\documentclass[12pt]{ociamthesis}          % OCIAM thesis class

\usepackage{graphicx}                      % needed for \includegraphics
\renewcommand{\logo}{%
  \includegraphics[width=60mm]{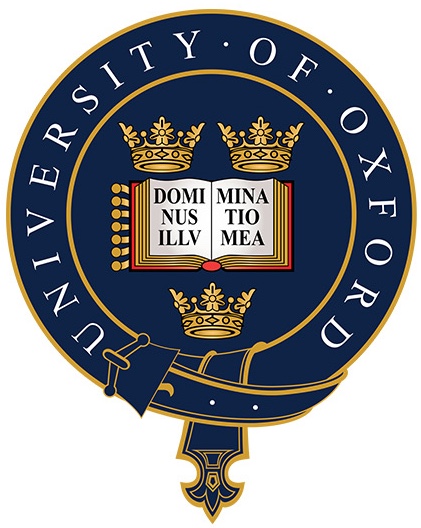}% ← adjust width
}

\usepackage{etoolbox}                      % for \patchcmd
\makeatletter
\patchcmd{\maketitle}{\vspace*{40mm}}{\vspace*{15mm}}{}{}
\patchcmd{\maketitle}{\vspace*{25mm}}{\vspace*{10mm}}{}{}
\patchcmd{\maketitle}
  {\Huge}
  {\LARGE}{}{}
\makeatother

\usepackage{amssymb}
\usepackage{xcolor}
\usepackage{amsmath}
\usepackage{braket}
\usepackage{bbm}
\usepackage{array}
\usepackage{amsthm}
\usepackage{enumitem}
\usepackage{qcircuit}
\usepackage{comment}
\usepackage{float}
\usepackage[hidelinks]{hyperref}
\usepackage{bibentry}
\usepackage{caption}
\nobibliography*                 % initialise bibentry package

\usepackage{setspace}
\newlength\poemparindent
\newenvironment{poemcompact}{%
  \begin{spacing}{0.9}\small\setlength{\parskip}{0pt}%
}{%
  \end{spacing}\setlength{\parindent}{\poemparindent}%
}

\newtheorem{theorem}{Theorem}

\title{Counterfactuals in \\Macroscopic Quantum Physics:\\
       Irreversibility, Measurement and Locality}

\author{Maria Violaris}
\college{Magdalen College}
\degree{Doctor of Philosophy}
\degreedate{Trinity 2024}

\begin{document}

\baselineskip=18pt plus1pt

%set the number of sectioning levels that get number and appear in the contents
\setcounter{secnumdepth}{3}
\setcounter{tocdepth}{2}

\maketitle                  % create a title page from the preamble info
\begin{abstract}
    
    Can quantum theory be applied on all scales? While there are many arguments for the universality of quantum theory, this question remains a subject of debate.  It is unknown how far the existence of macroscopic irreversibility can be derived from or reconciled with time-reversal symmetric quantum dynamics. Furthermore, reasoning about quantum measurements can appear to produce surprising and even paradoxical outcomes. The classical outcomes of quantum measurements are in some contexts deemed to violate the fundamental principle of locality, in particular when considering entanglement and Bell non-locality. Therefore measurement, irreversibility and locality can all appear to challenge the universality of quantum theory. In this thesis we approach these problems using counterfactuals — statements about the possibility and impossibility of transformations. Using the principles of constructor theory and quantum information theory, we find novel features of quantum thermodynamics relating to irreversibility, information erasure and coherence. We also develop tools to quantify the full implications of non-commutativity of quantum operators in settings where quantum theory is applied universally to measurement devices. This reveals new ways of characterising the quantum information stored in entanglement and quantum branching structure. Our results reinforce the ability of universal quantum theory to consistently describe both microscopic and macroscopic observers and thermodynamic systems.
\end{abstract}          % include the abstract
\chapter*{Acknowledgements}

I thank Chiara Marletto, who has been an inspiring collaborator and mentor throughout my DPhil. Thank you for showing me the significance of creativity in scientific research; providing constant encouragement and motivation to try new things and set ambitious goals; and for the many insightful research discussions, comments and ideas. I thank Vlatko Vedral, I have learnt many new perspectives from our enlightening discussions and collaborations. Thanks also to Artur Ekert for his help and to Tristan Farrow and David Deutsch for interesting discussions. I am grateful to have been able to spend the last four years in an environment with the freedom, openness and enthusiasm to explore exciting ideas. 

Thank you to all my collaborators that have contributed to the projects I present in this thesis: Anna Beever, Gaurav Bhole, Jonathan Jones, Samuel Plesnik, Abdulla Alhajri and Samuel Kuypers. I also thank those with whom I have ongoing research discussions, including Nicetu Tibau Vidal, Simone Rijavek, Charles Alexandre Bédard and Nick Ormrod, and my experimental collaborators at INRIM: Fabrizio Piacentini, Ivo Pietro Degiovanni, Alessio Avella, Marco Gramegna, Enrico Rebufello, Francesco Atzori and Marco Genovese. 

One of the most important parts of my DPhil life, both for my research development and day-to-day experience, has been being part of the Quantum Frontiers research group. I am lucky to have joined such a friendly and fun group. I thank all the past and present group members and visitors, including: Simone Rijavec, Lodovico Scarpa, Lucia Vilchez Estevez, Giuseppe Di Pietra, Amy Searle, Jiaxuan Zhang, Mattheus Burkhard, Michele Minervini, Samuel Hagh Shenas, Antonia Weber, Salvatore Raia, Nicetu Tibau Vidal, Samuel Kuypers, Abdulla Alhajri, and Aditya Iyer. I will miss all our times in the office, lunches, discussions, memes and much more!

I thank those in the Mathematical Physics group; the broader Oxford quantum community; the mathematical and quantum community in Bristol, who welcomed me during my visits there; students and staff at Magdalen College, where I have spent the last eight years including my undergraduate degree; and the huge range of people across the world I have met at conferences. Thank you all for widening my perspectives and embracing the opportunity to forge new connections.

Thank you to everyone I worked with in the IBM Quantum team on the ``Quantum Paradoxes" video series and other projects.

My friends are consistently a core part of my life, including during my four years doing this DPhil. Thank you for being amazing people! 

I thank all of my loving family. I thank my sister, Elena Violaris, with whom I have been analysing education and research on a meta-level since we were children. We have found consistent connections between our seemingly vastly different research interests in physics and literature, and even both reference paradoxical Penrose figures in our theses! I thank my little brother, Glafkos Nicholas Violaris, who never fails to entertain me with his decidedly Gen-Z perspectives and obsession with throwing ``quantum homogenizer" into conversation ever since he overheard me use the word in a presentation in 2020. 

Finally I thank my parents, Georgia Violaris and Patroklos Violaris, who unfailingly provide invaluable support and wisdom. 

%(supervisors, collaborators, colleagues, Oxford people, Bristol people, IBM people, conference people, Heilbronn, Magdalen, teachers, friends and family).
   % include an acknowledgements.tex file
% publications.tex
\chapter*{Publications}

\begingroup
  \sloppy
  \setlength\emergencystretch{3em} 
  \begin{enumerate}
    \item \bibentry{violaris2021transforming}
    \item \bibentry{violaris2022irreversibility}
    \item \bibentry{beever_comparing_nodate}
    \item \bibentry{plesnik2024impossibility}
    \item \bibentry{violaris2024penrose}
    \item \bibentry{violaris2024hardy}
  \end{enumerate}
\endgroup

\noindent The following publications in quantum education research were influenced by the thesis results, though not included in the thesis: 

\begin{enumerate}
    \item \bibentry{violaris2023physics}
    \item \bibentry{violaris2024physics}
\end{enumerate}      % include the publications.tex file
\begin{center}
  {\Large\scshape Alice and Bob:}\\[0.5\baselineskip]
  {\large\scshape The Scientists that Don’t Exist}\\[1\baselineskip]
  \hrule height 0.4pt\vspace{0.5\baselineskip}
  \hrule height 0.4pt
\end{center}

\bigskip

\begin{poemcompact}

  \flushleft
    “Not again!” cries Alice, as machines start to whir.
  \endflushleft

  \begin{center}
    Bob closes the heavy lab doors unperturbed.\\
    Another day, new experiments, for A and B to trial,
  \end{center}

  \flushright
    “We’re advancing science!” says Bob with a smile.
  \endflushright

  \flushleft
    “But Bob – we don’t exist, we’re just words on a page,\\
    Placeholders when scientists can’t think of a name.”
  \endflushleft

  \flushright
    “But Alice – we bring to life their theories, so vague,\\
    With a body, a story, a play on a stage.”
  \endflushright

  \flushleft
    “Exactly, a play, a toy, a rhyme!\\
    Testing fantasies of secret codes or quantum time.”
  \endflushleft

  \flushright
    “Our experiments are models, and even so,\\
    Fantasies today make technology tomorrow.”
  \endflushright

  \flushleft
    “Am I simply A character? A, with lice deleted?\\
    A tool like A test‐tube that’s cleaned and repeated?”
  \endflushleft

  \flushright
    “Alice, you’re a character of dimensions all kind,\\
    In paper after paper and mind after mind.”
  \endflushright

  \begin{center}
    Alice gathers her coins, her phone, her watch.\\
    Things were so simple before she met Bob.\\
\end{center}
    \begin{center}
    But perhaps they are no more real than her,\\
    These scientific theories, great and small.
    \end{center}

  \flushleft
    “Bob – if fact was always fiction once,\\
    Then science is fiction after all!”
  \endflushleft

  \bigskip
  \flushright
    \textit{— Maria Violaris, 2018}
  \endflushright

\end{poemcompact}

\begin{figure}[H]
    \centering
    \includegraphics[width=1\linewidth]{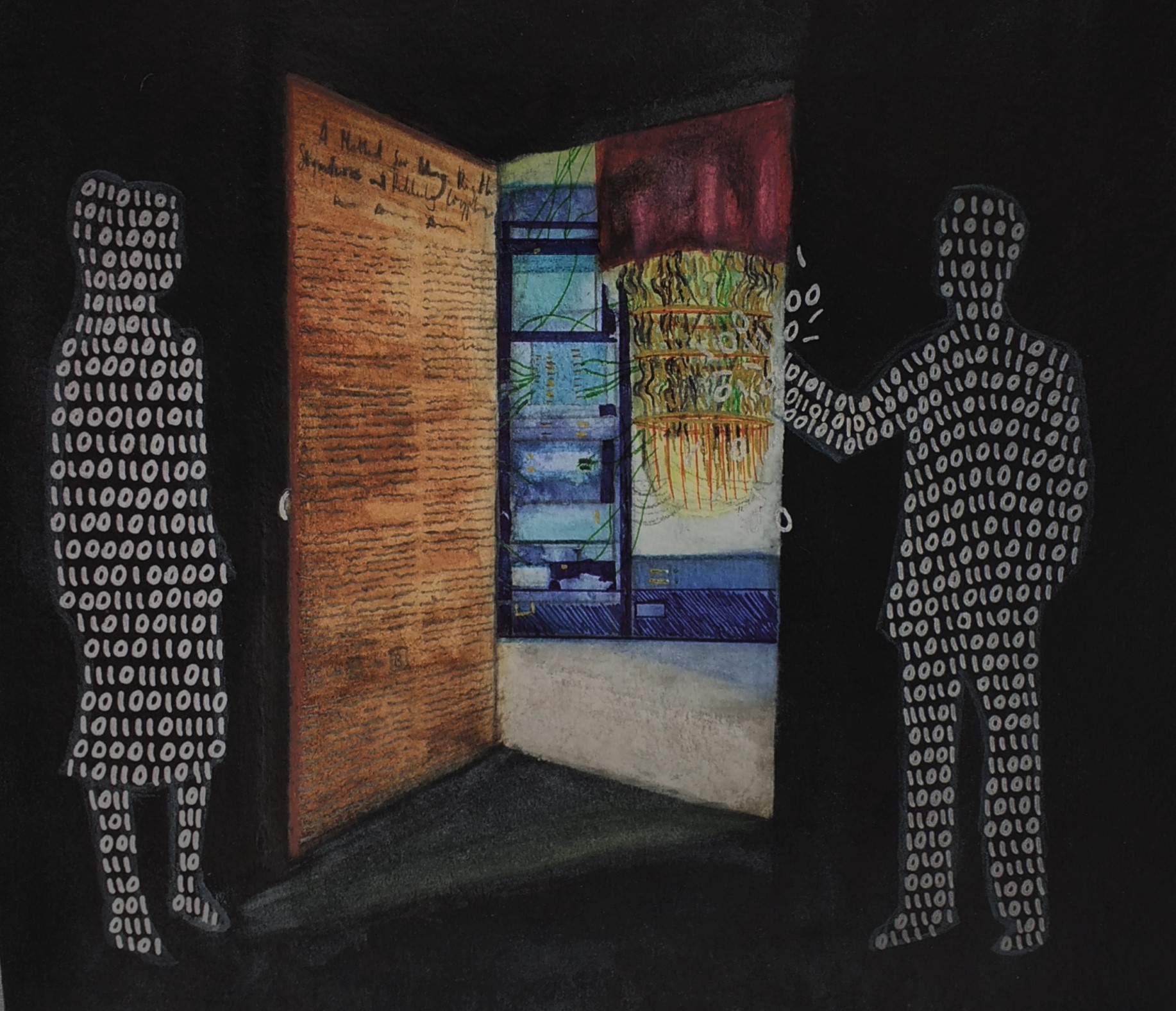}
    \captionsetup{labelformat=empty}
    \caption{Alice and Bob: Artwork response to the poem, by Maria Kostylew, 2020}
    \label{fig:alice-bob}
\end{figure}

\begin{romanpages}          % start roman page numbering
\tableofcontents            % generate and include a table of contents
%\listoffigures              % generate and include a list of figures
\end{romanpages}            % end roman page numbering

%now include the files of latex for each of the chapters etc
\chapter{Introduction}

Quantum theory is one of the most incredibly well-tested theories in the history of science. While it is often said to describe reality on the smallest scales, it is also essential to explaining macroscopic, everyday observations, such as why the sky is blue and the solidity of a table. Despite the range of effects on different scales explained by quantum theory, there remain open questions regarding its effectiveness to describe certain macroscopic phenomena, with respect to thermodynamics, observers and locality. Understanding the full domain of applicability of quantum theory is essential for distinguishing the cases where we expect to observe approximately classical behaviour from non-trivial quantum effects. This distinction is central to realising the full implications of quantum theory for foundational questions, as well as the limits of quantum technologies which depend on the validity of quantum theory in wide-ranging regimes. In this thesis, we examine these issues using tools from quantum information theory and an approach from counterfactuals. 

Quantum mechanics was first developed as a theory of dynamics, as encapsulated for instance by the Schrödinger equation. In the late 1900s, a new understanding of the underlying principles and scope of its applications was established with the emergence of quantum information theory. As a field, quantum information theory merges quantum mechanics with information theory, showing that information can be processed in an entirely different way when encoded in a system that obeys quantum dynamics, as opposed to a system that only obeys the dynamics of classical physics. Developments in quantum information theory led to huge advances in quantum science and technology, all the way from important foundational insights to the development of the theory of quantum computation. Today, there is a global effort working towards useful applications of quantum information processing technologies. Yet, the precise source of advantage when processing quantum information over classical information remains elusive. Furthermore, there are many open questions regarding how to characterise the behaviour of quantum systems in thermodynamic and macroscopic regimes.   

A promising way to approach these problems is by taking counterfactuals as a fundamental aspect of the laws of physics obeyed by quantum systems. Information is inherently counterfactual in nature; for a system to carry information, it must have the counterfactual potential to be in a different state to the one that it is in. For example, a coin must have the counterfactual ability to be in the Heads or Tails state in order to encode information. Recently, an approach called Constructor Theory has been developed, which takes the counterfactual nature of information to be part of the fundamental laws of physics, ultimately more fundamental than the associated dynamical laws \cite{deutsch_constructor_2013, deutsch2015constructor, marletto_constructor_2015}. More broadly, the theory conjectures that all fundamental laws of physics can be expressed using counterfactual principles about what can and cannot be made to happen, termed \textit{possible} and \textit{impossible} tasks. Dynamical laws such as those of quantum theory are supplementary to the constraining principles. 

This formulation enables us to express physical laws that cannot be expressed purely in terms of dynamical laws. A key example is that we can express physical laws about information, such as the principle of \textit{interoperability of information}, which means that information can be transferred between media. Furthermore, it provides a unifying, dynamics-independent and scale-independent expression of classical and quantum information. When physical principles are expressed in a dynamics-independent way, they can also be applied to constraining future dynamical theories of physics, such as a theory of quantum gravity, without specifying the (currently unknown) dynamics of such a theories. For example, the principles of interoperability and locality underlie a proposal for testing for non-classical features of gravity \cite{marletto_witness_2017}. The power of formulating laws in terms of scale-independent, non-probabilistic principles is particularly evident in formulations of thermodynamics, which is naturally expressed in terms of possible and impossible transfers of energy, work and heat. By applying tools from the constructor theory of information, thermodynamic laws could be expressed in an exact, scale-independent way that applies to quantum systems \cite{marletto_constructor_2017}. Building on this approach, in this thesis we show how apparently non-unitary macroscopic processes, such as measurement and information erasure, can be reconciled with underlying unitary quantum dynamics. This is done without the need for approximations or probabilistic statements. 

A central aspect of understanding macroscopic phenomena using quantum theory is to take seriously the consequences of non-commutativity of quantum operators, applied to systems on all scales. Non-commutativity is intertwined with the impossibility of single-shot distinguishability of quantum states. In the constructor theory of information, the impossibility of distinguishing non-orthogonal states is generalised to a dynamics-independent setting. Additionally, distinguishability can be defined in a precise way that avoids the circularity introduced in standard information-theoretic definitions \cite{deutsch2015constructor}. 

In Part I of this thesis, we use tools from this axiomatic, exact formulation of thermodynamics to analyse models for thermodynamic phenomena in quantum theory. We analyse protocols such as information erasure and work extraction, and the relevance of resources such as coherence. A useful qubit toy-model for investigating these phenomena are quantum homogenization machines, which implement an information spreading process via a collision model. Our results connect axiomatic constraints, such as those on distinguishability of states, with results from analysing quantum dynamics using resource theories.  

In Part II of this thesis, we consider how distinguishability and non-commutativity are manifested in quantum branching structure, and how they apply to quantum thought experiments involving reasoning about measurement outcomes. By applying quantum theory universally to measurement devices, we show how apparent contradictions can be fully explained using quantum theory, and provide a new way of quantifying the information stored in entanglement within the Heisenberg picture.  

Our results show new implications of applying quantum theory to macroscopic phenomena, giving novel insights into thermodynamics, measurement and locality. The approach from counterfactuals provides a unifying way to analyse these problems. This shows how underlying information-theoretic properties can help us resolve apparent contradictions and develop new understanding of both quantum phenomena, and those of the future theories that will supersede quantum mechanics. 

\section{Outline}

\noindent \textbf{Part I: Irreversibility, Erasure and Coherence}\\

\noindent In Part I, we investigate problems in the field of quantum thermodynamics using the tools of counterfactuals. In particular, we analyse models for an exact form of irreversibility; quantum information erasure in a quantum analog of the Maxwell's Demon thought experiment; and the usefulness of coherence for work extraction and spreading information via collision models. We use the qubit-based quantum homogenizer as a model for exploring these phenomena, and include results from an experimental collaboration implementing this qubit model on an NMR platform.\\

\noindent \textbf{Chapter 2 — Background: Quantum Thermodynamics}\\

\noindent We outline the context, notation and key results in quantum thermodynamics that our results build on, including: irreversibility, information erasure, constructor-theoretic thermodynamics, resource theories, and the quantum homogenizer.\\

\noindent \textbf{Chapter 3 — Axiomatic, Resource- and Constructor-Theoretic approaches}\\

\noindent Here we compare the axiomatic, resource-theoretic and constructor-theoretic formulations of thermodynamics, highlighting their domains of applicability, relevance of catalysts, and the key ordering relations involved in expressing the second law of thermodynamics.\\

\noindent \textbf{Chapter 4 — Quantum Model for Constructors and Irreversibility}\\

\noindent We analyse a model of ``constructor-based" irreversibility, which is exact, scale-independent and compatible with the time-reversal symmetric dynamics of quantum theory \cite{violaris2022irreversibility, marletto2022emergence}. An entity which can implement a task in a cycle arbitrarily well is called a constructor. Constructor-based irreversibility follows from the observation that a task can be possible, and the transpose task be impossible, even with reversible underlying dynamics. We provide a quantum model for a constructor, demonstrating how the scale-independent and dynamics-independent laws of Constructor Theory can be understood within the specific dynamics of quantum theory. Using this model, we give a general proof showing how constructor-based irreversibility is consistent with quantum theory, by considering the properties of a recursively defined limit of sets.\\

\noindent \textbf{Chapter 5 — Additional Cost to Information Erasure}\\

\noindent We demonstrate an example implementation of constructor-based irreversibility based on the quantum homogenizer. We consider how far the quantum homogenizer can be used in a cycle to transform a qubit from a mixed to a pure state in a cycle, and the transpose task. The former task is the erasure step of a quantum Szilard's engine. We show that the quantum homogenizer cannot be used in a cycle to transform a qubit from a mixed to a pure state, but it can be used in a cycle to perform the transpose task. Hence, this example demonstrates that constructor-based irreversibility is consistent with quantum dynamics. This new formulation of irreversibility in preparing pure qubits suggests there is an additional cost to the erasure of quantum information, and hence to constructing a reliable implementation of Szilard's engine. \\

\noindent \textbf{Chapter 6 — The CSWAP Quantum Homogenizer as an Incoherent Quantum Information Eraser}\\

\noindent Here we present a model for a universal quantum homogenizer that does not have coherence between the system and reservoir qubits, based on a controlled-swap (\textsc{cswap}) operation instead of a partial-swap (\textsc{pswap}). The added control qubit prevents interference between the homogenizer qubits, such that the \textsc{cswap} homogenizer is closer to a classical implementation of homogenization. We compare the homogenizers through analytical calculations of reduced states, numerical simulations and deriving bounds on resources required for homogenization. Then we analyse how far the coherent and incoherent homogenizers can be re-used to perform state transformations. The results suggest that our findings on constructor-based irreversibility based on homogenization machines can be generalised to incoherent models for thermalization and information erasure.\\

\noindent \textbf{Chapter 7 — Experimental Implementation of the Quantum Homogenizer using NMR}\\

\noindent Here we explain the results of an experimental collaboration to implement an NMR quantum homogenizer, using the partial swap on a system of four qubits \cite{violaris2021transforming}. In our protocol, a first system qubit interacts with two reservoir qubits via partial swap operations, and then a second system qubit interacts with the two reservoir qubits via partial-swap operations. Then the states of all four qubits are determined. The experiment is repeated for a range of coupling strengths of the partial-swap, and is performed for two initial conditions: the homogenization of system qubits in a pure state with reservoir qubits in a mixed state, and vice-versa. The results are consistent with the theoretical symmetric evolution of the pure and mixed qubit states when compared using standard measures such as von Neumann entropy.\\

\noindent \textbf{Chapter 8 — The Impossibility of Universal Work Extractors from Coherence}\\

\noindent It has been shown that for a known coherent state, more work can be extracted from the coherent state than from a thermal state with the same energy, when given access to a coherent catalyst \cite{korzekwa2016extraction}. There are also constructor-theoretic arguments that more work cannot be extracted deterministically from a coherent state than that which can be extracted from its basis states \cite{marletto2022information}. We demonstrate the consistency of these results by proving that the coherent catalyst that enables additional work extraction from coherent states must vary with the input coherent state. Hence, there is no deterministic work extractor that can extract additional work from an arbitrary coherent state. Our analysis connects the exact constructor-theoretic thermodynamic laws with resource-theoretic results in quantum thermodynamics, and we identify an interesting role played by constructor-based irreversibility via the quantum homogenizer.\\

\noindent \textbf{Part II: Measurement, Locality and Non-Commutativity}\\

\noindent In Part II, we demonstrate the full implications of applying quantum theory universally to measurement devices, identifying new features of quantum branching structure and developing tools to quantify information within the local account of quantum theory. We consider various thought experiments that involve reasoning about measurement outcomes, and show how they are self-consistent when the full quantum state of the measurement devices is accounted for. Then we focus on Hardy's paradox, which demonstrates Bell non-locality without inequalities. We give a local account of the thought experiment, and provide a new understanding of how the outcomes maximally deviate from classical physics for partial entanglement rather than maximum entanglement. With this setting as a toy model, we show that the locally inaccessible information present in entanglement in the Heisenberg picture is quantified by the usual measure of entanglement, von Neumann entropy.\\

\noindent \textbf{Chapter 9 — Background: Measurement Paradoxes and Non-Locality}\\

\noindent We summarise Bell non-locality using a simple quantum circuit example, and show how the local account of quantum theory explains the flow of information during measurements. Then we formally explain the key features of the local account of quantum theory using the Heisenberg picture.\\

\noindent \textbf{Chapter 10 — Incompatibility of Non-Commutativity and Classical Logic}\\

\noindent Here we show how classical reasoning about quantum measurement outcomes violates the key axiomatic principle that non-orthogonal quantum states cannot be distinguished by a single measurement. We present this result using a low-depth quantum circuit. The classical reasoning is essentially the chaining together of individually consistent statements when they refer to the definite state of a qubit in a different basis. This demonstrates the subtleties involved in reasoning about non-commuting measurements. Additionally, we show how this underlies a class of apparent paradoxes, which can all be simplified when presented using comparable quantum circuits. A common feature is that they involve reasoning about measurements that could have been made and were not. The underlying logical structure is analogous to the geometry of a Penrose triangle: just as the triangle has locally consistent corners that are chained together to create a globally inconsistent shape, quantum theory can sustain locally consistent statements that form a globally inconsistent structure when chained together.\\

\noindent \textbf{Chapter 11 — Local Account of Hardy’s Paradox}\\

\noindent In the Heisenberg picture, it is made explicit that Bell-type non-locality can be explained through locally inaccessible information. This information resides locally in the subsystems of a composite system but cannot be recovered through measurements performed on any isolated subsystem. Locally (in)accessible information has been qualitatively defined in a dynamics-independent way with the constructor theory of information \cite{deutsch2015constructor}.  In this Chapter, we explain how the distinguishing feature of Hardy’s thought experiment from the EPR thought experiment is that systems in the former contain both locally accessible- and inaccessible-information. Identifying the key role of the non-commutativity of operators used to retrieve incompatible pieces of information from the partially-entangled state, we derive the entangled state which maximises the violation from the classical expectations for the measurement results.\\

\noindent \textbf{Chapter 12 — Quantifying Locally Inaccessible Information}\\

\noindent We prove that the correct measure to quantify how much of the information stored in the system is locally (in)accessible is based on von Neumann entropy. We find boundary conditions on locally (in)accessible information in an overall pure, bipartite system, and relate the reduced density matrix of a system with a Heisenberg picture representation of its sharpest observable. Our formulation shows how the sum of locally inaccessible and locally accessible information remains constant under unitary evolution, and shows how partial bits of locally (in)accessible information can reside in quantum systems.\\

% Introduction

\part{Irreversibility, Erasure and Coherence} % Part I

\chapter{Background: Quantum Thermodynamics} \label{background_q_thermo}

Here we introduce some background context about the aspects of quantum thermodynamics relevant to this thesis, and introductory material on the specific ideas and results which this thesis builds on. 

\section{The problem of irreversibility}

Thermodynamics began as a phenomenological theory to quantify and bound the efficiency of large-scale exchanges of energy. The field of statistical mechanics arose as an attempt to put thermodynamics on firmer footing, showing how the large-scale laws given by thermodynamics arise from microscopic interactions between particles. There remains debate surrounding how far statistical mechanics provides an explanation for phenomenological thermodynamics, though statistical mechanics provides an accurate foundation for analysing the average behaviour of many systems on large scales.

There is particular controversy surrounding the status of the 2$^{\textrm{nd}}$ law of thermodynamics. While it is widely held that the 2$^{\textrm{nd}}$ law is one of the least likely to be violated in physics \cite{murphy_nature_1930}, it is difficult to find an exact derivation from underlying microphysics. It has been argued that irreversibility is actually impossible to derive exactly from reversible dynamics \cite{grnitz_temporal_1992}. The formulation of the 2$^{\textrm{nd}}$ law that ``the entropy of an isolated system is unlikely to decrease" leaves open the question of whether the 2$^{\textrm{nd}}$ law is then fundamentally probabilistic and an approximation that applies at some arbitrarily defined scale, or whether there exists an exact formulation of irreversibility in nature. This is related to the controversy over how far entropy is a subjective or objective notion \cite{rex_maxwells_2017}. Having full knowledge of all degrees of freedom as a system evolves leads to no change in entropy with time, causing some to argue that it is associated with the knowledge of the observer manipulating the system, rather than a property of a system itself \cite{jaynes1957information, jaynes1992gibbs}.

The problems in defining an exact 2$^{\textrm{nd}}$ law mean that it is unclear from traditional thermodynamics and statistical mechanics how to apply the law in the case of microscopic systems, possibly behaving in a fully quantum way, without using approximations. To account for situations where fluctuations are significant, the field of stochastic thermodynamics has emerged. This has particular applications for modelling non-equilibrium dynamics in biochemistry, for instance in DNA and molecular motors \cite{seifert_stochastic_2012}. Similarly the field of quantum thermodynamics has become established to investigate the thermodynamic laws when quantum effects are significant. Quantum information has provided key insights in the field of quantum thermodynamics \cite{goold_role_2016}. These range from demonstrating how work can be extracted from entanglement, to using quantum resource theories to define quantum laws of thermodynamics \cite{maruyama_morikoshi_vedral_2005, chitambar_quantum_2019}.

The emergence of these fields has led to progress in knowing how to treat specific physical systems at scales where classical thermodynamics and statistical mechanics no longer apply. However the question of whether there is a fundamental irreversibility in nature, or if irreversibility requires approximations, coarse-graining and probabilistic treatment, remains open. Furthermore, attempts to define a 2$^{\textrm{nd}}$ law of thermodynamics, or to define notions of work and heat at these scales, are generally specific to a system under particular constraints \cite{brandao_second_2015, muller_correlating_2018}. It could be that work, heat and the 2$^{\textrm{nd}}$ law will remain having many forms, and that future researchers creating practical quantum thermal machines will choose the particular definition that works best for modelling the system that is in question. Alternatively, if thermodynamics is truly a fundamental part of physics as opposed to an emergent and solely operational theory, then one might expect a unifying definition of work, heat and the 2$^{\textrm{nd}}$ law to exist.

In addition, in contrast to quantum theory, classical mechanics and electromagnetism, thermodynamics exists on a level that can apply to different theories without committing to particular underlying dynamical laws. This invites the question of whether there exists a formulation of thermodynamics that works for small and quantum scales, but retains the property of being independent of the dynamical theory used \cite{marletto2022information, marletto_constructor_2017}. One approach to the problems of defining a 2$^\textrm{nd}$ law and definitions of heat and work for quantum systems is offered by constructor theory, as we will now explain. We will also later discuss a different approach to quantum thermodynamics from resource theories, and how these approaches compare. 

\section{Constructor-theoretic approach to thermodynamics}\label{Intro:CT-thermo}

\textbf{Motivations and background}\\

\noindent Constructor theory aims to reformulate the laws of physics in terms of statements about whether transformations are \textit{possible} or \textit{impossible}, and why. The central conjecture of the research program is that these statements are fundamental principles, with dynamical laws (such as the dynamics of quantum theory) emerging as subsidiary theories.

When applied to thermodynamics, constructor theory lends itself to formulating general, axiomatic principles that underlie both classical and quantum systems, and systems whose dynamics may be partially unknown \cite{deutsch_constructor_2013,deutsch2015constructor,marletto_constructor_2017}. As we expand in Chapter \ref{resource_comparison}, resource theories take existing dynamical laws of physics as a starting point and derive constraints on what can be done under those laws. By contrast, constructor theory has its own laws of physics, including the \textit{interoperability of information} and the \textit{principle of locality}. These laws are conjectured to be more fundamental than dynamical laws.

In constructor theory, a task being \textit{possible} means that the laws of physics place no limit, short of perfection, on how well the task could be performed in a cycle. This means that the task could be performed to arbitrary accuracy, an arbitrary number of times. If there is however some limit imposed by the laws of physics to how well a task can be performed in a cycle, then it is \textit{impossible}. For example, consider the phenomenon whereby all the particles in an isolated room move to one corner. It is a dynamically allowed trajectory, but an impossible task, because there is no reliable way to cause the transformation in a cycle (without causing an irreducible change to the environment). 

These principles about what is possible and impossible are all exact, without dependence on probabilities or approximations. They are also scale-independent, meaning that they do not refer to a particular microscopic or macroscopic regime; and dynamics-independent, meaning they can be expressed without committing to a particular set of dynamical laws, such as those of quantum theory. The dynamics-independent properties of the principles have proved useful for devising tests for non-classical features of systems with unknown or intractable dynamics \cite{marletto2020witnessing, di2022temporal}. Meanwhile the scale-independent form of the principles means they can be used to formulate an exact form of irreversibility, whereby a task is possible and the opposite task is impossible \cite{marletto_constructor_2017}. In Chapters \ref{CT-irreversibility} and \ref{erasure_model} of this thesis, we will show that this form of irreversibility is compatible with reversible dynamical laws, and demonstrate this explicitly for quantum theory using a qubit model \cite{violaris2022irreversibility, marletto2022emergence}.\\

\noindent \textbf{Formulation and notation}\\

\noindent Individual systems are called \textit{substrates} and are characterised by their \textit{attributes}, which are sets of all states with a given property. These properties can be changed by physical transformations. A \textit{task} is a transformation of a substrate from an input attribute to an output attribute. A \textit{variable} is a set of disjoint attributes, $i.e.$ a set of distinct, independent properties that a system can have. For a system $\mathbf{S_1}$ with attribute $\mathbf{a}$ and $\mathbf{S_2}$ with attribute $\mathbf{b}$, the combined system $\mathbf{S_1} \oplus \mathbf{S_2}$ has attribute $\mathbf{(a, b)}$. An important principle is the \textit{principle of locality}, which states that if a transformation only operates on substrate $\mathbf{S_1}$, then only attribute $\mathbf{a}$ can change \cite{deutsch2015constructor}. 

The basic objects in constructor theory are \textit{tasks} $\mathbf{T}$, defined as sets of ordered pairs of input and output attributes specifying a transformation on a physical system, or substrate.  A \textit{constructor} for a task is a system capable of performing the task to arbitrarily high accuracy, while maintaining its ability to cause the transformation again. In Chapter \ref{CT-irreversibility}, a constructor in quantum theory is defined as being the limit set, $\mathbf{C}$, of a sequence of sets of quantum systems that implement $\mathbf{T}$ on the substrate. 

A task is possible ($\mathbf{T}^{\checkmark}$) if it can be performed by a constructor, and impossible ($\mathbf{T}^{\times}$) otherwise. For a possible task, there is no fundamental restriction on how closely a perfect constructor could be approximated by a sequence of (imperfect) actual machines. For a possible task, a perfect constructor can be approximated with arbitrarily good precision by a sequence of ever-improving approximate constructors. A task being impossible means that the accuracy to which it can be performed is fundamentally limited by the laws of physics.\\

\noindent \textbf{Work and heat}\\

\noindent In this setting, information can be defined in an exact way as part of physical laws, and quantum and classical information can be defined under a single local framework, without requiring approximations \cite{deutsch2015constructor}. An \textit{information medium} is defined by the transformations possible on that medium (the possibility of copying from one medium to another, and the possibility of permuting each state to any other state). Then \textit{heat media} and \textit{work media} are defined as information media with different additional properties \cite{marletto_constructor_2015}. All these definitions are scale- and dynamics-independent, and hence can hold exactly even if the dynamics of quantum theory are modified in a post-quantum, successor theory. Furthermore, the exact statements circumvent the need for coarse-graining or approximations, which can limit the domain of applicability of laws.

Clausius' definition of the 2$^{\textrm{nd}}$ law of thermodynamics is stated using cycles: we cannot entirely transform heat into work in a cycle, but we can entirely transform work into heat in a cycle. The limitations of this statement for small and quantum scales comes with the ambiguous definitions of work and heat on these scales. Hence, the constructor-theoretic scale- and dynamics-independent definitions of work and heat can be used in the spirit of Clausius' formulation of the 2$^{\textrm{nd}}$ law, in order to formulate it in exact terms. 

The ability to formulate such a 2$^{\textrm{nd}}$ law hinges on a conjecture that reversible dynamics are consistent with a transformation being possible in a cycle, and the reverse transformation being impossible in a cycle. A motivation for this conjecture is that there is no clear reason that the reversibility of dynamical laws should imply or not imply the existence of irreversibility based on cycles. This is because there is no reason for the dynamical time-reverse of a machine able to repeatedly perform a task to result in a machine that is able to repeatedly perform the reverse task. We prove the conjecture using a fully unitary quantum toy-model in Chapter \ref{CT-irreversibility}.\\

\noindent \textbf{Deterministic work extractors}\\

\noindent A \textit{deterministic work extractor} can transform the attributes of a substrate to those of a \textit{work variable}, where a work variable is the set of distinct attributes of a work medium \cite{marletto2022information}. Let's consider the task of work extraction from a quantum system as an example. The system $\rho_S$ (which is the substrate $\mathbf{S}$) can be coupled to an ancillary battery $\rho_B$ (which is the work medium $\mathbf{Q}$). Then an energy-conserving unitary transformation $U$ can transfer energy between the system and the battery, such that the final state of the battery is $\rho_B^\prime =  \operatorname{tr}_S U (\rho_S \otimes \rho_B) U^\dagger$. The work medium $\mathbf{Q}$ could for instance be an atom, whose corresponding attributes are energy levels that can get excited and de-excited following an interaction with another substrate $\mathbf{S}$. In this example, $\mathbf{S}$ could be a magnetic field or another atom. If both media are atoms, and both their variables are sets of energy eigenstates of the atom, then the deterministic work extractor maps each energy eigenstates of the system $\rho_S$ to a unique energy eigenstate of the battery $\rho_B$.

In quantum theory, single-shot distinguishability of states is possible if they are orthogonal. Using constructor theory, the same kind of single-shot distinguishability can be defined without referring to properties of dynamics specific to quantum theory, such as orthogonality \cite{deutsch2015constructor}. The physical meaning of distinguishability as defined in conventional information theory is circular: two states are distinguishable if they cause some measurement device to end up in a distinguishable pair of states, conditional on the state that was measured. So, the distinguishability of states that a system can be in is defined in terms of the distinguishability of states of the measurement device. The constructor theory of information solves this problem, giving a condition for distinguishability that does not itself refer to distinguishability, and instead only refers to the possible and impossible tasks on substrates that can instantiate information (information media).

The constructor-theoretic definition of distinguishability generalises the distinguishability of orthogonal states in quantum theory. In \cite{marletto2022information}, the constructor-theoretic formulation of distinguishability is used to prove that work variables are distinguishable, Theorem \ref{theorem:distinguish}:

\begin{theorem}
\label{theorem:distinguish}
  A work variable $\mathbf{W}$ is a distinguishable variable.
\end{theorem}

In general, a deterministic work extractor transforms each attribute of the substrate $\mathbf{S}$ to exactly one attribute of the work medium $\mathbf{Q}$. Hence, since $\mathbf{Q}$'s variable is distinguishable, $\mathbf{S}$ must also have a distinguishable variable \cite{deutsch2015constructor}.\\

\noindent \textbf{Adiabatic possibility and the 2$^{\textrm{nd}}$ law}\\

\noindent An axiomatic approach to thermodynamics was developed by Lieb and Yvnangson, formulated around the property of \textit{adiabatic accessibility}: the key relation between two equilibrium states is whether or not one is adiabatically accessible from the other, where adiabatic accessibility is defined operationally using a set of axioms \cite{lieb1999physics}. The overall idea is that a state is adiabatically accessible from another state if it is possible for the transformation to be implemented by some auxiliary system that returns to its original state, with the only side-effect being the lifting or lowering of a weight in a gravitational field. 

In the constructor theory of thermodynamics, the property of adiabatic accessibility is reformulated as \textit{adiabatic possibility}, where adiabatic possibility is defined solely in terms of possible and impossible tasks on a generalised form of states \cite{marletto_constructor_2015}. Specifically, the adiabatic possibility condition is: \textit{A task \{\textbf{x} → \textbf{y}\} is adiabatically possible if the task} \{(\textbf{x}, \textbf{w$_1$}) → (\textbf{y}, \textbf{w}$_2$)\} \textit{is possible for some two work attributes \textbf{w$_1$}, \textbf{w}$_2$ belonging to a work variable}. Then, the 2$^{\textrm{nd}}$ law can be stated as: \textit{There are tasks that are adiabatically possible, whose transpose is not adiabatically possible}.

Adiabatic accessibility is defined with reference to a weight in a gravitational field, which is deemed to represent any kind of work storage system. However, on microscopic and quantum scales, there are a variety of specific models for work storage systems depending on the context. Adiabatic possibility is not defined with reference to a weight in a gravitational field, but instead with reference to the scale- and dynamics-independent \textit{work medium}. In turn this makes adiabatic possibility a scale-independent and dynamics-independent property, avoiding ambiguity regarding whether or not some specific system counts as a work storage medium. It also enables us to reason about weights and work storage media that may not obey the dynamical laws of classical or quantum mechanics. 

\section{Information erasure, Landauer's principle, and Maxwell's Demon}\label{intro:erasure}

Erasure is essential for information processing and thermodynamics. The clearest realisation of the connection between erasure and thermodynamics is Bennett's application of Landauer's principle to solve the Maxwell's Demon paradox \cite{Bennett_Landauer_1982}, which was proposed to suggest how the 2$^{\textrm{nd}}$ law of thermodynamics could be violated by a light-fingered demon able to sort disordered particles into two distinct groups, increasing overall order \cite{bub_maxwells_2002, rex_maxwells_2017}. The solution to the seeimingly paradoxical thought experiment was first thought to be that the demon increases the entropy of its surroundings when it measures the properties of the particles in order to sort them. However it was later shown by Bennett that gaining this information can be done reversibly with no entropy cost. But the demon needs to store the information it gathers about the particles, and it cannot store an infinite amount of information. Therefore, at some point the information stored in its memory will have to be erased \cite{bennett_notes_2003, landsman_verified_2019}. The erasure of information is logically irreversible, and by Landauer's principle, it requires an irreducible entropy increase in the environment.

\begin{figure}
    \centering
    \includegraphics[width=0.4\linewidth]{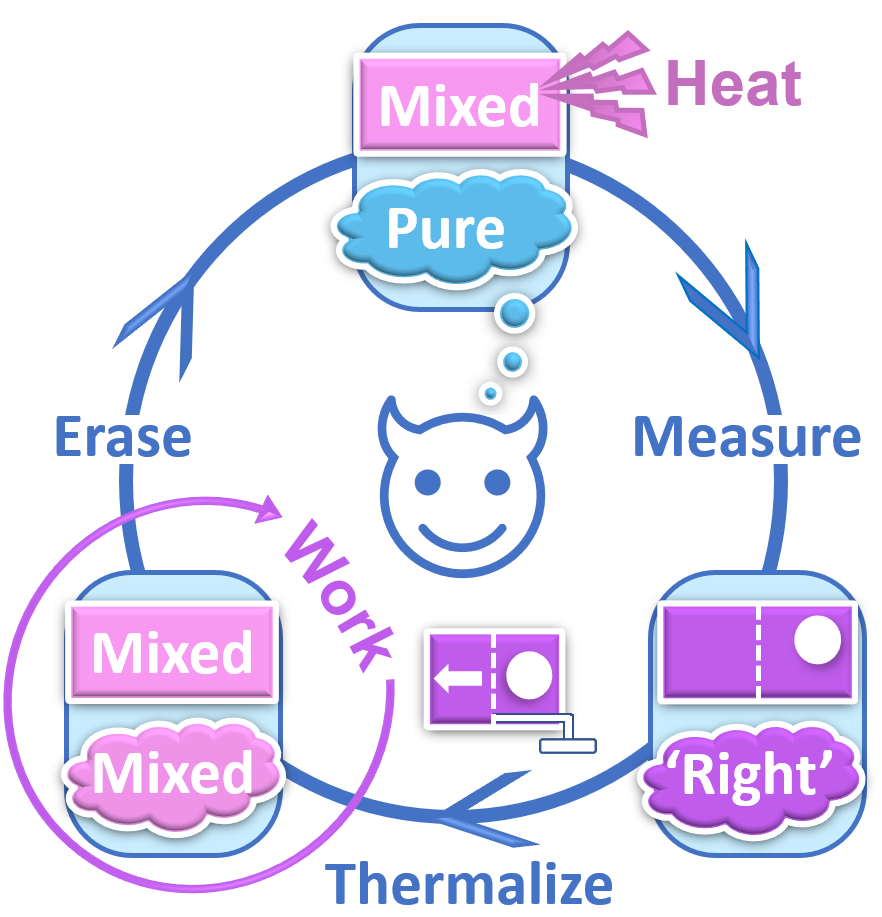}
    \caption{Visualisation of the Szilard engine cycle.}
    \label{fig:szilard}
\end{figure}

The irreducible entropy cost of erasing the demon's information is most manifest in the case of the notorious {\sl Szilard engine} (figure \ref{fig:szilard}). A Szilard engine follows a similar reasoning to Maxwell's Demon, but with a single particle. In the classical thought experiment, there is a particle in a box. Then a partition divides the box in two, and the demon measures which side of the box the particle is on. Depending on the outcome, the demon attaches a weight to that side of the partition. Then as the particle collides with the partition, it moves and pulls up the weight. Hence, it appears that we can convert the initial ``heat" of the single particle into work, without an additional cost. As with Maxwell's Demon, the resolution to the violation of the 2$^\textrm{nd}$ law is that the demon's memory contains the information about which side of the box the particle was on. For the demon to operate in a cycle, this information must be erased, which has an irreducible entropy cost -- see \cite{sep-information-entropy}.

We can then consider the quantum analog of this thought experiment. We can model the initial particle in the box as a qubit in a maximally mixed state. Then the demon's memory is initially in a pure state. When the demon measures the state of the particle, the demon's memory and the particle's position in the box become entangled. Then the process of work extraction involves the demon's memory becoming decorrelated from the particle, such that they are both maximally mixed states. For the cycle to repeat, the demon's memory must be reset from a maximally mixed state to a pure state \cite{vedral_landauers_2000}. This is a non-unitary process, whence the irreducible entropy cost of this step. It involves a decrease in the von Neumann entropy of the qubit, and since total von Neumann entropy is conserved in unitary interactions, the entropy of the environment must increase by the same amount. This is the quantum manifestation of Landauer's principle. Understanding the full quantum manifestation of Landauer's principle and potential modifications to the limit is an active area of research in quantum thermodynamics. There are also multiple quantum variations of the Szilard engine thought experiment, such as a ``demonless" engine and an analysis where all steps of the engine are unitary \cite{aydin_landauers_2020, maroney_information_2002}. 

Recent studies have investigated various additional costs to the entropy cost of Landauer erasure. For example, an additional entropy cost due to finite-time erasure has been studied in the classical case \cite{proesmans_ehrich_bechhoefer_2020} and quantum case \cite{van2022finite, PhysRevLett.125.160602}, where additionally quantum coherence induces heat dissipation that is not present in classical erasure. Modifications to Landauer's principle have also been considered in terms of collision models with non-Markovianity \cite{PhysRevA.99.042106}. In addition, entropy changes and Landauer's principle have been studied in the quantum regime to account for finite systems and correlations and entanglement, related through an equality improving on the Landauer bound \cite{Reeb_2014, esposito2010entropy}. In other studies the von Neumann entropy has been argued to be insufficient to characterise erasure \cite{egloff2015measure}, and there remain different measures to cost quantum erasure processes in different contexts.

In Chapter \ref{erasure_model}, we will consider a system where it is more difficult to perform a transformation of a qubit from a mixed to a pure state in a cycle, than from a pure to a mixed state. This asymmetry is not captured by the symmetric changes in von Neumann entropy in the two situations. Hence we conjecture that our results could indicate an additional cost to information erasure to that given by Landauer's principle. This would also suggest a connection between the type of exact irreversibility that emerges in constructor theory and the limits of quantum information erasure. 

\section{Resource theory of quantum thermodynamics} \label{background:resource}

Resource theories are a general way of approaching problems by drawing a distinction between resources which are finite and valuable, and those which are available in large numbers and hence modelled as unlimited or cheap \cite{chitambar_quantum_2019}. Applying these ideas to quantum information theory led to the emergence of quantum resource theories \cite{gour2015resource}. By varying which physical quantities are designated as \textit{free} (unlimited) or \textit{resources} (finite), different quantum resource theories can be constructed for different settings. For example, resource theories have been developed for entanglement, coherence, and Bell non-locality, and for each one there is a different set of states that are categorised as free or resources \cite{brandao2013resource, ng2019resource}. The basic formulation of a resource theory involves the \textit{free states}, which are free to create, and \textit{free transformations}, which are transformations that only require free states to be used. Then one can study which state transformations are allowed and disallowed under these conditions. A general formulation of resource theories has been constructed using process theory \cite{coecke2016mathematical}.

The resource theory of quantum thermodynamics has been used to derive a variety of useful results \cite{horodecki2013fundamental,ng2015limits}. These include deriving a family of free energies that underlie a version of the 2$^{\textrm{nd}}$ law of thermodynamics for quantum systems \cite{brandao2013resource,brandao2015second}, and limits on work extraction from quantum systems \cite{aaberg2013truly,skrzypczyk2014work,renes2014work}.\\

\noindent \textbf{Resource theories and the 2$^{\textrm{nd}}$ law}\\

\noindent In the context of thermodynamics, the relevant resource is work. Hence, the only \textit{free states} are thermal states, since no work can be extracted from an arbitrary number of such states. This constraint is a quantum version of the 0$^\textrm{th}$ law of thermodynamics. The allowed transformations are energy-conserving unitary operations, in analogy with the 1$^\textrm{st}$ law of thermodynamics. Then from these constraints, one can derive a family of quantum 2$^{\textrm{nd}}$ laws — each imposing that state transitions which decrease a type of free energy are disallowed \cite{brandao_second_2015}. In the thermodynamic limit, this family of free energies converges to the standard Helmholtz free energy used to define the 2$^{\textrm{nd}}$ law. This family of 2$^{\textrm{nd}}$ laws is derived assuming that the thermal machine performing the state transition is returned to an uncorrelated state with the system it transformed. By relaxing this assumption, it has been shown that the quantum 2$^{\textrm{nd}}$ law can be characterised by a single free energy, rather than a family of 2$^\textrm{nd}$ laws \cite{muller_correlating_2018}. How undisturbed a thermal machine should be after a transformation to count as a catalyst also affects the resulting quantum 2$^{\textrm{nd}}$ law(s). 

These examples illustrate how the approaches to defining quantum thermodynamics using resource theories are focused on the operational use of thermodynamics. We can arrive at different formulations of the 2$^{\textrm{nd}}$ law depending on what constraints we add to our analysis, with the constraints arising from the particular physical situation in question. One noted benefit of the approach is ``the advantage of separating out laws of fundamental physics, e.g. that evolution be unitary and energy conserving, from those of thermodynamics" \cite{brandao_second_2015}. \\

\noindent \textbf{Protocol to extract work from coherence}\\

\noindent The resource-theoretic approach is used to analyse work extraction from coherent states in \cite{skrzypczyk2014work}, where it is shown that the average work extracted from a single copy of a mixture of energy eigenstates can be equal to the free energy of the state. For coherent states, if no external sources of coherence are used, collective actions on multiple copies of the state are needed to extract work equal to the free energy. However, for an individual coherent state, no more work can be extracted than that for its incoherent counterpart of the same energy, despite the former having more free energy. This phenomenon is known as \textit{work-locking}.

It is then shown in \cite{aaberg2014catalytic} that a coherent resource can be repeatedly used as a catalyst for tasks involving a change in coherence of a system. Extending this set up to the problem of extracting work from coherent states, in \cite{korzekwa2016extraction} it is shown that thermal machines can extract single-shot work from coherent states with arbitrarily small failure probability, though the work extracted is non-optimal when using a bounded coherent catalyst.  Hence, adding a coherent resource enables one to bypass the constraints of work-locking. 

Here we review the assumptions and setting of work-locking and the protocol in \cite{korzekwa2016extraction}. The proposal in \cite{korzekwa2016extraction} has three central assumptions:\\

\noindent 1) Allowed transformations are energy-conserving unitary operations, imposed by the requirement that they commute with the overall Hamiltonian.  \\

\noindent 2) The average work that can be extracted from a system $\rho_S$, which is incoherent in the energy eigenbases, is given by the free energy difference: 

\begin{equation}
    \langle W \rangle (\rho_S) = \Delta F(\rho_S) = F (\rho_S) - F(\gamma_S)
\end{equation}

\noindent where $F(\rho) = \operatorname{tr}(\rho H_S) - kTS(\rho)$, and $S(\rho) = -\operatorname{tr}(\rho\textrm{ ln } \rho)$ is the von Neumann entropy; $T$ is the temperature of the heat bath that is coupled to the system; $H_S$ is the system Hamiltonian; and $\gamma_S = e^{-H_S/kT}/Z_S$ is the thermal Gibbs state with partition function $Z_S$.\\

\noindent 3) The work extractable from $\rho_S$ with a single shot is:
\begin{equation}
    W_{ss}^{\epsilon}(\rho_S) = \Delta F_{0}^\epsilon (\rho_S) = F_{0}^\epsilon (\rho_S) - F(\gamma_S)
\end{equation}
where $\epsilon$ denotes a small failure probability and $\rho_S$ is again incoherent in the energy eigenbasis. 

Then a general work extraction protocol can be expressed as a transformation of the form $\rho_S \otimes \rho_B \rightarrow \gamma_S \otimes \rho_B^\prime$, where $\rho_B$ and $\rho_B^\prime$ are initial and final incoherent battery states respectively. This is a thermal operation. It is useful to consider the effect of a dephasing channel $\mathcal{D}$ \cite{brandao2013resource}, which removes the coherence of a system and commutes with thermal operations: 

\begin{equation}
    \mathcal{D}(\rho) = \sum_i \operatorname{tr}(\Pi_i \rho)\Pi_i,
\end{equation}
where $\Pi_i$ are projectors onto the energy eigenspace. Applying the three assumptions then leads to the following conclusion (see \cite{korzekwa2016extraction} for the proof and more context): when there is no additional source of coherence, the maximum work extractable from a coherent state is equal to that of the decoherent state of the same energy, despite the former having more free energy. Specifically:

\begin{equation}
 \qquad \langle W \rangle (\rho_S) \leq \langle W \rangle (\mathcal{D}(\rho_S)).
\end{equation}
and similarly for single-shot work: 

\begin{equation}
    W_{ss}^{\epsilon}(\rho_S) \leq W_{ss}^{\epsilon}(\mathcal{D}(\rho_S))
\end{equation}

In this sense, there is some work ``locked" in the coherent state, that cannot be accessed without coherent resources. The question considered in \cite{korzekwa2016extraction} is then, how far can all this locked-up free energy be accessed by introducing a coherent resource?

It is shown that this ``locked work" can be accessed by introducing an ancilla system, which acts as a thermal machine. This is termed the \textit{reference system} $\rho_R$. It has the ladder system Hamiltonian $H_R = \sum_{n=0}^\infty n \ket{n}\bra{n}$, and its quality as a machine for the work extraction task is characterised by the coherence measure $ \langle \bar{\Delta} \rangle$ defined as: 

\begin{equation}
    \langle \bar{\Delta} \rangle = \operatorname{tr}_R (\rho_R \bar{\Delta}) = \frac{1}{2} \operatorname{tr}_R (\rho_R [ \Delta + \Delta^\dagger]).
\end{equation}
with $\Delta = \sum_{n=0}^\infty \ket{n+1}\bra{n}$ being the shift operator. The reference state $\rho_R$ is coupled to the system state $\rho_S$, from which work will be extracted, and a battery system $\rho_B$. The Hamiltonian of the overall system is $H = H_{S}\otimes H_R \otimes H_B$. 

Work can then be extracted from the coherence in the system by following a three-step protocol, involving $\rho_S$, $\rho_B$, and $\rho_R$. The pre-processing step applies a desired channel to $\rho_S$ using $\rho_R$ as an ancilla, to map the system to a state that is optimal for work extraction; the work extraction step involves extracting work from $\rho_S$ into $\rho_B$; and finally to ensure repeatability, the reference is ``repumped" to be close to original state, via a joint operation on $\rho_S$, $\rho_B$ and $\rho_R$: \\

\noindent 1) Pre-processing \\

\noindent The pre-processing operation involves rotating the state $\ket{\psi}$,
\begin{equation}
\label{eq:CoherenceInputstate}
    \ket{\psi} = \sqrt{1-p}\ket{0} + \sqrt{p}e^{-i\varphi}\ket{1},
\end{equation}
parameterised by $p$ and $\varphi$, into a coherent Gibbs state $\ket{\gamma}$ with thermal distribution $(1-r,r)$,
\begin{equation}
\label{eq:coherentGibbsState}
    \ket{\gamma} = \sqrt{1-r}\ket{0} + \sqrt{r}\ket{1}.
\end{equation}
This rotation is defined by the following quantum channel:
\begin{equation}
    \ket{\psi} \rightarrow \ket{\gamma}: \qquad \qquad \operatorname{tr}_R \left[ V \left( \ket{\psi}\bra{\psi} \otimes \rho_R \right) V^\dagger \right] = \ket{\gamma}\bra{\gamma},
\end{equation}
where $V(U)$ is the energy-preserving unitary. $V(U)$ is used to approximately induce the transformation $U$ on $\ket{\psi}$, 
\begin{align}
    V(U) &= \ket{0}\bra{0} \otimes \ket{0}\bra{0} + \sum_{l=1}^\infty V_l(U) \\
    V_l(U) &= \sum_{n,m = 0}^1  \bra{n} U \ket{m} \ket{n}\bra{m} \otimes \ket{l-n}\bra{l-m}.
\end{align}

The unitary $U$ is selected such that it rotates state $\ket{\psi}$ into $\ket{1}$,
\begin{equation} \label{eq: U}
    U = \begin{pmatrix} \sqrt{p} & -\sqrt{1-p} \\ \sqrt{1-p} & \sqrt{p} \end{pmatrix}.
\end{equation}
The system and reference states following the pre-processing step are denoted $\rho_S^\prime$ and $\rho_R^\prime$, respectively. \\

\noindent 2) Work extraction\\

\noindent The work extraction step is implemented by applying the transformation $W(\ket{\gamma}\bra{\gamma}\otimes \rho_{\textrm{battery}})W^\dagger$. This operation transfers an amount of energy $w$ to the battery with success probability $1-r$, and probability $r$ of failing to extract work. \\

\noindent 3) Resetting the reference state\\

\begin{figure*}[ht]
    \centering
    \includegraphics[width=5.9in]{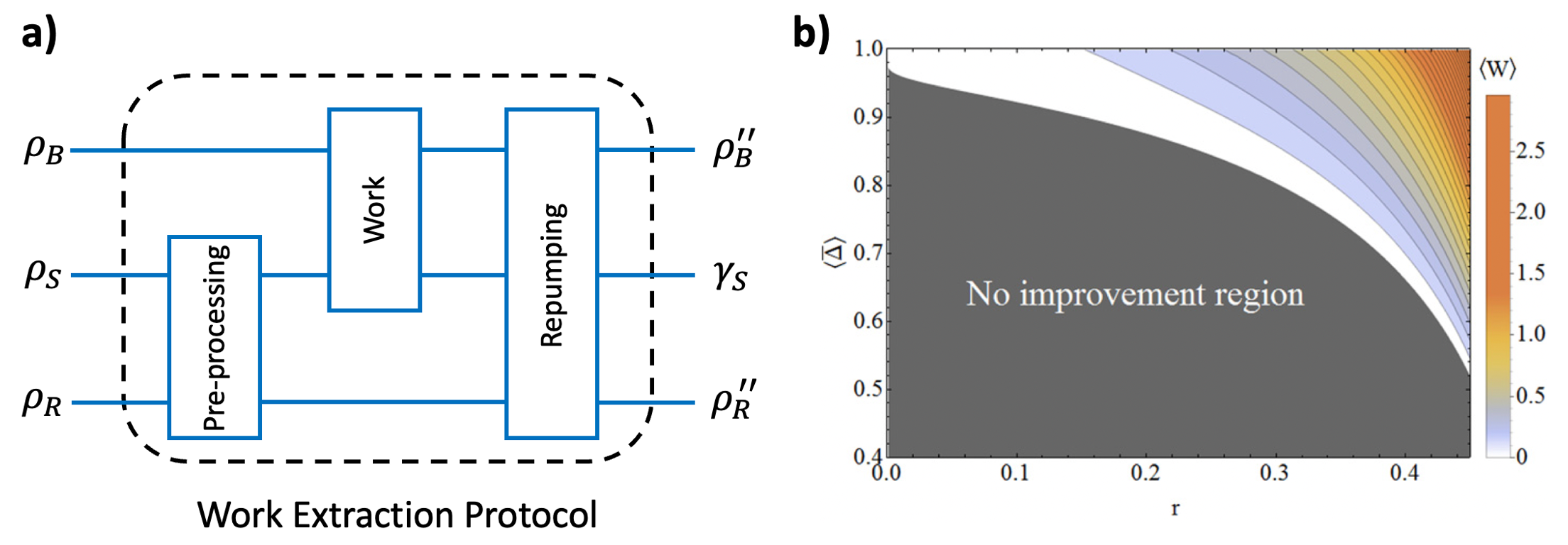}
    \caption{a) Quantum circuit diagram schematic of the work extraction protocol from coherence, where $\rho_B$ and $\rho_B^{\prime \prime}$ are initial and final states of the battery; $\rho_S$ and $\gamma_S$ are initial coherent state and final Gibbs thermal state; $\rho_R$ and $\rho_R''$ are the initial and final states of the reference, respectively. b) Figure reproduced from \cite{korzekwa2016extraction}, figure 4. A phase diagram showing regions where the coherent work extraction protocol provides improvement and is capable of extracting more work from a coherent state than from an incoherent state, where $r$ is the output Gibbs state coherent state parameterisation and $\langle \bar{\Delta} \rangle$ reference quality parameter.}
    \label{fig:korzewaPaper1}
\end{figure*}

\noindent Finally, for the protocol to be repeatable, the reference state $\rho_R^\prime$ must be returned back to a state where it can enable the transformation to occur again. This is implemented via a ``repumping" step, involving a joint unitary on the system, battery and reference states. During this process, one unit of work is used from the battery. The steps are summarised in figure \ref{fig:korzewaPaper1}, which also shows the regime for which a coherent catalyst offers an advantage in the work extraction task. 
 \\

\noindent \textbf{Caveats to coherent catalysis}\\

\noindent The ability to repeatedly use a coherent resource to allow transformations that change coherence of a system, without degradation of the resource, contrasts with the typical degradation undergone by quantum resources. Various caveats to the repeatability of coherent resources have been explored, arguing that coherence is not really being used repeatedly and these machines should not be termed as catalysts. In particular, it has been shown in \cite{vaccaro2018coherence} that while the reduced states of the outputs of the machine achieve the desired transformation, globally the outputs from multiple uses of the machine are in a very different state to that which they would be in as independent systems prepared in those reduced states. This is due to correlations between the output states affecting their global state, when they have interacted with the same machine. 

We note that for the purposes of our analysis of work extraction in Chapter \ref{Chapter:work}, the independence of the outputs of the machine is not a concern. We are comparing the effectiveness of the work extraction protocol for performing a task on an individual system state, and only the reduced state of the output is significant for our analysis. The same constraint, where only the reduced state of the output matters, is used in our constructor-theoretic analysis of quantum tasks in Chapters \ref{CT-irreversibility} and \ref{erasure_model}. This is because the specific state transformation that we are considering the possibility of has a well-defined input subspace and output subspace, defined on a single copy of the system. The difference between the collection of correlated outputs and a product state of independent systems in the same reduced state does not affect the possibility of the task in question. The dependence on correlations would be significant for tasks relating to what can be done to a collection of systems prepared in a certain subspace, which is a different class of tasks. 

Instead, in Chapter \ref{Chapter:work} we focus on a different caveat to using the catalytic properties of coherence as a thermal machine for extracting work: the impossibility of augmenting this proposal with the universality expected for typical thermodynamic machines. 

\section{The quantum homogenizer}\label{intro:homogenizer}

A particularly useful class of tools in quantum thermodynamics are known as \textit{collision models} \cite{ciccarello2022quantum}. This refers generally to a set up where there is a system that interacts with a reservoir of many discrete parts. The system typically interacts with the reservoir subsystems in the \textit{weak coupling} regime, such that the interactions cause the states of the system and reservoir subsystems to become slightly more similar. Thus collision models tend to be useful for modelling a process whereby information in a system gets spread out amongst a reservoir, which particularly lends itself to modelling thermalization processes, when additional assumptions regarding energy and thermal states are imposed \cite{scarani_thermalizing_2002, strasberg_quantum_2017}.

\begin{figure}
    \centering
    \includegraphics[width=0.8\linewidth]{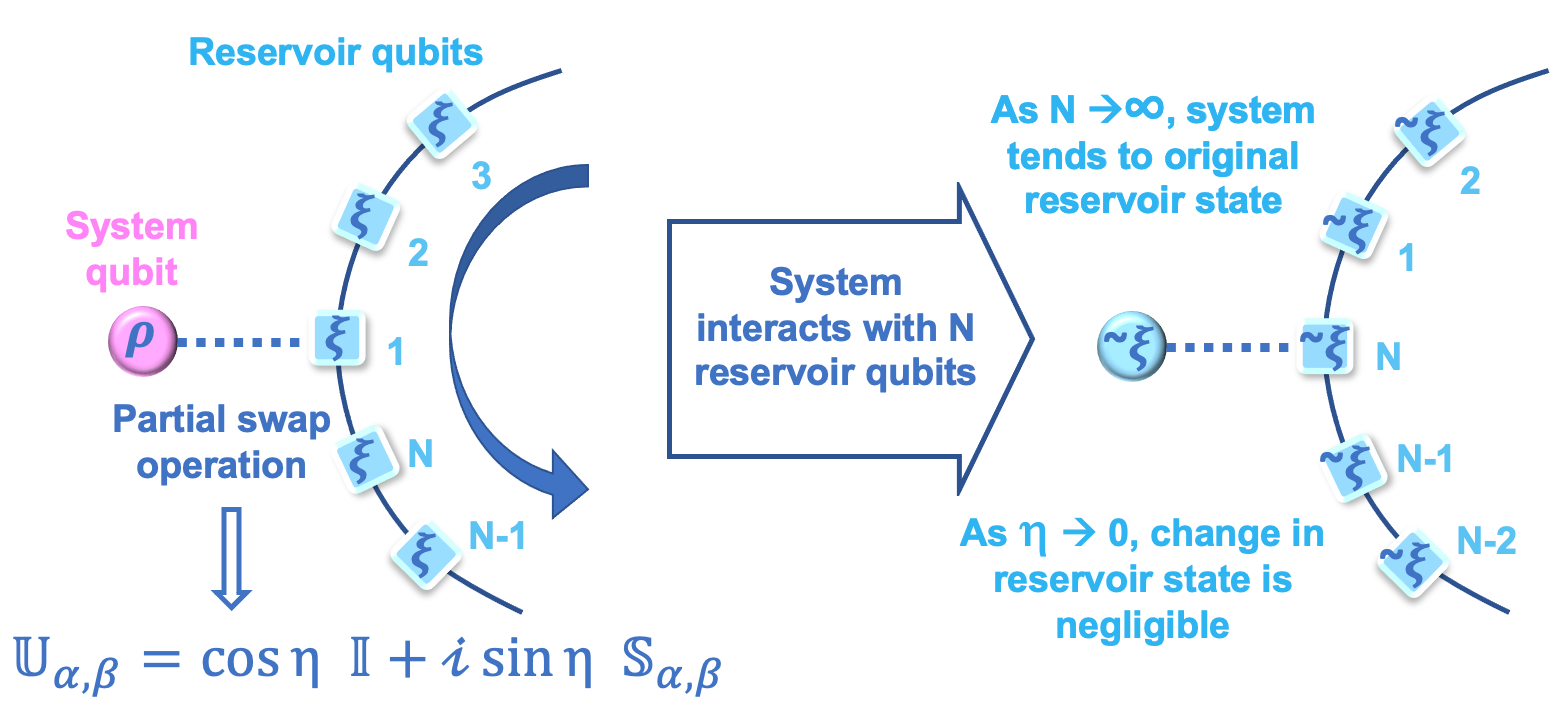}
    \caption{Depiction of the quantum homogenizer, which consists of $N$ identical reservoir qubits initially in the state $\xi$. By weakly interacting one by one with a system qubit in the state $\rho$ via a partial swap interaction $U$, the system qubit converges to the state of the reservoir qubits, while they remain almost unchanged.}
    \label{fig:QHomogenizer}
\end{figure}

One such collision model, that has a particularly simple implementation, is the quantum homogenizer. The quantum homogenizer is a machine consisting of $N$ identical reservoir qubits (figure \ref{fig:QHomogenizer}). These each interact, one by one, with the  system qubit (the qubit whose state is to be transformed) via a unitary {\sl partial swap}: 

\begin{equation} \label{partial swap}
U = \text{cos}\eta \mathbbm{1} + i\text{sin}\eta \mathbbm{S}.
\end{equation}

\begin{figure}
    \centering
    \includegraphics[width=0.5\linewidth]{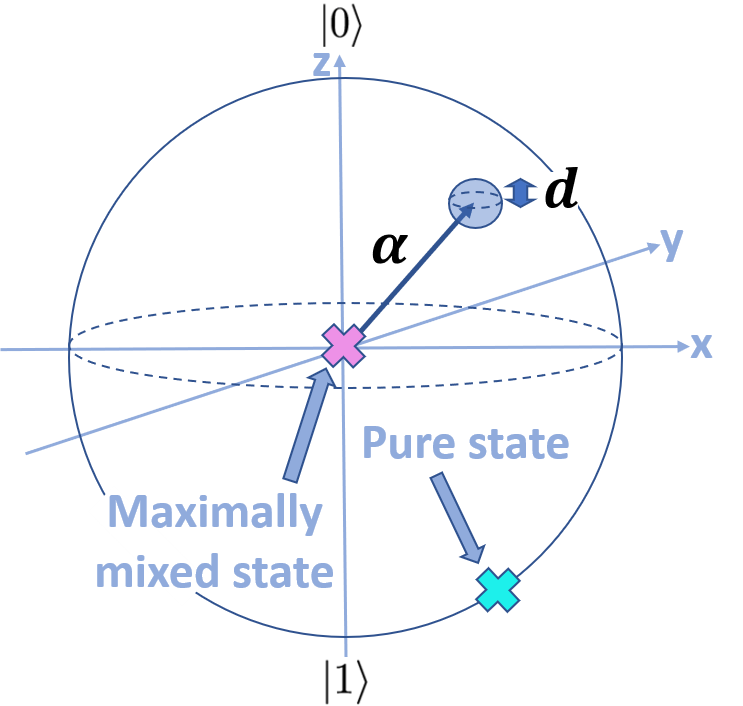}
    \caption{The Bloch sphere and a Bloch vector of size $\alpha$, pointing to the original reservoir qubit state. Following homogenisation, all the reservoir qubits, and the homogenized system qubit, are within some distance $d$ of the original reservoir qubit state, where $d$ can be made arbitrarily small as the size of the homogenizer is made arbitrarily large.}
    \label{fig:BlochSphere}
\end{figure}

The partial swap is a combination of the identity, $\mathbbm{1}$ and swap operation, $\mathbbm{S}$, weighted by the coupling strength parameter $\eta$. It has been shown that if the system qubit interacts with $N$ reservoir qubits via the partial swap, then as $N \to \infty$, the system qubit state converges to the original state of the reservoir qubits, for {\sl any} coupling strength $\eta \neq 0$ \cite{ziman_quantum_2001}. Furthermore, all of the reservoir qubits after the interaction are within some distance $d$ of their original state (see figure \ref{fig:BlochSphere}), which can be made arbitrarily small as coupling strength $\eta \to 0$. It was shown in \cite{ziman_quantum_2001} that in the limit of the best possible homogenization, any system qubit $\rho$ is sent to the reservoir qubit state $\xi$, with all the reservoir qubits remaining unchanged:

\begin{equation}
    U^{\dagger}_N...U^{\dagger}_1 (\rho\otimes \xi^{\otimes N})U_1 ...U_N\approx\xi^{\otimes N+1}
\end{equation}

\noindent where $U_k := U \otimes (\otimes_{j\neq k} \mathbbm{1}_j)$ denotes the interaction between the system qubit and the $k^{\text{th}}$ reservoir qubit. The information about the original system qubit state is seemingly erased, despite all the interactions being unitary and thus information-preserving. The information has actually become stored in the infinitesimal entanglement between infinitely many reservoir qubits, which sums to a finite value \cite{ziman_quantum_2001}.

\chapter{Axiomatic, Resource- and Constructor-Theoretic approaches} \label{resource_comparison}

\textit{The contents of this Chapter are partially based on a section of the preprint publication \cite{plesnik2024impossibility}, done in collaboration with Samuel Plesnik.} \\

\section{Introduction}

In this Chapter we give some brief comments on the aims and formulation of different approaches to quantum thermodynamics. In particular, we consider axiomatic approaches, resource theories and constructor theory. 

As discussed in section \ref{background:resource}, resource theories provide a general approach to formalizing scientific fields in terms of available physical states and processes. Constructor theory, introduced in section \ref{Intro:CT-thermo}, instead provides a general, axiomatic approach to unifying the underlying principles governing classical and quantum systems, and systems whose dynamics may be partially unknown \cite{deutsch_constructor_2013,deutsch2015constructor,marletto_constructor_2017}. While resource theories take existing dynamical laws of physics as a starting point and derive constraints on what can be done under those laws, constructor theory has its own laws of physics, for instance the \textit{interoperability of information} and the \textit{principle of locality}. These laws are conjectured to be more fundamental than dynamical laws. In the context of thermodynamics, constructor theory conjectures that thermodynamic laws are fundamental and can be formulated exactly, rather than being dependent on the constraints of a particular practical situation. It may therefore be possible to state the 2$^{\textrm{nd}}$ law of thermodynamics within constructor theory such that it is of equal status as the conservation of energy, rather than an emergent operational outcome that varies depending on the constraints applied.

\section{Dynamics-independence}

Laws governing quantum thermodynamics derived using resource theories are embedded within the dynamics of quantum theory, and make some scale-dependent assumptions about thermodynamics — for instance, in the unlimited access to thermal states, and in defining states relative to a large heat bath. By contrast, constructor theory has more general laws, formulated in terms of principles about possible and impossible transformations on sets. These principles are used to define a unified and generalised form of classical and quantum information, and physical laws regarding information. Building on the constructor theory of information \cite{deutsch2015constructor}, the constructor theory of thermodynamics introduces further constraints of possible and impossible tasks to define entities such as work media and processes such as work extraction \cite{marletto_constructor_2017}, again without committing to specific dynamics. The overarching aim in constructor theory is to abstract from scale-dependent assumptions and eventually from the dynamics of quantum theory, to state thermodynamic laws in a dynamics- and scale- independent form. 

However, to investigate how the form of irreversibility claimed by constructor theory manifests itself in quantum systems, in Chapters \ref{CT-irreversibility} and \ref{erasure_model} we consider how far a quantum machine can transform  the state of a system qubit in a cycle. This toy-model has parallels with models for thermal machines considered in the resource-theoretic analysis. In this setting, existing results and tools from quantum resource theories may have interesting connections with the constructor-theoretic possibility of thermodynamic tasks. It could be that catalysts, as defined in resource theory, can be framed as a special type of constructor. Therefore an interesting problem is how far results from resource theory and the toy-model for constructor-based irreversibility are related and could inform each others' development.

The approach to quantum thermodynamics using resource theories has been extended to the framework of general probabilistic theories (GPTs) \cite{barnum__2015, chiribella_scandolo_2015}. GPTs consider theories that relax restrictions from quantum theory, but can still violate principles of classical physics. There is a similarity with constructor theory in that studying GPTs can enable us to draw conclusions about thermodynamics that are not limited to the specific dynamics of quantum theory. However, the approaches differ in multiple ways. GPTs assume the existence of probabilities and that states can be completely characterised by measurements \cite{janotta_hinrichsen_2014}. By contrast, constructor theory is non-probabilistic and aims to formulate physical principles in terms of deterministic statements. Furthermore, the analog of states in constructor theory are attributes, which may or may not be directly measurable \cite{deutsch2015constructor}. In addition to this, constructor theory imposes new principles of physics, such as the interoperability of information. One could consider how far the two approaches lead to the same conclusions about thermodynamics as an area for future research. It may for instance turn out that the approach to generalising thermodynamic laws from GPTs and constructor theory is broadly analogous to the relationship between statistical mechanics and thermodynamics respectively.

\section{State transformations vs possible tasks 
}

The possibility of a task in constructor theory depends on a task being implemented in a cycle ($i.e.$ with no irreducible changes in the environment), with the machine transforming the system in a cycle called the constructor \cite{deutsch2015constructor}. The constructor should be approximately unchanged after causing the desired change in the system, and there is no limit to how well a perfect constructor for the task can be approximated. 

There are various ways of defining the conditions for a catalyst in resource theories, for instance regarding the potential for correlations between the system and catalyst or arbitrarily small errors in the catalyst's return to its original state (e.g. \cite{rubboli2022fundamental}). A key aspect of this approach is therefore the reusability of catalysts without significant alterations to their state. 

Contrastingly, constructors are a wider class of entities that cause transformations than catalysts. The key property of a constructor is its capacity to repeatedly perform a given task, without necessarily returning to its exact initial state or a state close to it. The possibility of a constructor for a task depends on how far a system in some input subspace can be repeatedly transformed to a state in an output subspace. This relates to the indirect way in which constructors emerge in constructor theory: the actual laws of physics, stated as possible and impossible tasks, do not refer to specific properties of the entities that implement them (the constructors). The principles instead constrain the limits of how closely an ideal entity that repeatedly performs a task can be approximated by physical machines. These properties mean that constructors are naturally formulated in terms of mappings between subspaces that enable a task, rather than retaining the same or approximately the same state. 

Recently, the implications of constructor-theoretic principles of thermodynamics have been explored within the specific dynamics of quantum theory \cite{violaris2022irreversibility, marletto2022emergence}. In this setting, existing results and tools from quantum resource theories may have interesting connections with the constructor-theoretic possibility of thermodynamic tasks, after accounting for the subtleties regarding the above mentioned characterisation of constructor-theoretic possibility and the generalisation of catalysts to constructors. An interesting open problem is to extend and define the notion of catalysts used in quantum resource theories to connect with the physical laws captured by constructors in constructor theory. Recent work has drawn connections between constructor theory and the formalism of process theories, which could help towards this goal \cite{gogioso2023constructor}.\\

\section{Connection to axiomatic thermodynamics
}

It is enlightening to consider both approaches in terms of their connection to the axiomatic classical thermodynamics formulated by Lieb and Yvnangson \cite{lieb1999physics}, which itself is based on and extends works by Giles and Carathéodory \cite{giles, caratheodory1907variabilitatsbereich}. 

Lieb and Yvnangson's approach is formulated around the property of \textit{adiabatic accessibility}: the key relation between two equilibrium states is whether or not one is adiabatically accessible from the other, where adiabatic accessibility is defined operationally using a set of axioms \cite{lieb1999physics}. A state is adiabatically accessible from another state if it is possible for the transformation to be implemented by some auxiliary system that returns to its original state, with the only side-effect being the lifting or lowering of a weight in a gravitational field.

The \textit{resource theory of noisy operations} coincides with the axioms of Lieb and Yvangson applied to quantum theory, leading to \textit{majorization} being the mathematical criterion for computing the adiabatic accessibility of one quantum state from another \cite{weilenmann2016axiomatic}. Using this framework, the von Neumann entropy can be derived as the necessary quantity for ordering quantum states under these operations, forming the basis of a 2$^{\textrm{nd}}$ law. Resource theories are more general than Lieb and Yvnangon's axiomatic formulation of thermodynamics, because they are capable of describing a family of theories that do not obey those axioms. 

Meanwhile, the constructor theory of thermodynamics extends the adiabatic accessibility condition to one of \textit{adiabatic possibility}, defined in terms of tasks rather than states, such that it is dynamics-independent and scale-independent \cite{marletto_constructor_2017, marletto2022emergence}. Adiabatic accessibility is defined in reference to a weight in a gravitational field, to generally represent any kind of work storage system. However, on microscopic and quantum scales, there are a variety of specific models for work storage systems depending on the context. Adiabatic possibility is not defined with reference to a weight in a gravitational field, but instead with reference to a work medium, where a work medium is itself fully defined in terms of possible and impossible tasks. This makes adiabatic possibility a scale-independent and dynamics-independent property, avoiding ambiguity regarding whether or not some specific system counts as a work storage medium. It also enable us to reason about ``weights" and work storage media that may not obey the dynamical laws of classical or quantum mechanics. 

Table \ref{table:thermo_relations} gives a summary of the approaches to ordering states according to the 2$^{\textrm{nd}}$ law in thermodynamics, and their associated domain of applicability. This implies the following relation between the ordering of states in resource theory and constructor theory: the constructor theory of thermodynamics may coincide with a constructor-theoretic generalisation of the resource theory of noisy operations. Then the constructor-theoretic generalisation of majorization could be adiabatic possibility, and a constructor-theoretic generalisation of the resource theory of noisy quantum operations could give rise to a generalised analog of von Neumann entropy. These are interesting connections for future investigation. 

\begin{table*}[ht]
\centering
\begin{tabular}{|>{\raggedright\arraybackslash}p{0.2\linewidth}|>{\raggedright\arraybackslash}p{0.15\linewidth}|>{\raggedright\arraybackslash}p{0.65\linewidth}|}
\hline
\textbf{Approach} & \textbf{Ordering relation} & \textbf{Domain of applicability} \\ \hline
Classical axiomatic thermodynamics & Adiabatic accessibility & Defined in terms of the behaviour of a weight in a gravitational field. Scale-dependent. \\ \hline
Constructor theory of thermodynamics & Adiabatic possibility & Defined entirely using possible and impossible tasks, can be applied to settings of systems instantiating classical and/or quantum and/or more general forms of information. Scale- and dynamics-independent. \\ \hline
Resource theory of noisy quantum operations & Majorization & Defined for quantum states, with a ``weight" understood as a work storage medium independent of a specific model. Scale- and dynamics-dependent.\\ \hline
\end{tabular}
\caption{Comparison of how states are ordered in different approaches to thermodynamics.}
\label{table:thermo_relations}
\end{table*} % A Quantum Model for Constructors and Scale-Independent, Exact Irreversibility
\chapter{Quantum Model for Constructors and Irreversibility}\label{CT-irreversibility}

\textit{The contents of this Chapter are based on part of the publication \cite{violaris2022irreversibility} done in collaboration with Chiara Marletto.} \\

\section{Introduction}

Here we give a general model for a constructor in quantum mechanics, in terms of the limit of a recursive sequence of sets. Then we apply this to give a general proof of the compatibility of constructor-based irreversibility and time-reversal symmetric laws. This shows that an exact form of irreversibility is compatible with unitary quantum dynamics, and with other potential theories based on reversible dynamics. We also compare this approach to irreversibility with recent results from resource theories. The background information for this section was introduced in section \ref{Intro:CT-thermo}.

\section{Quantum model for a constructor} \label{constructor-limit}

We first introduce a general quantum model for a constructor. Consider a pair of quantum systems $C$ and $S$, such that their overall Hilbert space is ${\cal H}={\cal H}_C\otimes {\cal H}_S$. The systems evolve via an overall unitary $U$, which is the law of motion governing their interaction.

Let's consider the attribute of a quantum system associated with its density operator. The dynamics-independent definition of attributes was given in section \ref{Intro:CT-thermo} as a property of substrates that can be changed. For our quantum model, the attribute of a quantum system is its density operator, which is one element in the set of all possible density operators the substrate could have. This set is the quantum substrate's variable in constructor-theoretic terminology.

Then the task of transforming a quantum system from density operator $\rho_{x}$, with attribute $\textbf{x}$, to $\rho_{y}$, with attribute $\textbf{y}$, is denoted $T= \{{\bf x} \rightarrow {\bf y}\}$. Here we consider the attributes being independent of time, which means that they will only change if the substrate is acted upon. Such attributes correspond to stationary states in quantum theory, whose observables are completely time-independent. However we expect the results shown here to also apply more generally for non-stationary states. 

Now we define the set of density operators of the substrate $C$, as follows:
\begin{equation}
 \Sigma^{(1)}_T=\{\rho_C \in {\cal H}_C \;:\;{\textrm{tr}}_{C}(U(\rho_C \otimes \rho_x)U^{\dagger}) = \rho_y\}.\\
\end{equation}

Every substrate $C$ in the set has the property that, when presented with a substrate $S$ that has attribute ${\bf x}$, it will return the substrate with attribute ${\bf y}$. Hence, every substrate $C$ is able to cause the task $T= \{{\bf x} \rightarrow {\bf y}\}$ once. The constraint on the output attribute only requires that the isolated substrate $S$ has the attribute ${\bf y}$, $i.e.$ that its reduced state is $\rho_{y}$. Therefore, the reduced state of $C$ may have changed from $\rho_C$, and $S$ may be entangled with $C$. This asymmetry between the inputs being necessarily in a product state, and the outputs being potentially entangled, will prove significant in our proof of constructor-based irreversibility being compatible with quantum theory. 

Any attribute in $\Sigma^{(1)}_T$ should universally cause the task $T$, without the laws of motion $U$ having a dependence on which attribute it is. This universality is captured by the property of being a convex set, which imposes that arbitrary mixtures of attributes in the variable $\Sigma^{(1)}_T$ are also an attribute in that variable. We can check that $\Sigma^{(1)}_T$ is a convex set by considering how a convex combination of N density operators $\sum_{i=1}^{N}\lambda_i \rho_{C_i}, \sum_{i=1}^{N}\lambda_i = 1$, where $\rho_{C_i}$ are in $\Sigma^{(1)}_T$, cause a system state $\rho_x$ to evolve under the unitary $U$:

\begin{align}
\begin{split}
    & \operatorname{tr}_C(U(\sum_{i=1}^{N}\lambda_i \rho_{C_i})\otimes \rho_x U^\dagger )\\
    & = \sum_{i=1}^{N}\lambda_i \textrm{tr}_C(U \rho_{C_i}\otimes \rho_x U^\dagger )\\
    & = \sum_{i=1}^{N}\lambda_i \rho_{y} = \rho_{y}.
\end{split}
\end{align}\\
Therefore the convex combination $\sum_{i=1}^{N}\lambda_i \rho_{C_i}$ is also in the set $\Sigma^{(1)}_T$, meaning that $\Sigma^{(1)}_T$ is a convex set.

\begin{figure}[t]
\includegraphics[width=\columnwidth]{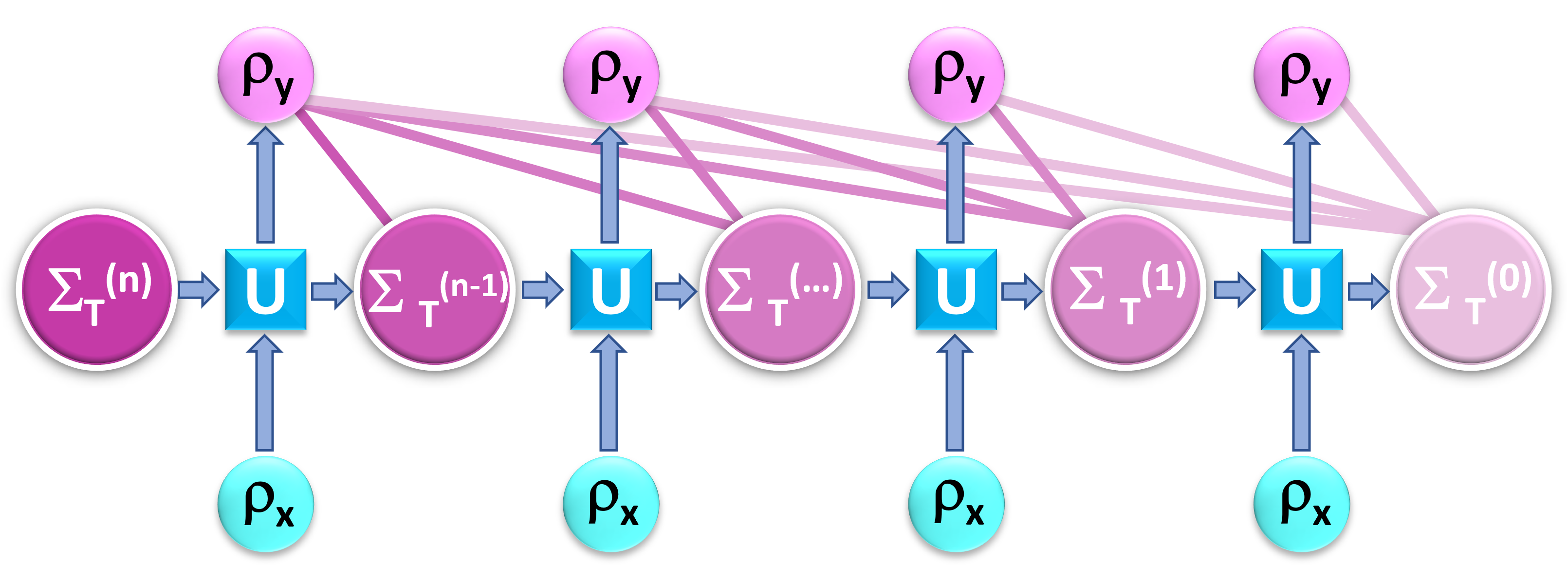}
\caption{\label{constructor_chain} Every time a machine in the set $\Sigma_T^{(i)}$ is evolved using unitary $U$ to transform a system from $\rho_x$ to $\rho_y$, it is left in a state in the set $\Sigma_T^{(i-1)}$. The lines between each $\rho_y$ and $\Sigma_T^{(i)}$ indicate that they may be entangled.}
\end{figure}

Now we can give a recursive definition of a family of sets: 

\begin{equation}
    \Sigma^{(n)}_T\subseteq \Sigma^{(n-1)}_T
\end{equation}

\noindent for all $n > 1$, with the property that $\forall \rho \in \Sigma_T^{(n)}$, setting $\rho_{out}=U(\rho\otimes \rho_x)U^{\dagger}$, we have that $\textrm{tr}_S\{\rho_{out}\} \in \Sigma_T^{n-1}$ and $\textrm{tr}_C\{\rho_{out}\} = \rho_y$.  The set $\Sigma^{(n)}_T$ is the set of states of $C$ that can perform the task $T$ consecutively $n$ times, as depicted in figure \ref{constructor_chain}. After $n$ times $C$ may lose its ability to cause the task once again. A constructor is a special kind of machine which keeps this ability indefinitely. As such, it must be defined as the limit point of the sequence of sets $\{\Sigma^{(n)}_T\}_n$. A necessary condition for $C$ to be a {\sl constructor} for the task $T$ under the law of motion $U$ is that the set
\begin{equation} \label{limit}
    \Sigma_{C_T}\doteq \lim_{n\rightarrow \infty} \{\Sigma^{(n)}_T\}_n
\end{equation}
exists and it is non-empty. That the task $T$ is possible implies, in quantum theory, that there exists a non-empty set $\Sigma_{C_T}$ with the above properties. 
\\

\noindent \textbf{The limit of a constructor}\\

\noindent A non-empty set $\Sigma_{C_T}$ defines a perfect constructor for a task: one which is able to perform the task arbitrarily well, an arbitrary number of times. In practice, if this machine is physically realized, it will not be a perfect constructor due to its inevitable deterioration from interactions with its environment. However, the existence of the non-empty set $\Sigma_{C_T}$ indicates that there is no limit to how well a perfect constructor for task $T$ can be approximated by a physical machine. 

For example, a CNOT gate and a qubit in the $\ket{1}$ state together form a  constructor for the task of transforming a state $\ket{0}$ to a state $\ket{1}$. This can be done by re-using the qubit in the $\ket{1}$ state as the control qubit with any number of target qubits in the state $\ket{0}$. However, physical CNOT gates cannot be implemented with no errors, and qubits cannot be prepared in the $\ket{1}$ state without errors. The transformation $\ket{0}$ to $\ket{1}$ is possible because there is no limit to how far the errors can be reduced in the CNOT implementation and $\ket{1}$ state preparation.

\section{Constructor-based irreversibility}

We now introduce the notion of constructor-based irreversibility. Consider a possible task $T$, that sends some input attribute $\textbf{x}$ to an output attribute $\textbf{y}$. Its transpose $T^\sim $ is defined to have the inputs and outputs of $T$ switched, so it has input attribute $\textbf{y}$ and output attribute $\textbf{x}$. Then the task $T$ is possible, while its transpose $T^\sim $ need not be possible, even with time-reversal symmetric dynamics \cite{marletto2022emergence}:

\begin{equation} 
T_{\checkmark} = \{\textbf{x} \to \textbf{y}\},\hspace{0.5cm} T^{\sim}_{\times} = \{\textbf{y} \to \textbf{x}\}.
\end{equation}

The existence of a machine that can approach a perfect constructor for a task $T$ {\sl does not imply} that such a machine exists for the task $T^{\sim}$ in the reverse direction. This is because a machine approximately capable of performing a task in a cycle is not necessarily able to perform the transpose task in a cycle to the same degree of approximation, simply by having its dynamics reversed. The asymmetry holds even if the underlying dynamical laws are time-reversal symmetric, like those of quantum theory. A model was recently proposed to illustrate this kind of irreversibility, based on unitary quantum theory \cite{marletto2022emergence}. Specifically, transforming a qubit from a pure state to a mixed state is possible, while transforming a qubit from a mixed to a pure state is not necessarily possible, even if this transformation is done via a series of unitary, time-reversal symmetric interactions. Here we will discuss this irreversibility in full generality within the formalism of quantum theory, and then in Chapter \ref{erasure_model} we will demonstrate it when the task is performed by the quantum homogenizer \cite{ziman_quantum_2001}.\\

\noindent \textbf{Constructor for transpose task}\\

\noindent We shall now argue that under quantum theory's time-reversal symmetric laws, the 
existence of a non-empty constructor set $\Sigma_{C_T}$ does not imply that a non-empty constructor set for the transpose task $T^{\sim}$,
$\Sigma_{C_{T^{\sim}}}$, must also exist. This statement holds even under the time-reversal $U^{\dagger}$ of the dynamical law. 

Consider again the task $T= \{{\bf x} \rightarrow {\bf y}\}$. Assume now that, according to time-reversal symmetry, all unitaries and their transposes are allowed. Assume also that $T$ is possible, so that the set $\Sigma_{C_T}$ exists and is non-empty under a given unitary $U$. This fact does not imply that the set $\Sigma_{C_{T^{\sim}}}$ must also be non-empty under $U^{\dagger}$, the time-reversal of $U$. For the inverse evolution $U^{\dagger}$ applied to $C$ prepared in $\Sigma_{C_T}$ and to the substrate $S$ initialized in the attribute ${\bf y}$, does not necessarily retrieve the substrate in the attribute ${\bf x}$. Hence we have the key result that $C$ initialized in the attribute $\Sigma_{C_T}$ is not necessarily a constructor for the task $T^{\sim}$ under the inverse unitary $U^\dagger$.

This result can be understood as follows. If it is possible to perform the task $T= \{{\bf x} \rightarrow {\bf y}\}$, then in particular it must be possible to perform the task $T$ consecutively on $n$ systems for arbitrary $n$ -- meaning that the set $\Sigma_T^{(n)}$ is non-empty. One might then think that it must be possible to perform $T^{\sim}= \{{\bf y} \rightarrow {\bf x}\}$ consecutively on $n$ systems for arbitrary $n$ by reversing the unitary interaction -- meaning that as a result, the set $\Sigma_{T^{\sim}}^{(n)}$ is also non-empty. This is false. As depicted in figure \ref{constructor_chain}, there can be entanglement (or more generally, correlations) between the system and machine following their interaction via $U$. Therefore to dynamically reverse the process in figure \ref{constructor_chain} to convert $n$ systems in the state $\rho_y$ back to $n$ systems in the state $\rho_x$ using $U^\dagger$, we need to prepare a very special initial state on the machine subspace. This initial state must not only have a reduced state in $\Sigma_T^{(0)}$ (the final state of a machine after performing the forwards task $T$ $n$ times), but also be entangled in a particular way with each of the $n$ systems to be transformed (all the systems $\rho_y$ shown in the top row of figure \ref{constructor_chain}). 

However for the task $T^{\sim}= \{{\bf y} \rightarrow {\bf x}\}$ to be possible, we require that \textit{any} system having the attribute $\textbf{y}$ interacting with any machine state in $\Sigma_{C_{T^\sim}}$ will be transformed to a state having the attribute $\textbf{x}$, independently of whether the system is initially entangled with the machine or not. In other words, the transformation should work for any system whose reduced state is initially $\rho_y$, whether or not it is initially entangled with the machine, but this is not the case. If we applied the inverse unitary between the machine in figure \ref{constructor_chain} with a system which was in the correct reduced state $\rho_y$ but not entangled with the machine in the correct way, then the inverse unitary would not in general transform the system to have the reduced state $\rho_x$. For let us denote by $U(\rho_C \otimes \rho_x)U^{\dagger} = \rho_{\textrm{out}}$ the global state of the machine and system after the interaction $U$. We know that, by unitarity, $U^{\dagger}\rho_{\textrm{out}}U = \rho_C \otimes \rho_x$. However, in general $U^{\dagger}(\rho_C \otimes \rho_y)U \neq \rho_C \otimes \rho_x$. Unlike the particular state $\rho_{\textrm{out}}$, a generic joint system and machine state, with reduced density operators $\rho_y$ on the system and $\rho_C$ on the machine, may not have the right correlations to allow $U^\dagger$ to lead the joint state to $\rho_C\otimes\rho_x$. Hence while $\rho_C$ in $\Sigma_{C_T}$ may well be a state that represents a constructor for $T$, its existence does not imply that a constructor should exist for the transpose task $T^{\sim}$, even when dynamical laws are unitarily reversible. 

This argument shows that irreversibility in constructor theory, stating that a task is possible but its transpose need not be, is compatible with time-reversal symmetric laws, in an exact way. This is because of the fundamental difference between a task being possible and a dynamical law being allowed. In Chapter \ref{erasure_model}, we look at a particular implementation of this constructor-based irreversibility within an eraser, using a qubit model for homogenization. 

\section{Possible connections with resource theories}

The notion of re-using quantum machines to perform a task has been studied in a variety of contexts in the field of quantum thermodynamics, including in quantum resource theories, as discussed in Chapter \ref{resource_comparison}. It is interesting to consider possible connections between our results and these approaches. 

In the resource theory of thermodynamics, measures such as R\'enyi divergence are used to assess the feasibility of an operation \cite{ng2018resource}. In constructor theory, the existence and non-emptiness of the set in equation \ref{limit} is the fundamental expression for the possibility of a task — this is a qualitative, not quantitative criterion. The relative deterioration is an expression for the feasibility of pure-to-mixed and mixed-to-pure tasks using homogenization machines, and could form the basis for constructing a similar expression for a wider class of transformations. Implementing a similar model in the resource theory of thermodynamics and comparing an analogue of relative deterioration to R\'enyi divergence would be an interesting part of a future project translating between the constructor-theoretic approach and the resource theory of thermodynamics. 

In resource theories, there is a fundamental distinction between which states and operations are available in unbounded number, and which are not. In the general case we are considering, there are no constraints on the unitary that can be applied between the system and environment, hence all operations are ``free” in the sense of resource theories. We also draw no distinction in the prior availability of pure states and maximally mixed states, since the homogenizer can be either initialised in pure states or maximally mixed states depending on the direction of the transformation being implemented. However the deterioration of using the homogenizer constrains the size of homogenizer required to achieve a given error on the system qubit. Hence, all states are treated as ``resources”. In this sense, translating our approach to a quantum resource theory would involve having all operations being free and all states being resources. One promising area for future work is to compare our results to work done in the resource theory of nonuniformity \cite{gour2015resource}, where all operations are free and the free resources are maximally mixed states, and translate the quantum homogenizer to a catalyst as defined in that resource theory (in light of the considerations discussed in Chapter \ref{resource_comparison}).  

We also note the difference between our approach and the resource theory of quantum thermodynamics: we have not introduced thermal states, or infinite thermal baths, anywhere in our approach. In a similar vein to the analysis in \cite{gour2015resource}, this enables us to consider the irreversibility arising in our model on an information-theoretic level, separately from studying the specific implications for energetics. 
 % An Additional Cost to Information Erasure Using the Quantum Homogenizer
\chapter{Additional Cost to Information Erasure} \label{erasure_model}

\textit{The contents of this Chapter are based on part of the publication \cite{violaris2022irreversibility}, done in collaboration with Chiara Marletto.} \\

\section{Introduction}

Erasure is fundamental for information processing, and key in connecting information theory and thermodynamics, as it is a logically irreversible task (introduced in section \ref{intro:erasure}). Here we use a model of erasure based on the quantum homogenizer (introduced in section \ref{intro:homogenizer}) to demonstrate how the constructor-based irreversibility discussed in Chapter \ref{CT-irreversibility} can occur in a particular case. We quantify the irreversibility using a quantity called the \textit{relative deterioration} of the homogenizer, and use this to note an additional cost to erasure, that is not captured by standard results such as Landauer's principle. In particular we argue that when performing erasure via quantum homogenization there is an additional cost to performing the erasure step of the Szilard's engine, because it is more difficult to reliably produce pure states in a cycle than to produce maximally mixed states. We also discuss the implications of this result for the cost of erasure in more general terms. 

Specifically, we analyse how far the quantum homogenizer can approximately realise the mixed-to-pure transformation needed for erasure, and approximately remain unchanged, so that it can work in a cycle. Various open-system dynamics can implement this transformation \cite{ng_limits_2015}, at the cost of causing changes to the environment in various ways. The quantum homogenizer can be used to implement this transformation by partially swapping the mixed qubit to be erased with a qubit from a large reservoir of initially pure qubits. In Chapter \ref{chapter:NMR}, we present an experimental implementation the protocol for small numbers of qubits using NMR, and it has also been implemented for photons \cite{violaris2021transforming, marletto2022emergence}.

In this Chapter we show that while a quantum homogenizer could in theory be constructed to perform a pure-to-mixed transformation to arbitrary accuracy — crucially, while working {\sl in a cycle} — such a homogenizer cannot be constructed to perform a mixed-to-pure transformation in a cycle. Specifically, in the mixed-to-pure case, the homogenizer deteriorates too quickly to perform the task indefinitely in a cycle. We conjecture that this captures an additional cost to operating Szilard's engine with the homogenizer, which goes beyond the entropy cost traditionally associated with erasure. We point out that this cost is not directly captured by a traditional analysis of entropy changes in terms of von Neumann entropy. The fact that erasure is a task that may not necessarily be performable in a cycle, while its transpose is, suggests that erasure may have an additional link to irreversibility via the constructor-based irreversibility discussed in Chapter \ref{CT-irreversibility}, beyond its traditional connection to the 2$^{\textrm{nd}}$ law via Landauer's principle and Maxwell's Demon. 

\section{Erasure and Landauer's principle} \label{Landauer}

The typical statement of Landauer's principle is that when a system stores information, erasing that information causes an irreducible entropy increase in the non-information-bearing degrees of freedom of the environment. This entropy increase can be physically manifested in the form of heat dissipation in the environment, or another form of disorder in the environment such as a rearrangement of spins, if angular momentum is a conserved quantity rather than energy \cite{vaccaro2011information, barnett2013beyond}. More generally, in a quantum setting, the Landauer bound on entropy changes to a system and environment holds also for entropy changes in the opposite direction. If instead of erasing information in a system, we require that its entropy increases, then this has an irreducible entropy decrease in the environment, which could be in the form of reduction in heat. This follows intuitively from the fact that the von Neumann entropy is conserved in unitary evolution, so a change in entropy of a system requires the same total change in entropy in the environment and system-environment correlations. Hence, the entropy changes associated with information erasure and the reverse process are symmetric, as are the associated physical implementations of those entropy changes on the environment via some form of heat dissipation.

Various experiments have verified classical and quantum versions of Landauer's principle \cite{hong2016experimental, yan2018single}. In these experiments information is erased by letting the system storing the information interact with a reservoir. These experimental implementations focus on accounting for the entropy changes in a single erasure, and do not consider what happens if the same reservoir is reused to erase further systems.

However, a physical implementation of erasing information of a system or performing the reverse process can never be performed perfectly with zero error. Hence, if a given environment is used to imperfectly erase a system, and then reused, it is unclear how these errors will combine with successive use. In principle the environment could be redesigned to be more robust to errors, improving its reusability. There is a tension between the improved robustness to errors of a modified environment, and the degradation in its ability to erase when reused. If modifying the environment reduces the error more than the degradation increases it, then successively improved environments could be used an unbounded number of times. Whereas if the error reduction achieved by modifying the environment does not compensate for the error caused by its degradation, then it has limited reusability. We formalized this distinction between whether there exists a limiting environment that could be reused an unbounded number of times in section \ref{constructor-limit}. Here we use this to compare the limiting reusability of a specific environment, the quantum homogenization machine, for performing erasure and the reverse process. By contrast to the Landauer bound, the reusability of the quantum homogenizer is not symmetric for erasure and the reverse process.

One of the strengths of our model is that it is not directly related to temperature. We quantify reusability using the quantity of relative deterioration, which depends only on the input and output states of the system. These may not be equilibrium states, therefore may not be associated with a particular temperature. One can also remove the dependence on temperature in the Landauer erasure case \cite{vaccaro2011information, barnett2013beyond, egloff2015measure}.

This asymmetry in erasure using the homogenizer is a toy-model for constructor-based irreversibility, and is fundamentally different to the irreversibility typically associated with information erasure. Traditionally, irreversibility can only be introduced in a quantum system by neglecting some information, causing the overall evolution to be described by some non-unitary process. As discussed in Chapter \ref{CT-irreversibility}, constructor-based irreversibility is not related to whether or not the dynamical trajectory of a process can be reversed, and hence is consistent with unitary, information-preserving dynamics. 

\section{Relative deterioration} \label{RelDet}

Here we show that in the weak coupling limit ($\eta \ll 1$), the quantum homogenizer can only be used as a constructor for the task of transforming a qubit from a pure state to a mixed state (pure-to-mixed task), and not for turning a mixed state into a pure state (mixed-to-pure task). To conclude whether or not a task can be performed to arbitrary accuracy in a cycle by the homogenizer, one can consider the evolution of the relative deterioration \cite{marletto2022emergence}. This accounts for both the \textit{error} in performing the task, to quantify the potential for arbitrary accuracy, and the \textit{robustness} of the machine with multiple iterations, to quantify the potential for performing the task again. 

In the following we denote the reduced state of the $I^{\textrm{th}}$ system qubit that has interacted with $j$ reservoir qubits as $\rho^{I}_{j}$. We denote the reduced state of the $j^{\textrm{th}}$ qubit in the reservoir that has been used $I$ times as $\xi^{I}_{j}$.

The accuracy in performing the task can be quantified as the fidelity of the output system qubit with the original state of the reservoir qubits. If the fidelity is close to 1, then the homogenization has been performed to high accuracy. Therefore one can define the error $\epsilon^n_N$ as this quantity subtracted from 1:

\begin{equation} 
\epsilon^n_N = 1 - F(\rho^n_N, \xi^0_j),
\end{equation}

where $n$ and $N$ are the total number of iterations and reservoir qubits respectively. Here $\rho^n_N$ is the state of the $n^{\textrm{th}}$ system qubit that has interacted with all $N$ reservoir qubits, and $\xi^0_j$ is the original state of the reservoir qubits.

Similarly, the robustness of the reservoir as it is used multiple times is the fidelity of the final reservoir state with the original reservoir state: 

\begin{equation} \label{robustness}
\delta^n_N = F(\xi^n_{N_{\textrm{tot}}}, \xi^{0 \otimes N}_j).
\end{equation}

Here $\xi^n_{N_{tot}}$ is the joint state of the $N$ reservoir qubits after the reservoir has been used $n$ times, and $\xi^{0 \otimes N}_j$ is the original joint state of the reservoir qubits, which is a product state of the original reservoir qubit states.

The robustness $\delta^n_N$ defined in this way is 1 if the reservoir is unchanged after homogenization. The effects of error and robustness on performance of the task can be summarised using their ratio, namely relative deterioration $R^n_N$: 

\begin{equation} 
R^n_N = \frac{\epsilon^n_N}{\delta^n_N} = \frac{1 - F(\rho^n_N, \xi^0_j)}{F(\xi^n_{N_{\textrm{tot}}}, \xi^{0 \otimes N}_j)}.
\end{equation}

Relative deterioration scales with error and robustness such that in the limit of large number of iterations and large number of reservoir qubits (the condition for best accuracy), then there are two cases: 

\begin{equation} \label{Possibility definitions}
\lim_{n\to\infty}\lim_{N\to\infty} R^n_N = \begin{cases} 0 & \text{Task is possible,} \\
\infty & \text{Task need not be possible.}
\end{cases}
\end{equation}\\

If $R^n_N \to 0$ in the limit $\lim_{n\to\infty}\lim_{N\to\infty}$ then the deterioration with number of iterations increases slower than the error decreases with reservoir size. This means the task {\sl can} be performed to arbitrary accuracy in a cycle, so a constructor exists for this task. If $R^n_N$ tends to $\infty$ in the limit $\lim_{n\to\infty} \lim_{N\to\infty} $ then deterioration with number of iterations increases faster than error decreases with reservoir size. This means the task {\sl cannot} be performed to arbitrary accuracy in a cycle by the homogenizer. Therefore, the homogenizer is not a constructor for this task. 

Note that we first take the limit of large homogenizer $N$, then large number of cycles $n$. In section \ref{constructor-limit} we defined a constructor existing if there is a non-zero set in the infinite limit of sets of machines able to perform a task some number of times. That treatment was idealised in the sense that it was assumed that the task can be performed with zero error $n$ times, whereas with the homogenizer the task cannot even be performed once with zero error. Hence, the definition of this limit becomes more subtle. We first take $N$ to infinity to consider the best possible homogenizer. Then we consider what happens when it is used an unbounded number of times. A third option is that the relative deterioration tends to a constant value. In this case the task also need not be possible, because the machine does not satisfy the condition of being able to perform the task to arbitrary accuracy an arbitrary number of times.

The fidelity in the robustness expression is difficult to calculate analytically. Instead, the entire reservoir state can be approximated as the tensor product of the individual reservoir qubit states,  

\begin{equation} 
\xi^n_{N_{\textrm{tot}}} \approx \xi^n_1 \otimes ... \otimes \xi^n_N, 
\end{equation} 

where each $\xi^n_j = \operatorname{tr}_1(U(\rho^n_{j-1} \otimes \xi^{n-1}_j)U^\dagger)$ is the local state of the $j^{\text{th}}$ reservoir qubit after $n$ iterations. Physically, the approximation amounts to neglecting the entanglement between the reservoir qubits. This is justified in the weak coupling limit, since inter-qubit entanglement becomes negligible for low $\eta$ \cite{ziman_quantum_2001}. This approximation was used successfully in \cite{marletto2022emergence}, and we discuss it further in section \ref{approximation}.

Then the expression for the denominator of the relative deterioration, which measures the robustness of the machine, becomes: 

\begin{equation} \label{Fidelity product}
\begin{split}
F(\xi^n_{N_{tot}}, \xi^{0 \otimes N}_j)
& \approx F(\xi^n_1 \otimes ... \otimes \xi^n_N, \xi^{0 \otimes N}_j)\\
& = F(\xi^n_1, \xi^0_j)...F(\xi^n_N, \xi^0_j).
\end{split}
\end{equation} 

Now we will derive expressions for system and reservoir states, error, robustness and relative deterioration for the pure-to-mixed and mixed-to-pure transformations, approximating to the weak-coupling limit.

\subsection{Recurrence relations for system and reservoir states} \label{appendix recurrence}

To derive these expressions, we first show how the states of system and reservoir qubits are affected by the partial swap operation, leading to recurrence relations for the states. We begin by calculating the effect of the partial swap operation between some reservoir qubit in the state $\xi$ and system qubit $\rho$, with Bloch vectors parallel along some direction $B$: 

\begin{eqnarray}
\xi &=& \frac{1}{2} (\mathbbm{1} + \alpha \boldsymbol{\sigma} \cdot \textbf{B}), \label{res state 1} \\
\rho &=& \frac{1}{2} (\mathbbm{1} + \beta \boldsymbol{\sigma} \cdot \textbf{B}). \label{sys state 1}
\end{eqnarray}

The overall resulting state from a unitary operation $U$ acting on the two qubits is given by $U(\rho \otimes \xi)U^\dagger$. Here $U$ is the partial swap (equation \ref{partial swap}), which generally entangles the two qubits, and was discussed in section \ref{intro:homogenizer}. We will only be interested in the reduced density operators, defined as: 

\begin{eqnarray}
\xi' = \operatorname{tr}_1(U(\rho \otimes \xi)U^\dagger),
\\
\rho' = \operatorname{tr}_2(U(\rho \otimes \xi)U^\dagger),
\end{eqnarray}

where $\operatorname{tr}_i$ denotes the partial trace over system $i$, here labelling the 1$^{\text{st}}$ or 2$^{\text{nd}}$ state in the tensor product. Using these reduced states in subsequent calculations means that entanglement between the two qubits is neglected. This approximation is dealt with by specialising to the limit of weak coupling, where the entanglement is negligible \cite{ziman_quantum_2001}. 

Computing the partial traces gives the local reservoir qubit state $\xi' = \frac{1}{2} (\mathbbm{1} + \alpha' \boldsymbol{\sigma} \cdot \textbf{B})$ and system qubit state $\rho' = \frac{1}{2} (\mathbbm{1} + \beta' \boldsymbol{\sigma} \cdot \textbf{B})$, where

\begin{eqnarray} \label{recurrence states}
\alpha' &=& \text{s}^2 \beta + \text{c}^2 \alpha, \label{alpha prime}
\\
\beta' &=& \text{c}^2 \beta + \text{s}^2 \alpha, \label{beta prime}
\end{eqnarray}

 with c $\equiv$ cos$(\eta)$ and s $\equiv$ sin$(\eta)$. 

Now we can generalize equations \ref{recurrence states} to arbitrary interactions across the reservoir and over many iterations of the quantum homogenizer. When a reservoir qubit interacts with a new system qubit, the number of iterations $I$ is increased by one (since each reservoir qubit only interacts with a system qubit once per iteration). The index $j$, labelling the position of the reservoir qubit within the entire reservoir of $N$ qubits, stays the same for the interactions of a given reservoir qubit. Hence, if $\beta \equiv \beta^I_j$ then $\beta' = \beta^{I+1}_j$. On the other hand, if $\alpha \equiv \alpha^I_j$ then the system qubit after the partial swap has interacted with the next reservoir qubit in the same iteration, so $j$ is increased by 1 and $I$ stays constant: $\alpha' = \alpha^I_{j+1}$. This means the states for general number of iterations $I$ and reservoir qubit position $j$ can be written: 

\begin{equation} \label{Res recur}
\begin{split}
\alpha^I_j & = \text{s}^2 \beta^I_{j-1} + \text{c}^2 \alpha^{I-1}_j \\
 & = \text{s}^2 \sum_{l=1}^{I}\beta^l_{j-1}+\text{c}^{2I}\alpha^0_I
\end{split}
\end{equation}

and

\begin{equation} \label{Sys recur}
\begin{split}
\beta^I_j & = \text{c}^2 \beta^I_{j-1} + \text{s}^2 \alpha^{I-1}_j \\
& = \text{c}^{2j}\beta^\text{I}_0 + \text{s}^2 \sum_{k=1}^{j}\alpha^{I-1}_{k}
\end{split}
\end{equation}

where we have solved the recurrence relations in terms of the initial system and reservoir states. The equations can be de-coupled by substituting expression \ref{Sys recur} into \ref{Res recur}, such that the recurrence relation for the reservoir qubit state becomes entirely in terms of previous reservoir qubit states. The same can be done for the system qubit recurrence relation by substituting \ref{Res recur} into \ref{Sys recur}. The resulting recurrence relations are: 

\begin{equation}
\beta^I_j = \text{c}^{2j}\Big(\beta^I_0 + \text{s}^2\text{c}^{2(I-1)}\sum_{k=1}^{j}\text{c}^{-2k}\Big(\alpha^0_k + \text{s}^2\sum_{l=1}^{I-1}\text{c}^{-2l}\beta^l_{I-1}\Big)\Big),
\end{equation}

\begin{equation}
\alpha^I_j = \text{c}^{2I}\Big(\alpha^0_j + \text{s}^2\text{c}^{2(j-1)}\sum_{l=1}^{I}\text{c}^{-2l}\Big(\beta^l_0 + \text{s}^2\sum_{k=1}^{j-1}\text{c}^{-2k}\alpha^{l-1}_{k}\Big)\Big).
\end{equation}

Now the initial conditions for the pure-to-mixed and mixed-to-pure transformations can be substituted into the expressions to derive specific recurrence relations. For the pure-to-mixed case, the initial conditions are a pure system qubit at the start of each iteration ($\beta^I_0 = 1$) and every reservoir qubit being maximally mixed before the first iteration ($\alpha^0_j = 0$). Then the recurrence relations for the pure-to-mixed case are: 

\begin{equation} \label{beta1'}
\beta^I_j = \text{c}^{2j}\Big(1 + \text{s}^4\text{c}^{2(I-1)}\sum_{k=1}^{j}\sum_{l=1}^{I-1}\text{c}^{-2(k+l)}\beta^l_{k-1}\Big),
\end{equation}

\begin{equation} \label{alpha1'}
\alpha^I_j = \text{c}^{2(j-1)}\Big(1 - \text{c}^{2I}+ \text{s}^4\text{c}^{2I}\sum_{l=1}^{I}\sum_{k=1}^{j-1}\text{c}^{-2(l+k)}\alpha^{l-1}_{k}\Big).
\end{equation} \\

Similarly, for the mixed-to-pure case we use the initial conditions $\alpha_0^I = 0$ and $\beta^0_j = 1$ to find: 

\begin{equation} \label{beta2'}
\tilde\beta^I_j = \text{c}^{2(I-1)}\Big(1 - \text{c}^{2j}+ \text{s}^4\text{c}^{2j}\sum_{k=1}^{j}\sum_{l=1}^{I-1}\text{c}^{-2(l+k)}\beta^{l}_{k-1}\Big),
\end{equation}

\begin{equation} \label{alpha2'}
\tilde\alpha^I_j = \text{c}^{2I}\Big(1 + \text{s}^4\text{c}^{2(j-1)}\sum_{l=1}^{I}\sum_{k=1}^{j-1}\text{c}^{-2(l+k)}\beta^{l-1}_{k}\Big).
\end{equation}
\begin{figure*}[!htb]
\minipage{0.49\textwidth}  
\includegraphics[width=\linewidth]{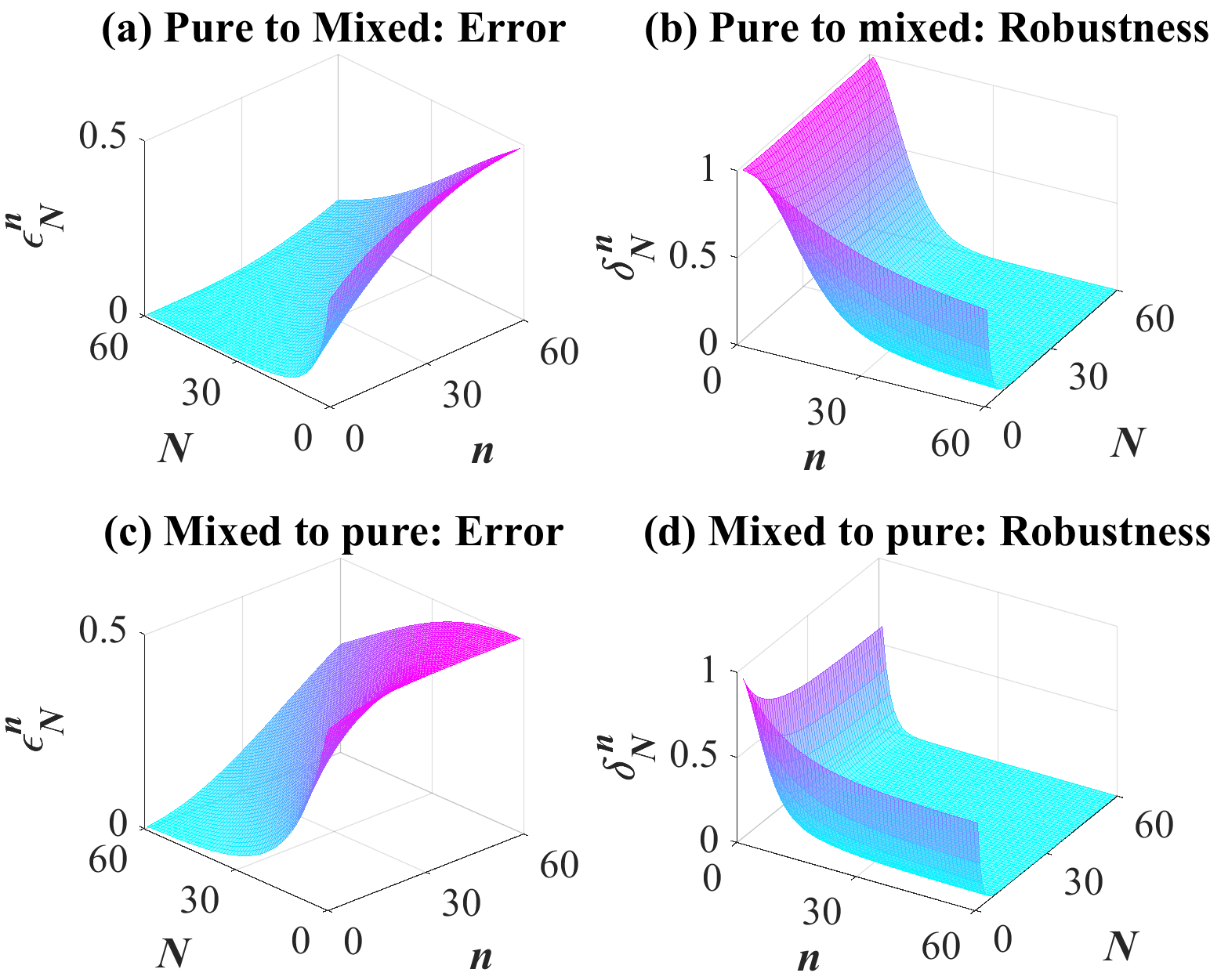} 
\caption{\label{error and robustness 0.1} Evolution of error and robustness for a reservoir of size $N$ used $n$ times, with coupling strength $\eta = 0.5$.}
\endminipage\hfill 
\hspace{0.02\textwidth}  
\minipage{0.49\textwidth}  
\includegraphics[width=\linewidth]{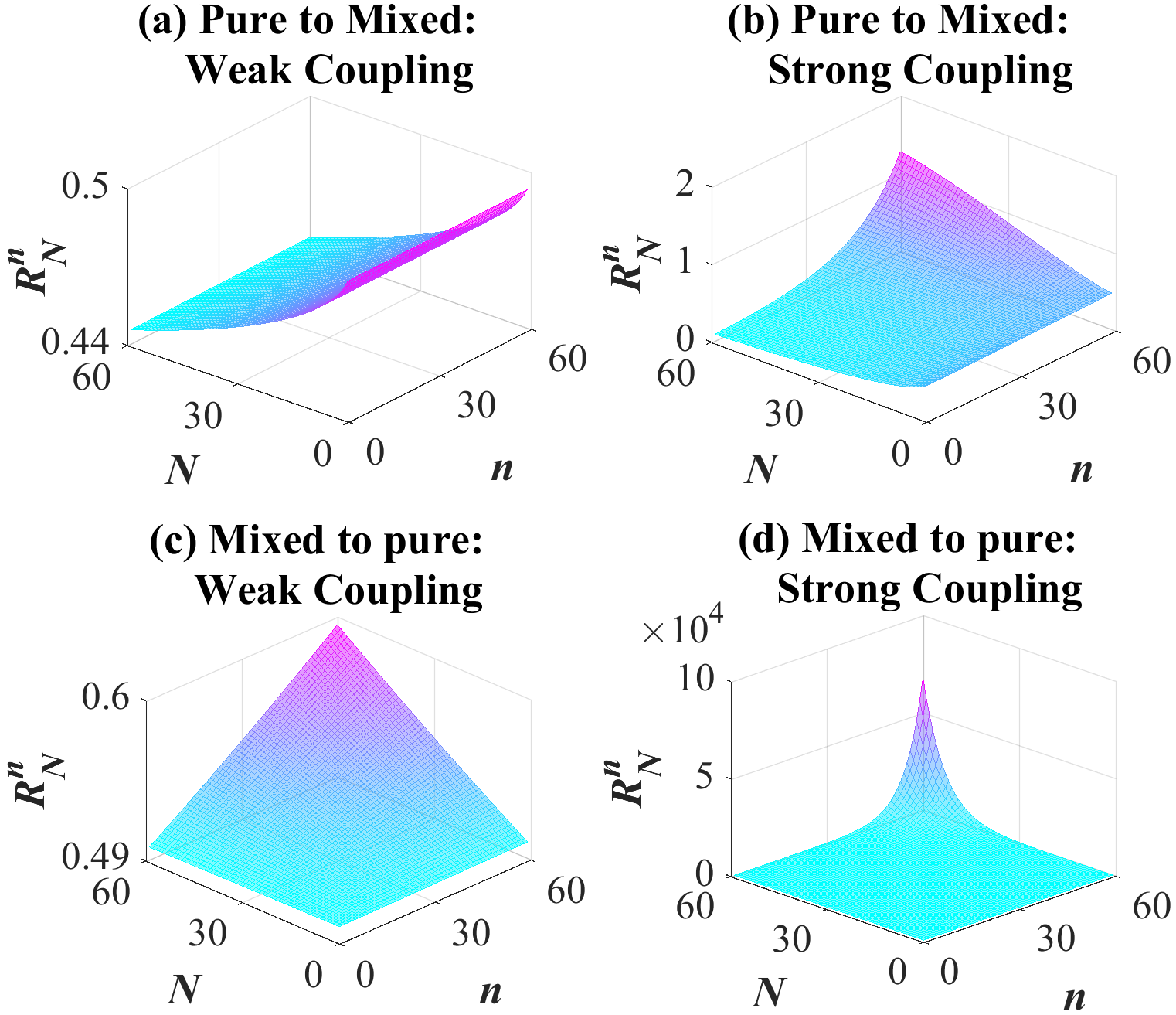} 
\caption{\label{relative deterioration} Surfaces of relative deterioration for a reservoir of size $N$ used $n$ times, with $\eta = 0.01$ for weak coupling and $\eta = 0.1$ for strong coupling.}
\endminipage
\end{figure*}

\subsection{Symmetries between states} \label{Appendix Symmetry}

The information to derive all four states is in fact contained within each one, so e.g. equation \ref{alpha1'} is enough to compute the other three. This is due to three independent symmetries constraining the Bloch vectors: 

\begin{eqnarray}
\alpha^I_a + \beta^a_j = 1, \label{Symmetry1}\\
\alpha^a_b = \tilde\beta^b_a,\hspace{0.5cm} \tilde\alpha^a_b = \beta^b_a, \label{Symmetry2}
\end{eqnarray}

where the states with and without a tilde denote the mixed-to-pure and pure-to-mixed tasks respectively. Equation \ref{Symmetry1} can be derived by summing equations \ref{alpha prime} and \ref{beta prime}, with careful attention to the indices, and using the initial condition that the Bloch vector sizes sum to 1 since one of them is 1 and the other is 0. Then equation \ref{Symmetry1} can be derived by induction as follows. 

The sum of Bloch vector sizes for the reservoir qubits on their 0$^{\text{th}}$ iteration and system qubits on their 0$^{\text{th}}$ interaction with the reservoir is always 1: 

\begin{equation}
\alpha^0_j + \beta^I_0 = 1.
\end{equation}

Also, expressing equations \ref{recurrence states} in terms of the indices, and summing them together, gives the general relation: 

\begin{equation} 
\alpha^a_b + \beta^c_d = \alpha^{a+1}_{b} + \beta^{c}_{d+1}.
\end{equation}

Then by induction, one can deduce: 

\begin{equation} 
1 = \alpha^0_j + \beta^I_0 = \alpha^1_j + \beta^I_1 = ... = \alpha^x_j + \beta^I_x.
\end{equation}

For the special case where $j = I$, then: 

\begin{equation} 
\alpha^a_b + \beta^b_a = 1,
\end{equation}

as required. 

Equations \ref{Symmetry2} can be inferred by direct substitution of the indices into the equations \ref{beta1'} to \ref{alpha2'}. They also follow logically when one keeps track of which qubits have interacted with one another. homogenization for a pure system qubit interacting with many mixed reservoir qubits (large $j$) is equivalent to deterioration for a pure reservoir qubit interacting with many mixed system qubits (large $I$).

\subsection{Error, robustness and relative deterioration} \label{Appendix error}

Here we derive the expressions for error, robustness and relative deterioration in terms of the Bloch vector sizes $\alpha$ and $\beta$, providing the expressions shown in table \ref{errors_summary}.

\begin{table}[ht]
\centering % Centers the table
\begin{tabular}{c||c c}
\hline \hline
 & Pure to Mixed & Mixed to Pure \\
\hline
$\epsilon^n_N$ & $\frac{1}{2}(1-\sqrt{1-(\beta^n_N)^2})$ & $\frac{1}{2}(1-\tilde{\beta}^n_N)$ \\
$\delta^n_N$ & $\prod_{j=1}^{N} \frac{1}{2}(1+\sqrt{1-(\alpha^n_j)^2})$ & $\prod_{j=1}^{N} \frac{1}{2}(1+\tilde{\alpha}^n_j)$ \\
\hline \hline
\end{tabular}
\caption{Expressions for error and robustness for the pure-to-mixed and mixed-to-pure tasks.}
\label{errors_summary}
\end{table}

These quantities are all defined in terms of the fidelity between quantum states, which is a measure of similarity between states. It is 0 for orthogonal states, 1 for identical states, and for two pure states it simply reduces to the inner product. A general result for fidelity of two qubits is \cite{jozsa_fidelity_1994}: 

\begin{equation} \label{Jozsa fidelity}
F(\rho,\xi) =  \operatorname{tr}(\rho\xi) + 2(\text{det}\rho\text{det}\xi)^{1/2}.
\end{equation}

\noindent In terms of Bloch vector sizes, the fidelity is:

\begin{equation} \label{Fidelity}
F(\rho,\xi) = \frac{1}{2}\big(1 + \alpha\beta+\sqrt{(1-\alpha^2)(1-\beta^2)}\big).
\end{equation} 

\noindent Hence, using equations \ref{beta1'} to \ref{alpha2'}, one can compute the fidelity between any system and reservoir qubits for any values of $I$ and $j$. 

\noindent The error is given by: 

\begin{equation}
    \epsilon^n_N = 1 - F(\rho^n_N, \xi).
\end{equation}

\noindent The fidelity of a system qubit with the original reservoir qubit can be calculated using equation \ref{Fidelity}. For the pure-to-mixed case, $\alpha^0_j = 0$ giving fidelity:

\begin{equation}
    F = \frac{1}{2}\bigg(1+\sqrt{1-(\beta^n_N)^2}\bigg).
\end{equation}

\noindent For the mixed-to-pure case, $\alpha^0_j = 1$ giving fidelity:

\begin{equation}
    F = \frac{1}{2}\bigg(1+\beta^n_N\bigg).
\end{equation}

\noindent Subtracting these from 1 to find the error, one deduces that the error for the pure-to-mixed case is: 

\begin{equation}
    \epsilon^n_N = \frac{1}{2}\bigg(1-\sqrt{1-(\beta^n_N)^2}\bigg).
\end{equation}

\noindent and for the mixed-to-pure case is: 

\begin{equation}
    \epsilon^n_N= \frac{1}{2}\bigg(1-\beta^n_N\bigg).
\end{equation}

\noindent The robustness is given by: 

\begin{equation}
    \delta^n_N = \prod^N_{j=1}F(\xi^0_j,\rho^n_j)
\end{equation}

\noindent For the pure-to-mixed case, $\alpha^0_j$ = 0, giving:

\begin{equation}
    F = \frac{1}{2}\bigg(1+\sqrt{1-(\alpha^n_j)^2}\bigg).
\end{equation}

\noindent For the mixed-to-pure case, $\alpha^0_j$ = 1, giving:

\begin{equation}
    F = \frac{1}{2}\bigg(1+\alpha^n_j\bigg). 
\end{equation}

\noindent Inserting these fidelities into the expression for robustness gives

\begin{equation}
    \delta^n_N = \prod^N_{j=1}\frac{1}{2}\bigg(1+\sqrt{1-(\alpha^n_j)^2}\bigg).
\end{equation}

\noindent for the pure-to-mixed case, and 

\begin{equation}
    \delta^n_N = \prod^N_{j=1}\frac{1}{2}\bigg(1+\alpha^n_j\bigg). 
\end{equation}

\noindent for the mixed-to-pure case. 

Putting these expressions together, one finds the relative deterioration for the pure-to-mixed case is: 

\begin{equation}
    R^n_N = \frac{\frac{1}{2}\bigg(1-\sqrt{1-(\beta^n_N)^2}\bigg)}{\prod^N_{j=1}\frac{1}{2}\bigg(1+\sqrt{1-(\alpha^n_j)^2}\bigg)}.
\end{equation}

and in the mixed-to-pure case is: 

\begin{equation}
    R^n_N = \frac{\frac{1}{2}\bigg(1-\beta^n_N\bigg)}{\prod^N_{j=1}\frac{1}{2}\bigg(1+\alpha^n_j\bigg)}.
\end{equation}
\section{Analysis}

\subsection{Patterns in relative deterioration. 
}
We analysed the pure-to-mixed and mixed-to-pure transformations with all the pure qubits initialised in the state $\ket{0}\!\bra{0}$ and mixed qubits initialised in the maximally mixed state $\frac{\mathbbm{1}}{2}$. The error and robustness for $N = n =$ 60, $\eta =$ 0.5 are shown in figure \ref{error and robustness 0.1}. Error decreases with increasing reservoir size, and increases with number of iterations, as expected. Conversely, robustness decreases with number of iterations as the reservoir deteriorates from its initial state with repeated use. The robustness also decreases with increasing reservoir size, as more reservoir qubits deteriorate from their initial state. The error for the mixed-to-pure case increases more with $n$ than for the pure-to-mixed case, and robustness for the mixed-to-pure case decreases more quickly and sharply with $n$ than for the pure-to-mixed case. These patterns indicate that the pure-to-mixed task can be performed {\sl more reliably} than the mixed-to-pure task. 

The relative deterioration reveals an asymmetry in the possibility of the two tasks. For a weak coupling of $\eta$ = 0.01, the surfaces of relative deterioration for the two directions of the task are shown in figures \ref{relative deterioration}(a) and (c). For large $N$ and $n$, $R^n_N \to 0$ for the pure-to-mixed case and $R^n_N \to \infty$ for the mixed-to-pure case. Using equation \ref{Possibility definitions}, the pure-to-mixed task is possible while the mixed-to-pure task is not possible when using the quantum homogenizer. 

\begin{figure}
\centering
\begin{minipage}{0.48\textwidth}
\includegraphics[width=\linewidth]{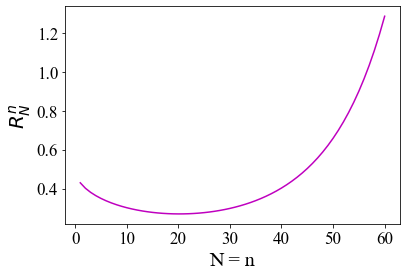} 
\caption{Approximate relative deterioration for a pure-to-mixed homogenizer used \(N=n\) times, with coupling strength \(\eta = 0.1\).}
\label{large N rel det 0.1}
\end{minipage}
\hfill 
\begin{minipage}{0.48\textwidth}
\includegraphics[width=\linewidth]{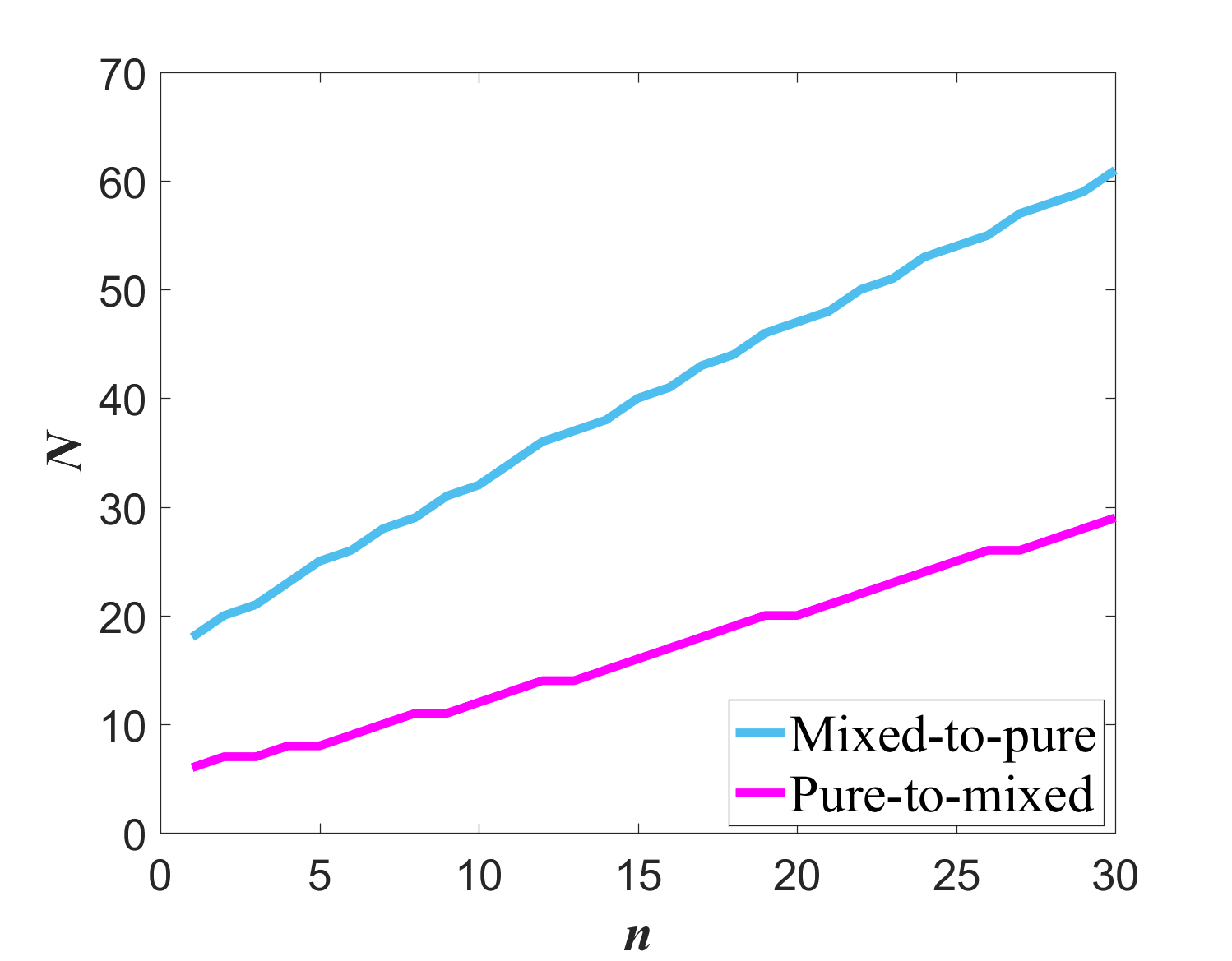}
\caption{Comparison of size of homogenizer required (\(N\)) to perform a given number of homogenizations (\(n\)), for a fixed accuracy (error 0.1) and coupling strength (0.3).}
\label{resources}
\end{minipage}
\end{figure}

The second column of figure \ref{relative deterioration} shows that for the strong coupling regime, the relative deterioration remains significantly higher in the mixed-to-pure case than the pure-to-mixed case. However, the relative deterioration for large $n$ and $N$ is increasing and not tending towards 0 for the pure-to-mixed transformation. As can be seen more clearly in figure \ref{large N rel det 0.1}, the relative deterioration decreases until around $N=n=20$, then begins increasing for larger $N=n$. 

This remains consistent with the pure-to-mixed transformation being possible using the homogenizer, since it need not be possible for all coupling strengths in order to be a possible task. It makes sense that the homogenizer has greatest reusability for small coupling strengths, as it is in the weak-coupling limit that the reservoir qubits are left almost unchanged after interacting with the system qubit. Additionally, in the strong coupling regime there is significant entanglement between the reservoir qubits, and between reservoir and system qubits that has been neglected due to the approximations in equation \ref{Fidelity product} and in deriving the recurrence relations. 

This has increasing impact as $N=n$ increases, demonstrated by the analysis of von Neumann entropy in section \ref{approximation}. This entanglement must be accounted for to determine conclusively how the relative deterioration behaves in the large $n$ and $N$ limit for strong coupling. 

\subsection{Resources required for homogenization}

Now we will compare the resources required for the pure-to-mixed and mixed-to-pure homogenizations. We can consider the size of the homogenizer needed to achieve a certain number of homogenizations all to a given accuracy, for a fixed coupling strength. Plotting the number of qubits required against number of qubits that can be homogenized to the desired accuracy (number of cycles) we find that the mixed-to-pure transformation requires consistently larger homogenizers than the pure-to-mixed transformation, indicating that the mixed-to-pure transformation requires more resources for repeated use (see figure \ref{resources}). Furthermore, the mixed-to-pure resources increase more steeply than pure-to-mixed, showing that the increase in resources required become more significant for a greater number of homogenizations. 

Similarly we can invert this relationship to conclude that the lifetime of a mixed-to-pure homogenizer of a given size is consistently lower than that of a pure-to-mixed homogenizer, since it can be used fewer times to achieve the same accuracy. This relationship reinforces the conclusion drawn from the asymmetry in relative deterioration: that the mixed-to-pure transformation is more difficult to perform in a cycle than the reverse process. \\

\begin{figure*}[!htb]
\centering  
\minipage{0.49\textwidth} 
\includegraphics[width=\linewidth]{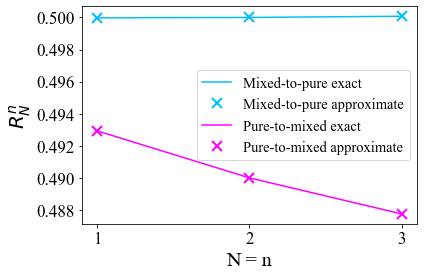}
\caption{\label{rel det 0.01} Comparison of exact (lines) and approximate (crosses) relative deterioration for pure-to-mixed (pink, bottom) and mixed-to-pure (blue, top) transformations for coupling of 0.01, showing good agreement.}
\endminipage\hfill
\minipage{0.49\textwidth}
\includegraphics[width=\linewidth]{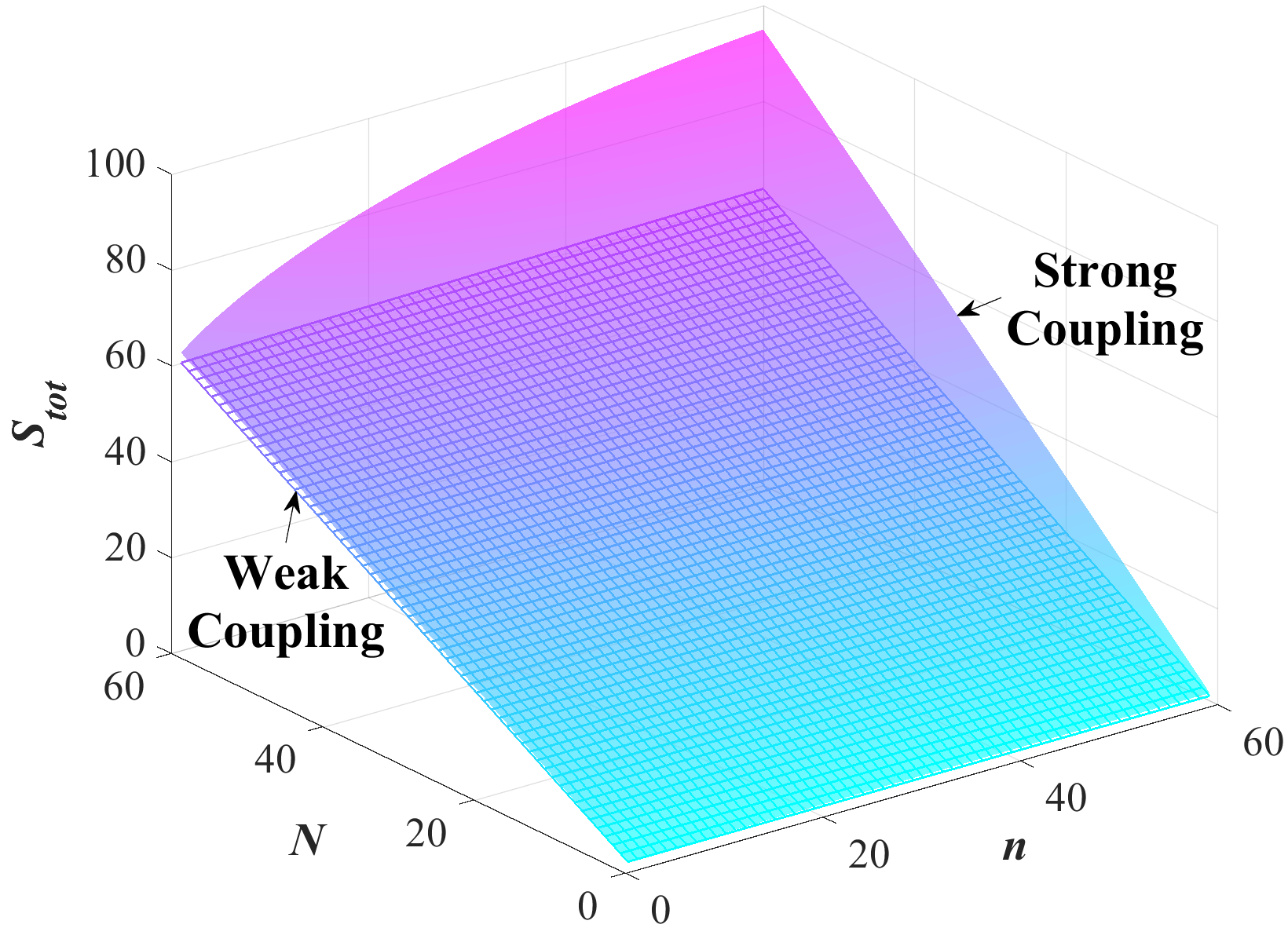}
\caption{\label{total entropies} Total von Neumann entropy of the reservoir qubits and system qubits in the pure-to-mixed transformation, with $\eta = 0.01$ for weak coupling (flat, meshed surface) and $\eta = 0.1$ for strong coupling (curved, solid surface). $N$ is the number of qubits in each reservoir, and $n$ is the number of iterations of each reservoir.}
\endminipage
\end{figure*}

\subsection{Comparing approximate and exact results}\label{approximation}

In the analytic calculation of robustness, we approximated the fidelity of the used reservoir with the original reservoir by treating the entangled reservoir as a product state of its composite qubits (equation \ref{Fidelity product}). We justified this approximation using the fact that the entanglement between reservoir qubits tends to zero for the limit of weak coupling. However, since we use relative deterioration to demonstrate an asymmetry between the mixed-to-pure and pure-to-mixed transformations, one may consider whether the asymmetry could be related to the approximation rather than being a fundamental feature of the transformations. A motivation for this could be that neglecting entanglement seems like a natural way to introduce irreversibility in a system: it amounts to a form of coarse-graining or neglecting information, as is typical for standard derivations of irreversibility. \\

\textbf{1. Conceptual difference in irreversibility.} In response to the latter argument, we note that the irreversibility suggested by the asymmetry in relative deterioration is of a different sort to the irreversibility from neglecting information about a system. Constructor-based irreversibility comes from the possibility to perform a transformation arbitrarily well in a cycle, which does not imply that the reverse transformation can be performed arbitrarily well in a cycle; this irreversibility does not emerge by reversing the dynamics of the system. By contrast, neglecting information about a system (or coarse-graining) leads to irreversibility emerging in an approximate or statistical way, such that all evolution of a system is ultimately dynamically reversible if all information about the system is known, and not necessarily reversible if there is neglected information. Therefore, constructor-based irreversibility is not of the type of irreversibility that would naturally emerge from neglecting entanglement in the reservoir. \\

\textbf{2. Symmetry of approximation.} Another reason that we would not expect the approximation to be the cause of asymmetry is that it is used in the same way for both the mixed-to-pure and pure-to-mixed transformations, neglecting entanglement between reservoir qubits in both cases, hence one would expect that it also affects the patterns in relative deterioration in a similar way for both cases.  \\

\textbf{3. Exact simulations.} To furthermore verify the validity of our approximation, we performed exact simulations of the quantum homogenizer and compared these to the approximation. Since the exact simulations become computationally large quickly with number of qubits, we have limited the simulations to maximum of a 3x3 homogenizer, where three system qubits are transformed by a three-qubit reservoir. Specialising to the case where N = n (number of reservoir qubits = number of cycles), we compared the robustness and relative deterioration for the pure-to-mixed and mixed-to-pure transformations, with and without the approximation. We found good agreement between the approximate and exact results for both coupling strengths 0.01 and 0.1. The relative deterioration for coupling 0.01 up to a size 3x3 homogenizer is shown in figure \ref{rel det 0.01}, demonstrating that the predicted asymmetry in relative deterioration holds for an exact calculation and is consistent with the approximation. \\

\textbf{4. Negligible entanglement.} An alternative way to consider the validity of the approximation for a large number of qubits is to use von Neumann entropy to quantify the entanglement building between the qubits. Figure \ref{total entropies} shows the total sum of von Neumann entropies of the individual reservoir qubits and system qubits, for strong and weak coupling. The plot is the same for the mixed-to-pure and pure-to-mixed transformations due to the symmetry of reservoir qubits in one transformation with system qubits in the reverse transformation. Specifically, $S_{tot}$ is defined: 

\begin{equation}
S_{tot}(N,n) = \frac{1}{\textrm{ln}2} ( \sum_{j=1}^{N} S(\xi^{n}_{j}) + \sum_{I=1}^{n} S(\rho^{I}_{N}) )
\end{equation}

where $S(\xi^{n}_{j})$ is the von Neumann entropy of the $j^{th}$ reservoir qubit after it has been used $n$ times, and $ S(\rho^{I}_{N})$ is the von Neumann entropy of the $I^{th}$ system qubit after it has interacted with $N$ reservoir qubits. These are computed using the states calculated in section \ref{appendix recurrence}. 

Now by unitarity, we know that the initial von Neumann entropy of all the system and reservoir qubits must equal the final entropy. So given a homogenizer of $N$ maximally mixed reservoir qubits and $n$ pure system qubits, the total entropy must be $N$ln$2$ before and after the $n$ system qubits are homogenized. Similarly, given a homogenizer of $N$ pure reservoir qubits and $n$ maximally mixed system qubits, the total entropy must be $n$ln$2$ before and after the $n$ system qubits are homogenized. 

Now correlations between the qubits contribute negative entropy, hence the difference between the total of the individual qubit entropies and the initial joint entropy indicates the amount of entanglement between the qubits that is neglected in considering only the joint states. Note that there are two contributions to the missing entropy: part from neglecting system-reservoir qubit entanglement in the approximation used to calculate the reduced states, and part from the actual entanglement that has built up between system qubits, reservoir qubits, and system-reservoir qubits. 

From figure \ref{total entropies}, we can see that for the pure-to-mixed transformation up to $N = 60$ and $n = 60$, the weak coupling $S_{tot}$ scales linearly with $N$ indicating negligible entanglement, while the strong coupling $S_{tot}$ is significantly greater, indicating that there is significant entanglement being neglected. The plot for the mixed-to-pure transformation is identical with $n$ and $N$ switched, such that the weak coupling $S_{tot}$ scales linearly with $n$. Therefore the approximation of neglecting entanglement between qubits can be used for weak coupling to demonstrate the asymmetry in relative deterioration.  

\section{Discussion}

\noindent We have shown that erasure of a mixed to a pure state using the quantum homogenizer is more difficult than the reverse process. A pertinent question is whether or not this is a special case or a signature of a more general phenomenon, which makes it difficult to generate pure states. One way to assess this would be to consider a different machine that can approximate the task and be approximately resed, and consider the behaviour of a measure of relative deterioration for such a machine.\\ 

\noindent \textbf{Comparison of homogenizer with SWAP}\\

\noindent A candidate for this could be a large reservoir of initially identical qubits, where each interaction of the system with the reservoir is a random swap -- the system qubit is swapped with a randomly selected reservoir qubit. Let's call this a probabilistic SWAP machine. Then as the reservoir becomes large, the system qubit becomes increasingly unlikely to be mistakenly swapped with a previous system qubit rather than one of the original reservoir qubits. There are some key differences between the homogenizer and probabilistic SWAP machine implementations of erasure. Firstly, with the homogenizer, there is no intrinsic limit on how many times the eraser can be resed, while still performing erasure to some effectiveness. By contrast the probabilistic SWAP machine will either perform perfect erasure or no erasure at all. While the homogenizer deterministically performs erasure with some accuracy, the probabilistic SWAP machine would probabilistically perform either perfect or no erasure. It may be that there are certain conditions under which one of the homogenizer and SWAP erasers is more optimal than the other, rather than one being conclusively the best eraser to consider for a task. 

Another such machine we can consider is a deterministic SWAP machine, which has a fixed reservoir, and reservoir qubits are swapped in order one by one with system qubits each time the task is performed. This means the task is performed with 100$\%$ accuracy, for a fixed number of times determined by the size of the reservoir. We can compare this machine to a battery for the task of charging a system: if we have a reservoir of high energy systems, then when a low energy system is presented to the machine, we can simply swap it with a high energy one to ``charge" it again. However changing the energy of a system is an impossible task in constructor theory, ensuring the fundamental principle of conservation of energy holds. A key reason that the battery does not class as a constructor is that each performance of the task leads to an irreducible degradation of the battery in terms of its ability to perform the task again, because the battery uses up a fixed amount of energy. Whatever knowledge is brought to bear in modifying the battery, there is no way to reduce the battery's degradation in a single task. By analogy, the deterministic SWAP machine faces a similar issue, in that each performance of the task results in an irreducible degradation of the machine, since one of the target states is used up. Unlike the homogenizer, there is no parameter such as coupling strength that can be varied to reduce the degradation in performing the task. An interesting problem for future research is to generalise the formulation of relative deterioration such that it can be applied to the probabilistic and deterministic SWAP machines, and more general protocols. \\ 

\noindent \textbf{Comparison of quantum and classical erasure}\\ 

\noindent Furthermore we could investigate a toy-model for classical erasure and consider whether the same difficulty of erasure arises. This way we can deduce whether the additional cost of erasure is specific to quantum theory or a more general feature of erasure that works across different regimes. In section \ref{chapter:cswap}, we analyse an incoherent version of the quantum homogenizer: a controlled swap eraser. We find that the output reduced states of the relevant systems resulting from the controlled swap eraser are very close to and sometimes identical (for certain special cases) to the quantum homogenizer, showing that our results hold for more general implementations than the specific one considered here. 

Having said this, there are particular motivations for paying special attention to the quantum homogenizer as an eraser. Different erasers can be used as models for different types of physical systems. The homogenizer is closely related to the collision model commonly used to model thermalization processes, with the weak-coupling regime being physically motivated to model systems of interest. By exploring and comparing the limitations and constraints on different forms of erasure, we can reveal how far the limits of erasure vary depending on the environment and physical model being considered.\\

\noindent \textbf{Interpretation of the additional cost of erasure}\\

\noindent Our work suggests that there may be an additional cost to erasure when using the quantum homogenizer. While we can conjecture that this cost holds for other forms of erasure, this broader conclusion is not shown by these results. If the cost does not hold for all forms of erasure, then that invites further questions about the space of erasers that are bound by the additional cost. For instance, it may be that the cost appears for quantum but not classical models, or for deterministic but not probabilistic models, or for models with weak coupling between the system and eraser but not for strong coupling. This outcome would also leave open the possibility for other processes or transformations to exhibit constructor-based irreversibility, perhaps connected by a common property of the homogenizer that is different to information erasure. 

Alternatively, if the cost does hold for other forms of erasure, then this suggests a link between constructor-based irreversibility and information erasure. The cost would have practical implications for any implementation of Szilard engines and programmable nanomachines that depend on repeated information erasure. We could then consider whether the irreversibility can be formulated as a constraint on erasure in a more general setting than quantum theory, in terms of constructor-theoretic principles. 

\section{Summary}

The traditional cost of erasure, and the reverse process, via Landauer's principle is an entropy change in the environment. This necessitates a minimum thermodynamic change in heat (or disorder of a conserved quantity other than energy). Here we have considered instead an asymmetry between performing erasure and the reverse process with the quantum homogenizer, characterised by the non-existence of a homogenizer for performing erasure in a cycle. This asymmetry is manifested in the size of homogenizer required to perform the two processes, as shown in figure \ref{resources}.      

A natural future development of this work is to better characterise the additional cost of erasure, for instance by expressing the relative deterioration in terms of a type of entropy. Furthermore, while the relative deterioration is adequate for considering the reusability of the homogenizer to perform tasks with a fixed coupling strength, an alternative analysis could have the coupling strength as a variable which is taken to the zero limit in a modified relative deterioration. 
 % The CSWAP Quantum Homogenizer as an Incoherent Quantum Information Eraser
\chapter{The CSWAP Quantum Homogenizer as an Incoherent Quantum Information Eraser}\label{chapter:cswap}

\textit{The contents of this Chapter are based on the preprint publication \cite{beever_comparing_nodate}, done in collaboration with Anna Beever, Chiara Marletto and Vlatko Vedral.} \\

\section{Introduction}

Here we investigate the role of quantum interference in the quantum homogenizer. In the original quantum homogenizer protocol, reviewed in section \ref{intro:homogenizer} and further developed in Chapter \ref{erasure_model}, a system qubit converges to the state of identical reservoir qubits through partial swap (\textsc{pswap}) interactions that allow interference between reservoir qubits. In the \textsc{pswap} homogenizer, each interaction between the system qubit and the reservoir qubits is unitary, resulting in a web of interference between system and reservoir qubits that have interacted. This raises the question as to how far the properties of the homogenizer are affected by the coherence of the unitary \textsc{pswap}, and whether the homogenizer's properties result from non-trivial quantum phenomena. 

We propose a new universal quantum homogenizer, which is an incoherent variation of the \textsc{pswap} homogenizer. We introduce an additional control qubit for each reservoir qubit in the protocol, and replace the \textsc{pswap} by a controlled swap (\textsc{cswap}) interaction, conditioned on the control qubit with a system and reservoir qubit as targets. The \textsc{cswap} gate has been previously investigated in a variety of contexts, including studies on comparing entangling power of \textsc{pswap} and controlled unitary gates \cite{balakrishnan2008entangling}, experimental implementations (e.g. \cite{ono2017implementation}), comparing quantum states, and detecting entanglement \cite{foulds2021controlled}. We show that our incoherent homogenizer satisfies the essential conditions for homogenization, being able to transform a qubit from any state to any other state to arbitrary accuracy, with negligible impact on the reservoir qubits' states. 

Mediating the interaction via a control qubit prevents interference between the system and reservoir qubits. We place an upper bound on the difference between the system qubit convergence achieved using the \textsc{cswap} and \textsc{pswap} homogenizers for arbitrary system and reservoir states, demonstrating that the difference tends towards zero as the size of the homogenizer increases. Furthermore, we identify a number of cases where the homogenizations are identical. We reinforce our conclusions with numerical simulations. Our analysis shows that the states in the two protocols differ in their paths to converging to a state, and also have major differences in the joint entropy of the system and environment qubits, but these aspects do not affect the homogenization properties. 

In addition, we derive new results regarding the reusability of both homogenizers. By calculating lower bounds on the resources needed to homogenize a general number of system qubits, we conclude that there always exists a protocol for approximately homogenizing $n$ system qubits to within a given error $\Delta$, with $N$ reservoir qubits remaining $\Delta$ close to their initial state. This requires making $N$ larger and the coupling strength weaker than the equivalent constraints for performing only a single homogenization within some error. Our analysis of the \textsc{cswap} is more general than that for the \textsc{pswap} as the lack of coherent terms simplifies the analysis, leading to tighter bounds for that protocol. We demonstrate that both homogenizers are universal for any number of homogenizations, for an increased resource cost.

Our results also suggest that the results explained in Chapters \ref{CT-irreversibility} and \ref{erasure_model} about a new form of irreversibility and information erasure in quantum homogenization machines are not dependent on non-trivial quantum coherence in the homogenizer, making them more generally applicable. 

\subsection{CSWAP quantum homogenizer}\label{incoherent}

There are two conditions that must be satisfied for homogenization. For any distance $\delta$, defined according to some distance measure between quantum states such as trace norm, the system qubit must become at least $\delta$ close to the initial reservoir qubit state, with all the reservoir qubits also at least $\delta$ close to their initial state. Formally, for some distance measure $D(\rho_1, \rho_2)$ and number of reservoir interactions $N$: 

    \begin{equation}
        D(\rho_N,\xi) \leq \delta
    \end{equation}
    and
    \begin{equation}
        D(\xi_j,\xi) \leq \delta ~ \forall ~ j, j \leq N
    \end{equation}
    
    for arbitrarily small $\delta$. Here $\rho_{j}$ denotes the state of the system qubit after interacting with $j$ reservoir qubits, and the $j^{\textrm{th}}$ reservoir qubit to interact with the system is denoted by $\xi_{j}$.

    It was shown in \cite{2002paper} that the quantum homogenizer based on the \textsc{pswap} satisfies these conditions, for any initial state of the system and reservoir qubits. Furthermore, the \textsc{pswap} is the only unitary operator that satisfies the conditions, meaning it uniquely determines the universal quantum homogenizer. 

    We will now define a universal quantum homogenization protocol based on the \textsc{cswap} instead of the \textsc{pswap}, removing the coherence between the system qubits and reservoir qubits. The \textsc{cswap} operation is a three-qubit gate, where the two-qubit swap operation is applied to the 2$^{\textrm{nd}}$ and 3$^{\textrm{rd}}$ qubits if the 1$^{\textrm{st}}$ (control) qubit is a $\ket{1}$, and they are left alone if the control qubit is a $\ket{0}$: 

    \begin{equation} \label{control swap}
    U = \frac{1}{2} ( \ket{0}\bra{0} \otimes \mathbbm{1} + \ket{1}\bra{1} \otimes  \mathbb{S} ).
    \end{equation}
    
    In our protocol, the control qubit begins in the state $\ket{c}$, a weighted superposition of $\ket{0}$ and $\ket{1}$, parameterized by a coupling strength $\eta$:
    
    \begin{equation}
        \ket{c} = \cos \eta \ket{0} + \sin \eta \ket{1}
    \end{equation}
    
    Consider a system qubit and a reservoir qubit with initial states $\rho_0 = \frac{\mathbbm{1} + \vec{\beta} \cdot \vec{\sigma}}{2}$ and $\xi_1 = \frac{\mathbbm{1} + \vec{\alpha} \cdot \vec{\sigma}}{2}$ respectively. Here the subscript zero indicates that the system qubit has interacted with zero reservoir qubits. Table \ref{state_table1} shows the results of letting the two qubits interact with a control qubit $c$, initially in the state $\rho_c = \ket{c}\bra{c}$, via the \textsc{cswap} interaction. The table shows the final joint state of the system and reservoir qubits, and the final state of the system qubit. Table \ref{state_table2} shows the corresponding states when the two qubits instead interact via a \textsc{pswap} interaction. The key difference between the final joint states in the two cases is that there are additional terms in the final joint state and final state of the system qubit when the \textsc{pswap} is used instead of the \textsc{cswap}. These additional terms indicate coherence between the qubits, and by comparing the \textsc{pswap} and \textsc{cswap} we investigate how far they impact the convergence and reusability properties of a quantum homogenization protocol.

        \begin{table}[H]
    \caption{Initial and final states after \textsc{cswap}}
    \centering
    \begin{tabular}{ |c|c| }
    
     \hline
      & Controlled Swap \\ 
     \hline
     Initial States & $\rho_0 = \frac{\mathbbm{1} + \vec{\beta} \cdot \vec{\sigma}}{2}$, $\xi = \frac{\mathbbm{1} + \vec{\alpha} \cdot \vec{\sigma}}{2}$, $\rho_c = \ket{c}\bra{c}$ \\ 
     \hline
     Final Joint State & $\rho_{\textrm{s+r}}^{\textsc{cswap}} = c^2 (\rho_0 \otimes \xi) + s^2 (\xi \otimes \rho_0)$ \\ 
     \hline
     Final System State & $\rho_1 = \frac{\mathbbm{1}}{2} + \frac{s^2}{2} \vec{\alpha} \cdot \vec{\sigma} + \frac{c^2}{2} \vec{\beta} \cdot \vec{\sigma}$\\
     \hline
    \end{tabular}
    \label{state_table1}
    \end{table}
    
    \begin{table}[H]
    \caption{Initial and final states after \textsc{pswap}}
    \centering
    \begin{tabular}{ |c|c| }
    
     \hline
      & Partial Swap \\ 
     \hline
     Initial States & $\rho_0 = \frac{\mathbbm{1} + \vec{\beta} \cdot \vec{\sigma}}{2}$, $\xi = \frac{\mathbbm{1} + \vec{\alpha} \cdot \vec{\sigma}}{2}$ \\ 
     \hline
     Final Joint State & $\rho_{\textrm{s+r}}^{\textsc{pswap}} = c^2 (\rho_0 \otimes \xi) + s^2 (\xi \otimes \rho_0)$ \\
      & - $\frac{cs}{8} (\vec{\beta} - \vec{\alpha}) \cdot (\sigma \otimes \mathbbm{1} \wedge \mathbbm{1} \otimes \sigma)$ \\
      & - $\frac{cs}{8} (\vec{\beta} \wedge \vec{\alpha}) \cdot (\sigma \otimes \mathbbm{1} - \mathbbm{1} \otimes \sigma)$ \\ 
     \hline
     Final System State & $\rho_1 = \frac{1}{2} [\mathbbm{1} + c^2 \vec{\beta} \cdot \vec{\sigma} + s^2 \vec{\alpha} \cdot \vec{\sigma} $ \\
     & $ + \frac{cs}{4} (\vec{\beta} \times \vec{\alpha}) \cdot \vec{\sigma}]$ \\
     \hline
    \end{tabular}
    \label{state_table2}
    \end{table}
 
\begin{figure}
    \centering 
    \includegraphics[width=0.7\linewidth]{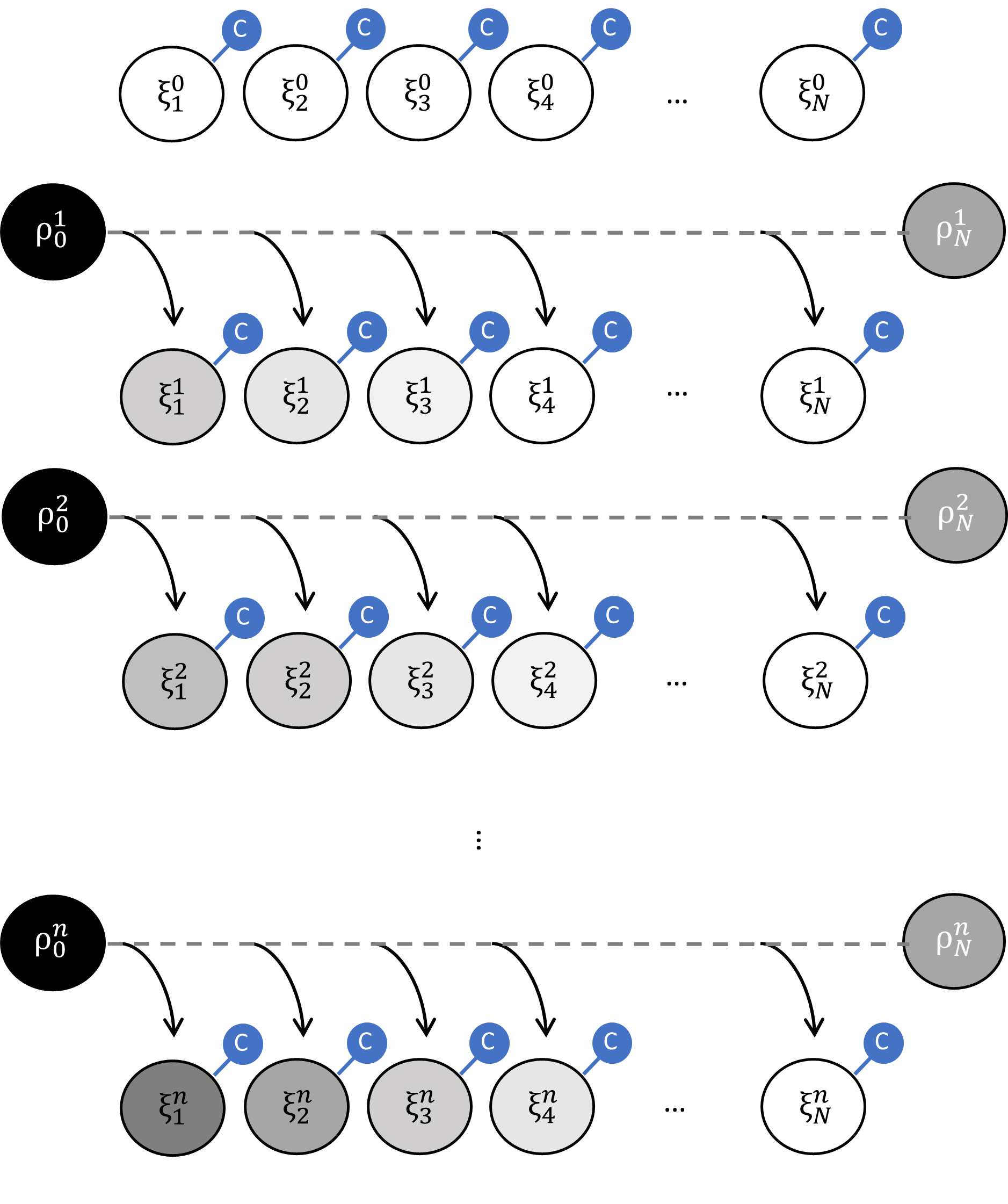}
    \caption{The \textsc{cswap} homogenizer. Control qubits are labelled c. The $j^{\textrm{th}}$ reservoir qubit after interaction with the $i^{\textrm{th}}$ system is in the state $\xi_j^i$, and the $i^{\textrm{th}}$ system after interaction with the $j^{\textrm{th}}$ reservoir qubit is in the state $\rho_j^i$.}
    \label{protocol_diagram}
\end{figure}

    Our incoherent quantum homogenization protocol involves a system qubit, a reservoir of identical environment qubits, and a set of control qubits. The system qubit interacts sequentially with each environment qubit through a \textsc{cswap} gate, moderated by a new control qubit in the state $\ket{c} = \cos \eta \ket{0} + \sin \eta \ket{1}$. The protocol is shown in figure \ref{protocol_diagram}, for a reservoir of $N$ qubits. 

\section{Convergence}\label{convergence}

    Here we demonstrate that the \textsc{cswap} homogenizer has the same convergence properties as the \textsc{pswap} homogenizer, meaning that convergence is not affected by the coherence terms. We will show that the two homogenizers achieve the same convergence using fidelity as a measure of distance, first using analytic calculations and supported by a Qiskit simulation \cite{ibmq}. 
        
        The aim of a homogenization protocol is to approximate $F(\rho_N,\xi) = 1$ as closely as possible, where $\rho_N$ is the system qubit state after interacting with $N$ reservoir qubits, $\xi$ is the initial state of the reservoir qubits, and $F(\rho_N,\xi)$ is the fidelity between $\rho_N$ and $\xi$. For a system starting with Bloch vector $\vec{\beta}$ and reservoir qubit with Bloch vector $\vec{\alpha}$, with the shorthand $c = \cos \eta$, $s = \sin \eta$, we derive the reduced states of the system qubit and reservoir qubit with the \textsc{CSWAP} homogenizer as follows.

\subsection{\textsc{CSWAP} reduced states derivation}

        Let the starting states of a control qubit, system qubit and reservoir qubit be $\rho_c^i$, $\rho_s^i$ and $\rho_r^i$ respectively, with Bloch vectors $\vec{c}$, $\vec{s}$ and $\vec{r}$. We have initial states:
        
        \begin{equation}
            \rho_c^i = \frac{\mathbbm{1} + \vec{c} \cdot \vec{\sigma}}{2},
        \end{equation}
        
        \begin{equation}
            \rho_s^i = \frac{\mathbbm{1} + \vec{s} \cdot \vec{\sigma}}{2},
        \end{equation}
        
        \begin{equation}
            \rho_r^i = \frac{\mathbbm{1} + \vec{r} \cdot \vec{\sigma}}{2}.
        \end{equation}
        
\noindent       Then let the controlled swap operator be $U$ from equation \ref{control swap}, and act on these states: 
        \begin{equation}
            U \{ \rho_c \otimes \rho_s \otimes \rho_r \} U ^\dagger .
        \end{equation}
        
        \noindent We then obtain final states of $\rho_c^f$, $\rho_s^f$ and $\rho_r^f$, where $c_x$, $c_y$ and $c_z$ are the x, y and z components of the Bloch vector $\vec{c}$ :
        
        \begin{equation}\label{3}
            \rho_c^f = \frac{\mathbbm{1}}{2}  +  \left( 1 + \vec{r} \cdot \vec{s} \right) \frac{c_x \sigma_x}{4} + \left( 1 + \vec{r} \cdot \vec{s} \right) \frac{c_y \sigma_y}{4} + \frac{c_z \sigma_z}{2},
        \end{equation}
        
        \begin{equation}
            \rho_s^f = \frac{\mathbbm{1}}{2} + \frac{(1-c_z)}{4} \vec{e} \cdot \vec{\sigma} + \frac{(1+c_z)}{4} \vec{s} \cdot \vec{\sigma},
        \end{equation}
        
        \begin{equation}
            \rho_r^f = \frac{\mathbbm{1}}{2} + \frac{(1+c_z)}{4} \vec{r} \cdot \vec{\sigma} + \frac{(1-c_z)}{4} \vec{s} \cdot \vec{\sigma}.
        \end{equation}
        
        \noindent Since the final states of system and reservoir depend only on 
        \begin{math}
        c_z
        \end{math}
        we can let 
        \begin{math}
        c_z = 2\cos^2{\eta} - 1
        \end{math}
        so that the controlled swap is parameterized by the swap strength 
        \begin{math}
        \eta
        \end{math}
        \cite{2002paper}. Then we see the system and reservoir final states can be written simply as 
        \begin{equation}
            \rho_s^f = \frac{\mathbbm{1}}{2} + \frac{\sin^2{\eta}}{2} \vec{r} \cdot \vec{\sigma} + \frac{\cos^2{\eta}}{2} \vec{s} \cdot \vec{\sigma},
        \end{equation}
        
        \begin{equation}
            \rho_r^f = \frac{\mathbbm{1}}{2} + \frac{\cos^2{\eta}}{2} \vec{r} \cdot \vec{\sigma} + \frac{\sin^2{\eta}}{2} \vec{s} \cdot \vec{\sigma}.
        \end{equation}
    
        \noindent Also, because the final states of the system and reservoir only depend on $c_z$, which remains unchanged after a \textsc{cswap} as shown in equation \ref{3}, the control can be reused as long as it is with a different reservoir and system qubit each time, avoiding interference terms between the control and its target qubits. Note that this relies on the number of reservoir qubits being greater than the number of system qubits being homogenized for there to be different control qubits used in each interaction, which is consistent with a large reservoir being the typical regime in which homogenization is used to transform system states. The final states after one \textsc{cswap} interaction are summarized in tables \ref{state_table1} and \ref{state_table2}.         

        \subsection{State Fidelity}\label{fidelity}
        
        Table \ref{state_table1} shows that for the \textsc{cswap}:
        \begin{equation}
            \vec{\beta}_1 = c^2\vec{\beta} + s^2\vec{\alpha},
        \end{equation}
        
        \noindent and for the \textsc{pswap}
        \begin{equation}
            \vec{\beta}_1 = c^2\vec{\beta} + s^2\vec{\alpha} + \frac{cs}{4} \vec{\beta} \times \vec{\alpha},
        \end{equation}

        \noindent where the subscript indicates that the system qubit has interacted with one reservoir qubit.
        
        The fidelity between the system and reservoir state $\vec{\alpha}$ for the incoherent homogenizer, using the \textsc{cswap}, is 
        \begin{equation} \label{cswap_fid}
            F_{\text{inc}} = \frac{1}{2}(1+c^2 \vec{\beta} \cdot \vec{\alpha} + s^2) + \frac{1}{2} \sqrt{(1-|c^2\vec{\beta} + s^2\vec{\alpha}|^2)(1 - |\vec{\alpha}|^2)}
        \end{equation}
        and for the coherent homogenizer, using the \textsc{pswap}, is
        \begin{align*}
            F_{\text{coh}} = \frac{1}{2}(1+c^2 \vec{\beta} \cdot \vec{\alpha} + s^2) + ~~~~~~~~~~~~~~~~~~~~~~~~~~~~~~~~~~~~~~~~~~
        \end{align*}
        \begin{equation}\label{pswap_fid}
            \frac{1}{2} \sqrt{(1-|c^2\vec{\beta} + s^2\vec{\alpha} + \frac{cs}{4} \vec{\beta} \times \vec{\alpha}~|^2)(1 - |\vec{\alpha}|^2)}
        \end{equation}

        The additional term introduced in the \textsc{pswap} fidelity is zero if $|\vec{\alpha}| = 1$, $\vec{\beta} \parallel \vec{\alpha}$, $\vec{\alpha} = 0$ or $\vec{\beta} = 0$. Even at its maximum, the additional term has a significantly smaller contribution to the fidelity than the other terms. Specifically, we can derive an upper bound on the difference between the fidelities as follows: 

        \subsection{Bounding fidelity difference} \label{comparison}

Here we bound the difference between the magnitudes of the system qubit's fidelity with the target state in the \textsc{pswap} and \textsc{cswap} homogenizers, hence the accuracy of the homogenization. We show that the ratio of the magnitudes of the additional term in the \textsc{pswap} fidelity to the \textsc{cswap} fidelity is much less than one. The ratio is: 

\begin{equation}
\frac{\delta F}{F_{\text{inc}}} = \frac{F_{\text{inc}} - F_{\text{coh}}}{F_{\text{inc}}},
\end{equation}

where $F_{\text{inc}}$ and $F_{\text{coh}}$ are defined in equations \ref{cswap_fid} and \ref{pswap_fid}. To bound the maximum value of this ratio, we consider the case where $\vec{\alpha}$ is perpendicular to $\vec{\beta}$, such that $\vec{\alpha} = \alpha \hat{z}$, $\vec{\beta} = \beta \hat{x}$, and $\vec{\beta} \times \vec{\alpha} = \alpha \beta \hat{y}$, maximising the difference between the two fidelities. Then the upper bound on $\frac{\delta F}{F_{\text{inc}}}$ is: 

\begin{equation}
\frac{\sqrt{1-\alpha^2} \left( \sqrt{1-c^4 \beta^2 + s^4 \alpha^2} - \sqrt{1-c^4 \beta^2 + s^4 \alpha^2 - \frac{c^2 s^2 \alpha \beta}{2}}\right)}{1 + s^2 + \sqrt{1-\alpha^2}\sqrt{1-c^4 \beta^2 + s^4 \alpha^2}}.
\end{equation}

From the form of the extra term in the coherent fidelity, the difference will be maximised for $\beta=1$ and $c = s = \frac{1}{\sqrt{2}}$. Then we can further simplify the bound, solely in terms of $\alpha$:

\begin{equation}
\frac{\delta F}{F_{\text{inc}}} \leq \frac{\sqrt{1-\alpha^2} (\sqrt{3 - \alpha^2} - \sqrt{3 - \alpha^2 - \frac{\alpha}{2}})}{3 + \sqrt{1-\alpha^2}\sqrt{3 - \alpha^2}}.
\end{equation}

The maximum of the RHS as $\alpha$ varies between $0$ and $1$ is $\approx 0.0208$ at $\alpha \approx 0.805$. Hence, at a maximum, $F_{\text{coh}}$ has an approximately $2\%$ deviation from $F_{\text{inc}}$.

Now let's consider the effect that subsequent interactions of the homogenization protocols have on this difference in fidelity. For our worst-case upper-bound we initialised $\vec{\alpha}$ and $\vec{\beta}$ to be perpendicular. From the convergence properties of the \textsc{pswap}, for subsequent interactions with reservoir qubits, the Bloch vectors of the system and reservoir states will no longer be perpendicular and will tend towards the same direction. The deviation between the coherent and incoherent fidelities will be scaled down by a factor of $\textrm{sin}{(\theta)}$ due to the contribution from the cross-product of the vectors $\vec{\alpha}$ and $\vec{\beta}$, where $\theta$ is the angle between the system and reservoir qubit Bloch vectors. This scaling factor will tend towards zero, the more reservoir interactions are included in the homogenization protocol. This means that the convergence properties of the \textsc{pswap} and \textsc{cswap} fidelities in this limit are equivalent. 

In summary, there is a $\approx 2\%$ upper bound on the fidelity deviation of the coherent homogenizer from the incoherent homogenizer for a finite number of system interactions with the reservoir, and in the limit of a large reservoir, the difference between the fidelities tends towards zero. 
        
\begin{figure}[H]
            \centering
            \includegraphics[width=0.7\linewidth]{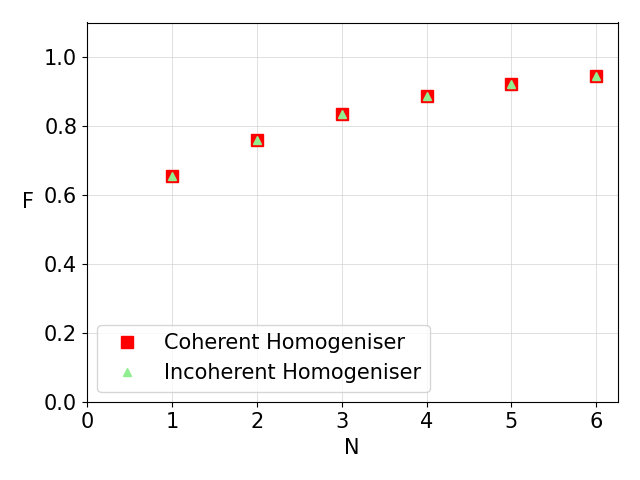}
            \caption{State fidelity $F$ against number of system-reservoir interactions $N$ for transforming $\ket{0}$ to $\ket{+}$.}
            \label{fidelitygraph}
        \end{figure}

  Hence, the convergence properties of the fidelities, for a large reservoir size, will be equivalent. Therefore there is a close agreement between the state fidelity outcomes achieved by the two homogenization protocols, for arbitrary system and reservoir states. A Qiskit simulation of both protocols transforming a system originally in the $\ket{0}$ state to the $\ket{+}$ state is shown in figure \ref{fidelitygraph}. 

        The incoherent \textsc{cswap} homogenizer therefore achieves the same accuracy as the coherent \textsc{pswap} homogenizer, up to a small correction which tends to zero in the limit of a large homogenizer. This demonstrates that the coherence introduced by the \textsc{pswap} is not contributing to the homogenization properties.

    \subsection{Trace Distance}\label{tracedistance}
    Another way of demonstrating an equivalence between the homogenizers is calculating the minimum number of system-reservoir interactions required so that the trace distance between the final system state and original reservoir state is below some error:
    \begin{equation}\label{s_delta}
        D(\rho_{N},\xi) \leq \delta 
    \end{equation}
    
    \noindent whilst also having the distance of every environment state with the original reservoir state being below that error: 
    \begin{equation}\label{e_delta}
        D(\xi_i, \xi) \leq \delta ~ \forall i, i \leq N.
    \end{equation}
    
    \noindent Here $\rho_{N}$ is a system state that has interacted with $N$ reservoir qubits, $\xi$ is the original reservoir qubit state, and $\xi_i$ is the state of the $i^{\textrm{th}}$ reservoir qubit. Following the method in \cite{2002paper} we use the trace distance as a measure of the distance between two states, corresponding to the distance on the Bloch sphere between the two states' Bloch vectors. Since $\xi_1$ interacts first with the system qubit it will be furthest from the reservoir state, so as long as this satisfies equation \ref{e_delta} all other reservoir qubits also satisfy equation \ref{e_delta}. Using table \ref{state_table1}:
        \begin{equation}
            \xi_1 = \frac{\mathbbm{1}}{2} +  \frac{s^2}{2} \vec{\beta} \cdot \vec{\sigma} + \frac{c^2}{2} \vec{\alpha} \cdot \vec{\sigma}.
        \end{equation}

    \noindent We will consider the initial system state and initial reservoir states having an absolute difference between their Bloch vectors given by: 
        
        \begin{equation}
            d = |\vec{\beta} - \vec{\alpha}|.
        \end{equation}

    \noindent This makes our analysis initially more general than the bound derived in \cite{2002paper}, where the extreme case of a distance between the states of 2 (where the initial system and reservoir states are orthogonal) is assumed from the beginning. The relevant trace distance for the \textsc{cswap} case is simpler analytically than the \textsc{pswap}: for \textsc{cswap}, the cross-terms in Table \ref{state_table1} are zero for any $d$, but for the \textsc{pswap}, the cross-terms in Table \ref{state_table2} are zero for $d=2$ (initially orthogonal qubits) but not in general. Hence, we can use the general distance $d$ to find tighter bounds on the resources needed for the \textsc{cswap} protocol. Later we will specialise to $d=2$ to compare to the \textsc{pswap} homogenizer results for the worst-case homogenization. In our general \textsc{cswap} case, the trace distance between the first reservoir qubit after it has interacted with the system with its original state is:  
        
        \begin{equation}
            D( \xi_1, \xi ) = d s^2.
        \end{equation}
        So the limit for satisfying equation \ref{e_delta} is
        \begin{equation}
             s^2 = \frac{\delta}{d}.
        \end{equation} 
        
        The system state after N system-reservoir interactions is
        \begin{equation}
            \rho_N = \frac{\mathbbm{1}}{2} + \frac{c^{2N}}{2}  \vec{\beta} \cdot \vec{\sigma} + \frac{(1 - c^{2N})}{2}  \vec{\alpha} \cdot \vec{\sigma},
        \end{equation}
        so that
        \begin{equation}\label{1}
            D( \rho_N, \xi ) = d c^{2N}.
        \end{equation}
        Using $s^2 = \frac{\delta}{d}$ we get
        \begin{equation}
            D( \rho_N, \xi ) = d \left[1 - \frac{\delta}{d}\right]^N.
        \end{equation}
        To satisfy equation \ref{s_delta} we require 
        \begin{equation}
            D( \rho_N, \xi ) \leq \delta.
        \end{equation}
        Solving for $N$, 
        \begin{equation} \label{NBound}
            N \geq \frac{\ln \frac{\delta}{d}}{\ln(1 - \frac{\delta}{d})}.
        \end{equation}
        This is the minimum number of gates required to achieve convergence to within $\delta$ for a system qubit and all the reservoir qubits.
        
        In \cite{2002paper} it is shown that for the \textsc{pswap} homogenizer the number of gates required to achieve convergence within $\delta$ for the case of two orthogonal pure states is
        
        \begin{equation}
            N_{\delta} \geq \frac{\ln \frac{\delta}{2}}{\ln(1 - \frac{\delta}{2})}.
        \end{equation}
        
        \noindent This is the same as our result for the \textsc{cswap} homogenizer, where for orthogonal pure states 
        
        \begin{equation}
        d = |\vec{\beta} - \vec{\alpha}| = 2,
        \end{equation}
        so that 
        \begin{equation}
                N \geq \frac{\ln \frac{\delta}{2}}{\ln(1 - \frac{\delta}{2})}.
            \end{equation}
        
        \noindent Therefore we have derived an equivalent upper bound on the number of reservoir qubits needed for a successful homogenization using the \textsc{cswap} homogenizer as for the \textsc{pswap} homogenizer.

    \subsection{Differences between homogenizers}

    Despite the similarity in convergence properties of the two homogenizers, we also found significant differences between them regarding how the Bloch vector of the system qubit evolves during homogenization, and in the joint system-reservoir von Neumann entropy.\\

\noindent \textbf{Evolution of Bloch vectors}\\

\begin{figure}[t]
                \includegraphics[width=\linewidth]{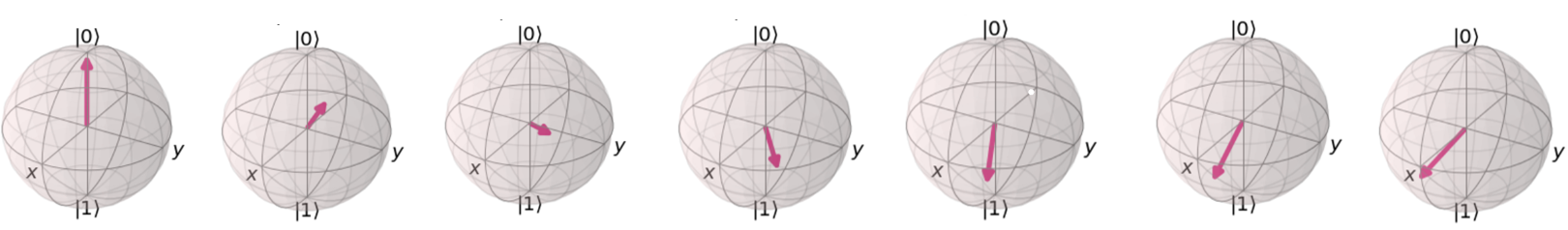}
                \caption{System Bloch vector evolution for the coherent homogenizer with initial state $\ket{0}$ and reservoir state $\ket{+}$, simulated using Qiskit \cite{ibmq}.}
                \label{pswap_series}
            \end{figure}

\begin{figure}[t]
                \includegraphics[width=\linewidth]{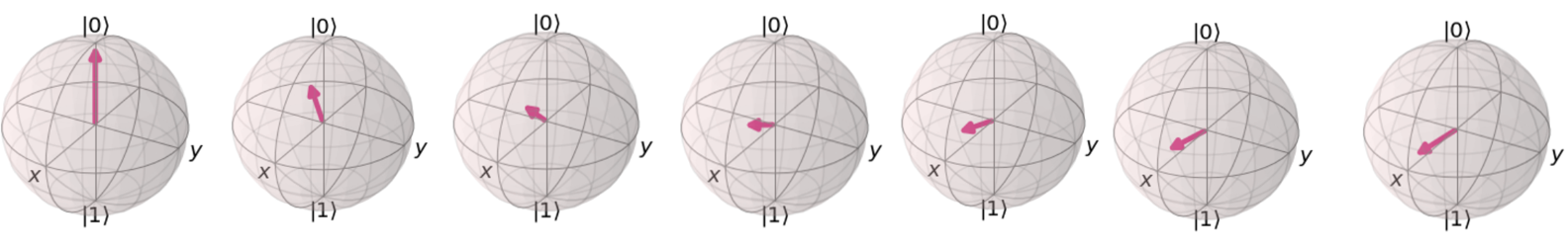}
                \caption{System Bloch vector evolution for the incoherent homogenizer with initial state $\ket{0}$ and reservoir state $\ket{+}$, simulated using Qiskit \cite{ibmq}.}
                \label{cswap_series}
            \end{figure}

\noindent Despite the similarity in fidelities computed for the two homogenization protocols, figures \ref{pswap_series} and \ref{cswap_series} show that there is a significant difference in how the states are evolving on the Bloch sphere. In the \textsc{cswap} case, the Bloch vector remains in the X-Z plane throughout its evolution. The coherence term in the \textsc{pswap} case changes the path the Bloch vector takes but not the final state.\\
    
\noindent \textbf{Joint system-reservoir entropy}\\

\noindent Here we show the significant difference in joint von Neumann entropy of the system and reservoir qubits for the \textsc{pswap} and \textsc{cswap} homogenizers, which nonetheless does not affect the homogenization properties. Specifically the joint von Neumann entropy is $S = - \operatorname{tr}(\rho_{s+r} \log \rho_{s+r})$ where $\rho_{s+r}$ is the joint state of the system and reservoir qubits (with the control qubit traced out for the \textsc{cswap}).
            
        With the coherent \textsc{pswap} homogenizer, all interactions between the system and reservoir qubits are unitary, and hence the overall von Neumann entropy is constant. By contrast, the incoherent \textsc{cswap} homogenizer involves a control qubit which is traced out to find the joint system and reservoir state. Therefore we expect that the system-reservoir von Neumann entropy in general changes with number of interactions. Specifically, since the entanglement of the system-reservoir qubits with the control qubit contributes negatively to the von Neumann entropy, we might intuitively expect that the joint system-reservoir von Neumann entropy increases with number of interactions. 
    
\begin{figure}[t] 
                \centering
                \includegraphics[width=0.7\linewidth]{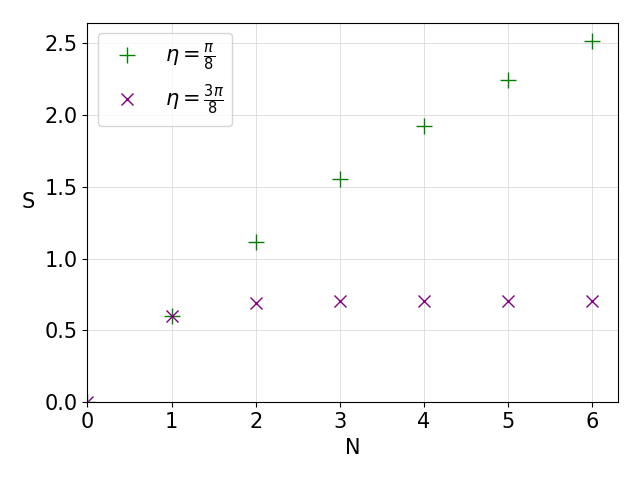}
                \caption{Von Neumann entropy $S$ as a function of the number of system-environment interactions $N$ with coupling strengths $\eta = \frac{\pi}{8}$ and $\eta = \frac{3\pi}{8}$}.
                \label{entropy}
            \end{figure}

        When we compute numerical simulations of the von Neumann entropy for the joint system-reservoir state with \textsc{cswap} interactions, we indeed find that it increases with interactions, and then reaches a plateau, which happens sooner for strong coupling than weak coupling, though at a smaller value of maximum von Neumann entropy. This can be understood in terms of the system being homogenized quicker in the strong coupling case (leading to a plateau in joint system-reservoir von Neumann entropy) but there is also more negative entropy contributed by the entanglement with the control qubit (leading to a smaller maximum value of von Neumann entropy), shown in figure \ref{entropy}.  
            
    \section{Repeated homogenization}\label{reusability}
            
            Now we will consider the reusability of both homogenizers, which was not considered in \cite{2002paper}. Reusability is of interest in quantum thermodynamics for assessing whether transformations can be enabled using catalysts. For instance, analysis using resource theories has found that catalysts can drastically increase performable state transformations \cite{ng2015limits, ng2019resource, brandao2015second, lipka2021all}. 
            Furthermore, reusability is important for assessing whether a transformation can be performed reliably, as in the constructor theory approach to thermodynamics discussed in section \ref{Intro:CT-thermo} \cite{deutsch_constructor_2013, marletto_constructor_2017, marletto2022information}. 
            
            First we discuss how the similarity we derived earlier between the \textsc{pswap} and \textsc{cswap} homogenizers shows how those results can be generalized to a wider class of incoherent protocols. Then, we use a different approach to investigating the reusability of homogenization machines, bounding the number of reservoir qubits $N$ required and number of times $n$ the homogenization task can be performed, in order to satisfy the conditions for homogenization. \\

\noindent \textbf{Generalising constructor-based irreversibility}\\

\noindent A pertinent question is whether the convergence and irreversibility properties of the quantum homogenizer presented in Chapters \ref{CT-irreversibility} and \ref{erasure_model} are dependent on the web of interference between system and reservoir qubits that arises due to the coherent partial swap interactions. Those Chapters demonstrate an asymmetry in reusing homogenization machines to transform qubits from mixed to pure states and the opposite process. If the coherence is important, this would suggest that the homogenizer's properties are a non-trivial quantum effect, whereas if the properties are independent of the coherence, this suggests the homogenizer's properties can be generalized to a wider class of incoherent protocols, which are closer to classical implementations of thermalization and information erasure. The similarity we derived between the \textsc{pswap} and \textsc{cswap} homogenizers supports the latter case.
        
            The results in Chapters \ref{CT-irreversibility} and \ref{erasure_model} were derived using a quantity called the relative deterioration, which is a function of fidelities between qubits and their target states. We showed in section \ref{fidelity} that the difference between the \textsc{pswap} and \textsc{cswap} fidelities of system states with the target state is small and tends to zero in the limit of a large reservoir. Since relative deterioration is quantified using fidelities, the equivalence of the convergence properties of the homogenizers suggests that the additional repeatability cost of coherently erasing information also applies for the incoherent model. Hence our result expands the range of applicability of the repeatability cost of erasure to a wider class of models. By contrast with protocols in recent studies of coherence as a resource for quantum information processing tasks \cite{streltsov2017colloquium}, it is not a resource for the homogenization and information erasure protocols presented here. \\

\noindent \textbf{Bounds on reusing homogenization machines}\\

\noindent For a reusable homogenizer, we require that all homogenizer qubits remain arbitrarily close to their original states, but we extend the requirement on the system qubit such that all $n$ system qubits must be homogenized arbitrarily close to the original reservoir qubits' state. 
            
            For both the \textsc{cswap} and \textsc{pswap}, we find that there is a finite number of homogenizer qubits required such that a given homogenizer is able to transform a given number of system qubits from an initial state to a final state, within a specified error, with all homogenizer qubits also being within a certain error from their original state. This is a natural extension of the conditions for homogenization introduced in \cite{2002paper} to a setting where the homogenizer needs to also be reused to transform some finite number of systems. Note that for the \textsc{cswap} homogenizer, after the first system qubit has been homogenized the control qubits are reordered, so that for the next system qubits no reservoir qubit interacts with the same control. This prevents quantum interference terms developing between control and reservoir qubits. 
            
            The first reservoir qubit deteriorates most rapidly as this is first to interact with each new system qubit, so when this state is within the required distance from $\xi$ so are the other reservoir qubits.
            
            The state of the first reservoir qubit after $n$ interactions with a fresh system qubit is:
            \begin{equation}
                \vec{\alpha}_1^n = (1 - c^{2n}) ~ \vec{\beta} + c^{2n} \vec{\alpha_1^0}
            \end{equation}

            \noindent where the superscript indicates the number of system qubits the reservoir qubit has interacted with.
        
            As in the previous section, for homogenization we require $D( \xi^{n}_{1}, \xi ) \leq \delta$. Hence, $1 - c^{2n} \leq \frac{\delta}{d}$, such that
            
            \begin{equation}\label{reuse_constraint}
                c^{2n} \geq 1 - \frac{\delta}{d}.
            \end{equation}
           This can be rearranged to find the constraint on the number of qubits that can be homogenized for a given error and coupling strength: 

            \begin{equation}
                n \leq \frac{\textrm{ln}(1-\frac{\delta}{d})}{2 \textrm{ln(c)}}.
            \end{equation}

            Now we can consider constraining the system qubits such that every system qubit is within a distance $\epsilon$ from the most-deteriorated reservoir qubit, namely the first one, which is in state $\xi^{(n-1)}_1$ before the final homogenization. Therefore in a worst-case scenario, we could use a homogenizer entirely composed of qubits in the state of the most deteriorated one, $\xi_1^{(n-1) \otimes N}$, to transform the state of a system qubit. The lower bound on the number of reservoir qubits needed for the system qubit to be within a distance $\epsilon$ of the reservoir qubits' states is of the same form as the original bound when the homogenizer was used once, in equation \ref{NBound}:

            \begin{equation} \label{NBound2}
                N \geq \frac{\ln \frac{\epsilon}{d'}}{\ln(1 - \frac{\epsilon}{d'})}.
            \end{equation}

            \noindent Here $d' = |\vec{\beta}^0 - \vec{\alpha}^{(n-1)}_1|$ is the distance between the first reservoir qubit after interacting with $n-1$ system qubits, and the original system qubit state on the Bloch sphere. Using $d = |\vec{\beta}^0 - \vec{\alpha}^0_1|$, we find $d - d' = |\vec{\alpha}^0_1 - \vec{\alpha}^n_1|$. Simplifying this expression leads to the following relation between $d$ and $d'$: 

            \begin{equation}
                d' = c^{2n}d.
            \end{equation}
            
            \noindent Now the distance of the worst-case reservoir qubit $\xi^{(n-1)}_1$ from the target state is $d(1-c^{2(n-1)})$, from equation \ref{reuse_constraint}. Therefore the distance $\epsilon$ must satisfy the condition $\Delta = \epsilon + d(1-c^{2(n-1)})$, for all system qubits to be within $\Delta$ of the target state. Substituting the resulting expression for $\epsilon$ into equation \ref{NBound2}, along with the expression for $d'$, the bound can be rewritten as: 

            \begin{equation} \label{NBound3}
                N \geq N_{\min} = \frac{\ln(1 - \frac{d-\Delta}{dc^{2(n-1)}})}{\ln \frac{d-\Delta}{dc^{2(n-1)}}}.
            \end{equation}

            \noindent Now if the conditions in equation \ref{NBound3} and equation \ref{reuse_constraint} are both satisfied, then $N$ reservoir qubits and $n$ system qubits are a maximum distance $\Delta$ from the original reservoir qubits' state. Since the bound comes from a worst-case approximation, the minimum $N$ needed for specific transformations will be smaller than $N_{\min}$. We can therefore always homogenize $n$ qubits, with all system and reservoir qubits within an error $\Delta$, for any $n$ and $\Delta$, by making $\eta$ sufficiently small and $N$ sufficiently large. For the single-use homogenizer, reducing the desired error $\Delta$ requires $\eta$ to decrease and $N$ to increase. For our reusable homogenizer, we have the added condition that imposing $n$ to be greater also requires $\eta$ to decrease and $N$ to increase, further constraining the conditions for homogenization. Note that setting $n=1$ and $d=2$ reproduces the constraints on $\eta$ and $N$ derived in \cite{2002paper}. 

            We derived the conditions on $N$ and $n$ for a general initial distance $d$ between the initial system qubit state and initial homogenizer qubits' state. This general expression holds for the \textsc{cswap} homogenizer. By considering the worst-case scenario where the initial reservoir state and initial system states are a distance $d=2$ apart (orthogonal pure states), then we have conditions for reusable partial swap homogenization.
  
            We also note that the bounds we derived on $\eta$ and $N$ in this Chapter assumed a worst-case homogenization. Hence, we cannot directly use such bounds to compare the resource costs of more specific tasks, such as transforming pure states to mixed states and the opposite process. In addition, the asymmetry shown between pure and mixed state homogenization is in a different physical context to \cite{violaris2022irreversibility}, where the coupling strength $\eta$ is first fixed, and then it is considered how far the homogenization machines can be reused to perform a homogenization within a given error, while remaining close to their original state. 

\section{Summary}

    We proposed a model for a universal quantum homogenizer that does not have coherence between the system and reservoir qubits, based on a \textsc{cswap} operation instead of \textsc{pswap}. We computed an upper-bound on the difference between the reduced states of the system and reservoir qubits of the \textsc{cswap} homogenizer compared to the \textsc{pswap}, showing that it tends to zero in the limit of a large reservoir, and simulated an example where the homogenization protocols are equivalent. Then we derived a bound on the resources needed for an arbitrarily good \textsc{cswap} homogenization, showing that it satisfies the required convergence conditions for homogenization. Our result is more general than that previously derived for the \textsc{pswap}, showing the dependence of resources required on the distance between the initial system and reservoir qubit states. We also contrasted the \textsc{cswap} and \textsc{pswap} homogenizers in terms of the von Neumann entropy of the joint system-reservoir qubits. 

    Then we analysed how far the coherent and incoherent homogenizers can be re-used to perform state transformations, deriving constraints on the resources needed to repeatedly perform imperfect homogenizations. Our analysis also suggests that the demonstrations of a new kind of irreversibility based on homogenization machines in Chapters \ref{CT-irreversibility} and \ref{erasure_model} can be generalised to incoherent models for thermalization and information erasure. 

    Future work could investigate connections between the general bounds on repeated homogenizations found here with approaches to modelling catalysts in quantum resource theories, building on the connections discussed in Chapter \ref{resource_comparison}. Another interesting avenue is to investigate in more detail how entanglement builds up in the two homogenizers, building on recently-proposed approaches to describe quantum correlations in collision models (e.g. \cite{sergey2022entanglement}). Entanglement may be distributed differently in the coherent and incoherent homogenizers, despite the negligible differences in the ultimate convergence properties.  % Experimental Implementation of the Quantum Homogenizer using NMR
\chapter{Experimental Implementation of the Quantum Homogenizer using NMR}\label{chapter:NMR}

\textit{The contents of this Chapter are based on the publication \cite{violaris2021transforming}, done in collaboration with Gaurav Bhole, Jonathan Jones, Chiara Marletto and Vlatko Vedral. Gaurav Bhole and Jonathan Jones performed the experiment.} \\

\section{Introduction}

Here we present an experimental implementation of the quantum homogenizer. We use a four-qubit NMR system to compare the homogenization of a pure state to a mixed state with the opposite process. After accounting for the effects of decoherence in the system, we find the experimental results to be consistent with the theoretical symmetry in how the qubit states evolve in the two cases. We analyse the implications of this symmetry by interpreting the homogenizer as a physical implementation of pure state preparation and a mechanism for spreading information. We also explain how the entropy changes in quantum homogenization are consistent with unitary quantum theory.

\begin{figure*}[tb]
\centering
\includegraphics[width=0.7\linewidth]{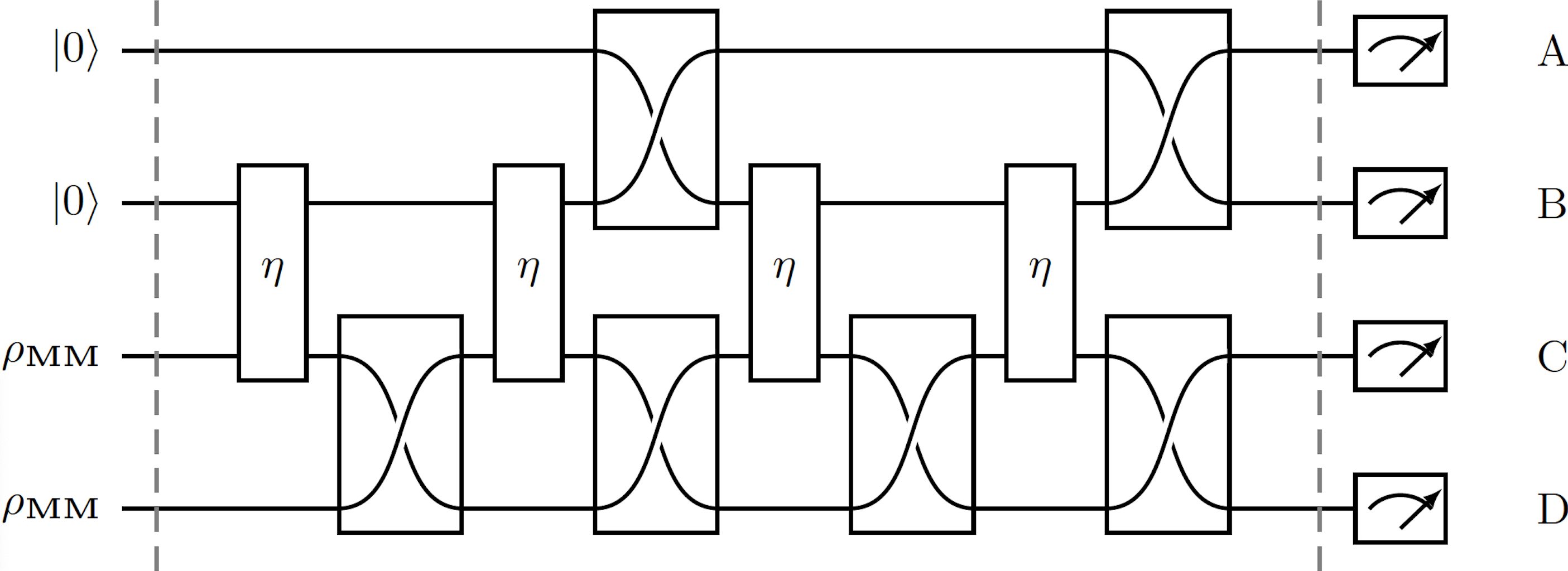}
\caption{A quantum homogenizer with two system qubits and two reservoir qubits designed to operate on a linear chain system where two qubit gates are possible only between adjacent qubits. homogenization is achieved using the partial swap gates, labelled as $\eta$, which connect the two middle qubits, while full \textsc{swap} gates are used to rotate the system and reservoir registers to bring other qubit pairs into contact.}\label{nmr-circuit}
\end{figure*}

\section{Experimental Simulation}

The quantum circuit used for the simulation is shown in figure \ref{nmr-circuit}. The two system qubits are initialised in the pure state $\ket{0}$, and the two reservoir qubits are initialised in the maximally mixed state $\rho_{\rm MM}$. The circuit is designed to be implemented on a linear chain where nearest-neighbour two-qubit gates are enabled between qubits. By using full \textsc{swap} gates, it is possible to achieve indirect contact between any pair of qubits. The full experimental details of the simulation using NMR can be found in the associated paper, \cite{violaris2021transforming}. Here we focus on the significance of the protocol that is implemented and the theoretical interpretations of the results. 

The final state of the four qubits is in general an entangled state. However the readout stage in the experiment implicitly involves a partial trace over the other three qubits, resulting in four single-qubit states which all lie on the z-axis of their Bloch spheres. Therefore their states can be characterised completely through z-basis measurements. The state of each qubit can be written as:
\begin{equation} \label{general state}
\rho=\frac{\mathbbm{1}}{2}+f(\eta)\times\frac{\sigma_z}{2}
\end{equation}
with 
\begin{equation}\label{fZ}
 f={\rm tr}(\rho\sigma_z)   
\end{equation}
lying in the range $\pm 1$. This value corresponds to the difference between the probabilities of finding $\ket{0}$ or $\ket{1}$ when measuring a qubit in the computational basis. The value of $f$ must lie between 0 and 1, and for our implementation, we find for the individual qubits:
\begin{equation} \label{form B}
f_B=\cos^4(\eta)
\end{equation}
and $f_C=1-f_B$. Therefore $f_B$ decreases smoothly from 1 to 0, whereas $f_C$ increases in the opposite way to qubit $B$. 

Similarly
\begin{equation} \label{form A}
f_A=4\cos^2(\eta)-9\cos^4(\eta)+8\cos^6(\eta)-2\cos^8(\eta)
\end{equation}
also decreases from 1 to 0, but following a more complex path. Finally, $f_D=1-f_A$, so it again increases in the opposite way to qubit $A$.

The homogenization becomes more efficient as the coupling strength $\eta$ increases. Hence, when $\eta=\pi/2$, circuits with identical numbers of system and reservoir qubits become symmetrical in that the system and reservoir qubits can be interchanged. In this case, the homogenization can be viewed either as a process that randomizes the pure qubits or causes the mixed qubits to become polarised (as they become pure). 

\begin{figure}[tb]
\centering
\includegraphics[width=85mm]{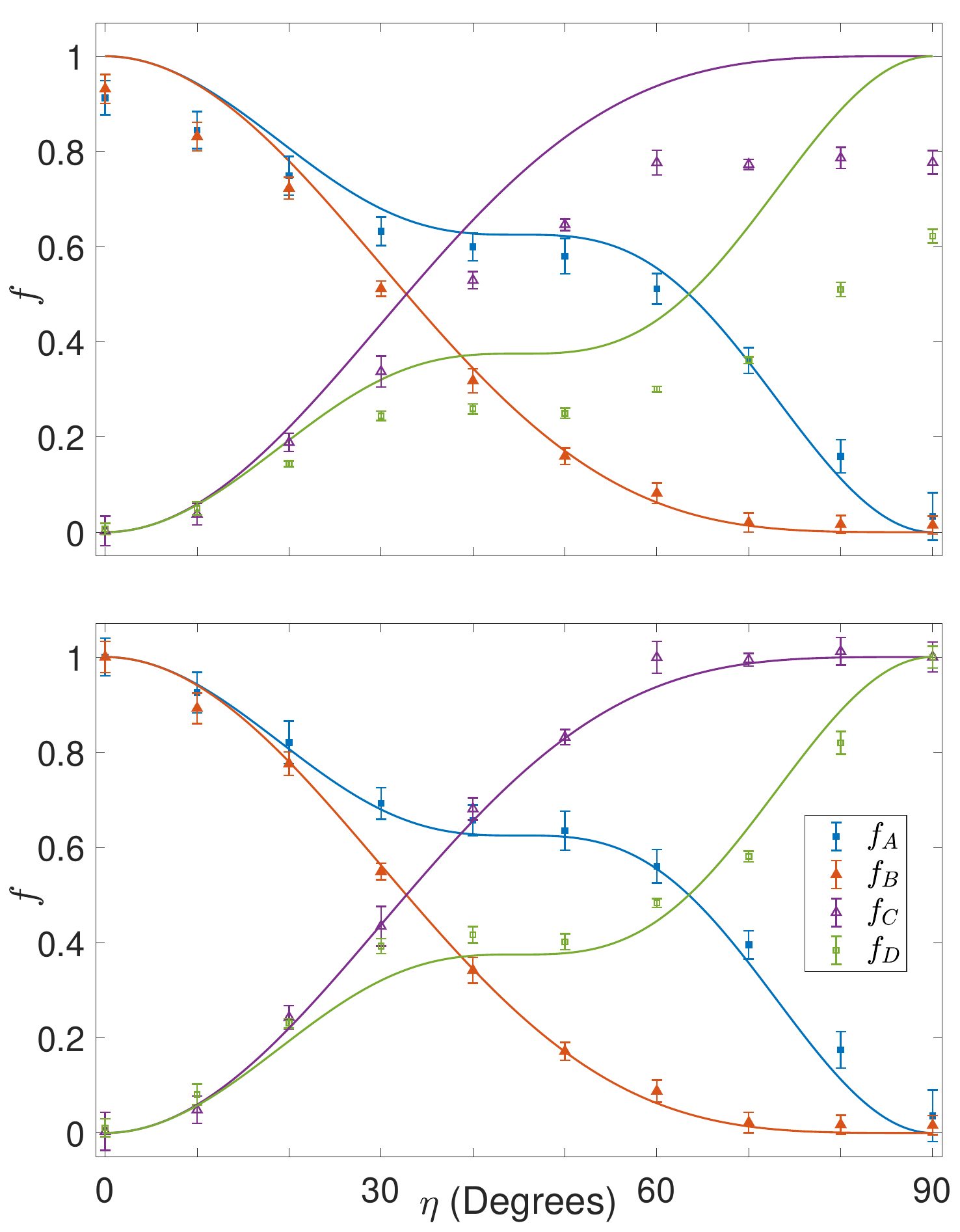}
\caption{Experimental results and theoretical predictions for the quantum homogenizer. The data points show the results from ten repetitions of the experiment, while solid lines show theoretical calculations using equations \ref{general state}--\ref{form A}. In the upper image, signal strengths were normalised against the average intensity from spins A and B in the initial state.  In the lower image, each spin's signal is normalised against a reference signal chosen for that spin.}
\label{experiment}
\end{figure}

The experimental results are shown in figure \ref{experiment}. Each experiment was repeated ten times, and the standard deviation around the mean is indicated by the error bars. The upper and lower panels show different normalisations of the data. In the upper image, the data is systematically lower than theoretical predictions due to experimental decoherence, which leads to the signal being lost. This has a greater effect for mixed states being homogenized to pure states than the opposite direction, since the decoherence acts towards making states more mixed. In the lower image, the data has been normalised to remove the effects of loss of signal. This approximately restores the symmetry that is expected between the homogenizations. 

\section{Interpretations}

The experimental data show the convergence of the system qubit to the state of the reservoir qubits. This has been tested for the limiting cases of transforming a qubit from a mixed state to a pure state and from a pure to a mixed state, demonstrating that the homogenization is effective regardless of the initial states of the system or reservoir qubits. After accounting for the bias towards mixed states caused by decoherence, the experimental results are consistent with the theoretical symmetry in the evolution of how the states evolve, where the states for the pure-to-mixed homogenization vary inversely compared to the states for the mixed-to-pure homogenization.\\

\noindent \textbf{Pure state preparation}\\

\begin{figure}[tb]
\centering
\includegraphics[width=85mm]{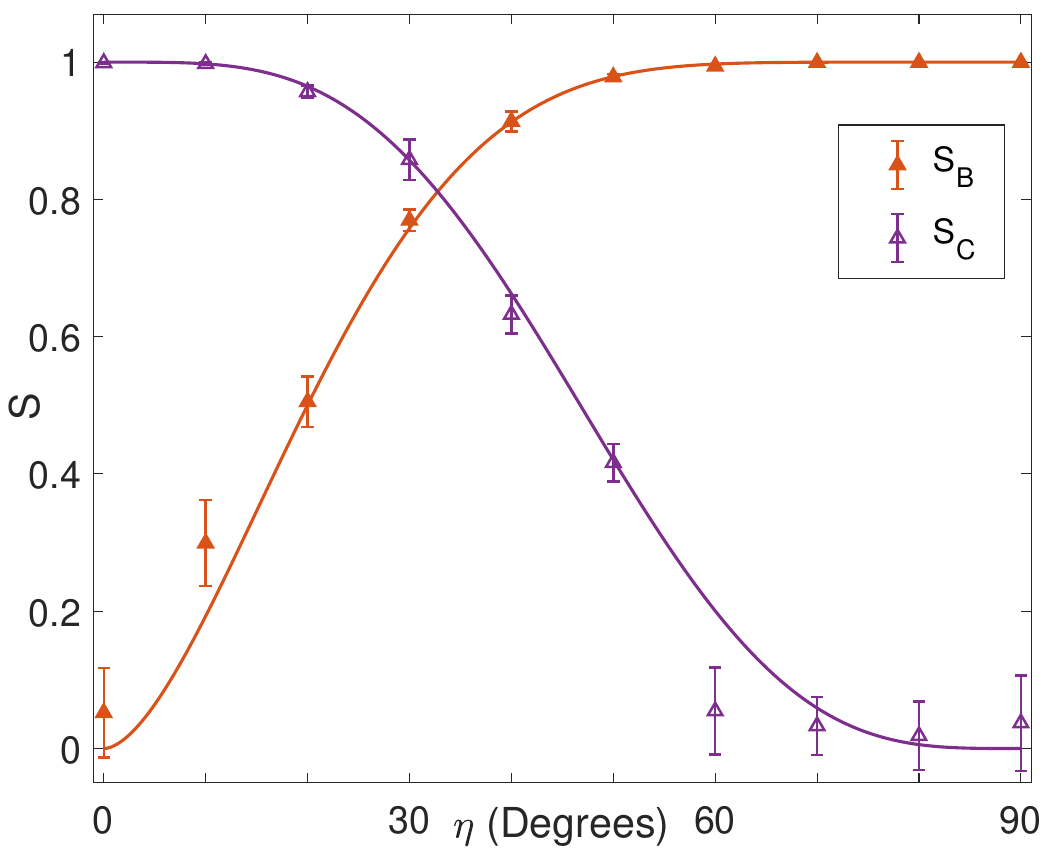}
\caption{Von Neumann entropies of qubits B and C against coupling strength, calculated from the experimental results and plotted against the theoretical predictions.}
\label{VN_B_and_C}
\end{figure}

\noindent Figure \ref{VN_B_and_C} plots the von Neumann entropy:
\begin{equation}
S=-\left(\frac{1+f}{2}\right)\log_2 \left(\frac{1+f}{2}\right)-\left(\frac{1-f}{2}\right)\log_2 \left(\frac{1-f}{2}\right)
\end{equation}
of the theoretical and experimental qubit states against coupling strength, where $f$ was defined in equation \ref{general state}. Theoretical curves were calculated using equation \ref{form B} for $f_B$ and $f_C=1-f_B$. The qubit B is the system qubit for the pure-to-mixed homogenization, while C is the system qubit for the mixed-to-pure homogenization. As expected, the qubit being transformed from a mixed to a pure state decreases in entropy, with the effect being strongest for strong coupling, while the qubit being transformed from a pure to a mixed state increases in entropy. Since all the interactions are unitary, the total von Neumann entropy of the combined homogenizer and system must remain constant. Hence it can be deduced that the entropy decrease of the mixed-to-pure system qubit C must be accompanied by an increase in the entropy of the homogenizer (qubits A and B), which is the irreducible entropic cost associated with preparing a pure state.\\

\noindent \textbf{Information spreading}\\

\begin{figure}[tb]
\centering
\includegraphics[width=85mm]{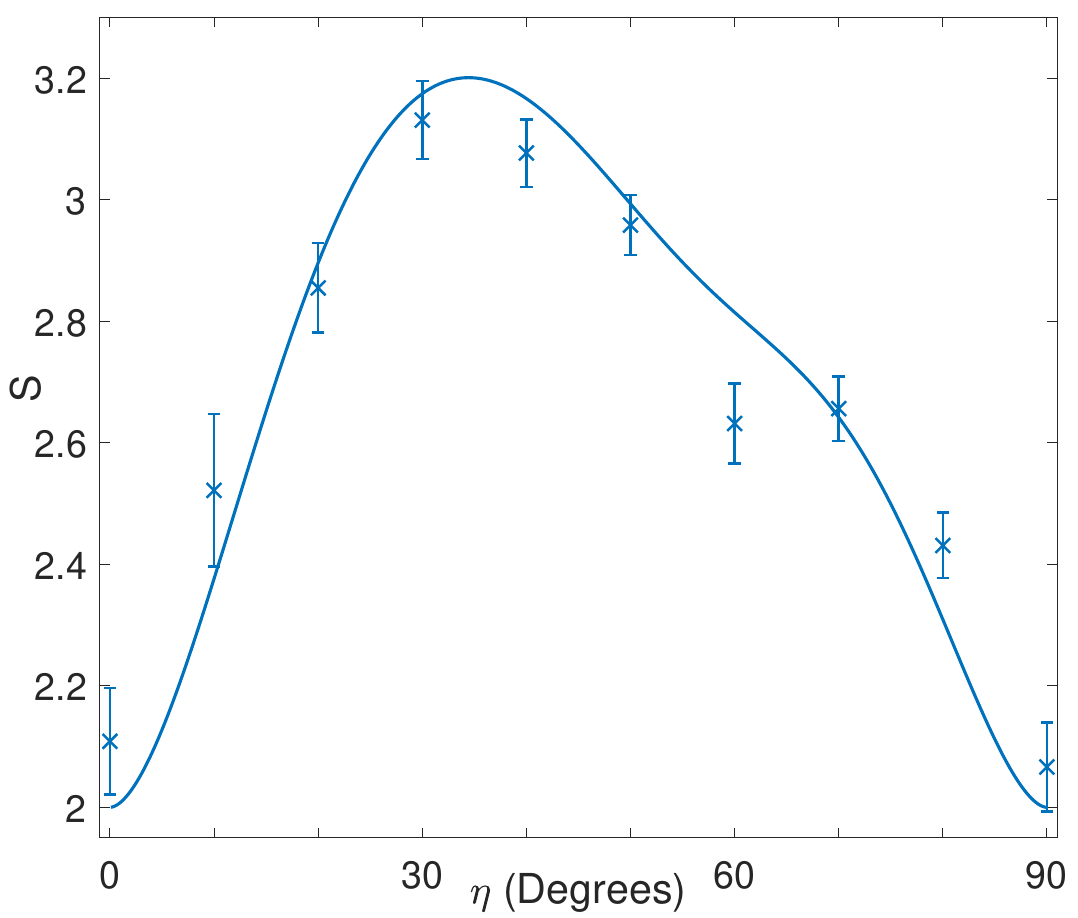}
\caption{Total von Neumann entropies of the four individual qubits, calculated from the experimental results and plotted against the theoretical predictions.}
\label{Total VN}
\end{figure}

\noindent The initial pure or mixed state of a system qubit becomes indistinguishable from the original state of the homogenizer qubits. In this sense, the homogenization protocol has similar features to the interesting phenomenon of quantum information scrambling, by spreading out information in entanglement in such a way that is not retrievable from local measurements on subsystems \cite{touil2024information}. As explained in \cite{ziman_quantum_2001}, the information about the system's initial state becomes hidden in mutual correlations between the homogenizer qubits. If there were no mutual correlations, one would expect the sum of the von Neumann entropies of the four qubits to equal two for all coupling strengths. The actual sum of the von Neumann entropies is in figure \ref{Total VN}. While this is two for the cases of an identity or a SWAP operation, for intermediate coupling strengths it is larger. This indicates that the negative contribution to von Neumann entropy from mutual correlations has been unaccounted for, which is due to considering only reduced density operators to describe the qubit states. \\

\noindent \textbf{Reusability}\\

\noindent The results in \cite{ziman_quantum_2001} show that for the regime of weak coupling and a large reservoir, the reservoir qubits remain almost unchanged by the interaction. We can therefore hypothesise that the same homogenizer could be reused for a second time (or more) to successfully homogenize more system qubits, relating to the theoretical models discussed in Chapters \ref{CT-irreversibility} and \ref{erasure_model}. Whilst the system we have tested here only has a small homogenizer, we can explore the hypothesis by comparing the states of the four qubits in the weak coupling regime.

Qubit A can be interpreted as interacting with a homogenizer that has already been ``used" once, to homogenize B. Similarly, qubit D can be interpreted as being homogenized by a homogenizer that has already been used to homogenize C. These second system qubit states are closely aligned with the first system qubit counterparts for weak coupling, and diverge for strong coupling, in figure \ref{experiment}. This indicates that the homogenizer is minimally changed from its original state in the weak coupling regime, allowing it to perform just as effective a homogenization on the second system qubit as it did on the first. Hence, the homogenizer can be reused to some extent to give the same incremental changes in system qubit state ($i.e.$ the qubit being homogenized), in the weak coupling regime. 

\section{Summary}

We have performed an NMR demonstration of the quantum homogenizer, using the partial swap on a set of four qubits. This demonstrates the principle behind a machine that can perform processes such as information erasure and the preparation of pure states using entirely unitary interactions. The experiments show the homogenization of a pure state and of a mixed state. The asymmetry in the evolution of the states with coupling strength can be explained by the decoherence within the experiment. After accounting for this, the results are consistent with the theoretical symmetric evolution of the pure and mixed qubit states.

Our experiment was limited to showing the principles behind a homogenization machine with a small number of NMR qubits.  Expanding the experimental demonstration of the homogenizer to different regimes and quantum technologies could give an additional insight into the fundamental limits to homogenising pure and mixed states by physical machines. % The Impossibility of Universal Work Extractors from Non-Orthogonal Coherent States
\chapter{The Impossibility of Universal Work Extractors from Coherence}\label{Chapter:work}

\textit{The contents of this Chapter are based on part of the preprint publication \cite{plesnik2024impossibility}, done in collaboration with Samuel Plesnik.} \\

\section{Introduction}

We compare how the impossibility of a universal work extractor from coherence arises from different approaches to quantum thermodynamics: an explicit protocol accounting for all relevant quantum resources (introduced in section \ref{background:resource}), and axiomatic, information-theoretic constraints imposed by constructor theory (introduced in section \ref{Intro:CT-thermo}). We first explain how the impossibility of a universal work extractor from coherence is directly implied by the constructor-theoretic theorem based on distinguishability, which is scale- and dynamics- independent. Then we give an explicit demonstration of this result within quantum theory, by proving the impossibility of generalising the proposed quantum protocol for deterministically extracting work from coherence. We demonstrate a new connection between the impossibility of universal work extractors and constructor-based irreversibility, which was shown using the quantum homogenizer in Chapter \ref{erasure_model}. Finally we discuss additional avenues for applying the constructor-theoretic formulation of work extraction to quantum thermodynamics, including the irreversibility of quantum computation and thermodynamics of multiple conserved quantities. 

Specifically, we investigate a case-study to show how the resource theory and constructor theory approaches reach intersecting conclusions: the impossibility of a universal work-extractor from coherence. By analysing a proposed protocol for extracting work from coherence \cite{korzekwa2016extraction}, we demonstrate within quantum mechanics that there cannot be a generalisation of the protocol to achieve a universal work extractor. This forbids deterministic work extraction of different amounts of work from any pair of non-orthogonal input states, including coherent and incoherent states of the same energy. We explain how this constraint on work extraction directly follows from a recent dynamics-independent theorem derived in constructor theory \cite{marletto2022information}. Our results show the information-theoretic origins of the impossibility of a universal work extractor from coherence, for quantum theory and its potential successors. During our analysis, we explain how the phenomenon of ``work-locking", whereby the additional free energy in a coherent state is not accessible, emerges as a consequence of constructor-theoretic distinguishability principles being applied to quantum theory. 

Furthermore, we find a new connection between the impossibility of universal work extractors and the recent constructor-theoretic analysis of the quantum homogenizer. In Chapter \ref{erasure_model}, we used the quantum homogenizer to demonstrate constructor-based irreversibility, which is a novel, exact form of irreversibility consistent with unitary quantum dynamics \cite{violaris2022irreversibility, marletto2022emergence}. Here we show that the constructor-theoretic impossibility of a universal work extractor in quantum theory relies on the constructor-theoretic impossibility of transforming a qubit from a mixed state to a pure state, which has been shown to hold when the task is implemented by the quantum homogenizer.

We also comment on additional promising avenues for connecting constructor-theoretic results with other approaches in quantum thermodynamics. This could help generalise existing results in quantum thermodynamics to potential successor theories of quantum mechanics, and elucidate the principles from which these phenomena originate. We conjecture an information-theoretic foundation for the additional entropy dissipation that makes quantum computation fundamentally irreversible \cite{bedingham2016thermodynamic}, and suggest how constructor-theoretic results could be extended to multiple conserved quantities. This could lead to new connections between conservation laws and thermodynamic-type laws. 

\section{Constructor-theoretic impossibility of work from coherence}

\begin{figure}[t]
\centering
\includegraphics[width=\linewidth]{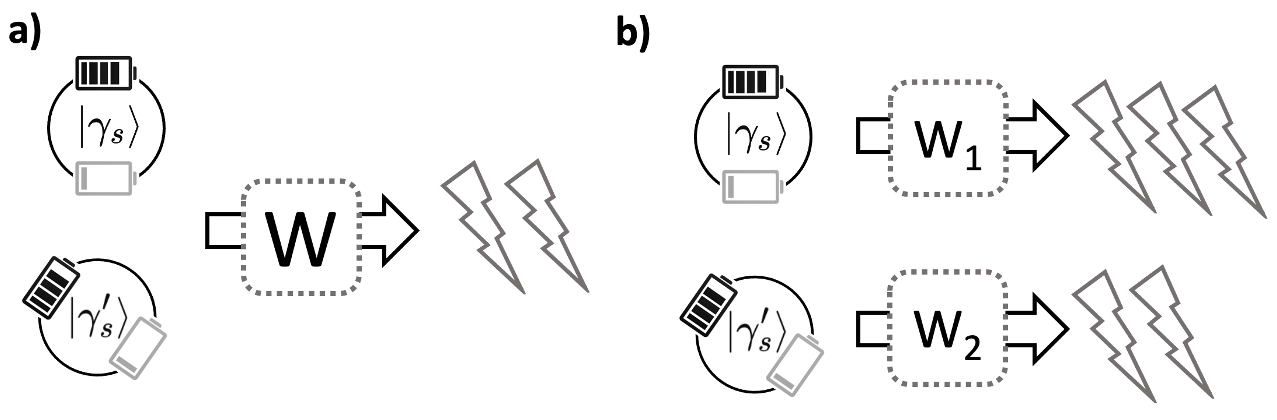}
\caption{Consider deterministic work extraction from non-orthogonal coherent states, $\ket{\gamma_s}$ and $\ket{\gamma'_s}$. a) A given machine could extract the same amount of work from both inputs. b) Extracting different amounts of work requires special-purpose machines.}
\label{work_coherence}
\end{figure}

In section \ref{Intro:CT-thermo}, we discussed how distinguishability is formulated in constructor theory. When applied to quantum theory, theorem \ref{theorem:distinguish} means that it is possible to deterministically extract different amounts of work from two quantum states only if they are orthogonal. If two quantum states are non-orthogonal, then it could be possible to deterministically extract the same amount of work from the two states. However, the theorem rules out having a universal deterministic work extractor able to extract different amounts of work from unknown, non-orthogonal states, visualised in figure \ref{work_coherence}. This theorem has interesting consequences for (in)coherent inputs, since a coherent state and incoherent state of the same energy are in general non-orthogonal, and so not perfectly distinguishable in a single shot. Hence, the theorem forbids a work extractor that can deterministically extract more work from a coherent state than the corresponding incoherent state of the same energy, despite the former having greater free energy. 

Specifically, consider a system with two energy eigenstates, $\ket{0}$ and $\ket{1}$. The system can be in a coherent state, $\ket{\psi} = \alpha \ket{0} + \beta \ket{1}$, with $\rho_{\text{coherent}} = \left( \begin{smallmatrix} |\alpha|^2 & \alpha\beta^* \\ \beta\alpha^* & |\beta|^2 \end{smallmatrix} \right)$ or the system can be in an incoherent state of the same energy, with $\rho_{\text{incoherent}} = \left( \begin{smallmatrix} |\alpha|^2 & 0 \\ 0 & |\beta|^2 \end{smallmatrix} \right)$.\\

Let's denote the task of extracting a quantity of work $w$ from a quantum state $\rho$ as $\mathbf{T_w}( [\rho, w])$. Assuming the task is possible, let's denote a constructor capable of causing the task as $\mathbf{C_w}([\rho, w])$. The task of deterministically extracting amounts of work $w_1$, $w_2$ from a pair of states $\rho_1$, $\rho_2$ respectively is $\mathbf{T_w}([\rho_1, w_1], [\rho_2, w_2])$. A consequence of the constructor-theoretic theorem is that $\mathbf{T_w}$ is possible ($\mathbf{T_w^{\checkmark}}$) if and only if $\rho_1$ and $\rho_2$ are perfectly distinguishable with a single shot measurement, i.e. are orthogonal quantum states. These orthogonal states could be energy eigenstates, or a pair of orthogonal states coherent in the energy basis. The theorem therefore does not permit a constructor $\mathbf{C_w}([\rho_1, w_1], [\rho_2, w_2]))$ for extracting different amounts of work from non-orthogonal states. This includes both arbitrary pairs of non-orthogonal coherent states, and pairs of coherent  and incoherent states of the same energy. Consequently, there cannot be a universal work extractor capable of distinguishing between and extracting different amounts of work from such states, highlighting a fundamental limitation in the physics of work extraction from quantum coherence. 

We demonstrate this explicitly using the coherent work extraction protocol proposed in \cite{korzekwa2016extraction}. We focus on the pre-processing step, where knowledge of the input state is important, and show how this fails to generalise for sets of non-orthogonal inputs. This demonstrates the physical significance of prior knowledge of input states for constraining extractable work, which has been explored in various quantum settings (e.g. \cite{vsafranek2023work}).

\section{Impossibility of universal quantum work from coherence}

\subsection{Connections between constructor-theoretic distinguishability and work-locking}

In  section \ref{background:resource} we discussed the phenomenon of work-locking. We now consider how work-locking emerges as a natural consequence of the constructor-theoretic theorem \ref{theorem:distinguish}, including limits on its domain of applicability. Work-locking refers to the phenomenon where without access to an external source of coherence, the same work is extractable from both a coherent and incoherent state of the same energy. The additional free energy of the coherent state cannot be converted into work. For the case of deterministic work extraction, a related phenomenon is implied more generally by theorem \ref{theorem:distinguish}: if different amounts of work were deterministically extractable from coherent and incoherent inputs, this would violate the theorem, as the work-extractor would act as a perfect distinguisher of inputs that are not perfectly distinguishable. 

However, the constraint of work-locking can be evaded to extract work that harnesses the free energy of the coherent state, in a regime where an asymptotic number of systems are processed (shown in section IV. b) ``Extracting work arbitrarily close to the free energy difference" in \cite{korzekwa2016extraction}). Theorem \ref{theorem:distinguish} is based on the impossibility of distinguishing states with a single shot, but in the limit of processing an asymptotic number of systems, any quantum states become perfectly distinguishable. This is the principle behind quantum tomography, expressed more generally as the principle of asymptotic distinguishability in \cite{marletto2022information}. Hence, the fact that the full free energy of the coherent state can be extracted when an asymptotic number of system states are input into the work extractor intersects with the domain of applicability of the constructor-theoretic theorem. In the next section, we discuss the more subtle situation whereby systems are processed in a single shot (not in the asymptotic limit), and external catalytic resources are allowed. 

\subsection{Proof protocol is not universal}

The step in the work-from-coherence protocol where knowledge of the system's state is significant is the pre-processing step and application of the rotation $U$ (eq. \ref{eq: U}). The transformation $U$ is required to rotate the state $\ket{\psi}$ to $\ket{1}$, enabling extraction of maximal work \cite{kammerlander2016coherence}. The protocol uses an initial state coherent in the energy eigenbasis, where $\ket{0}$ has lower energy than $\ket{1}$. The ability to rotate the state from which work is to be extracted to the higher energy $\ket{1}$ state is crucial to extract the optimal amount of work. 

\begin{figure*}[ht]
    \centering
    \includegraphics[width=5.9in]{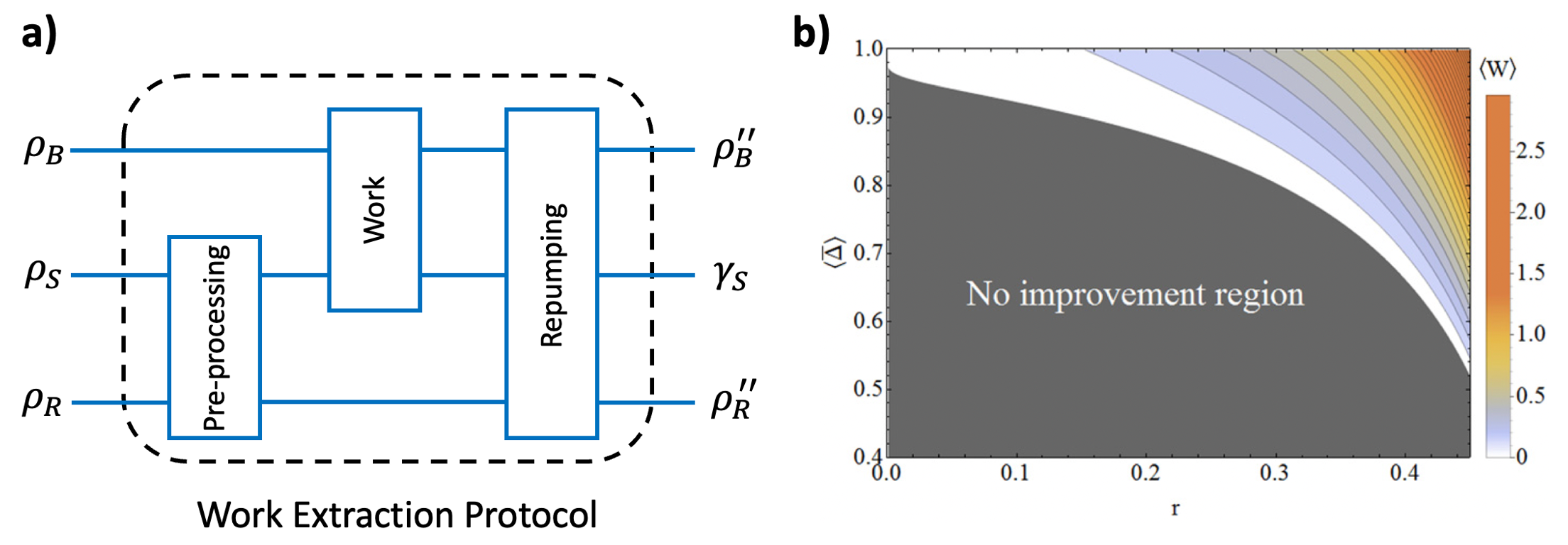}
    \caption{a) Quantum circuit diagram schematic of the work extraction protocol from coherence, where $\rho_B$ and $\rho_B^{\prime \prime}$ are initial states of the battery, $\rho_S$ and $\gamma_S$ are initial coherent state and final Gibbs thermal state and $\rho_R$ is the reference. b) Figure reproduced from \cite{korzekwa2016extraction}, figure 4. A phase diagram showing regions where the coherent work extraction protocol provides improvement and is capable of extracting more work from a coherent state than from an incoherent state, where $r$ is the output Gibbs state coherent state parameterisation and $\langle \bar{\Delta} \rangle$ reference quality parameter.}
    \label{fig:korzewaPaper2} (Note this figure is the same as figure \ref{fig:korzewaPaper1}, and is replicated here for clarity.)
\end{figure*}

This is because the quality of the resulting reference state in the protocol affects the amount of work that can be extracted, as the reference quality is defined by the distance of the rotated state from $\ket{1}$. It was shown in \cite{korzekwa2016extraction} that for each coherent Gibbs state parametrised by $r$ (equation \ref{eq:coherentGibbsState}), there exists a minimal quality of the reference state that ensures an advantage in extracting work from a coherent state compared to its incoherent counterpart. In the single-shot case, this advantage is a smaller probability of failure. The failure rate can be made arbitrarily small as the quality of the reference state is improved, tending towards deterministic work extraction. The phase diagram in figure \ref{fig:korzewaPaper2} b) shows the state and reference quality combinations that result in improved work extraction by the protocol. Therefore, mapping the initial state to $\ket{1}$ in the energy eigenbasis is a crucial step in the coherent work extraction protocol, to ensure that the reference state will fall in a region where an improvement in work extracted from the coherent state can be seen. \\

\textbf{Pure, distinct and non-orthogonal input states.} Let's first consider attempting to generalise the protocol so that it can take as input two non-orthogonal, pure coherent states. It is clearly impossible to choose $U$ such that $U\ket{\psi} = \ket{1}$ for two distinct inputs $\ket{\psi}$, as there will be an irreducible distance between their output states. A key property of fidelity $F$ (distance between states) is its invariance under unitary transformations on the state space \cite{jozsa1994fidelity}. For example, the fidelity between $U\ket{0}$ and $U\ket{+}$ is $\frac{1}{\sqrt{2}}$ for any $U$. More generally, for two pure states, consider a unitary transformation $U$ such that $U\ket{\psi_1} = \ket{1}$ and $U\ket{\psi_2}=\ket{1}$ exists for arbitrary two states $\ket{\psi_1}, \ket{\psi_2}$ where $\ket{\psi_1} \neq \ket{\psi_2}$. Then we can write: $1 = \braket{1 | 1} = \bra{\psi_1} U^\dagger U \ket{\psi_2} = \braket{\psi_1 | \psi_2}$. Since $\braket{\psi_1 | \psi_2} = 1$ is true only if $\ket{\psi_1} = \ket{\psi_2}$, this contradicts our initial assumption, so this $U$ cannot exist. \\

\textbf{Orthogonal input states.} A key exception to this is for perfectly distinguishable states. If the two inputs are deterministically, single-shot distinguishable, e.g. $\ket{0}$ and $\ket{1}$, the fidelity of $U\ket{0}$ and $U\ket{1}$ will be 0. However, a measurement can be done in the basis in which the inputs are perfectly distinguishable, and then the appropriate $U$ applied conditionally to the input, depending on its state. \\

\textbf{Allowing an error tolerance.} The impossibility of generalising the protocol is not quite so clear cut, since there is some error tolerance built in: $V(U)$ only needs to approximately implement $U$ on $\ket{\psi}$ to gain an advantage in work extraction from coherence, giving a final system state:

\begin{equation}
    \rho_S^\prime \approx U (\ket{\psi}\bra{\psi})U^\dagger = \ket{1}\bra{1}.
\end{equation}

This means that there is some tolerance for the protocol to work for multiple, non-orthogonal input states. The input states must be close enough that $V(U)$ approximately maps the system to $\ket{1}\bra{1}$, with the accuracy of the approximation lower bounded by the distance from $\ket{1}\bra{1}$ admissible for work to be extracted from coherence. Specifically, the effectiveness of applying $V(U)$ is quantified by a parameter $q$ (see equation 11 in \cite{korzekwa2016extraction}), with $q = 1$ indicating an ideal channel mapping the system to $\ket{1}\bra{1}$. If $U\ket{\psi} = \ket{1}$, then $q$ can be made arbitrarily close to 1.  If $U\ket{\psi} = \epsilon \ket{0} + \sqrt{1-\epsilon}\ket{1}$ for some small $\epsilon$, then the final system state $\rho'_s$ has an $O(\epsilon)$ correction, leading to $q$ having an $O(\epsilon)$ correction. So, the protocol can be applied for multiple non-orthogonal inputs for the constrained set of states where coherent work extraction is possible within an $O(\epsilon)$ approximation.  

By definition, a universal machine must work for any input state. There will always be possible input states that have a large distance between them (e.g. those that are close to orthogonal). Yet, non-orthogonal inputs cannot be perfectly distinguished in a single-shot measurement, meaning that the unitary cannot be made conditional on the input state; it must operate with no specific knowledge of the input state being retrieved. The step of the protocol where $V(U)$ must approximately map any input $\ket{\psi}$ to a state close to $\ket{1}$ will have a fundamentally restricted set of input states, forbidding universality of such a machine. Therefore, this is the specific part of the work extraction protocol that embodies the constraint introduced by the quantum instance of the constructor-theoretic theorem, where $\mathbf{T_w}([\rho_1, w_1], [\rho_2, w_2])$ is possible only if $\rho_1$ and $\rho_2$ are perfectly distinguishable. \\

\textbf{Mixed states as input.} We considered the impossibility of unitarily mapping two arbitrary pure states to a region within the $\ket{1}\bra{1}$ state required for the deterministic work extraction from coherence. However, to deduce the possibility of a deterministic work extractor that can take in both a coherent pure state and an incoherent thermal state of the same energy as input, we must consider the effect of inputting a mixed state to the protocol. In general, two arbitrary, distinct input mixed states $\rho_1$, $\rho_2$ cannot be unitarily transformed to $U\rho_1U^\dagger = U\rho_2U^\dagger = \ket{1}\bra{1}$, again since $1 = F(U\rho_1U^\dagger, U\rho_2U^\dagger) = F(\rho_1, \rho_2) \rightarrow \rho_1 = \rho_2$ , contradicting the assumption of distinct $\rho_1$ and $\rho_2$.

However, there is a more pertinent problem with attempting to use the pre-processing step on the incoherent state corresponding to a coherent state of the same energy: it involves mapping a mixed state to a pure state. This cannot be done unitarily, since it involves an irreducible change in von Neumann entropy. We now consider approximating the mapping of mixed states to pure states using external resources. 

\subsection{Role of catalysis}

We have shown that $V(U)$ cannot directly implement a universal channel for mapping arbitrary input systems to the set of states close to $\ket{1}\bra{1}$. However, there are many quantum channels that become available by the use of catalysts or approximate catalysts, as demonstrated in quantum thermodynamics and quantum resource theories \cite{ng2015limits, lipka2021all}. In particular, various channels are made approximately possible with the introduction of catalysts. 

As noted in Chapter \ref{resource_comparison}, certain aspects of catalysis are generalised by that of a constructor. We will now draw upon recent results that demonstrate the constructor-theoretic impossibility of transforming mixed states to pure states ($i.e.$ impossibility of performing the transformation in a cycle), to show that the coherent work extraction protocol cannot be made universal even when taking advantage of a candidate constructor for the task. 

\section{Constructor-theoretic impossibility of generalising quantum protocol}

As discussed in Chapter \ref{erasure_model}, a candidate constructor for enabling the approximate transformation of a mixed state to a pure state is the universal quantum homogenizer. There it was shown that the task of transforming a qubit from a maximally mixed to a pure state via the quantum homogenizer is constructor-theoretically impossible, meaning that it deteriorates too rapidly with repeated use to be a constructor for the task \cite{violaris2021transforming,violaris2022irreversibility}. This has direct implications for the constructor-theoretic impossibility of performing the pre-processing step of the work extraction protocol for a universal input. If the input is an arbitrary, unknown pure state, then it is equivalent to a maximally mixed state, and cannot be reliably mapped to the pure $\ket{1}\bra{1}$ state via the quantum homogenizer. Similarly, if the input is an incoherent state other than $\ket{0}$ and $\ket{1}$ (which is always the case in the context of distinct coherent and incoherent states that have the same energy), then it is mixed, and again is constructor-theoretically impossible to map to $\ket{1}\bra{1}$ using the quantum homogenizer. Hence, the coherent work extraction protocol in \cite{korzekwa2016extraction} is impossible to generalise to a universal machine, even given access to powerful quantum homogenization machines.  Specifically, in the protocol for extracting work from coherence, we need to transform a pure state with coherence in the computational basis (equation \ref{eq:CoherenceInputstate}) arbitrarily close to $\ket{1}$, also a pure state. \\

\noindent \textbf{Pure, distinct and non-orthogonal input states}\\

\noindent Assume that some operation can transform an arbitrary pure state, into a fixed pure state, let's say $\ket{1}$. Since the operation must work for any pure input state, without prior knowledge of which state it is, the input is equivalent to a maximally mixed state. For example, if the input could be either $\ket{0}$ or $\ket{1}$, with no prior information about which of those states it is, then we describe the input by the mixture $\rho = \frac{1}{2}(\ket{0}\bra{0}+ \ket{1}\bra{1})$. By extension, if the input could be any single-qubit pure state, with no information about which state it is, it is also described by the maximally mixed state. 

Hence, the possibility of mapping any pure state to a fixed pure state reduces to the possibility of transforming a maximally mixed state to a fixed pure state, i.e. the task $\tilde{T} = \{ \text{\textit{mixed state}} \rightarrow \text{\textit{pure state}} \}$. However, it has been shown that the quantum homogenizer is not a constructor for this task, meaning that the task is not constructor-theoretically possible using the quantum homogenizer. 

Therefore universal work extraction from coherence cannot be reliably implemented with any known constructor, and is ruled out by the conjecture that the mixed-to-pure task is constructor-theoretically impossible. A universal work extractor from coherence using this protocol is constructor-theoretically possible if and only if the task $\tilde{T} = \{ \text{\textit{mixed state}} \rightarrow \text{\textit{pure state}} \}$ is also possible. 

As the impossibility is conjectured to hold for any task transforming a mixed state to a less mixed state, the impossibility also applies for a mapping of an arbitrary pure state to a state within the vicinity of a fixed pure state (which has an equivalent description as a non-maximally mixed state). \\

\noindent \textbf{Incoherent state}\\

\noindent When a coherent and incoherent state are distinct and have the same mean energy, the incoherent state is mixed ($i.e.$ when the incoherent state is not $\ket{0}$ or $\ket{1}$). Therefore, using the quantum homogenizer to map the state to within the vicinity of the pure state $\ket{1}$ is also a constructor-theoretically impossible task. Hence, the coherent work extraction protocol in \cite{korzekwa2016extraction} is impossible to generalise to a universal machine even when given access to powerful quantum homogenization machines. 

\section{Possibility of extracting identical work from non-orthogonal inputs}

What if the same amount of work is extracted from both input states? Then, the possibility of deterministic work extraction is not ruled out by the constructor-theoretic theorem. Let's imagine that the work $w_1$ that can be extracted from $\rho_1$ is less than the work $w_2$ that can be extracted from $\rho_2$, where $\rho_1$ and $\rho_2$ are arbitrary states that may or may not be perfectly distinguishable (i.e. orthogonal). Then there may exist a constructor that is able to extract the same amount of work, $w_1$, from either of the two states, without prior knowledge of the states, i.e.: $\mathbf{T_w^{\checkmark}}([\rho_1, w_1], [\rho_2, w_1])$ for arbitrary $\rho_1$ and $\rho_2$, given that $\mathbf{T_w^\checkmark}( [\rho_1, w_1])$, $\mathbf{T_w^\checkmark}( [\rho_2, w_2])$, and $w_1 \leq w_2$. More generally, the task:
\begin{equation}
\bigcup_{i}\mathbf{T_w}( [\rho_i, w_j]) 
\end{equation}
could be possible if:

\begin{equation}
\forall i: \mathbf{T_w^{\checkmark}}( [\rho_i, w_i]) \textrm{  and  } w{_i} \geq w{_j}
\end{equation}

This result does not mean that even this minimum amount of identical work is necessarily possible to extract from all the states. Let's denote the smallest amount of optimal work that can be extracted from any individual state as $w_\textrm{smallest}$. A constructor for extracting $w_\textrm{smallest}$ from that state may not necessarily be able to extract $w_\textrm{smallest}$ from the other possible input states; the result does not forbid such a protocol existing, but neither does it require it to exist. It could be interesting to demonstrate these results explicitly with a quantum protocol in future work. 

\section{Further information-theoretic foundations of quantum thermodynamics}

We now discuss how constructor-theoretic work extraction and thermodynamics could extend to other settings and connect with results from resource theories. \\

\noindent \textbf{Irreversibility of non-classical computation}\\

\noindent We conjecture that the constraint on deterministic work extraction only being possible for distinguishable inputs demonstrates, and in some respects generalises, the ``irreversibility" of quantum computation. Here the term irreversibility is used differently from the constructor-based irreversibility referred to earlier in the paper, hence we will refer to the use of irreversibility here from now on as entropic-irreversibility. 

It was shown in \cite{bedingham2016thermodynamic} that general quantum operations have an additional irreducible entropy cost to the environment as compared to classical operations. In quantum mechanics, a general quantum state $\rho$ can be considered as a distribution of states. Extending the classical definition of Shannon entropy to the quantum case, we get the von Neumann entropy $S(\rho) = - \operatorname{tr}(\rho \log \rho)$. Then the entropy cost of the quantum operation $\rho \rightarrow \rho '$ is based on the difference in von Neumann entropy \cite{schumacher1995quantum}, $\Delta E \geq \textrm{kTln} 2 [S(\rho) - S(\rho ')]$.

This constraint is the quantum generalisation of Landauer's principle. However, when the input quantum system can be one of at least two non-orthogonal states, and each input has a unique output, there is in general an additional entropy dissipation to the environment. A computation which saturates the Landauer bound is ``reversible", in the sense that there is no limit to how closely a complete cycle could be approached where the overall entropy change is zero (i.e. the entropy gain in the environment is perfectly balanced by the entropy change of the system). By contrast, a process that has a minimum entropy change exceeding the Landauer bound is fundamentally ``irreversible", meaning that a complete cycle cannot approach zero entropy change. The additional irreducible entropy cost to the environment for quantum operations therefore makes quantum computation entropically-irreversible.  

Similarly, theorem \ref{theorem:distinguish} from \cite{marletto2022information} shows that there are unavoidable changes in the environment for the specific task of extracting different amounts of work from non-orthogonal states. However the theorem is dynamics-independent. Understanding the possible connections between these constraints on non-classical information processing could give a dynamics-independent generalisation of the entropic-irreversibility of quantum computation to a more general form of non-classical computation. Further work in this direction could therefore lead to information-theoretic foundations of entropic-irreversibility in quantum computing, and computing with the future potential successor theories of quantum mechanics. \\

\noindent \textbf{General conserved quantities}\\

\noindent Theorem \ref{theorem:distinguish} was derived for the simplifying assumption of a single conservation law, that of energy conservation, as is typical in thermodynamics. Development of resource theories of coherence \cite{winter2016operational,baumgratz2014quantifying}, purity \cite{horodecki2013quantumness} and entanglement \cite{chitambar2019quantum} has shown parallels between systems conserving energy and systems with other conserved quantities. There are now generalised formulations of 1$^{\textrm{st}}$ and 2$^{\textrm{nd}}$ laws of thermodynamics for systems with one or more conserved quantities, and it has been argued that energy has no special status for these laws in this respect \cite{guryanova2016thermodynamics, sparaciari2020first, lostaglio2017thermodynamic, yunger2016microcanonical}. 

Interestingly, the constructor-theoretic thermodynamics formulation lends itself to being extended for additional conserved quantities. When a quantity other than energy is conserved, the associated work medium could be defined in exactly the same way, but with its attributes having different values of the conserved quantity, rather than different values of energy. This generalised form of a work medium could then be used as a basis for formulating constructor-theoretic 2$^{\textrm{nd}}$ laws in terms of adiabatic possibility, and a theorem regarding deterministic work extraction relying on distinguishable input attributes, in precisely the same way as it was for energy. This would create a natural connection between the results regarding thermodynamics for different conserved quantities in quantum theory, and the constructor-theoretic formulation of thermodynamics.

We could further consider how constructor-theoretic work media are defined under multiple conserved quantities. The thermodynamic properties of a work medium should not be dependent on which one or combination of conserved quantities are used to define work. Hence, the work medium definition could be extended to account for multiple quantities being conserved simultaneously. This could give rise to a substrate-independent formulation of emergent entropy that neatly unifies entropy changes in different quantities under one 2$^{\textrm{nd}}$ law. Furthermore, since a constructor-theoretic 2$^{\textrm{nd}}$ law (based on adiabatic possibility) relies on a conserved quantity being used to define a work medium, we could explore the converse property in quantum theory: whether the 2$^{\textrm{nd}}$ laws found for different quantum resources in quantum resource theories each correspond to some kind of conservation law. This could shed light on what the underlying properties of a resource are that enable a 2$^{\textrm{nd}}$ law to be formulated for it. Such questions are particularly interesting in light of recent results that surprisingly, no 2$^{\textrm{nd}}$ law can exist for entanglement, though it does exist for a variety of other quantum resources \cite{lami2023no}.

\section{Summary}

There are various approaches to the field of quantum thermodynamics, each giving insights into thermodynamics of systems in different regimes. Here we considered a case-study where there is an intersection in results from axiomatic thermodynamics, resource theories and constructor theory. We first gave a general review of these different approaches to thermodynamics, comparing some of their key features and domains of applicability. Then we analysed the problem of deterministically extracting work from coherence. We considered how the phenomenon of work being locked in coherence in a single-shot setting (work-locking), while being accessible using tomographic methods, relates to constructor-theoretic constraints from dynamics-independent laws about distinguishability. 

Building on these insights, we showed that an existing quantum protocol for deterministically extracting work from coherent states cannot be generalised to extract different amounts of work from different input states, if those inputs are non-orthogonal. This demonstrates the impossibility of a universal deterministic work extractor from coherence. We also explained how this result is implied by a recently found dynamics-independent, scale-independent theorem from constructor theory, which is based on the 1$^{\textrm{st}}$ law of thermodynamics and information-theoretic foundations. 

Additionally, we used a recently proposed quantum model for constructor-theoretically possible tasks to explicitly show the constructor-theoretic impossibility of a universal work extractor from coherence with a particular implementation, based on the quantum homogenizer.  We also considered the converse of the constructor-theoretic theorem, whereby deterministically extracting the same amount of work from non-perfectly distinguishable states can be possible. Finally we discussed potential extensions to our work, including a constructor-theoretic analysis of the entropic-irreversiblity of quantum computation, and expanding the formulation of the 2$^{\textrm{nd}}$ law based on adiabatic possibility to multiple conserved quantities. 

In general, the usefulness of resources such as coherence in quantum thermodynamics has many subtleties and a high sensitivity to the particular assumptions and regimes being considered. We hope that demonstrating intersecting conclusions will help unify and generalise different approaches and clarify the fundamental nature and domain of applicability of results in the field.  % A Constructor-Based Quantum Toy-Model for the Cosmological Fine-Tuning Problem

\part{Measurement, Locality and Non-Commutativity} % Part II

\chapter{Background: Measurement Paradoxes and Non-Locality}

Claims that quantum theory is universal have been historically objected to on grounds regarding problems of observation and measurement \cite{myrvold_philosophical_2018}. The classic measurement problem is the ambiguity of where the so-called ``Heisenberg cut" takes place, where we transition from a system which can sustain a superposition of states to one which can only be in one state, by measuring the system. The Wigner's friend thought experiment elucidates the problem: if Wigner's friend measures a quantum system's state, then Wigner measures his friend's state, did the collapse take place in the former or latter measurement \cite{mehra_remarks_1995}? This thought experiment can be explained by understanding the measurement process as being the creation of entanglement between the system and the measuring apparatus, or observer. Once the apparatus becomes entangled with the system, they enter an entangled superposition, where each state in the superposition has the apparatus consistently measuring a single outcome of the system. Hence measurement can be described by unitary quantum dynamics without ambiguity about the point where ``collapse" takes place.

An extension to the Wigner's friend thought experiment was proposed by Deutsch \cite{deutsch_quantum_1985}. Here an observer can deduce whether or not they were previously in a superposition. They make a measurement of a quantum system in superposition relative to the computational basis, tell another observer that they definitely made a measurement (without revealing the outcome) and then are reversed back into their original state, along with the system they measured. By testing that the quantum system has returned to a superposition in the computational basis, the observer can be assured that they definitely made a measurement, without it causing an irreversible collapse of the system they measured. They can even retain memory of the fact that they made a measurement and saw a single outcome, providing that they do not remember the value of the outcome. Hence, the subjective experience of a single observation can be consistent with an observer existing in a superposition of states \cite{vedral_observing_2016, vedral_observing_2018}.

Another question regarding the universality of quantum theory is whether or not it is possible to have classical and quantum systems that interact. This would mean quantum theory could describe some part of the universe, but another part behaves classically, without any quantum properties. However there are arguments that if a given subsystem has non-classical features, then if it is able to interact with another subsystem, that must also have non-classical features \cite{dewitt_global_2003, marletto2022quantum}. This suggests that non-classicality is ``infectious", in the sense that if we accept some microscopic system has non-classical features, this conclusion must extend to all systems it can interact with. The universal nature of non-classicality is key when applying unitary quantum theory to all scales: if we treat one part of a system as classical and another part as quantum, we can run into inconsistencies. Semiclassical models are very useful for many situations, however can cause problems when taken seriously as models for the foundations of quantum theory. Mistakenly treating part of a system as classical is therefore the underlying cause of many claimed problems with applying quantum theory to describe observations, as we will demonstrate in Chapter \ref{ch:classical_paradoxes}.

Classical measurement apparatus also causes a significant problem for the consistency of physics by violating Einstein's principle of locality. This phenomenon is the root of Bell non-locality: the assumption that measurements of quantum systems are entirely characterised by single outcomes leads to the conclusion that a measurement on one system can instantaneously influence a spatially separated system with which it is entangled, albeit the influence is unobservable and hence does not violate no-signalling. By contrast, when treating the measurement apparatus as a quantum system, the assumption of measurements retrieving single outcomes must be dropped. Modifying the real description of a local system to include its full quantum description restores the principle of locality, and this is made explicit when modelling the measurement process in the Heisenberg picture \cite{deutsch2000information}.

In this Part of the thesis, we show how universal quantum theory can be applied consistently in contexts that involve measurement and locality, by elucidating the role of non-commutative operators in causing deviations from classically expected outcomes. In Chapter \ref{ch:classical_paradoxes}, we present a novel analysis of quantum branching structure in the presence of a trio of measurements, analogous to a Penrose-triangle. This structure is core to many apparent paradoxes that emerge from reasoning about measurement outcomes, particularly those that involve postselecting on particular outcomes. In Chapter \ref{ch:Hardy}, we focus on Hardy's paradox, which gives a demonstration of Bell non-locality without inequalities, and show that locally inaccessible information retrieved from non-commuting measurements explains the root of the paradox. In Chapter \ref{ch:quantifying_info} we show how to quantify the locally inaccessible information stored in Heisenberg picture observables, and prove analytically the intuitive connection between locally inaccessible information and the information stored in entanglement as quantified by von Neumann entropy.

We now present the background for Bell non-locality, and the explicitly local formulation of quantum theory in the Heisenberg picture. We first give a high level explanation using a demonstrative quantum circuit to explain the implications of the local account, and then give a breakdown of the key aspects of the Heisenberg picture required to track information flow.  

\section{Locality with an entangled pair}

In 1935, Albert Einstein, Boris Podolsky and Nathan Rosen proposed a thought experiment that came to be known as the EPR paradox \cite{einstein1935can}. By analysing the properties of two entangled systems, they concluded that quantum mechanics is an incomplete theory, as otherwise the implications of entanglement seemed to conflict with core principles of Einstein's relativity theories. Later, John Bell showed a novel way of taking measurements of quantum systems to see if the outcomes violate statistical inequalities \cite{bell1964einstein}. If these inequalities are violated, it rules out a class of theories known as ``local hidden variable models", which are potential candidates for modifying or completing quantum theory. Experimental implementations to test out Bell inequalities, and indeed rule out the impossibility of local hidden variable models describing reality, were then carried out by John Clauser, Alain Aspect and Anton Zeilinger \cite{clauser1978bell, aspect1982experimental, zeilinger1989hidden}. These had such fundamental significance for physics that they won the 2022 Nobel Prize. To understand the nature of entanglement, we need to consider also the limitations of Bell's theorem, and the local formulations of quantum theory that are not ruled out by violations of Bell inequalities.

A common popular idea is that given two entangled quantum systems, quantum theory implies that measuring one of the systems has an instantaneous effect on the other system, however far apart they are. Often attributed to Einstein's phrase ``spooky action at a distance", such behaviour would violate Eintein's \textit{principle of locality}. Einstein's principle of locality says that, for any two systems A and B, ``the real factual situation of system A is independent of what is done with the system B, which is spatially separated from the former" (paraphrased from \cite{schilpp1959albert}). This rules out instantaneous communication via entanglement, and even more strongly rules out any instantaneous influence from one of the systems to the other. The criterion has been formalised to include the constraint that a complete mathematical description for the properties of an individual physical system must exist, meaning global properties of a composite system can be constructed from its parts, and local properties from the global ones \cite{raymond2021local}. Quantum theory has been shown to satisfy this principle with an appropriate definition of the state of a quantum system, which is made explicit in the Heisenberg picture, as we discuss in detail in section \ref{Heis-pic} \cite{deutsch2000information}. We will first give a demonstrative summary of the local account using a quantum circuit example of an entangled pair of qubits. This is all done without reference to the local hidden variable models that are ruled out by Bell's theorem and violations of Bell inequalities. 

\subsection{Quantum circuit demonstration}\label{epr-circuit}

\begin{figure}[htbp!]
\centerline{\includegraphics[width=0.4\textwidth]{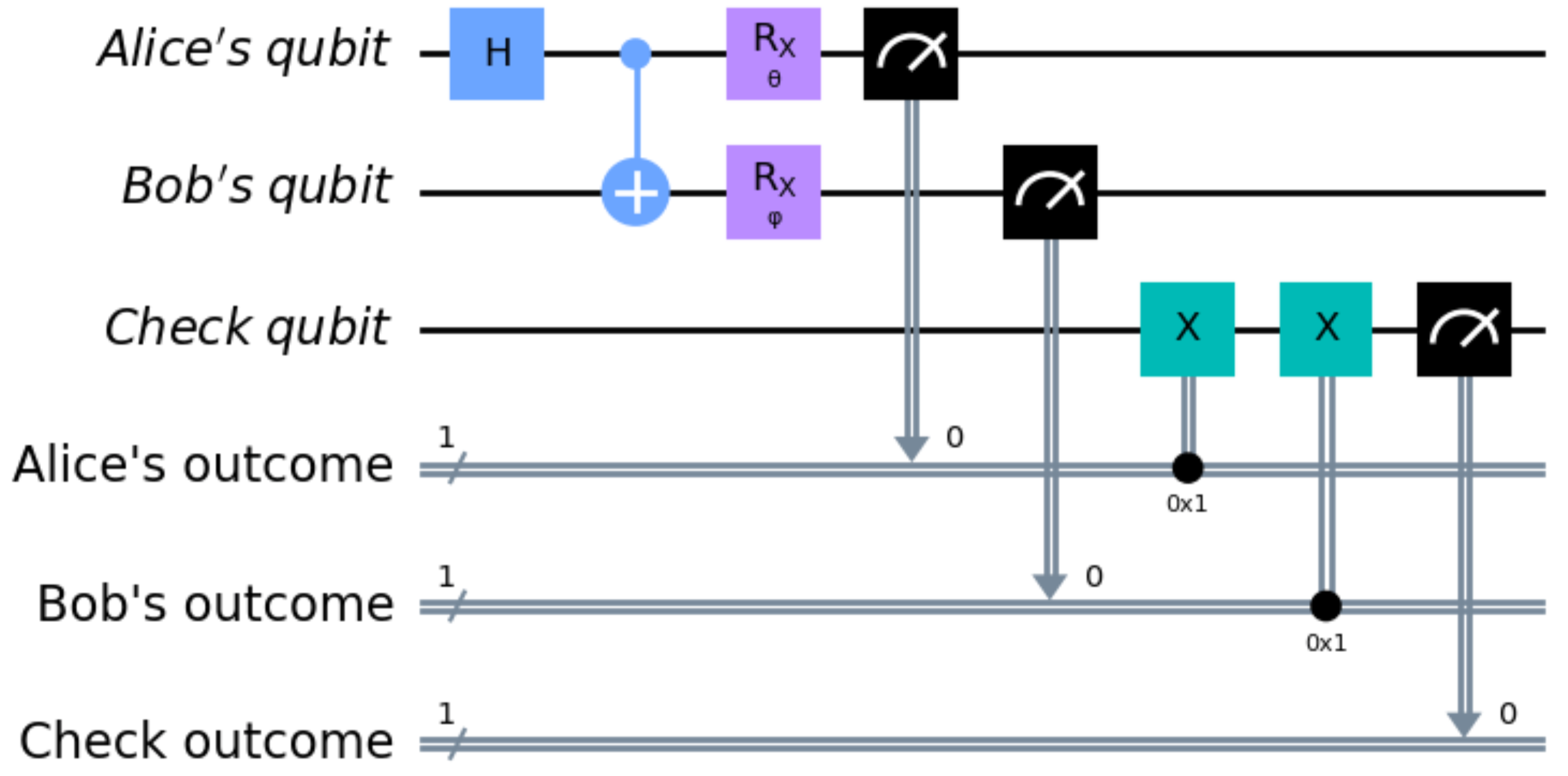}}
    \caption{Quantum circuit for measuring correlations of an entangled pair of qubits. Alice and Bob share an entangled pair; they each apply an arbitrary single-qubit X-rotation, paramaterised by $\theta$ and $\phi$ respectively; and they compare whether their outcomes are the same or different via a parity check using the check qubit.}
    \label{fig:epr-classical}
\end{figure}

Consider an entangled pair of qubits, where Alice has one of the qubits and Bob has the other. Alice and Bob can then make measurements on their qubits, and identify correlations between their measurement outcomes. The quantum circuit for this thought experiment is shown in figure \ref{fig:epr-classical}.  This and the other circuit figures in this section were created using Qiskit \cite{ibmq}.

Alice and Bob can each choose an arbitrary basis in which to measure their qubits, which we have captured by adding a paramaterised single-qubit gate before each of their $z$-measurements. For simplicity, we have made both these gates be X-rotation gates, with Alice's characterised by $\theta$ and Bob's by $\phi$. In general they could both be arbitrary unitary gates $U$, where each $U$ is defined by three parameters, and the same explanation will hold. 

We have then introduced a ``check qubit", which stores the outcome of a parity check on Alice and Bob's measurement outcomes, implemented by a pair of CNOT gates controlled on Alice and Bob's measurement outcomes and targeted on the check qubit. If Alice and Bob's measurement outcomes are the same, the check qubit will stay in the $\ket{0}$ state, while if Alice and Bob's measurement outcomes are different, it will change to the $\ket{1}$ state. \\

\noindent \textbf{Quantum and classical correlations}\\

\noindent The initial Hadamard and CNOT gates prepare Alice and Bob's entangled pair in the maximally entangled Bell state $\frac{1}{\sqrt{2}}(\ket{00}+\ket{11})$. It is well-known that whichever basis they now measure their qubits, if they choose the same basis, they will always get fully correlated outcomes. The surprising aspect is that information about the outcome of measuring a qubit e.g. in the $z$-basis and $x$-basis cannot be retrieved simultaneously, and yet measurement outcomes are correlated whichever of these bases is chosen. 

This leads to the common idea that  Alice's choice of measurement basis has some instantaneous (yet unobservable) effect on Bob's qubit, since Bob's qubit always gives the correct measurement outcome such that it is correlated with Alice's. This effect is manifest in the quantum circuit. Even though Alice and Bob are free to choose the basis in which they measure their own qubit independently, by varying the parameters $\theta$ and $\phi$, somehow their measurement outcomes conspire to be the same when the parameters coincide, such that the check qubit is always $\ket{0}$. \\

\noindent \textbf{Bell non-locality}\\

\noindent Bell's theorem is often quoted to suggest there is an influence on Bob's qubit caused by Alice's choice of measurement basis, or of Bob's choice of measurement basis on Alice's qubit's state. Bell's theorem considers a class of local theories, called ``local hidden variable models" \cite{brunner2014bell}. These assumptions place constraints on Alice and Bob's measurement outcomes, which are violated by quantum mechanics, showing that quantum mechanics cannot be a local hidden variable-model. The type of locality ruled out by ruling out this class of local theories is called Bell-non-locality. \\

\noindent \textbf{Alternative to Bell's local hidden variables}\\

\begin{figure}[htbp!]
\centerline{\includegraphics[width=0.5\textwidth]{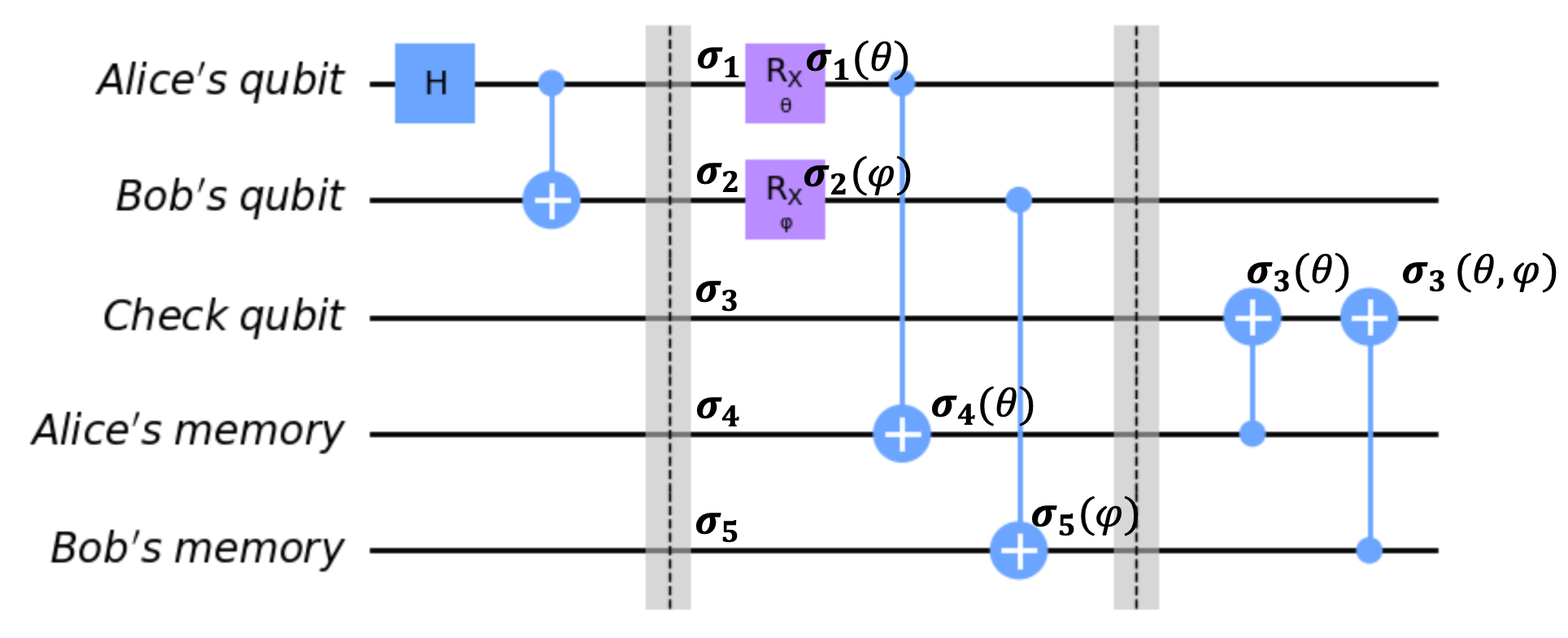}}
    \caption{Alice and Bob are treated as quantum systems, so all measurement operations are implemented via CNOT gates. Flow of the measurement basis information is shown using individual qubit descriptors.}
    \label{fig:info-flow}
\end{figure}

\noindent However, Bell's theorem only rules out one class of local realistic models for quantum theory \cite{deutsch2000information}. To explain the alternative local formulation, it is useful to consider an alternative quantum circuit, shown in figure \ref{fig:info-flow}. This is identical to the previous quantum circuit of figure \ref{fig:epr-classical}, except we have changed the classical registers of Alice and Bob's measurement outcomes to quantum registers. Then Alice and Bob's measurement operations become CNOT gates, such that if the control is $\ket{0}$, the memory will stay in the $\ket{0}$ state, and if the control is $\ket{1}$, the memory will be flipped to the $\ket{1}$ state, such that each memory qubit stores the outcome of measuring the entangled qubit. 

Switching measurements for CNOT gates at the end of a circuit makes no difference to the final distribution of measurement outcomes, known as the ``principle of deferred measurement" in quantum computing. However, a key difference compared with local hidden variable models is that here we do not assume that Alice and Bob have single, probabilistic measurement outcomes. Our overall description of events here is unitary, maintaining the overall entangled superposition of the multiple measurement outcomes that Alice and Bob could retrieve, freeing it from the assumptions of Bell's theorem. This is why the local account being explained here is fully consistent with measurement outcomes that violate Bell inequalities. \\

\noindent \textbf{Local and complete information from observables}\\

\noindent Now to explicitly see the local account of this thought experiment, we need to shift from the Schrödinger picture description of quantum states using a global statevector, to an explicitly local description of quantum states using observables in the Heisenberg picture. These two pictures are mathematically equivalent in the sense that they give the same predictions about the distributions of measurement outcomes in quantum mechanics, but only the Heisenberg picture makes the local formulation of quantum mechanics explicit. 

The properties of an individual quantum system can be fully described by the observables of that system. In section \ref{Heis-pic}, we precisely specify this description, but for now we will denote the object characterising a qubit \(\mathcal{Q}_i\) as the vector $\vec{\sigma}_i$, and refer to it as a \textit{descriptor}. The descriptor gives local and complete information about quantum states in the following sense: if we know the descriptor of two individual qubits, then we can recover the descriptor of their combined system, even if the two qubits are entangled. For example, if we know the descriptors of entangled qubits A and B individually, $\vec{\sigma_A}$ and $\vec{\sigma_B}$, we can recover their overall statevector $\ket{\psi}_{AB}$. Importantly, when quantum gates are applied to a system, the gates can only affect the descriptor of that quantum system; they cannot affect the descriptor of any quantum system on which the operation is not being directly applied \cite{deutsch2000information}. Hence, the existence of a local, complete description of quantum states using descriptors shows explicitly that quantum theory satisfies the powerful constraints imposed by Einstein's principle of locality. \\

\noindent \textbf{Tracking information flow for an entangled pair}\\

\noindent We can now apply this insight to the quantum circuit for Alice and Bob's measurements of the entangled pair of qubits, to track the flow of information in the thought experiment, as shown in Figure \ref{fig:info-flow}. 

The figure shows how the descriptor of each qubit comes to depend on $\theta$ and $\phi$ from local interactions. We can see from the quantum circuit that there is a direct quantum operation between Alice's qubit and Alice's memory, which means that the descriptor of Alice's memory qubit can depend on the parameter $\theta$, which captures the basis she chose to measure her qubit in. Similarly, the observables describing Bob's memory qubit can depend on his basis-measurement parameter $\phi$. 

Next we have the two CNOT gates that implement the parity check with the check qubit. One of them acts directly between Alice's memory qubit and the check qubit, so the descriptor of the check qubit can now depend on $\theta$. The other acts directly between Bob's memory qubit and check qubit, so the check qubit's descriptor can also depend on $\phi$.

Therefore the final outcome of the check qubit, which reveals whether or not Alice and Bob's qubits are correlated, can depend on Alice and Bob's individual, independent choices of measurement basis. We can explain this dependence in a fully local way, tracking the flow of information about which bases Alice and Bob chose to measure their qubits in all the way through the circuit. Alice's choice of measurement basis had absolutely no distant influence on any aspect of Bob's qubit, and vice-versa; the correlations between their outcomes are only present when information about their measurement basis choices has locally travelled from Alice and Bob's qubits to the check qubit. A detailed account of how Bell inequalities emerge with this formulation is presented in \cite{deutsch2000information} and \cite{bedard2024local}.

\section{Local account of quantum systems in the Heisenberg Picture} \label{Heis-pic}

Now we give a precise account of descriptors and the key features of describing quantum systems locally in the Heisenberg picture. \\

\noindent \textbf{Algebraic constraints on descriptors}\\

\noindent For any quantum system, we can define a set of operators that form a basis of local observables, describing only that system as it evolves over time. The overall set of operators is termed a \textit{descriptor}, since it gives complete information about the system. For some qubit \(\mathcal{Q}_1\), we can denote its descriptor as a vector of the component descriptors associated with the $x$, $y$ and $z-$bases: 
\begin{equation*}
    \hat{\mathbf{q}}_1(t) = (\hat{q}_{1x}(t),\hat{q}_{1y}(t),\hat{q}_{1z}(t)).
\end{equation*}
These components are Hermitian and obey the Pauli algebra, with the constraint: 
\begin{equation} \label{pauli}
    \hat{q}_{1i}(t)\hat{q}_{1j}(t) = \hat{1} \delta_{ij} + i \epsilon_{ij}^{\ \ k}\hat{q}_{1k}(t) \qquad \qquad (i,j \in \{x,y,z\}).
\end{equation} 
The eigenvalues of the $x$, $y$ and $z$ observables are $+1$ and $-1$, and hence they each form a Boolean observable, whose eigenvalues can take one of two values by definition. We will discuss Boolean observables extensively in Chapter \ref{ch:quantifying_info}, in which we refer specifically to those that satisfy the constraint \(\hat{O}^2=\hat{1}\), and so are traceless and satisfy the Pauli algebra. For a network of qubits, an additional constraint on the algebraic constraints on the observables is that distinct qubits obey the following commutation relation:
\begin{equation}  \label{commutator}
    [ \hat{q}_{ai}(t), \hat{q}_{bj}(t) ] =0 \qquad \qquad (\forall i,j \in \{x,y,z\} ).
\end{equation}\\

\noindent \textbf{Heisenberg state}\\

\noindent When computing an expectation value from the descriptor components, one requires the initial state of the system, $\hat{\rho}$:
\begin{equation*}
    \braket{\hat{A}(t)} \stackrel{\text{def}}{=} \operatorname{tr}(\hat{\rho} \hat{A}(t)).
\end{equation*} 
The initial state $\hat{\rho}$ the \textit{Heisenberg state}, and is fixed (does not evolve over time). Without loss of generality, the Heisenberg state can be set such that every qubit is in the state $\ket{0}$: 
\begin{equation} \label{Heisenbergstate}
\hat{\rho} = \bigotimes_{a =1}^n \ket{0;t=0}_a\bra{0;t=0}_a,
\end{equation}
Then at time $t=0$, the $z-$component of each qubit's descriptor has expectation value $+1$
:\begin{equation} \label{expectation}
    \braket{\hat{q}_{az}(0)} = 1 \qquad \qquad (\forall a \in \{1,...,n\}).
\end{equation}\\

\noindent \textbf{Variance and sharpness}\\

\noindent A key quantity that can be defined in terms of the expectation value is the variance of an observable: 
\begin{align*}
\text{Var}(\hat{O}(t))& := \left \langle ( \braket{\hat{O}(t)} - \hat{O}(t))^2 \right \rangle,
\end{align*}
which simplifies to the alternative form, \(\text{Var}(\hat{O}(t))= \braket{\hat{O}(t)^2} - \braket{\hat{O}(t)}^2\). The variance quantifies the level to which an average measurement outcome of the observable deviates from the expectation value. The special case of zero variance occurs when a measurement can be made that will deterministically give one value as the outcome, namely an eigenvalue of the observable. Observables with this property are termed \textit{sharp}. For example, if  \(\text{Var}(\hat{q}_{az}(t)) = 0\), then equivalently $\braket{\hat{q}_{az}(t)}^2 = \braket{\hat{q}_{az}(t)^2}$, meaning that (since \(\hat{q}_{az}(t)^2 = \hat{1}\)) $\braket{\hat{q}_{az}(t)} = \pm 1$, and vice-versa. By contrast, if an observable has a maximal variance, then it is maximally non-sharp. \\

\noindent \textbf{Quantum gates and unitary evolution}\\

\noindent When a unitary gate \(\mathbf{G}\) is applied to a network of qubits, it evolves each descriptor component at time $t$ to one at time $t+1$, which is in general a function of the time $t$ descriptors of the qubits on which the gate acted. Denoting the action of the unitary gate by $U_G$, the descriptors of some qubit \(\mathcal{Q}_a\) in the network evolve as:
\begin{equation*}
   \hat{\mathbf{q}}_a(t+1) = U^{\dagger}_G (\hat{\mathbf{q}}_1(t), ..., \hat{\mathbf{q}}_n(t)) \hat{\mathbf{q}}_a(t)U_G(\hat{\mathbf{q}}_1(t), ..., \hat{\mathbf{q}}_n(t)).
\end{equation*} 
Importantly, the algebra of the qubits is preserved by the unitary evolution, as shown in \cite{bedard2021abc}. Hence the algebra is fixed over time, and is a characteristic property of the system. 

When working in the Schrödinger picture, gates are fixed unitary operators applied to the evolving statevector. By contrast, in the Heisenberg picture, the operators evolve with each time-step. Since the qubit algebra is fixed over unitary evolution, operators also have a fixed representation, which takes the input observables at time $t$ and maps them to output observables at time $t+1$. This is the \textit{functional representation} of a gate, and is equivalent to the unitary operator representation, such that one can map back and forth between the representations \cite{bedard2021abc}. For example, the unitary applying an X gate to a qubit \(\mathcal{Q}_a\) at time $t$ acts as: 
\begin{equation} \label{watergate}
\textbf{X:} \ \hat{\mathbf{q}}_a(t+1) = (\hat{q}_{ax}(t), -\hat{q}_{ay}(t), -\hat{q}_{az}(t)),
\end{equation}
irrespective of the details of the initial descriptor $(\hat{q}_{ax}(t), \hat{q}_{ay}(t), \hat{q}_{az}(t))$. \\

\noindent \textbf{Einstein's locality}\\

\noindent Recall that a way of stating Einstein's criterion for locality requires that a system's is independent from a spatially separated one, and that a complete description of a quantum system can be composed from that of its parts (and vice-versa) \cite{schilpp1959albert, raymond2021local}. We will now broadly outline the way in which this criterion is explicitly satisfied in the Heisenberg picture. 

The independence can be proved by showing that a unitary applied to a qubit \(\mathcal{Q}_a\) alone can only affect the descriptors of that qubit, and not those of any other qubit (even those entangled with it). This is evident from the fact that the unitary applied to \(\mathcal{Q}_a\) can be expressed purely as a function of \(\mathcal{Q}_a\)'s observables. For instance, consider a single qubit rotation $U_R(\hat{\mathbf{q}}_a(t))$. We can compute how this affects the observables of \(\mathcal{Q}_b\): 
\begin{align*}
   \hat{\mathbf{q}}_b(t+1) &= U_R(\hat{\mathbf{q}}_a(t))^{\dagger}\hat{\mathbf{q}}_b(t)U_R(\hat{\mathbf{q}}_a(t)),\\
   &= \hat{\mathbf{q}}_b(t) U_R(\hat{\mathbf{q}}_a(t))^{\dagger} U_R(\hat{\mathbf{q}}_a(t)), \\
   &= \hat{\mathbf{q}}_b(t).
\end{align*}
Since all the observables of \(\mathcal{Q}_a\) commute with those of \(\mathcal{Q}_b\), we find that the description of \(\mathcal{Q}_b\) is invariant under any unitary evolution of \(\mathcal{Q}_a\). 

We can also prove the completeness of the description, such that individual qubits fully describe the networks that they compose, and vice-versa. First consider the direction of constructing the complete descriptions of the individual systems from the description of the global system. The global system's description includes the global observables and the stationary Heisenberg state \(\hat{\rho}\), which as discussed in \eqref{Heisenbergstate} can be initialised to a product state without loss of generality. From these, all the expectation values can be calculated. Now the set of all observables on the network includes as a subset all of the observables of each qubit, and the associated expectation values. Hence the complete description of each qubit can be derived, in addition to its reduced density matrix.

The criterion also requires the opposite direction to hold, such that the global network's complete description can be constructed from the complete description of the qubits composing it. A complete description of an individual qubit includes its observables and their expectation values (the latter being summarised by the reduced density matrix). Now any observable of the global network can be expressed in terms of sums, products and scalar multiples of the observables of the component qubits. Additionally, from the principle of local tomography, the expectation values of individual qubits that form a network are sufficient to reconstruct the overall density matrix of that network. Hence, the descriptions of individual qubits can be used to reconstruct the description of the overall network, and satisfy the completeness criterion \cite{bedard2021abc}.  % Background: Measurement Paradoxes and Non-Locality

\chapter{Incompatibility of Non-Commutativity and Classical Logic} \label{ch:classical_paradoxes}

\textit{The contents of this Chapter are based on a publication in preparation, done in collaboration with Abdulla Alhajri, Chiara Marletto and Vlatko Vedral. In the publication, the theoretical work will be presented alongside a single-photon experimental demonstration with Marco Genovese's group in the INRIM laboratory.} \\

\section{Introduction}

Multiple thought experiments have been suggested to show paradoxical features that emerge when we attempt to understand deductions about quantum measurements. In particular, there is a class of thought experiments which have a common feature of referring to measurements that could have been made in a different basis, but were not made. These include Hardy's paradox, the quantum pigeonhole paradox, and more recently the Frauchiger-Renner paradox \cite{frauchiger_quantum_2018, hardy_quantum_1992, aharonov_quantum_2016}. 

Various resolutions to each of these paradoxes have been proposed. These range from arguing that they simply demonstrate the same ``strangeness" of quantum mechanics as Bell non-locality, like the EPR paradox, to suggesting that they can be made self-consistent, experimentally testable but still surprising by analysing in terms of weak measurements \cite{aharonov_revisiting_2002}. Others argue more radically that some such paradoxes mean we need to abandon some logical axioms or the unitarity of quantum theory \cite{nurgalieva_testing_2020}. 

In this Chapter we investigate a quantum circuit which demonstrates a feature of quantum theory that emerges when we use classical logic on the outcomes of quantum measurements. In particular, we find that chaining together logically consistent statements leads to an inconsistent conclusion. This conclusion enables us to distinguish non-orthogonal quantum states deterministically with a single measurement. Hence, the impossibility of distinguishing non-orthogonal states is used to identify where a violation takes place. This counterfactual statement is interlinked with other fundamental principles of quantum theory, such as Heisenberg's uncertainty principle. 

The root of the inconsistency in the quantum circuit is that the two individually consistent statements refer to the same qubit in a different basis. The underlying problem is therefore that the non-commutativity of measurements in different bases is not accounted for when chaining together the individually consistent statements.  We compare this property to the geometry of a Penrose triangle, where each corner of the triangle is consistent but they join to create a globally inconsistent structure. Hence the quantum circuit reveals a twisted geometry within quantum branching structure. 

This quantum circuit provides a model for the problems that occur in the aforementioned measurement paradoxes. When we abstract each of them to the level of a quantum circuit, we can see the common features and how they map onto the Penrose-triangle framework where individually consistent statements lead to globally inconsistent outcomes, demonstrating a feature of quantum branching structure. Understanding that contradictions arise due to mistakenly applying classical logic ensures that universal quantum theory is self-consistent, and does not contain any known paradoxes.  

\section{Distinguishing non-orthogonal quantum states} \label{distinguish}

Here we present a thought experiment whereby it can (misleadingly) appear that we can distinguish non-orthogonal quantum states deterministically with a single-shot measurement, resulting in an internal contradiction within quantum theory.

\begin{figure}
\begin{tabular}{c}
\Qcircuit @C=1em @R=.7em { 
&& \text{{\sl Qubit A}} &&&&&& \lstick{\ket{0}} & \gate{U} & \ctrl{1} & \qw \barrier[-1.7em]{1} & \gate{H} & \meter \\ 
&& \text{{\sl Qubit B}} &&&&&& \lstick{\ket{0}} & \qw & \gate{H} & \qw & \qw & \qw & \gate{H} & \meter \\
}
\end{tabular}
\caption{\label{original short} Quantum circuit which demonstrates the impossibility of using classical logic on measurement outcomes.}
\end{figure}
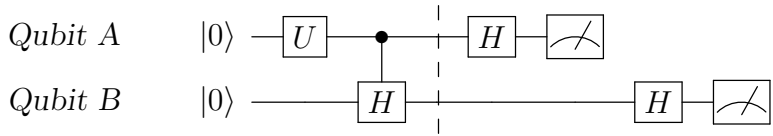

We will use the quantum circuit in Figure \ref{original short}. Let's consider the state of the two qubits as we move through the quantum circuit, for some arbitrary unitary gate $U$, where $U\ket{0} = \alpha \ket{0} + \beta \ket{1}$. The global state evolves as follows: 

\begin{align}
    & \ket{00}\\ \label{line 2}
    & \Rightarrow (\alpha \ket{0} + \beta \ket{1})\ket{0} \\ \label{line 3}
    & \Rightarrow \alpha \ket{00} + \frac{\beta}{\sqrt{2}}\ket{10} + \frac{\beta}{\sqrt{2}}\ket{11} \\ \label{line 4}
    & \Rightarrow (\frac{\alpha}{\sqrt{2}}+\frac{\beta}{2})\ket{00} + (\frac{\alpha}{\sqrt{2}}-\frac{\beta}{2}) \ket{10} + \frac{\beta}{2}\ket{01} - \frac{\beta}{2}\ket{11}\\ \label{line 5}
    & \Rightarrow (\frac{\alpha}{2}+\frac{\beta}{\sqrt{2}})\ket{00} + (\frac{\alpha}{2}-\frac{\beta}{\sqrt{2}})\ket{10} + \frac{\alpha}{2}\ket{01} + \frac{\alpha}{2}\ket{11}    
\end{align}

Now consider the case where Qubit A is measured to be in the state $\ket{1}$. The measurement reduces equation \ref{line 4} to having two terms, $\ket{10}$ and $\ket{11}$. If the coefficient of one of these terms is $0$, then we can deduce with certainty whether Qubit B is in a $\ket{0}$ or $\ket{1}$ state. Setting the coefficient of $\ket{11}$ to $0$ means $\beta = 0$, which would lead us to the trivial case where the global state is $\ket{00}$ until each qubit is measured in the $x$-basis. However we could alternatively set $\frac{\alpha}{\sqrt{2}}-\frac{\beta}{2} = 0$, such that $\alpha = \sqrt{\frac{1}{3}}$ and $\beta = \sqrt{\frac{2}{3}}$. In this case, we deduce with certainty that Qubit B must be in the $\ket{1}$ state upon measuring Qubit A in the $\ket{1}$ state. Now given that Qubit B is certainly in the $\ket{1}$ state, we can conclude from equation \ref{line 3} that before the measurement of Qubit A in the  $x$-basis, Qubit A was for certain in the $\ket{1}$ state. This suggests that stepping back to equation \ref{line 2}, only the $\ket{10}$ term could exist (given our final measurement of Qubit A as $\ket{1}$). Then, the Controlled-Hadamard gate must have definitely applied a Hadamard to Qubit B, and the measurement of Qubit B will for certain give $\ket{0}$ as an outcome. The outcome of measuring $\ket{11}$ as the final state is therefore forbidden.

Consider now an alternative input state, where instead of equation \ref{line 4} the state $\ket{--}$ is prepared before the measurements in the $x$-basis. Then, the application of $H\otimes H$ will clearly lead to the state $\ket{11}$ being deterministically measured in the computational basis at the end of the circuit. So, since $\ket{11}$ is forbidden as an outcome of the input state $\frac{1}{\sqrt{3}} \ket{00} + \frac{1}{\sqrt{3}}\ket{10} + \frac{1}{\sqrt{3}}\ket{11}$  by the logic above, and $\ket{11}$ is the deterministic outcome of the input state $\ket{--}$, a direct implication is that these two states can be distinguished by a single-shot measurement. Now $\bra{--}(\frac{1}{\sqrt{3}} \ket{00} + \frac{1}{\sqrt{3}}\ket{10} + \frac{1}{\sqrt{3}}\ket{11}) = \frac{\sqrt{3}}{2}$, hence the states are clearly non-orthogonal. The reasoning presented above therefore directly leads to the ability to perfectly distinguish non-orthogonal quantum states from a single measurement. 

Following the same logic for the case where the measurement of Qubit A gives $\ket{0}$, we instead set $\frac{\alpha}{\sqrt{2}}+\frac{\beta}{2} = 0$ to find $\alpha = \sqrt{\frac{1}{3}}$ and $\beta = -\sqrt{\frac{2}{3}}$. This leads to the same conclusion that the measurement of Qubit B will for certain give $\ket{0}$ as an outcome, such that measuring the state $\ket{01}$ is forbidden. Similarly to the previous example, one could prepare the state $\ket{+-}$ before the $x$-basis measurements, ensuring a deterministic outcome of $\ket{01}$. Then we can deterministically distinguish non-orthogonal states $\ket{+-}$ and  $\frac{1}{\sqrt{3}} \ket{00} - \frac{1}{\sqrt{3}}\ket{01} - \frac{1}{\sqrt{3}}\ket{11}$.

Therefore the seemingly logical deductions we made from equation \ref{line 4} enable us to fully distinguish two non-orthogonal states, violating an axiomatic principle of quantum theory whereby non-orthogonal states cannot be distinguished by a single measurement. This could appear to be an internal contradiction in quantum theory.

\section{A quantum Penrose triangle} \label{triangle}

Here we will explain why the deductions made from equation \ref{line 4} are false, and how this reasoning maps onto the logical structure of a Penrose triangle. Taking the case where Qubit A is measured to be $\ket{1}$, we argued that through a series of logical deductions, the state of Qubit A after the Controlled-Hadamard gate must have been $\ket{1}$. This claim is clearly faulty: we cannot draw a conclusion about the prior state of Qubit A because the Hadamard gate between the prior state and the later measurement does not commute with the measurement, which is in the $z$-basis.

The reason that deducing that Qubit A was in state $\ket{1}$ in the past can seem reasonable is that the claim can follow from two deductions that are each individually consistent. To see this, consider the set of deductions needed to conclude Qubit A's past state: 

\begin{enumerate}
    \item If Qubit A is measured to be $\ket{1}$, then Qubit B must be in the state $\ket{1}$ before it is measured.
    \item If Qubit B is in the state $\ket{1}$, then Qubit A must be in the state $\ket{1}$ before the Hadamard gate.
    \item Therefore, if Qubit A is measured to be $\ket{1}$, then before the Hadamard gate, Qubit A was in the state $\ket{1}$.
\end{enumerate}

\begin{figure}[h] 
\centering
\includegraphics[scale=0.4]{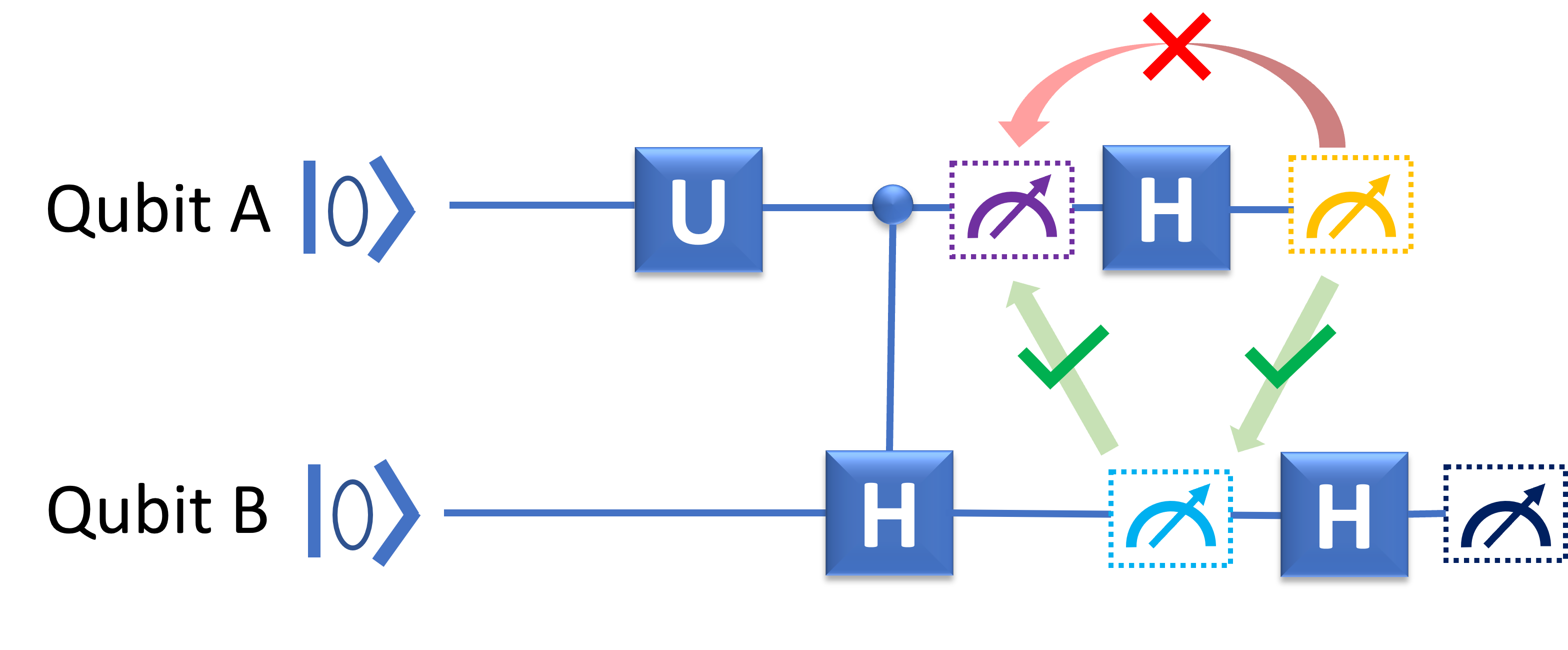}
\caption{Illustration of the individually consistent but incompatible deductions that can be drawn at different points of the quantum circuit. By measuring Qubit A to be $\ket{1}$, we can conclude that Qubit B is a $\ket{1}$. If Qubit B is a $\ket{1}$, then Qubit A was a $\ket{1}$ before the Hadamard gate. But this does not mean that we can conclude that if Qubit A is measured to be $\ket{1}$ then Qubit A was also $\ket{1}$ before the Hadamard, because it does not commute with the $z$-measurement.}
\label{circuit}

\end{figure}

\noindent The statement 1 is true: if Qubit A is measured to be in the state $\ket{1}$, then if Qubit B is measured in the $z$-basis at that moment, it will indeed be $\ket{1}$. Statement 2 is also true: if Qubit B is measured in the $z$-basis to give 1, then if Qubit A is measured in the $z$-basis before the Hadamard gate, Qubit A will also be in the state $\ket{1}$. The problem is that these logical deductions cannot be chained together: we cannot conclude that if Qubit A is measured to be $\ket{1}$, then Qubit B must have been in the state $\ket{1}$, \emph{therefore} Qubit A must have been in the state $\ket{1}$ before the Hadamard gate. This is because the two statements both refer to the state of Qubit A but in different bases, since there was a Hadamard gate between the two states of Qubit A. Having a defined state of Qubit A both before and after the Hadamard gate would violate the uncertainty principle, since we would simultaneously know the precise state of a qubit measured in both the X- and Z- bases. This is visualised in figure \ref{circuit}. 

\begin{figure}[h] 
\centering
\includegraphics[width=50mm,scale=0.5]{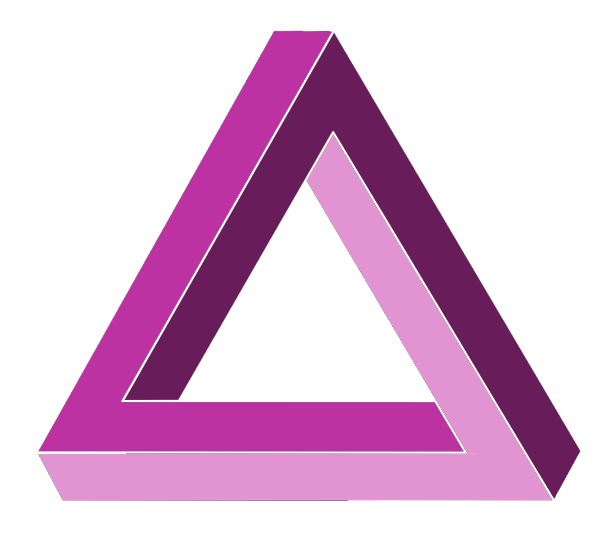}
\caption{A Penrose triangle. Each corner is consistent as the corner of a triangle. However the corners are chained together such that the global structure is inconsistent. This resembles how the logical deductions at different points of the quantum circuit are individually consistent, but cannot be consistently chained together.}
\label{penrose}

\end{figure}

We therefore find that quantum theory sustains a logical structure which resembles that of a Penrose triangle, shown in figure \ref{penrose}. The corners of the Penrose triangle are each consistent individually, but when chained together the global triangle is impossible. Similarly, the two logical deductions about the state of one qubit based on another are individually consistent, but impossible when chained together. This means that the reasoning which led to an ability to distinguish states in section \ref{distinguish} is invalid, because it relies on the forbidden chaining together of individually consistent statements. In a related publication, we demonstrate these results experimentally with high quality single photon qubits \cite{violaris2024penrose}. 

\section{Quantum paradoxes with twisted logic}\label{paradoxes}

Here we present three measurement-based paradoxes as a simple quantum circuit, and explain how they could map onto a Penrose-triangle structure.

\subsection{The Frauchiger-Renner paradox}\label{FR paradox}

The Frauchiger-Renner (FR) paradox is a thought experiment that was recently proposed to demonstrate that quantum theory cannot consistently describe users of quantum theory \cite{frauchiger_quantum_2018}. The authors suggest that the only solution is a radical departure from one of three natural axioms: the universality of quantum theory, the consistency of different observers' deductions, or that an observer only sees one outcome from their point of view.  

We will explain how the FR paradox and its resolution can be understood by comparing two quantum circuits. The first one is what is claimed to be the quantum circuit analysed in the proposal for a paradox, and maps onto the two-qubit quantum circuit discussed in section \ref{distinguish}. The first quantum circuit can be constructed by following the FR protocol step-by-step. \\

\begin{enumerate}
    \item A ``coin'' is put into a superposition of heads and tails. This is equivalent to applying a single-qubit rotation gate $U$ to a qubit initialized as $\ket{0}$, to rotate the state to $\sqrt{\frac{1}{3}} \ket{0} + \sqrt{\frac{2}{3}} \ket{1}$.
    
    \item Observer $\bar{F}$ measures the coin. The FR paper states that everything in the protocol is treated unitarily. Hence, the measurement is equivalent to a CNOT gate between a qubit representing $\bar{F}$ and the qubit representing the coin.
    
    \item $\bar{F}$ prepares a qubit in a state conditioned on her measurement result. If she measured $0$, then she prepares a $\ket{0}$ state, and if she measured $1$, then she prepares a $\ket{+} = \frac{1}{\sqrt{2}} (\ket{0} + \ket{1})$ state. This is equivalent to applying a Controlled-Hadamard gate with the control being the qubit representing $\bar{F}$ and the target being the qubit that $\bar{F}$ is preparing.
    
    \item The prepared qubit is sent to observer $F$, and $F$ measures the qubit. This is a CNOT gate controlled on the prepared qubit, with the target being the qubit representing $F$.
    
    \item Now observer $\bar{W}$ performs a Bell measurement on $\bar{F}$ and the coin. This involves a CNOT then Hadamard, then computational basis measurement of the $\bar{F}$ and coin qubits.
    
    \item Then observer $W$ performs a Bell measurement on $F$ and the prepared qubit. Again, this involves a CNOT and a Hadamard, followed by measurements in the computational basis.
\end{enumerate}

After running this circuit, we can compare the conclusions drawn by $\bar{W}$ and $W$ based on their measurements. The quantum circuit for the protocol is shown in figure \ref{original}. The statevector before the Bell measurements is:

\begin{figure}\centering
\begin{tabular}{c}
\Qcircuit @C=1em @R=.7em { 
&& \text{{\sl Coin}} &&&&&& \lstick{\ket{0}} & \gate{U} & \ctrl{1} & \qw & \qw & \qw \barrier[-1.7em]{3} & \gate{H} & \ctrl{1} & \meter \\ 
&& \text{$\bar{F}$} &&&&&& \lstick{\ket{0}} & \qw & \targ  & \ctrl{1} & \qw & \qw & \qw & \targ & \meter \\
&& \text{{\sl Qubit}} &&&&&& \lstick{\ket{0}} & \qw & \qw  & \gate{H} & \ctrl{1} & \qw & \gate{H} & \ctrl{1} & \meter \\
&& \text{{\sl F}} &&&&&& \lstick{\ket{0}} & \qw & \qw  & \qw & \targ & \qw & \qw & \targ & \meter \\
}
\end{tabular}
\caption{\label{original} Unitary implementation of the Frauchiger-Renner paradox quantum circuit. The first four gates are: the preparation of the coin; $\bar{F}$'s measurement of the coin; $\bar{F}$'s preparation of the qubit; and $F$'s measurement of the qubit. The final gates are $\bar{W}$ and $W$'s Bell measurements of $\bar{F}$ and $F$'s labs.}
\end{figure}
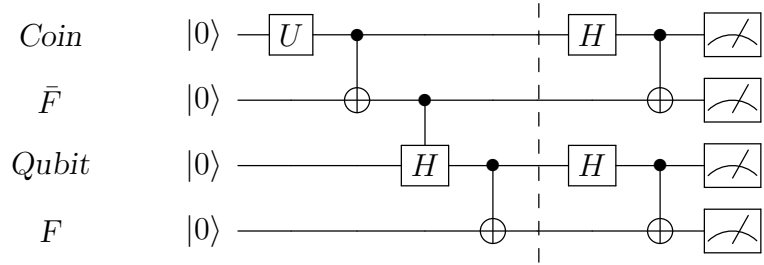

\begin{equation} \label{FR state}
    \sqrt{\frac{3}{4}} \ket{ \phi^{+}\phi^{+}} + \frac{1}{\sqrt{12}}\ket{ \phi^{+}\phi^{-}} -\frac{1}{\sqrt{12}}\ket{ \phi^{-}\phi^{+}} + \frac{1}{\sqrt{12}}\ket{\phi^{-}\phi^{-}}
\end{equation}\\

\noindent \textbf{The solution}\\

\noindent Now the FR paper claims that if the observers within the quantum circuit reason logically about each others' results, they conclude that there should be no probability of both $W$ and $\bar{W}$ measuring $\ket{\phi^{-}}$. Various resolutions have been proposed. These range from arguing that they simply demonstrate the same ``strangeness" of quantum mechanics as Bell non-locality, like the EPR paradox, to suggesting that such paradoxes mean we need to abandon some logical axioms or the unitarity of quantum theory \cite{nurgalieva_testing_2020}. 

Others argue the paradox arises due to a mistake: one observer's measurement is assumed to cause a ``collapse" rather than the observer entering an entangled superposition with a qubit \cite{sudbery_comment_2020}. We can see this argument as follows: the FR agents assume that when $\bar{F}$ measures the coin to be in the state $\ket{1}$ (or ``tails"), the branch where the coin is in the state $\ket{0}$ does not subsequently interfere with the $\ket{1}$ branch later on in the circuit. This is false. The branches do interfere later in the quantum circuit, so the agent $\bar{F}$ cannot disregard the existence of the $\ket{0}$ branch after measuring the outcome $\ket{1}$ of the coin.

\begin{figure}\centering
\begin{tabular}{c}
\Qcircuit @C=1em @R=.7em { 
&& \text{{\sl E}} &&&&&& \lstick{\ket{0}} & \qw & \targ & \qw & \qw & \qw & \qw & \qw & \qw & \meter \\
&& \text{{\sl Coin}} &&&&&& \lstick{\ket{0}} & \gate{U} & \ctrl{-1} & \ctrl{1} & \qw & \qw & \qw \barrier[-1.7em]{3} & \gate{H} & \ctrl{1} & \meter \\ 
&& \text{$\bar{F}$} &&&&&& \lstick{\ket{0}} & \qw & \qw & \targ  & \ctrl{1} & \qw & \qw & \qw & \targ & \meter \\
&& \text{{\sl Qubit}} &&&&&& \lstick{\ket{0}} & \qw & \qw & \qw  & \gate{H} & \ctrl{1} & \qw & \gate{H} & \ctrl{1} & \meter \\
&& \text{{\sl F}} &&&&&& \lstick{\ket{0}} & \qw & \qw & \qw  & \qw & \targ & \qw & \qw & \targ & \meter \\
}
\end{tabular}
\caption{\label{FR observers} Modification of the Frauchiger-Renner paradox quantum circuit to include an external observer $E$ who measures the coin after it is prepared. The outcomes attributed to the observers' reasoning in the Frauchiger-Renner paradox actually correspond to this modified quantum circuit.}
\end{figure}
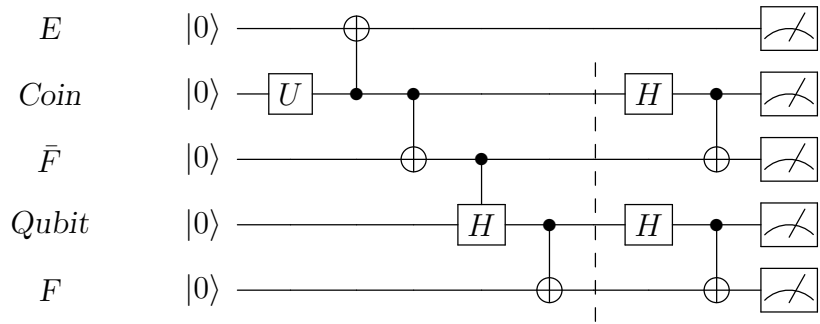

We will demonstrate this by constructing a different quantum circuit which leads to the version of events described by the FR agents. For this circuit, after the coin is prepared in a superposition, an external observer $E$ measures the state of the coin. $E$'s measurement is represented by a CNOT gate, with the coin as a control and a qubit representing observer $E$ as the target. Then the rest of the quantum circuit proceeds with the same gates as before. The resulting quantum circuit is shown in figure \ref{FR observers}.

The statevector before the measurements is now: 

\begin{equation}
    \frac{1}{\sqrt{12}} \ket{0} ( \ket{ \phi^{+}\phi^{+}} + \ket{ \phi^{+}\phi^{-}} + \ket{ \phi^{-}\phi^{+}} + \ket{ \phi^{-}\phi^{-}} ) + \frac{1}{\sqrt{3}} \ket{1} ( \ket{\phi^{+}\phi^{+}} - \ket{\phi^{-}\phi^{+}})
\end{equation}

Notice now that there is no term in the resulting statevector where the external observer measures $\ket{1}$ (``tails") on the coin, \textit{and} $\bar{W}$ and $W$ both measure the Bell states $\ket{\phi^{-}}$. This is precisely the forbidden term that appears in the FR paradox due to the observers' reasoning. Hence the circuit makes it clear that the reasoning claimed to be from the perspective of the observers in the FR paradox is referring to a different quantum circuit to the one described by the original thought experiment. Both versions of events can be described perfectly precisely within unitary quantum theory, as demonstrated by the two quantum circuits.

Note that labelling the extra qubit as an ``external observer" is only one interpretation of this circuit. We could also consider the extra qubit to represent an``environment" in which the coin decoheres. Alternatively we can consider the extra qubit to represent a permanent ``memory" of $\bar{F}$ that is not wiped out by any subsequent manipulations of $\bar{F}$.\\

\noindent \textbf{The twisted logic} \\

\noindent The false conclusion that the outcome $\ket{\phi^{-}\phi^{-}}$ can seem convincing due to the twisted logic described in section \ref{triangle}. To see how, notice that all the information contained in the circuit in figure \ref{original} remains if we remove the CNOT gates and $F$ and $\bar{F}$ qubits. This is because they cause the state of the control qubits to be copied onto the target qubits, so all the information is contained in just the control qubits. Then we get the quantum circuit of figure \ref{original short}, with Qubit A as the ``coin" and Qubit B as the ``qubit". The measurement of the coin is done by $\bar{W}$, and of the qubit is done by $W$. Now the reasoning made by the FR agents maps precisely to the statements in section \ref{triangle}: \\

\begin{enumerate}
\item If $\bar{W}$ measures the coin in the state $\ket{1}$, then $\bar{W}$ concludes that $F$ measured a $\ket{1}$ on the qubit. \\
\item If $F$ measures a $\ket{1}$ on the qubit, then $F$ concludes that $\bar{F}$ must have measured a $\ket{1}$ on the coin. \\
\item Therefore if $\bar{W}$ measured a $\ket{1}$, then $\bar{F}$ must have measured a $\ket{1}$ on the coin.
\end{enumerate}

The third conclusion means that $\bar{F}$ must have prepared the qubit sent to $F$ in the $\ket{+}$ state, and therefore $W$ must measure the state of the qubit to be $\ket{+}$ (giving an outcome of $\ket{0}$ at the end of the measurement circuit). Hence the measurement outcome for $W$ and $\bar{W}$ of $\ket{11}$ is forbidden (corresponding to the $\ket{\phi^{-}\phi^{-}}$ outcome in equation \ref{FR state}).

The statements 1 and 2 are each consistent individually, but cannot be chained together to give statement 3. This is because the first two statements refer to the state of the coin qubit in non-orthogonal bases. Therefore the FR paradox arises when using the twisted logic and Penrose-triangle structure identified in section \ref{triangle}.  

\subsection{Hardy's paradox} \label{hardy_paradox}

Hardy's paradox was proposed to show that quantum theory cannot be described by a local hidden variable model, through a different explanation than Bell inequalities and GHZ states \cite{hardy_quantum_1992, hardy_1993}. While the former relies on violating a statistical bound, and the latter requires at least three qubits, Hardy's paradox can show deterministically the Bell non-locality of quantum theory with only 2 qubits, when we post-select on a particular outcome of the experiment. There have been a variety of interpretations of the paradox over time and attempted demonstrations using weak measurements, e.g. \cite{aharonov_revisiting_2002}.  

The original presentation of Hardy's paradox involves two intersecting Mach-Zender interferometers. A particle is sent through one interferometer, while the corresponding antiparticle is sent through the other. The interferometers are arranged such that the bottom path of one intersects with the top path of the other, so a particle and anti-particle will annihilate if they both travel through this path. Then deductions about the paths of the particles can be made using measurements of the particles' paths after passing through their respective second beam-splitters. 

The quantum state of the particles before they reach their second beamsplitters can be expressed as an entangled state of the following form:

\begin{align} 
    \ket{\psi} &= \frac{1}{\sqrt{3}}(\ket{00} + \ket{10} + \ket{01}) \label{state}\\
    &= \sqrt{\frac{2}{3}}\ket{0}\ket{+} + \frac{1}{\sqrt{3}}\ket{10} \label{x1}\\
    &= \sqrt{\frac{2}{3}}\ket{+}\ket{0} + \frac{1}{\sqrt{3}}\ket{01} \label{x2} 
\end{align}

Then one can consider what happens when we measure each qubit in either the $x$-basis or the $z$-basis. If we measure the 1$^{\textrm{st}}$ qubit in the $z$-basis and get an outcome of 0, we know for sure that an $x$-basis measurement of the 2$^{\textrm{nd}}$ qubit will give $\ket{+}$, from equation \ref{x1}. Similarly, if we measure the 2$^{\textrm{nd}}$ qubit in the $z$-basis and get an outcome of 0, we know from equation \ref{x2} that if we measure the 1$^{\textrm{st}}$ qubit in the $x$-basis, it will give $\ket{+}$. We also know from equation \ref{state} that if we were to measure both the 1$^{\textrm{st}}$ and 2$^{\textrm{nd}}$ qubit in the $z$-basis, at least one of them would be $\ket{0}$. The false conclusion from these statements is that since at least one of the qubits would give $\ket{0}$ if measured in the $z$-basis, and if a qubit is $\ket{0}$ in the $z$-basis then the other qubit would be $\ket{+}$, then we know for sure that if we measure both qubits in the $x$-basis, at least one will be in the state $\ket{+}$. This conclusion is in contradiction with the conclusion from writing out the state in the $x$-basis: 

\begin{equation}
    \ket{\psi} = \frac{\sqrt{3}}{2}\ket{++} + \sqrt{\frac{1}{6}}\ket{+-} + \sqrt{\frac{1}{6}}\ket{-+} - \sqrt{\frac{1}{6}}\ket{--}
\end{equation}

From this state it is clear that there is a finite probability of measuring $\ket{--}$, an apparent contradiction to the reasoning from the potential $z$-measurements that could have been made before the $x$-measurements. This maps naturally onto the Penrose-triangle structure expressed in the quantum circuit we presented in section \ref{distinguish}. In this case, our three statements are:

\begin{enumerate}
\item If Qubit 1 is measured to be $\ket{0}$, then Qubit 2 will be a $\ket{+}$, and vice-versa.
\item At least one of Qubits 1 and 2 would be a $\ket{0}$ if they were measured in the $z$-basis.
\item Therefore at least one of Qubits 1 and 2 will be measured in the $\ket{+}$ state if both are measured in the $x$-basis. 
\end{enumerate}

\begin{figure}
    \centering
    \includegraphics[scale=0.5]{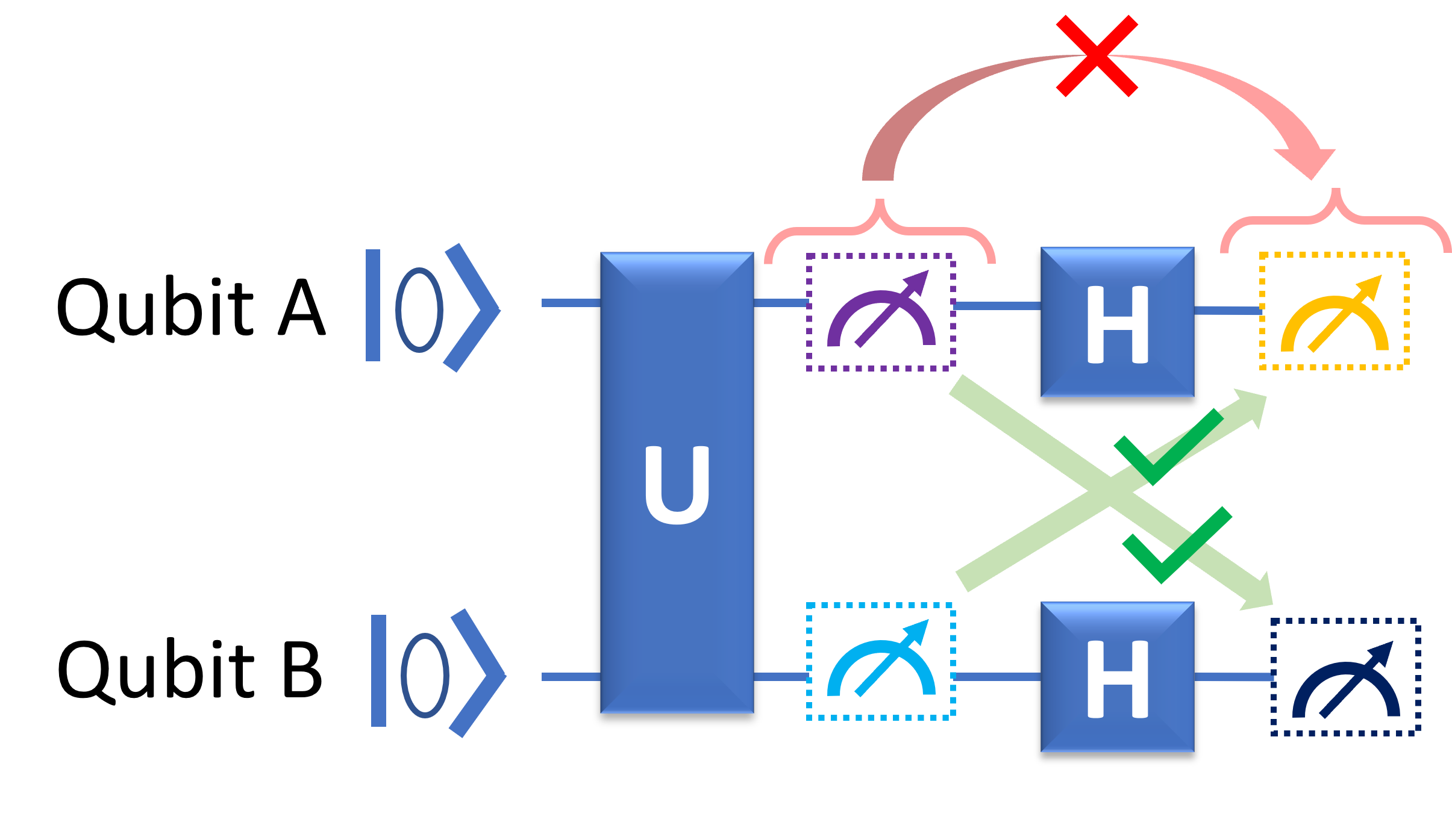}
    \caption{A depiction of Hardy's paradox as a quantum circuit.}
    \label{hardy circuit}
\end{figure}

Here statements 1 and 2 are each individually consistent, but they cannot be strung together to produce statement 3, because statement 3 involves conclusions about the definite state of the same qubit in a different basis - just as it did in the section \ref{distinguish} quantum circuit. This can be summarised by the quantum circuit in figure \ref{hardy circuit}. 

As before, we notice that the globally inconsistent statement arises when we are referring to the definite state of the same qubit before and after a Hadamard gate, as a result of reasoning about measurements that could have happened in a different basis, but did not.

\subsection{The quantum pigeonhole paradox}

The quantum pigeonhole paradox thought experiment claims to show that quantum mechanics violates the pigeonhole principle \cite{aharonov_quantum_2016}. Namely, a fundamental principle of number theory that if we put three objects into two boxes, at least two objects are in the same box. Since the pigeonhole principle is central to mathematics, this violation is claimed to have radical consequences for mathematics in the context of quantum theory. The solution and interpretation of the paradox has been expressed in a variety of ways, including claims that it is an unsurprising feature of quantum theory once we already accept Bell non-locality, and claims that the thought experiment shows a novel aspect of quantum theory that can be experimentally tested using weak measurements \cite{reznik_footprints_2020}. By simplifying the paradox to a quantum circuit, we will present an interpretation in terms of the Penrose-triangle logic we have elucidated above.

The thought experiment was proposed in the context of considering three quantum particles that can be in two boxes. Our initial state is a product state of each particle being in an equal superposition of both boxes. Then the particles are measured in the Y-basis, and we post-select on the outcome where all the Y-measurements give $\ket{0}$. This means the particles end up in a product state in the Y-basis. Then we note that if we project our post-selected state onto a state where any pair of particles were in the same box, the overlap with the initial state is $\ket{0}$. From this, we can conclude that none of the three qubits were ever both in the one of two boxes, violating the classical pigeonhole principle.

This scenario can be presented as a quantum circuit where the qubit state $\ket{0}$ indicates the 1$^{\textrm{st}}$ box and qubit state $\ket{1}$ indicates the 2$^{\textrm{nd}}$ box. The three qubits all begin in the $\ket{+}$ state, and are then measured in the y-basis. We post-select on the $\ket{+i +i +i}$ outcome. Then we deduce that we could have made a measurement on each pair of qubits, to deduce whether or not they are in the same state. There is no overlap between the projection of our outcome state $\ket{+i +i +i}$ onto a state where any pair of qubits is in the same state, with the initial $\ket{+++}$ state. In this case, our statements are:

\begin{enumerate}
\item The projections of the $\ket{+i +i +i}$ state where each pair of qubits is the same are orthogonal to the $\ket{+++}$ state.
\item The quantum state began as $\ket{+++}$ and became $\ket{+i+i+i}$ with no intermediate changes.
\item Therefore between the pre- and post- selection, no pair of the three qubits were in the same state, violating the pigeonhole principle. 
\end{enumerate}

\begin{figure}
    \centering
    \includegraphics[scale=0.4]{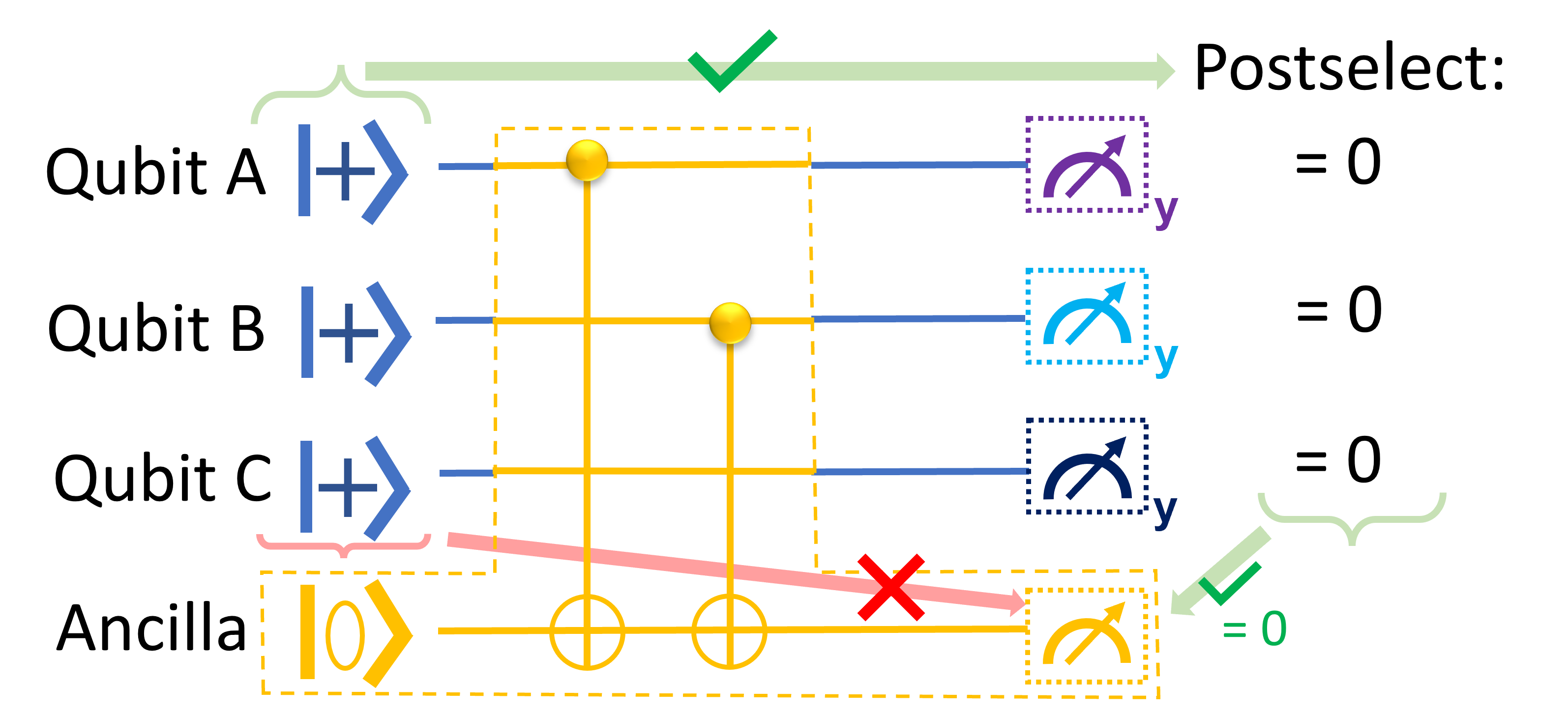}
    \caption{The pigeonhole paradox as a quantum circuit.}
    \label{pigeonhole circuit}
\end{figure}

The statements 1 and 2 are each individually consistent. However when we use them together, we once more make a definite claim about the state of a qubit in two different bases, causing an overall inconsistent statement. Once more this can be understood using a quantum circuit, as has been suggested in a paper commenting on the paradox using simulations on IBM quantum computers \cite{kunstatter2020escape}. Specifically, we begin with three qubits in the $\ket{+++}$ state. Then we measure the three qubits in the y-basis, by adding X-rotations followed by a $z$-measurement. After post-selecting on the $\ket{+i +i +i}$ outcome, we consider what would have happened if we had measured whether two qubits were in the same state. This can be done using CNOT gates between the pair of qubits and the same ancilla qubit, such that if the ancilla is measured to be $\ket{0}$ then the qubits must have been in the same state, while if the ancilla is $\ket{1}$ the qubits were in different states. If this measurement is made, and we find the pair of qubits were in the same state, then there is no possibility of the three qubits then being measured in the $\ket{+i +i +i}$ state. This is summarised in figure \ref{pigeonhole circuit}.  

\section{Discussion}

Our Penrose-triangle structure demonstrates how using classical logic in a quantum setting leads to contradictions. Classically, we are able to refer to the past state of a given system after its future state has been measured, whereas quantum theory prevents us from combining logical statements that refer to the same qubit at different points in time when it has been manipulated by operations which do not commute with the measurement.

Our set-up emphasises how the notion of ``branch structure" in unitary quantum theory can be misleading. In expositions of unitary quantum theory, when a qubit enters a superposition relative to some basis, it can be described as two ``branches". When it becomes entangled with another qubit, the qubits are correlated in each branch. However a key point that can be overlooked is that the branches are able to interfere, and hence the splitting of branches can change as a quantum circuit evolves. Deductions about Qubit A's ``past state" in the $z$-basis before a Hadamard cannot be made based on the outcome of measuring Qubit A in the $z$-basis after the Hadamard, as there is not a single quantum ``branch" connecting the past state with the measurement outcome. This occurs because the Hadamard gate does not commute with the $z$-basis measurement. Therefore we reinforce that a necessary consequence of unitary quantum theory is that branches are able to interfere.

We have drawn a parallel between deductions that can be made within unitary quantum theory and a geometrical structure, which raises the question as to how far we can interpret counterintuitive structures within quantum theory geometrically, as has been suggested in other quantum contexts \cite{isham_possible_2000}. A similar link between local possibility and global impossibility arising due to the non-commutativity of observables has been identified for local elements of reality, when investigating the nonlocality of a photon \cite{heaney_extreme_2011}. In this work the local elements of reality are mapped onto a Penrose square. Additionally there has been a comparison to Penrose-triangle structure in work on contextuality \cite{abramsky_contextuality_2020}. Further investigation into how non-commutativity in quantum theory maps onto impossible geometries when using classical logic could unify these examples of this property of quantum theory. 

\section{Summary}

We have discussed a mechanism by which it naively appears that we can perfectly distinguish non-orthogonal quantum states with a single measurement, violating a key axiomatic principle of quantum theory. We have demonstrated that this contradiction arises from applying classical logic to quantum deductions, and chaining together incompatible deductions that are individually consistent. The logical structure we have identified resembles the geometry of a Penrose triangle, with locally consistent and globally inconsistent statements paralleling the locally consistent corners of a triangle impossibly chained together. Understanding this twisted quantum logic gives insights into quantum branching structure, helps navigate apparent paradoxes, and links with other manifestations of impossible geometries arising from the non-commutativity of operators in quantum theory.  % Incompatibility of Non-Commutativity and Classical Logic 

\chapter{Local Account of Hardy's Paradox} \label{ch:Hardy}

\textit{The contents of this Chapter are based on a publication in preparation, done in collaboration with Samuel Kuypers.} \\

\section{Introduction}

There are a variety of thought experiments demonstrating the counterintuitive consequences of entanglement, which lead to experimental outcomes that deviate significantly from those allowed and expected by classical physics. These include the famous EPR paradox and subsequent Bell inequality violations, which demonstrate the impossibility of classical local hidden variable descriptions of reality. This effect has been termed Bell non-locality, due to the common view that the experiments demonstrate the impossibility of quantum theory satisfying local realism. If this is truly the case, then either quantum theory does not describe physical reality, or entangled systems can instantly influence each other, even if space-like separated. 

However, local hidden variable models are just one class of local models, and it is fully consistent for quantum theory to have a local description whilst not satisfying a classical local hidden variable model, at the cost of modifying our assumptions about which objects described in quantum theory are deemed to be physically real. It is shown in \cite{deutsch2000information} that contrary to the view of quantum theory violating local realism, it can be formulated in a fully local way, which is made explicit in the Heisenberg picture. This local account can be applied to tracking information flow in various quantum protocols, including a typical Bell pair scenario and the quantum teleportation protocol. Using this formulation, it can be explicitly proven that manipulating a quantum system can only affect that system, satisfying Einstein's principle of locality. These results have been further generalised and formalised in \cite{raymond2021local}. There, the properties of a local realistic theory are defined mathematically, and it is shown that quantum theory is consistent with those properties. 

This has important implications for physical laws about quantum information: since manipulating a quantum system can only affect that system alone, any information (classical or quantum) encoded in the system must reside in that local region of spacetime, even if the system is entangled with others. The local account shows that entanglement does not somehow enable information to be distributed across quantum systems, but rather entanglement enables quantum systems to act as lock-and-key pairs. When a pair of quantum systems becomes entangled, each one can now act as a ``key", for unlocking information ``locked" in the other. The locked-up information is stored in a ``locally inaccessible" way, meaning that local measurements (even quantum tomography) cannot retrieve the information; the information can only be retrieved following a local interaction between the entangled qubits. Information stored in a locally inaccessible way has the key property of being robust to decoherence, since decoherence is essentially unwanted measurements done on the system by the environment. Qubits storing information in a fully locally inaccessible way are ``effectively classical", since decoherence and measurements leave their locally observable properties unchanged. The precise meaning and physical relevance of this effectively classical behaviour has been made more precise in recent studies \cite{bedard2023teleportation, bedard2024local}. 

We discussed Hardy's paradox in section \ref{hardy_paradox}, but will review the background here for clarity. Hardy's paradox is a thought experiment which demonstrates Bell non-locality without using inequalities. Similarly to Bell inequality violations, the experiment provides a method to demonstrate the impossibility of local hidden variable models describing reality. Instead of violating a statistical inequality to show this conclusion, Hardy's paradox uses classical local hidden variables to conclude that a certain outcome should be forbidden, and thus forms an all-or-nothing test for Bell non-locality. If the classically-forbidden outcome is observed, then that rules out local hidden variable models. It is useful to also contrast it to the GHZ test for non-locality, which also does not require violating inequalities, however it requires a greater Hilbert space dimension, e.g. three qubits. Meanwhile, Hardy's paradox achieves a Bell non-locality demonstration without inequalities with only two qubits. 

In addition to the possibility of formulating a two-qubit test for Bell non-locality without inequalities, another interesting difference between the tests is that Hardy's paradox features the ``anomaly of non-locality" \cite{methot2006anomaly}. This effect demonstrates that entanglement and non-locality are not monotonically related, and has subsequently been found in a variety of other manifestations of Bell non-locality. It remains an actively discussed and surprising feature of quantum theory \cite{brunner2014bell}.

Given these significant differences in the Hardy's paradox construction when compared with Bell inequality violations, it is interesting to formulate the local account of Hardy's paradox, to gain insight into how locally inaccessible information is being distributed and the mechanism behind the false classical conclusion that causes the apparent paradox. Here we give a precise local account of the information flow in Hardy's paradox. We will show that the origin of Hardy's paradox is the ability to extract locally inaccessible information via joint measurements on observables that do not commute. The locally inaccessible information from non-commuting observables cannot be extracted simultaneously, and it is making the false assumption that it can be extracted simultaneously that leads to the paradox. In this picture, we can see the mechanism by which the paradox disappears for maximally entangled states (and product states). Based on this account, we find that the maximum violation of classical physics via Hardy's paradox is directly related to the incompatibility of the observables of these joint measurements, with maximum incompatibility coinciding with maximum classical violation. 

Interestingly, our account relies on a different measure of classical violation from that conventionally used in studies of Hardy's paradox. Instead of using $P_{Hardy}$, which is the probability of the classically forbidden outcome, we use a probability that we denote $P_{paradox}$, which includes the probabilities of measuring the other outcomes required to construct the paradox. Therefore we present both a new explanation for the maximum classical violation in terms of observables' incompatibility, and show the significance of the other post-selected outcomes involved in the construction of the paradox. This gives a novel way of understanding how the ``anomaly of non-locality" emerges in the paradox, demonstrating the mechanism by which the partially entangled state shows maximal violation, rather than the maximally entangled one. Our account shows how both locally inaccessible and locally accessible information are required to construct the paradox, with the key role of locally accessible information distinguishing it from Bell inequality violations. We also use relative descriptors to show explicitly how locally inaccessible information can be encoded in and retrieved from the partially entangled Hardy states. 

Furthermore, our analysis involves considering how information is distributed in its locally accessible and locally inaccessible forms during unitary evolution in a quantum circuit. In Chapter \ref{ch:quantifying_info}, we use the setting of two entangled qubits in a pure state as a toy-model for formulating a quantification of locally (in)accessible information, leading to a new connection with von Neumann entropy. 

\section{General Hardy's paradox}

Let's consider a generalised version of Hardy's paradox. Hardy showed that the reasoning to demonstrate this Bell non-locality can be extended to any two qubit entangled state apart from maximally entangled states \cite{hardy_1993, hardy_quantum_1992}. This generalisation was simplified in \cite{goldstein1994nonlocality}. We will now give a general version of the thought experiment, broadly following \cite{goldstein1994nonlocality}.

Any two-qubit state that is not a product state, and not maximally entangled, can be written in the following form:  
\begin{align}
    \ket{\psi} &= a\ket{v_{1}v_{2}} + b\ket{u_{1}v_{2}} + c\ket{v_{1}u_{2}} \label{Hardy_1}\\
    &= \sqrt{|a|^2+|c|^2}\ket{v_1 w_2} + b\ket{u_1 v_2} \label{Hardy_2} \\
    &= \sqrt{|a|^2+|b|^2}\ket{w_1 v_2} + c\ket{v_1 u_2} \label{Hardy_3}
\end{align}

\noindent given appropriate choice of bases $\ket{v_{1}}$ and $\ket{u_{1}}$ for the first qubit, $\ket{v_{2}}$ and $\ket{u_{2}}$ for the second qubit, and providing $abc \neq 0$. Here $\ket{w_1} = \frac{a\ket{v_1} + b\ket{u_1}}{\sqrt{|a|^2+|b|^2}}$ and $\ket{w_2} = \frac{a\ket{v_2} + c\ket{u_2}}{\sqrt{|a|^2+|c|^2}}$. \\

Then we can make the following statements: 

\begin{enumerate}
\item Either qubit A would be measured to be $\ket{v_1}$ or B measured to be $\ket{v_2}$ if measured in the $\{\ket{v_1},\ket{u_1}\}, \{\ket{v_2},\ket{u_2}\}$ bases respectively.
\item If qubit A is measured to be in the state $\ket{v_1}$ then B is in the state $\ket{w_2}$, and if B is measured to be in the state $\ket{v_2}$ then A is in the state $\ket{w_1}$.
\item Therefore, qubit A should be measured to be $\ket{w_1}$ and/or qubit B measured to be $\ket{w_2}$, so they cannot be measured to be $\ket{w_1^\perp w_2^\perp}$. 
\end{enumerate}

However, the state $\ket{w_1^\perp w_2^\perp}$ has a finite probability of being measured if qubits A and B are measured in the $\ket{w_1, w_1^\perp}, \ket{w_2, w_2^\perp}$ bases, respectively. Namely, the probability of the $\ket{w_1^\perp w_2^\perp}$ state is: $\frac{|b^2||c|^2(1-|b|^2-|c^2|)}{(1-|b|^2)(1-|c|^2)}$. For $|b|^2 = |c|^2 = \frac{1}{3}$, then this probability is $\frac{1}{12}$, as in the original state used for the thought experiment. The conclusions about the paradox described in section \ref{hardy_paradox} for a specific state also hold for this general case.

\section{Distinguishability and non-commutativity}

The incompatibility of the information gained from joint measurements on the relevant non-commuting observables is directly related to the distinguishability of the basis states used to formulate the paradox, $\ket{u_i}$ and $\ket{w_i}$. 

In the original paradox, these basis states are $\ket{u_i} = \ket{1}$ and $\ket{w_i} = \ket{+}$. Then two of the relevant non-commuting observables whose joint measurements are used to access locally inaccessible information are $Z_1Z_2$ and $Z_1X_2$. The inner product $\braket{u_i | w_i} = \frac{1}{\sqrt{2}}$, indicating the partial distinguishability of the states.

If instead the inner product is $\braket{u_i | w_i} = 1$, then the states are perfectly indistinguishable. In this case, all three ways of writing the global state $\ket{\psi}$ become equivalent (with $a = 0$), and without loss of generality, the joint observables all become the same, $Z_1Z_2$. In this case, we cannot extract incompatible pieces of locally inaccessible information to combine for a paradox. 

Now let's consider the converse case: the inner product is $\braket{u_i | w_i} = 0$, so that the states are perfectly distinguishable (\textit{i.e.} orthogonal). Then $\ket{w_i} = \ket{v_i}$, and again all ways of writing the global state $\ket{\psi}$ are the same, but it is a product state and the coefficient $c = 0$. Again, we have no incompatible joint observables to measure — but, even more problematically for the paradox, we cannot extract any locally inaccessible information through joint measurements. All information stored locally in the individual qubits adds up to the global information stored in both qubits. In Chapter \ref{ch:quantifying_info}, we demonstrate this explicitly in the Heisenberg picture. 

In summary: the states $\ket{u_i}$ and $\ket{w_i}$ must be partially distinguishable for the global state $\ket{\psi}$ to meet the condition that joint measurements on non-commuting observables can extract incompatible, semi-deterministic, locally inaccessible information. This condition is fulfilled only for and for all partially entangled states. Product states cannot store locally inaccessible information in the first place. Maximally entangled states have a degeneracy in the observables measured so that incompatible information cannot be extracted. 

We show in section \ref{section:incompatibility} how the distinguishability of these basis states is precisely the determining factor for maximising the probability of retrieving the measurement outcomes required to formulate Hardy's paradox. First, we review how entanglement can be quantified by von Neumann entropy in the thought experiment. 

\section{Von Neumann entropy in Hardy's paradox}

Given the general form of a two-qubit non-maximally entangled state in equation \ref{Hardy_1}, we can consider the von Neumann entropy of the individual qubits as the coefficients $a$, $b$ and $c$ vary. Note that, using $|a|^2 + |b|^2 + |c|^2 = 1$, we can characterise the state by only two of the coefficients.\\

\noindent The reduced density matrix of the first qubit is:
\(
\begin{pmatrix}
    1-|b|^2 & a^*b \\
    ab^* & |b|^2
\end{pmatrix}
\).\\

\noindent The eigenvalues of this density matrix are $\frac{1}{2}\pm\sqrt{\frac{1}{4}-|b^2||c|^2}$.\\

\noindent Then the von Neumann entropy of this qubit is:

\begin{equation}
  \begin{split}
    S_1 = & -\left(\frac{1}{2}+\sqrt{\frac{1}{4}-|b^2||c|^2}\right)
    \log_2\left(\frac{1}{2}+\sqrt{\frac{1}{4}-|b^2||c|^2}\right) \\
    & -\left(\frac{1}{2}-\sqrt{\frac{1}{4}-|b^2||c|^2}\right)
    \log_2\left(\frac{1}{2}-\sqrt{\frac{1}{4}-|b^2||c|^2}\right)
  \end{split}
\end{equation}\\

\noindent By symmetry, the second qubit has the same von Neumann entropy, so the total von Neumann entropy of the individual qubits is twice that of the first qubit: 

\begin{equation}
  \begin{split}
    S_{tot} = & -\left(1+\sqrt{1-4|b^2||c|^2}\right)
    \log_2\left(1+\sqrt{1-4|b^2||c|^2}\right) \\
    & -\left(1-\sqrt{1-4|b^2||c|^2}\right)
    \log_2\left(1-\sqrt{1-4|b^2||c|^2}\right)
  \end{split}
\end{equation}\\

\noindent Now when either $b = 0$ or $c = 0$, the two qubits are in a product state and $S_{tot}=0$. Meanwhile when $a = 0$ and $b = c = \pm\frac{1}{\sqrt{2}}$ the two qubits are maximally entangled, and $S_{tot}=2$. In all other cases, $0<S_{tot}<2$ and the qubits are non-maximally entangled. These are the cases in which Hardy's thought experiment can be used to show Bell non-locality. \\

\noindent \textbf{Comparing to probability of paradox}\\

\noindent We can compare how the sum of the individual von Neumann entropies of the two qubits (which is a measure of the entanglement between them) compares to the probability of successfully implementing the logic that leads to Hardy's paradox. In particular, this should account for the probabilities of post-selecting the appropriate states $\ket{v_1 w_2}$, $\ket{w_1 v_2}$, and $\ket{w_1^{\perp} w_2^{\perp}}$. These have probabilities $1-|b|^2$, $1-|c|^2$, and $\frac{|a|^2|b|^2|c|^2}{(1-|b|^2)(1-|c|^2)}$ respectively, meaning that the probability of post-selecting on the appropriate outcomes during the three types of measurements is their product, $P_{\textrm{paradox}} = |a|^2|b|^2|c|^2$. 

How would we expect $S_{tot}$ and $P_{\textrm{paradox}}$ to be correlated? $S_{tot}$ is a measure of the entanglement between the qubits $A$ and $B$, while $P_{\textrm{paradox}}$ is the probability of a successful post-selection for Hardy's paradox. Naively, we might expect that greater entanglement should lead to a greater likelihood of formulating Hardy's paradox, since more entanglement may be more non-classical. 

We compare $S_{tot}$ and $P_{\textrm{paradox}}$ for the simplified case of the Hardy state $\ket{\psi} = a\ket{00} + b\ket{01} + b\ket{10}$, for values of $a$ in range $[0,1]$, in figure \ref{fig:incompatibility}. Considering this simplified state is sufficient for capturing the behaviour of the relevant entropies and probabilities. We can consider the computational basis for the first qubit without loss of generality since the choice of basis is arbitrary. The paradox can be formulated for the second qubit in whichever basis it is expressed, and hence we can again set it to be the computational basis for simplicity, and absorb the basis transformation required for formulating the paradox into the form of the operators measured on the second qubit. 

Interestingly, we see that $P_{\textrm{paradox}}$ reaches a peak for a partially entangled state that has $S_{tot}$ at an intermediate value, between the product state and maximally entangled Bell state. Hence, the behaviour of $P_{\textrm{paradox}}$ cannot be explained by considering the amount of entanglement alone. 

\section{Incompatibility and paradox probability} \label{section:incompatibility}

We need an alternative explanation for the behaviour of $P_{\textrm{paradox}}$ for variations of the Hardy state, as measures of entanglement such as von Neumann entropy are insufficient. As mentioned above, formulating Hardy's paradox depends on extracting incompatible pieces of locally inaccessible information from the entangled state, which itself depends on the distinguishability of the basis states in which measurements are done to retrieve this information. 

It is useful to consider the entropic uncertainty relation, which is a basis-independent generalisation of Heisenberg's uncertainty principle for arbitrary pairs of observables \cite{deutsch1983uncertainty, maassen1988generalized, coles2017entropic}. Namely, given the Shannon entropy of some operator $V$ is \(H(V)\) and $W$ is \(H(W)\), then:
\begin{equation}
    H(V) + H(W) \geq 2\ln (\frac{1}{c})
\end{equation}
where \(c = \max_{i,j} \left| \langle v_i | w_j \rangle \right|^2\) is the maximum overlap of eigenbases $\ket{v_i}$ and $\ket{w_j}$ of \(V\) and \(W\) respectively. Since the right hand side of the expression lower bounds the total uncertainty of the operators, we can use it as a measure of incompatibility of the pair of operators. Let's denote incompatibility by $I(V,W) = 2\ln (\frac{1}{\max_{i,j} \left| \langle v_i | w_j \rangle \right|^2})$.

Then, given that the paradox is formulated by retrieving incompatible pieces of locally inaccessible information, the Hardy states providing maximum probability of paradox should coincide with those that have greatest incompatibility of the observables being measured to retrieve the required information. Recall that for the boundary cases where the two-qubit state is a product state and a maximally entangled state, the observables being measured to retrieve information become identical, meaning that they have zero incompatibility, and cannot be used to retrieve distinct, incompatible pieces of locally inaccessible information. 

We will now show that the paradox probability is maximised by the incompatibility of the observables, by comparing the two functions for two-qubit states characterised by a parameter that varies the level of entanglement, from product states to maximally entangled states. 

Let's again specialise to the state $\ket{\psi} = a\ket{00} + b\ket{01} + b\ket{10}$. Then, $\{\ket{v_1},\ket{u_1}\} = \{\ket{0},\ket{1}\}$. One pair of incompatible observables that we measure is $Z_1$ and $W_2$, where $W_2$ has corresponding basis states $\{ \ket{w_2},\ket{w_2^{\perp}}\}$, with $\ket{w_2} = \frac{a\ket{0} + c\ket{1}}{\sqrt{|a|^2+|c|^2}}$ and $\ket{w_2^{\perp}} = \frac{c^*\ket{0} - a^*\ket{1}}{\sqrt{|a|^2+|c|^2}}$. Another is $W_1$ and $Z_2$, where $W_1$ has corresponding basis states $\{ \ket{w_1},\ket{w_1^{\perp}}\}$, with $\ket{w_1} = \frac{a\ket{0} + b\ket{1}}{\sqrt{|a|^2+|b|^2}}$ and $\ket{w_1^{\perp}} = \frac{b^*\ket{0} - a^*\ket{1}}{\sqrt{|a|^2+|b|^2}}$. 

\begin{figure}
    \centering
    \includegraphics[scale=0.5]{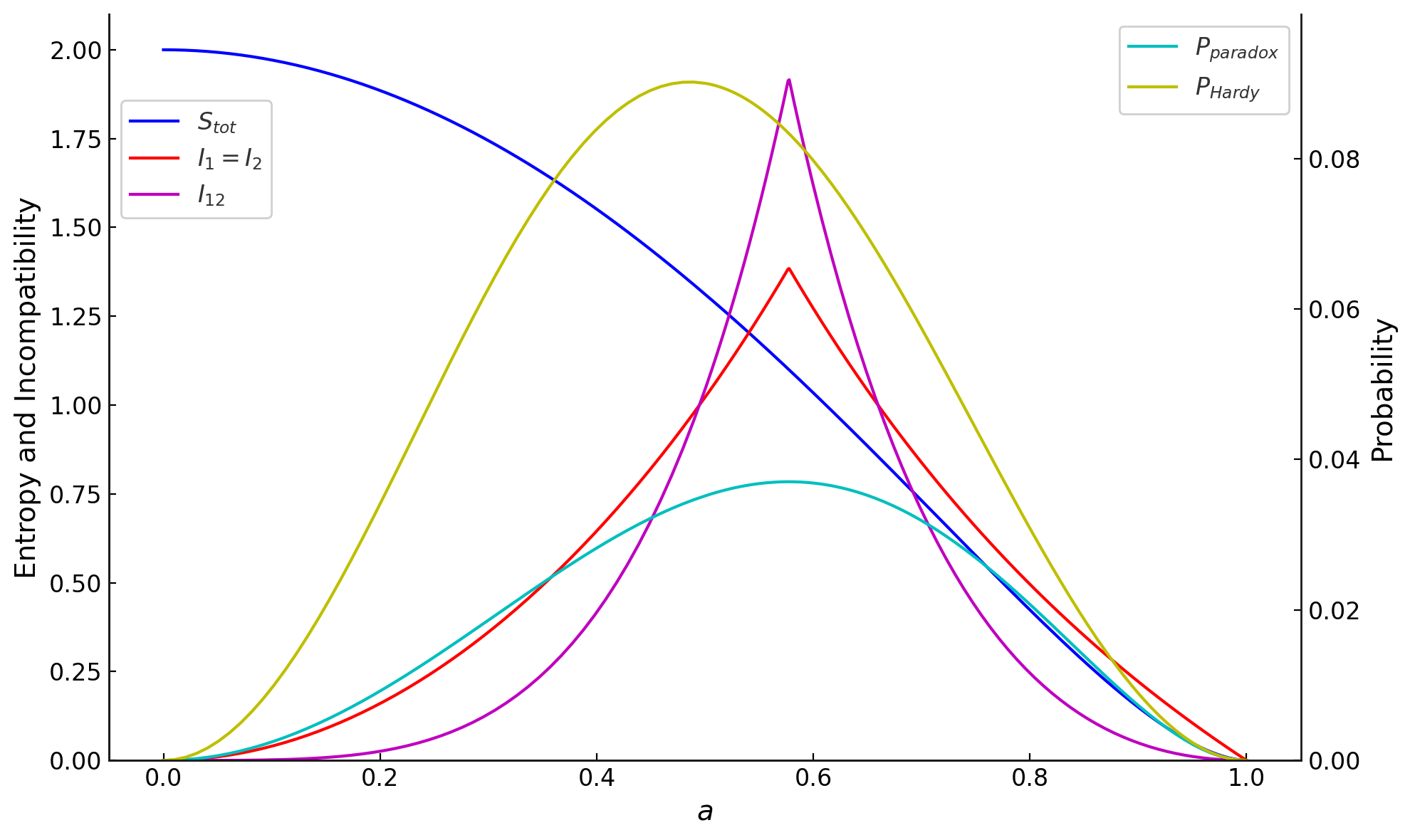}
    \caption{For the state $\ket{\psi} = a\ket{00} + b\ket{01} + b\ket{10}$: Sum of von Neumann entropy of reduced states; probability of post-selecting on states required to formulate a Hardy-type paradox; and incompatibility of relevant pairs of observables, as functions of $a$.}
    \label{fig:incompatibility}
\end{figure}

Now let $I_1 = I(Z_1, W_2)$, $I_2 = I(W_1,Z_2)$ and $I_{12} = I_1 I_2$. Plotting these as functions of $a$, we get figure \ref{fig:incompatibility}.  In this case, the symmetry of the state means that $I_1 = I_2$, and we see that the value of $a$ that maximises $I_1$, $I_2$, $I_{12}$ and $P_{\textrm{paradox}}$ is equal. Hence, the maximum incompatibility is a signature of the maximum probability of formulating the classically paradoxical logic with a given Hardy-type state.  

\subsection{Analytical proof}

We can prove that $I_{12}$ and $P_{\textrm{paradox}}$ are maximised by the same value of the parameter $a$. Since $I_1 = I_2$, we just need to consider the value of $a$ which maximises $I_1$. The general expression for $I_{1}$ is:  
\begin{equation}
I_1 = 2 \ln \left( \frac{1}{\max_{ij} |\langle v_i | w_j \rangle|^2} \right)
\end{equation}
Now expanding the expression, applied to the operators $\hat{Z}_1$ measured on the first qubit and $\hat{W}_2$ measured on the second qubit, the expression becomes:
\begin{equation}
I(\hat{Z}_1, \hat{W}_2) = 2 \ln \left( \frac{1}{\max \left\{ |\langle 0 | w_1 \rangle|^2, |\langle 1 | w_1 \rangle|^2, |\langle 0 | w_1^{\perp} \rangle|^2, |\langle 1 | w_1^{\perp} \rangle|^2 \right\}} \right)
\end{equation}

\begin{comment}

We can calculate each of the terms in the set to be maximised over:

\begin{align}
\langle 0 | w_1 \rangle &= \frac{a}{\sqrt{|a|^2 + |b|^2}} & \Rightarrow |\langle 0 | w_1 \rangle|^2 = \frac{|a|^2}{|a|^2 + |b|^2} \\
\langle 1 | w_1 \rangle &= \frac{b}{\sqrt{|a|^2 + |b|^2}} & \Rightarrow |\langle 1 | w_1 \rangle|^2 = \frac{|b|^2}{|a|^2 + |b|^2} \\
\langle 0 | w_1^{\perp} \rangle &= \frac{b^*}{\sqrt{|a|^2 + |b|^2}} & \Rightarrow |\langle 0 | w_1^{\perp} \rangle|^2 = \frac{|b|^2}{|a|^2 + |b|^2} \\
\langle 1 | w_1^{\perp} \rangle &= \frac{-a^*}{\sqrt{|a|^2 + |b|^2}} & \Rightarrow |\langle 1 | w_1^{\perp} \rangle|^2 = \frac{|a|^2}{|a|^2 + |b|^2}
\end{align}

\begin{align}
I(\hat{Z}_1, \hat{W}_2) &= 2 \ln \left( \frac{1}{\max \left\{ \frac{|a|^2}{|a|^2 + |b|^2}, \frac{|b|^2}{|a|^2 + |b|^2} \right\}} \right) \\
&= 2 \ln \left( \frac{|a|^2 + |b|^2}{\max \left\{ |a|^2, |b|^2 \right\}} \right)
\end{align}
\end{comment}

\noindent Calculating the terms, this becomes:

\begin{equation}
I(\hat{Z}_1, \hat{W}_2) = 2 \ln \left( \frac{|a|^2 + |b|^2}{\max \left\{ |a|^2, |b|^2 \right\}} \right)
\end{equation}

\noindent Given that \(a \in \{0, 1\}\) and $|a|^2 + 2|b|^2 = 1$, then we find the following relations: 
\begin{align}
\frac{1}{3} < |a|^2 \leq 1 &\Rightarrow |a|^2 > |b|^2 \\
0 \leq |a|^2 < \frac{1}{3} &\Rightarrow |a|^2 < |b|^2 \\
|a|^2 = \frac{1}{3} &\Rightarrow |a|^2 = |b|^2
\end{align}

\noindent The value of $I(\hat{Z}_1, \hat{W}_2)$ for \( |a|^2 > |b|^2 \), \( |a|^2 < |b|^2 \) and \( |a|^2 = |b|^2 = \frac{1}{3} \) is $2 \ln \left( 1 + \frac{|b|^2}{|a|^2} \right)$, $2 \ln \left( 1 + \frac{|a|^2}{|b|^2} \right)$ and $2 \ln 2$ respectively. Therefore $I_1$ is maximised when $|a|^2 = \frac{1}{3}$. Let's compare this to the $a$ which maximises the probability of paradox:

\begin{comment}
For \( |a|^2 > |b|^2 \),

\begin{align*}
I(\hat{Z}_1, \hat{W}_2) &= 2 \ln \left( \frac{|a|^2 + |b|^2}{|a|^2} \right) \\
&= 2 \ln \left( 1 + \frac{|b|^2}{|a|^2} \right)
\end{align*}

For \( |a|^2 < |b|^2 \),

\begin{align*}
I(\hat{Z}_1, \hat{W}_2) &= 2 \ln \left( 1 + \frac{|a|^2}{|b|^2} \right)
\end{align*}

For \( |a|^2 = |b|^2 = \frac{1}{3} \),

\begin{align*}
I(\hat{Z}_1, \hat{W}_2) &= 2 \ln 2
\end{align*}
\end{comment}

\begin{equation}
P_{\text{paradox}} = |a|^2 |b|^2 |c|^2 = |a|^2 |b|^4
\end{equation}
using $b = c$. We find:
\begin{align*}
\frac{dP_{\text{paradox}}}{da}
&= \frac{1}{2} \left( 2|a| - 8|a|^3 + 6|a|^5 \right) = 0 \\
&\Rightarrow |a| = 0, |a|^2 = 1 \text{ or } |a|^2 = \frac{1}{3}
\end{align*}

\noindent where $|a|^2 = \frac{1}{3}$ is the solution which maximises $P_{\text{paradox}}$, such that $P_{\text{paradox}}=\frac{1}{27}$ and $|a|^2 = |b|^2 = |c|^2$. Hence, we see that the value of $a$ that maximises $P_{\textrm{paradox}}$ is equal to that which maximises \(I_1\), namely $|a|^2 = \frac{1}{3}$.

\begin{comment}
\begin{align*}
P_{\text{Paradox}} &= |a|^2 |b|^2 |c|^2 \\
&= |a|^2 |b|^4 \quad
\end{align*}
because in our case $b = c$. We find:
\begin{align*}
\frac{dP_{\text{paradox}}}{da} &= \frac{d \left( |a|^2 \left( \frac{1 - |a|^2}{2} \right)^2 \right)}{da} \\
&= \frac{d}{da} \left( \frac{|a|^2 (1 - 2|a|^2 + |a|^4)}{2} \right) \\
&= \frac{d}{da} \left( \frac{|a|^2 - 2|a|^4 + |a|^6}{2} \right) \\
&= \frac{1}{2} \left( 2|a| - 8|a|^3 + 6|a|^5 \right) = 0 \\
&\Rightarrow |a| = 0 \text{ or } 1 - 4|a|^2 + 3|a|^4 = 0 \\
&\Rightarrow (3|a|^2 - 1)(|a|^2 - 1) = 0 \\
&\Rightarrow |a|^2 = 1 \text{ or } |a|^2 = \frac{1}{3}
\end{align*}
\begin{align*}
|a| = 0 &\Rightarrow P_{\text{paradox}} = 0 \text{ (minimum)} \\
|a|^2 = 1 &\Rightarrow P_{\text{paradox}} = 0 \text{ because } b = c = 0 \text{ (minimum)} \\
|a|^2 = \frac{1}{3} &\Rightarrow P_{\text{paradox}} = \frac{1}{3^3} = \frac{1}{27} \text{ because } |a|^2 = |b|^2 = |c|^2 \text{ (maximum)}
\end{align*}
\end{comment}

\subsection{Paradox probability vs Hardy probability}

An interesting feature of our analysis is that we use a different measure of classicality violation to the typical literature analysing the paradox \cite{hardy_1993, hardy_quantum_1992, goldstein1994nonlocality, brunner2014bell}. The violation is typically quantified by the ``Hardy probability", which is the probability of the classically forbidden state being measured. In our $P_{\textrm{paradox}}$, we also include the probabilities of measuring the other post-selected states required for formulating the paradox. When quantified via $P_{\textrm{Hardy}}$, it is well known that the maximal violation occurs when $a = \sqrt{\sqrt{5}-2}$, with a probability of $P_{\textrm{Hardy}}= \frac{5\sqrt{5}-11}{2}$ \cite{rabelo2012device}. Meanwhile, for $P_{\textrm{paradox}}$, we find the neat outcome that states of the original form of Hardy's paradox, where the three terms have equal amplitudes of $\frac{1}{\sqrt{3}}$, are in fact the ones that maximise the violation, giving  $P_{\textrm{paradox}} = \frac{1}{27}$. We compare $P_{\textrm{paradox}}$ and $P_{\textrm{Hardy}}$ directly in figure \ref{fig:incompatibility}. 

Hence the incompatibility of observables used for measurements characterises the trade-offs involved in retrieving information from the non-commuting measurements. One interpretation of the difference between the role of $P_{\textrm{Hardy}}$ and $P_{\textrm{paradox}}$ is that the latter integrates the significance of the non-commutativity for formulating the paradox, since it combines the probabilities of outcomes of measurements that do not commute. The relationship between measures of Bell non-locality and measures of entanglement is subtle and an open problem (indeed for multipartite systems, measures of either quantity alone become difficult to characterise), hence our derivation of the maximum of $P_{\textrm{paradox}}$ from incompatibility of observables provides an interesting way of giving firmer foundations to the measure of violation used in this setting, and sharpens the significance of certain properties for quantifying Bell non-locality.  

\section{Hardy's paradox in the Heisenberg picture}

In section \ref{Heis-pic}, we introduced the essential components of the local account of quantum theory using the Heisenberg picture. Here we analyse Hardy's thought experiment using the recently developed \textit{relative descriptors} \cite{kuypers2021everettian}, which were formulated to express Everettian relative states in the Heisenberg picture. This enables us to precisely state the post-selected measurement outcomes under which the paradox is formulated, in terms of the relevant observables, in the setting of fully universal quantum theory. We also see how locally inaccessible information can be encoded in and retrieved from the Hardy state. 

\begin{figure}[h]
  \centering
  \[
  \Qcircuit @C=1em @R=1em {
    \lstick{\ket{0}} & \gate{U_R(\phi)} & \ctrl{1} & \qw \\
    \lstick{\ket{0}} & \qw & \gate{H} & \qw
  }
  \]
  \caption{Quantum circuit to prepare a Hardy-type state parameterised by $\phi$.}
  \label{fig:quantum_circuit}
\end{figure}
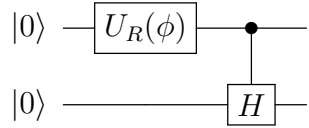

We start by preparing two qubits in the Hardy state. Let the two qubits of the entangled state be denoted by Qubit A, \(\mathcal{Q}_a\), and Qubit B, \(\mathcal{Q}_b\), respectively. Their initial state has the descriptors: 
\begin{align}
    \hat{\mathbf{q}}_a(0) & = (\hat{q}_{ax}, \hat{q}_{ay}, \hat{q}_{az}), \\
    \hat{\mathbf{q}}_b(0) & = (\hat{q}_{bx}, \hat{q}_{by}, \hat{q}_{bz}),
\end{align}
and their Heisenberg state is: \(\hat{\rho} = \ket{00;t=0}\bra{00;t=0}\). With this choice of Heisenberg state, the initial $z$-observables are all sharp. In the Schrödinger picture, a typical state analysed for Hardy's paradox is \ref{state}, $i.e.$ $\ket{\psi} = \frac{1}{\sqrt{3}}(\ket{00} + \ket{10} + \ket{01})$. To prepare this state, we can first apply the following unitary to \(\mathcal{Q}_a\): 
\begin{align}
    U_I = \frac{1}{\sqrt{3}}(\sqrt{2} \hat{1} +  i \hat{q}_x)
\end{align}
This evolves the observables of \(\mathcal{Q}_a\) to be: 
\begin{align}
    \hat{\mathbf{q}}_a(1) & = \left ( \hat{q}_{ax}, \frac{1}{3} \hat{q}_{ay} + \frac{2\sqrt{2}}{3} \hat{q}_{az}, \frac{1}{3} \hat{q}_{az} - \frac{2\sqrt{2}}{3} \hat{q}_{ay} \right ).
\end{align}
The preparation process can be generalised to a greater range of entangled states by introducing a parameter $\phi$ to the unitary rotation, for which the latter gate is a special case where \(\cos(\phi) = \frac{1}{3}\) and \(\sin(\phi) = \frac{2\sqrt{2}}{3}\):
\begin{align}
    U_R(\phi) = \cos(\phi/2) \hat{1} + i \sin(\phi/2) \hat{q}_{bx}
\end{align}
Next we need to apply a Controlled-Hadamard (C-H) gate with \(\mathcal{Q}_a\) as the control and \(\mathcal{Q}_b\) as the target. The unitary that implements a Hadamard gate is: 
\begin{align}
    U_H = \frac{1}{\sqrt{2}}(\hat{q}_x + \hat{q}_z)
\end{align}
The C-H gate evolves the descriptors to become: 
\begin{align}
    \hat{\mathbf{q}}_a(2) & = ( U_H \hat{q}_{ax}(1), \hat{q}_{az}(1)), \\
    \hat{\mathbf{q}}_b(2) & = (\hat{q}_{bx}(1) , \hat{q}_{bz}(1) )\Pi_{-1}(\hat{q}_{az}(1)) + (\hat{q}_{bz}(1) , \hat{q}_{bx}(1) )\Pi_{1}(\hat{q}_{az}(1))
\end{align}
where the descriptors now contain projectors that capture the conditional nature of the C-H gate on the eigenvalue of the $z$-observable of \(\mathcal{Q}_a\). The overall quantum circuit is shown in figure \ref{fig:quantum_circuit}.

Combining these expressions, the final state of the descriptors after the rotation and C-H gate are: 
\begin{align}
    \hat{\mathbf{q}}_a(2)  = & ( U_H \hat{q}_{ax}, \hat{q}_{az} \cos{\phi}  -  \hat{q}_{ay} \sin{\phi} ), \\
    \hat{\mathbf{q}}_b(2)  = &(\hat{q}_{bx},  \hat{q}_{bz} )\Pi_{-1}(\hat{q}_{az} \cos{\phi}  -  \hat{q}_{ay} \sin{\phi} ) \\ 
    & + (\hat{q}_{bz},  \hat{q}_{bx} )\Pi_{1}(\hat{q}_{az} \cos{\phi}  -  \hat{q}_{ay} \sin{\phi} ))  .
\end{align}
In Hardy's paradox, the first consideration (statement \ref{Hardy_1}) is: if we measure \(\mathcal{Q}_a\) in the $z$-basis and get an outcome of $\ket{0}$, \textit{i.e.} the +1 eigenvalue, what locally inaccessible information can we now retrieve from measuring \(\mathcal{Q}_b\) in the $x$-basis? We need to project the component $\hat{q}_{az}(2)$ to the +1 eigenstate, $\Pi_{1}(\hat{q}_{az}(2))$, and consider the information retrievable by performing a general measurement of \(\mathcal{Q}_b\) in the $x$-basis, conditioned on that outcome. The retrievable locally inaccessible information from the $x$-basis of \(\mathcal{Q}_b\), conditioned on \(\mathcal{Q}_a\)'s +1 $z$-eigenvalue, is then given by the expectation of the relative descriptor:
\begin{align}
\frac{ \braket{\hat{q}_{bx}(2) \Pi_{1}(\hat{q}_{az}(2)) }} {\braket{\Pi_{1}(\hat{q}_{az}(2)) }} & =   \frac{\braket{\hat{q}_{bz} }\braket{ \Pi_{1}(\hat{q}_{az} \cos{\phi}  -  \hat{q}_{ay} \sin{\phi}) }}{\braket{\Pi_{1}(\hat{q}_{az} \cos{\phi}  -  \hat{q}_{ay} \sin{\phi})}} = 1
\end{align}
Similarly, for the 2$^{\text{nd}}$ statement \ref{Hardy_2}, if \(\mathcal{Q}_b\) is measured to have a +1 eigenvalue in the $z$-basis and \(\mathcal{Q}_a\) is measured in the $x$-basis, the locally inaccessible information that can be retrieved from \(\mathcal{Q}_a\) is:  
\begin{align}
\frac{ \braket{\hat{q}_{ax}(2) \Pi_{1}(\hat{q}_{bz}(2)) }} {\braket{\Pi_{1}(\hat{q}_{bz}(2)) }} = 1.
\end{align}
Since the expectation of this relative descriptor is $\pm1$, it is sharp. To see explicitly how the relative descriptor reveals information, we need to consider a setting where it has multiple possible outcomes. For example, if we apply a general parameterised gate \(U_{\theta}\) on \(\mathcal{Q}_b\) before the C-H gate, then the relative descriptors have an expectation value exactly equal to \(\theta\):
\begin{align}
\frac{ \braket{\hat{q}_{ax}(2) \Pi_{1}(\hat{q}_{bz}(2)) }} {\braket{\Pi_{1}(\hat{q}_{bz}(2)) }} = \theta.
\end{align}

\begin{figure}[h]
  \centering
  \[
  \Qcircuit @C=1em @R=1em {
    \lstick{\ket{0}} & \gate{U_R(\phi)} & \ctrl{1} & \qw \\
    \lstick{\ket{0}} & \gate{U_\theta} & \gate{H} & \qw
  }
  \]
  \caption{Quantum circuit which stores the value $\theta$ in a partially locally inaccessible way.}
  \label{fig:theta}
\end{figure}
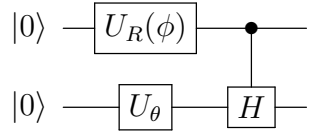

The corresponding preparation circuit is shown in figure \ref{fig:theta}. By varying $\theta$, the gate  \(U_{\theta}\) can be made into an Identity and into an X-gate. Depending on which of the two gates is implemented, the sign of \(\hat{q}_{bz}\) may or may not be inverted, such that the outcome will be $+1$ or $-1$. In this case, we gain exactly one bit of information from the measurement, by semi-deterministically obtaining exactly one of two possible outcomes in our measurement of \(\mathcal{Q}_a\) (semi-deterministic because we have a deterministic outcome conditioned on post-selecting the measurement of \(\mathcal{Q}_b\)). 

\noindent Finally the third statement for formulating Hardy's paradox, \ref{Hardy_3}, states that:
\begin{align}
 \braket{\Pi_{-1}(\hat{q}_{az}(2))\Pi_{-1}(\hat{q}_{bz}(2))} = 0 
\end{align}
with the corresponding sharp observable enabling us to retrieve locally inaccessible information being: 
\begin{align}
 \braket{ \hat{1} - \Pi_{-1}(\hat{q}_{az}(2)) \Pi_{-1}(\hat{q}_{bz}(2)) } = 1.
\end{align}
Then the classical reasoning about measurement outcomes leads to the false conclusion that: 
\begin{align}
 \braket{\Pi_{-1}(\hat{q}_{ax}(2))\Pi_{-1}(\hat{q}_{bx}(2))} = 0 
\end{align}
which is in contradiction with the fact that this expectation is not 0 when calculated directly. 

\section{Summary}

We have presented an account of Hardy's paradox in terms of the locally (in)accessible information that can be retrieved from non-commuting measurements. We found that the incompatibility of the observables used to formulate the paradox is key in determining the maximum violation from classical reasoning. In doing so, we reformulated the probability of violation from its typical form to include contributions from all the post-selected states required to formulate the paradox. Our analysis provides an explanation for why a partially entangled state is the one which maximises the classical violation in Hardy's paradox, giving a new insight into the ``anomaly of non-locality" whereby the maximum Bell non-locality does not correspond to maximum entanglement. Then we analysed the thought experiment in the Heisenberg picture, using relative descriptors, and showed explicitly how locally inaccessible information can be encoded in Hardy-type states.  In Chapter \ref{ch:quantifying_info}, we further develop this analysis by showing that the locally (in)accessible information stored in the Hardy-type partially entangled states is quantified by the von Neumann entropy. 

 % Local-Realistic Account of Hardy's Paradox

\chapter{Quantifying Locally Inaccessible Information} \label{ch:quantifying_info}

\textit{The contents of this Chapter are based on a publication in preparation, done in collaboration with Samuel Kuypers.} \\

\section{Introduction}

Information in quantum systems is typically quantified using the von Neumann entropy, which is the generalisation of Shannon entropy to quantum systems \cite{vedral2002role}. Von Neumann entropy is also used as a measure of entanglement \cite{plenio2007introduction}. Hence, entanglement can be understood as a store of quantum information. An example where this is explicitly evident is in the superdense coding protocol, where a qubit sent from Alice to Bob is able to transport two bits of classical information, provided they share an entangled pair of qubits \cite{bennett1992communication}. This works despite the fact that isolated qubits can only store at most 1 bit of classical information. Explanations of this protocol emphasise the qualitative and quantitative role of entanglement for information storage, saying for instance: ``the information is clearly being stored not in the particle but in the entanglement between the two particles" \cite{wootters1998quantum}, and ``the amount of entanglement determines exactly how much information Alice can convey to Bob" \cite{vedral2002role}. In this Chapter we will show how information residing locally in a particle quantitatively relates to the amount of entanglement.  

In the previous Chapter we discussed the local account of quantum theory, which is made explicit in the Heisenberg picture. In local accounts of quantum information processing protocols such as quantum teleportation, superdense coding, and demonstrations of Bell-type non-locality, the fundamental significance of the ability of quantum systems to carry locally inaccessible information is manifest. However, locally inaccessible information is only referred to qualitatively: we can identify its presence, e.g. in quantum teleportation, but how do we quantify the locally inaccessible information and locally accessible information residing in a system? Since locally inaccessible information can be stored using entanglement, intuitively it should be quantified by the von Neumann entropy. 

Building on the Hardy's paradox setting analysed in the previous Chapter, here we consider an overall pure, potentially entangled pair of qubits, and formulate precisely how the locally inaccessible, and locally accessible information residing in each qubit is uniquely quantified by von Neumann entropy. We provide a novel connection between von Neumann entropy and the local account of quantum theory as expressed in the Heisenberg picture. 

We express the conservation of total information by a constraint on the sum of locally inaccessible and locally accessible information. Our information measure in the Heisenberg picture is characterised by the \textit{sharpest observable} of a quantum system, which is the observable with the least variance, and is unique for all possible states except maximally mixed states. Interestingly, the eigenbasis of this observable is exactly the one in which the density matrix is diagonal. This leads to a direct connection between information characterised by the least-variance, sharpest observable of a quantum system, and the information accessible via its density matrix, and hence von Neumann entropy, which is a function of the density matrix. Using this formulation, we explain how systems can encode a fractional bit of information. 

\section{Forms of information in quantum systems}

\subsection{Bits in pure qubit observables} \label{bits}

Qubits can instantiate bits of information, simply by using their ability to be in the $\ket{0}$ and $\ket{1}$ states as the states for classical bits of information (or using any other orthogonal pair of states). Let's consider how bits are encoded in observables, in particular the descriptors used in the Heisenberg picture. 

A bit is a variable that can take one of two values. Let's consider those two values being either $+1$ or $-1$, which are the eigenvalues of the qubit states $\ket{0}$ and $\ket{1}$ respectively. This enables us to naturally encode a bit in an observable such as \(\hat{q}_{az}(t)\). More generally, when some observable \(\hat{O}\) is ``sharp", its expectation value can take exactly one of these two values, $i.e.$ \(\braket{\hat{O}} \in \{1,-1\}\). Hence, sharp observables form Boolean variables, and qubits with sharp observables can carry 1 bit of information. 

In addition to the potential for having two values, two more key properties of bits are that they can be copied, and that they are perfectly distinguishable \cite{deutsch2015constructor}. When an observable of a qubit is sharp, a single (``one-shot") measurement is sufficient to copy the value of the information stored to another qubit, for instance via a CNOT gate between the qubits. 

Specifically, a CNOT gate controlled on the qubit that initially stores the bit, and targeted on a measurement qubit initialised with the $z$-observable +1 at time $t$, affects the associated expectation values as follows:
\begin{align*}
\left\{ \begin{array}{c}
    \braket{\hat{O}(t)} = \pm 1 \\
    \braket{\hat{q}_z(t)} =  1
\end{array} \right\} 
\xrightarrow{\textbf{CNOT}} 
\left\{ \begin{array}{c}
    \braket{\hat{O}(t+1)} = \pm 1 \\
    \braket{\hat{q}_z(t+1)} =  \pm 1 
\end{array} \right\}.
\end{align*}
Importantly, this copying process is only possible for values of the observable that are sharp. For example, if its $z$-observable is sharp, then it is not possible to copy the $x$-observable, which cannot be sharp simultaneously with the $z$-observable, as a consequence of Heisenberg's uncertainty principle. Another key property of the values of the bit is that they are perfectly \textit{distinguishable} via a single measurement. Again, a non-sharp value of an observable does not satisfy this property. 

So far, we have an ``all-or-nothing" formulation of the ability for quantum systems to carry bits, in terms of sharpness of observables. An observable can either be sharp, in which case the qubit can instantiate a bit using the two values of that observable, or non-sharp, in which case the qubit cannot instantiate a bit using that observable. However, there is an entire continuum of ``sharpness" between a qubit being maximally mixed (maximally not sharp, meaning no information at all can be stored) and pure (sharp, meaning it can store a bit of information). In the former case of maximally non-sharp observables, there is a 50:50 chance of retrieving a value of $+1$ or $-1$, regardless of the observable measured, meaning that measurement results of the qubit are intrinsically indistinguishable from entirely random noise.  

The interesting case of information stored in partially mixed qubits happens when there are varying degrees of entanglement between two qubits, as demonstrated in the analysis of Hardy's paradox. Bits in this context require a quantitative, intermediate characterisation. Furthermore, in physical reality, a qubit never has a completely sharp observable, though we model it that way as an ideal abstraction. While there is no limit to how closely a sharp observable can be approximated, there is always some error causing a deviation from the ideal case. Considering the physical nature of information stored in a close to (but not exactly) sharp observable requires an understanding of how an approximation of a bit is stored in a qubit with a partially sharp observable. We define and explain approximately sharp observables in section \ref{measure}. 

\subsection{Locally accessible information} \label{sec:acc_info}

Qubits have much richer properties than merely instantiating a bit. We have just considered the possibility of storing a bit in the $z$-observable of a qubit. However, qubits have at least three distinct, Boolean observables ($x$, $y$ and $z$, that can each take two values of $+1$ or $-1$). In addition, there are an infinite number of linear combinations of these three leading to other Boolean observables. This leads us to naturally consider the how far these distinct observables can be used to store bits of information, $i.e.$ the question of how much information can be stored in a pure qubit. The solution elucidates what it means for a single qubit to store locally accessible information, and constraints on the nature of information that can be stored in this way. \\

\noindent \textbf{Single pure qubit}\\

\noindent Consider a qubit \( \mathcal{Q}_a\), which has the $x$, $y$ and $z$  observables \(\hat{\mathbf{q}}_a(t)\). At time $t$, the qubit is in a pure state, and its the expectation values are as follows: 
\begin{align} \label{pure}
\braket{\hat{\mathbf{q}}_a(t)} = (0,0,\theta) \quad \quad (\theta \in \{1,-1\}).
\end{align}
The observable \(\hat{q}_{az}(t)\) is sharp with value \(\theta\), so  \(\hat{q}_{az}(t)\) encodes one bit of information, which can be retrieved by a single-shot measurement, as described in section \ref{bits}. Information such as $\theta$ that can be retrieved from measurements on a qubit is termed \textit{locally accessible information}. 

Since a qubit has multiple, distinct Boolean observables, it is natural to ask whether multiple observables can be used to encode further information, including perhaps an additional fractional piece of information. Yet encoding any more than 1 bit in a single qubit is impossible, which can be seen as follows: when the $z$-observable is sharp, the $x$- and $y$- observables are maximally non-sharp, and so they cannot encode any information. 

To prove this, consider the general form of an arbitrary Boolean observable of a qubit: 
\begin{equation}
\hat{A}(t) \stackrel{\text{def}}{=} a \hat{q}_{ax}(t) + b\hat{q}_{ay}(t) + c\hat{q}_{az}(t),
\end{equation}
where \(a\), \(b\), and \(c\) are real-valued coefficients such that \((\hat{A}(t))^2=\hat{1}\).  Its expectation value is $\braket{\hat{A}(t)} = c\theta$, hence here $\theta$ is retrieved (approximately) from measuring this observable. It is approximate in the sense that $\theta$ is only retrieved in a single-shot measurement when $c=1$, and in all other cases, is retrieved probabilistically, which is effectively equivalent to a noisy measurement. In other words, retrieving $\theta$ by measuring an ideal bit that is subject to random errors. Hence we see that by measuring an arbitrary observable, we can at best retrieve a noisy version of the information that can be perfectly retrieved when measuring the $z$-observable. 

So far we have analyzed a qubit which is sharp in the $z$-observable, however in general a pure qubit could be sharp in any other observable, and that could be used to encode $\theta$ instead. Our analysis directly generalises for any sharp observable, because it can always be transformed to the $z$-observable by a single-qubit unitary gate that effectively implements a basis transformation, without affecting the information that is encoded ($i.e.$ $\theta$). Consider an observable  \(\mathbf{\hat{q}}_a(t)\), where the $z$-observable is sharp. A single-qubit gate is applied to the qubit. The new observable \(\mathbf{\hat{q}}_a(t+1)\) is composed of observables that are linear combinations of the components of \(\mathbf{\hat{q}}_a(t)\). Hence, a general form of $\hat{q}_{az}(t)$ is: 
\begin{align} \label{alwayssharp}
\hat{q}_{az}(t) = \sum_{j\in\{x,y,z\}} \operatorname{tr}\bigl(\hat{q}_{az}(t) \hat{q}_{aj}(t+1)\bigr)  \hat{q}_{aj}(t+1).
\end{align}
When the LHS of \ref{alwayssharp} is sharp, so is the RHS, and hence the RHS of this expression gives a sharp observable at time $t=1$, which stores the information $\theta$. Therefore, the proof that at most one bit of information can be stored in any pure qubit extends to any unitary evolution on that qubit. \\

\noindent \textbf{Multiple pure qubits}\\

\noindent We have demonstrated the familiar result that a pure qubit can store exactly one bit of information. This can be extended simply to a collection of multiple pure qubits; given $n$ pure qubits, they can store exactly $n$ bits of classical information. To demonstrate this, we need to prove that no additional information storage capacity becomes available via the joint observables of the two pure qubits, compared to the individual observables, which can each encode one bit. Consider two pure qubits \(\mathcal{Q}_a\) and \(\mathcal{Q}_b\), with observables \(\hat{\mathbf{q}}_a(t)\) and  \(\hat{\mathbf{q}}_b(t)\) respectively. Since they are each in a pure state, their joint state must be a product state. Hence, at time $t$, 
\begin{comment}
(check derivation)
\end{comment} 
\begin{equation}
   \braket{\hat{q}_{ai}(t)\hat{q}_{bj}(t)} = \braket{\hat{q}_{ai}(t)}\braket{\hat{q}_{bj}(t)} \quad (\text{for all} \ i,j\in\{x,y,z\}).
\end{equation}
We see that the information that can be retrieved from the joint observables of the two qubits is fully characterised by the information retrievable from the individual observables of each qubit. Hence, there is no additional information that can be stored in the joint state, compared to the individual states, as required. The proof straightforwardly generalises to $n$ qubits, such that $n$ qubits in pure states can store at most $n$ bits of information, specifically one bit of locally accessible information per qubit. 

\subsection{Locally inaccessible information}

So far we have considered the information that can be retrieved via measurements on a pure qubit, characterised by the information revealed in the expectation value of its observables. This is locally accessible information, since it can be retrieved by measurements on that qubit alone. However, the Heisenberg picture makes explicit another kind of information that can be stored in qubits, which is locally inaccessible information. This kind of information can be present when a qubit is in a mixed state, which occurs for instance when a qubit is entangled with another system. To characterise and quantify the nature of locally inaccessible information, here we will focus on the setting of two entangled qubits which are overall in a pure state. We give comments on generalising our analysis to settings where the overall state is not pure, and where there are more qubits involved, in section \ref{info_discussion}. \\

\noindent \textbf{Two-qubit pure state}\\

\noindent To see what locally inaccessible information means physically, consider an example of two qubits \(\mathcal{Q}_1\) and \(\mathcal{Q}_2\) with Heisenberg state \(\hat{\rho}= \ket{11;t =0}\bra{11;t=0}\). They then evolve by two gates (two time-steps), such that they have the following observables at time \(t=2\):
\begin{align} \label{maximally mixed}
   \hat{\mathbf{q}}_1(2) & = (\hat{q}_{1x}, \ \hat{q}_{1y}\hat{q}_{2x}, \ \hat{q}_{1z}\hat{q}_{2x}), \\
   \hat{\mathbf{q}}_2(2) & = (\hat{q}_{2z}\hat{q}_{1x}, \  -\hat{q}_{2y}\hat{q}_{1x}, \ \hat{q}_{2x}).
\end{align} 
Note that the observables $\hat{q}_{1i}$ and $\hat{q}_{2j}$ are the ones at $t=0$, though we have dropped the time label to simplify the notation. We see that both observables at time $t=2$ contain a mix of observables of the two qubits from $t=0$, which is a signature of the gates applied between $t=0$ and $t=2$ having caused the qubits to interact (there must have been 2-qubit gates). Specifically, the gates formed a Bell-state preparation circuit. The corresponding statevector in the Schrödinger picture at $t=2$ is: 
\begin{align} \label{Bell}
\frac{1}{\sqrt{2}}\left(\ket{1}_1\ket{1}_2 + \ket{-1}_1\ket{-1}_2 \right).
\end{align}
We can compute the expectation values of the qubits \(x\)-, \(y\)-, and \(z\)-observables, equivalently using the Schrödinger picture representation in \eqref{Bell} with the $t=0$ observables, or the Heisenberg picture presentation in \eqref{maximally mixed} with the $t=0$ Heisenberg state \(\hat{\rho}\). We find that all three observables are maximally non-sharp: 
\begin{align} \label{maximally non-sharp}
   \braket{\hat{\mathbf{q}}_1(2)} = \braket{\hat{\mathbf{q}}_2(2)} = (0,0,0).
\end{align}
This demonstrates the well-known property that the individual state of a qubit in a maximally entangled Bell pair is maximally mixed, and hence whichever observable is measured, the outcome will always be randomly $+1$ or $-1$. 

However, the interesting behaviour emerges when measuring joint observables. Since the overall state is pure, there exists for instance the two Boolean joint observables \(\hat{q}_{1z}(2)\hat{q}_{2z}(2)\) and \(\hat{q}_{1x}(2)\hat{q}_{2x}(2)\), which are both sharp with value 1: 
\begin{equation*}
    \braket{\hat{q}_{1z}(2)\hat{q}_{2z}(2)} = \braket{\hat{q}_{1x}(2)\hat{q}_{2x}(2)} = 1
\end{equation*}
A remarkable feature of entanglement is that the sharpness of the joint observables enables us to encode a classical bit \(\theta \in \{1,-1\}\) in one of the individual qubits (e.g. qubit \(\mathcal{Q}_1\)), despite the observables of qubit \(\mathcal{Q}_1\) alone all being maximally non-sharp. Explicitly, consider the case that we apply an X-gate to \(\mathcal{Q}_1\) if \(\theta=-1\) and an Identity gate if \(\theta=1\), $i.e.$ we apply the following gate to \(\mathcal{Q}_1\) between $t=2$ and $t=3$: 
\begin{align*}
U_\theta (\hat{\mathbf{q}}_1(t))\stackrel{\text{def}}{=} \frac{1}{2} (1+\theta)\hat{1} + \frac{1}{2} (1-\theta)U_{\text{not}}(\hat{\mathbf{q}}_1(t)) .
\end{align*}
Following the application of this gate, the qubit observables at $t=3$ are as follows: 
\begin{align*}
   \hat{\mathbf{q}}_1(3) & = (\hat{q}_{1x}, \ \theta \hat{q}_{1y}\hat{q}_{2x}, \ \theta \hat{q}_{1z}\hat{q}_{2x}), \\
   \hat{\mathbf{q}}_2(3) & = (\hat{q}_{2z}\hat{q}_{1x},\  -\hat{q}_{2y}\hat{q}_{1x}, \ \hat{q}_{2x}).
\end{align*} 
Evidently, \(\mathcal{Q}_1\)'s observables now depend on $\theta$, which is stored in its  \(y\)- and \(z\)-observables. Yet, the  observables of \(\mathcal{Q}_1\) remain maximally non-sharp, as are those of \(\mathcal{Q}_2\) which remained entirely unchanged from time $t=2$, $i.e.$ $\braket{\hat{\mathbf{q}}_1(3)} = \braket{\hat{\mathbf{q}}_2(3)} = (0,0,0)$. This means $\theta$ cannot be retrieved via local measurements on the individual qubits. By contrast, the expectation values of the joint observables have now gained a dependence on $\theta$ following the application of \(U_\theta \). Specifically, the expectation value of the Boolean \(\hat{q}_{1z}(3)\hat{q}_{2z}(3)\) has become:
\begin{align*}
   \braket{\hat{q}_{1z}(3)\hat{q}_{2z}(3)} = \theta.
\end{align*} 
The bit $\theta$ can therefore be retrieved from the joint measurement of the two qubits, despite not being retrievable from measuring either qubit in isolation. Note that the joint measurement can be achieved by measuring each individual qubit and bringing the measurement outcomes together, through entirely local interactions, to retrieve the value of the bit $\theta$, as required by the principle of local tomography. Therefore, a bit $\theta$ can be encoded in a qubit, in a locally inaccessible form \cite{deutsch2000information}. The pair of entangled qubits here forms a lock-and-key pair, whereby \(\mathcal{Q}_2\) acts as a key to unlock the bit of information $\theta$ that resides in \(\mathcal{Q}_1\). The two-qubit network can store up to two bits of information in a locally inaccessible form, while $n$ qubits can store up to $n$ qubits in a locally inaccessible form. 

\section{A measure of locally accessible and inaccessible information} \label{measure}

So far we have considered the extreme cases of either bits of locally accessible or locally inaccessible information are encoded in the qubits, corresponding to the pure two-qubit pair being either in a product state or maximally entangled respectively. These results are summarised in Table \ref{tableofinformation}: 
\begin{table}[ht]
\centering  
\begin{tabular}{|c|c|c|}  
\hline
 & Locally accessible & Locally inaccessible \\ \hline
\(n\) pure-state qubits & \(n\) bits & 0 bits \\ \hline
\(n\) maximally-entangled qubits & 0 bits & \(n\) bits \\ \hline
\end{tabular}
\caption{Comparison of locally accessible and locally inaccessible information capacity of a globally pure, $n$-qubit network. In the locally accessible case, each qubit is pure, while in the locally inaccessible case, the network is maximally entangled.}
\label{tableofinformation} 
\end{table}

However, in general, the qubits can be partially entangled, in which case they can contain a fraction of a bit of information in a locally accessible form, and a fraction in a locally inaccessible form. We consider this intermediate case, and identify an appropriate measure of locally (in)accessible information. This reveals an interesting connection between the Heisenberg description of quantum states, and the density matrix, and uniquely translates to von Neumann entropy, as we might intuitively expect given the central role of entanglement in enabling the encoding of locally inaccessible information. To prove these results, we need to consider the \textit{sharpest observable} of a qubit. 

\subsection{The sharpest observable}

Consider a qubit \(\mathcal{Q}_a\), which may be entangled, and hence may not have any sharp observable. We can instead consider its \textit{sharpest} observable, which for a given time $t$ we define as the Boolean observable of \(\mathcal{Q}_a\) with the least variance out of all the observables of the qubit, \(\hat{O}_a(t)\). We will prove that the sharpest observable is uniquely defined up to a minus sign, and can be expressed in terms of the $x$-, $y$-, and $z$- observables of \(\mathcal{Q}_a\) at time $t$ as follows:
\begin{equation}\label{sharpest}
   \hat{O}_a(t) \stackrel{\text{def}}{=}  \frac{\braket{\hat{q}_{ax}(t)}\hat{q}_{ax}(t)+\braket{\hat{q}_{ay}(t)}\hat{q}_{ay}(t)+\braket{\hat{q}_{az}(t)}\hat{q}_{az}(t)}{\sqrt{\braket{\hat{q}_{ax}(t)}^2+\braket{\hat{q}_{ay}(t)}^2+\braket{\hat{q}_{az}(t)}^2}}.
\end{equation}
The expectation value \(\braket{\hat{O}_a(t)}\) is:
\begin{align}
\gamma(t) \stackrel{\text{def}}{=}  \sqrt{\braket{\hat{q}_{ax}(t)}^2+\braket{\hat{q}_{ay}(t)}^2+\braket{\hat{q}_{az}(t)}^2},
\end{align}

\noindent which is the size of the qubit's Bloch vector, a measure of the mixedness for a qubit's state. The value of \(\gamma(t) \) can be in the range \(0 \leq \gamma(t) \leq 1\). For \(0 < \gamma(t) \leq 1\), $\hat{O}_a(t)$ is well-defined, but for \(\gamma(t) = 0\), the qubit is maximally mixed and $\hat{O}_a(t)$ is not well-defined. In this case, there is no sharpest observable, since all observables (except the unit) have an identical (maximum) variance and hence are equally (maximally) non-sharp. Taking \(0 < \gamma(t) \leq 1\), such that the sharpest observable is well-defined, the variance of the observable is:  
\begin{align}
   \text{Var}({\hat{O}_a(t)}) = 1-\braket{\hat{q}_{ax}(t)}^2 -\braket{\hat{q}_{ay}(t)}^2 -\braket{\hat{q}_{az}(t)}^2.
\end{align}
When the variance $\text{Var}({\hat{O}_a(t)})$ is small, \textit{i.e.} the sharpest observable is close to being sharp, then the qubit can imperfectly store information, with arbitrarily high accuracy as the variance approaches zero. Similarly as the sharpest observable $\hat{O}_a(t)$ approaches an observable that is sharp, it can encode arbitrarily close to one bit of locally accessible information. Hence, we require that the measure to quantify locally accessible information is a continuous function of its sharpest observable, \(\hat{O}_a(t)\), and its expectation value, \(\braket{\hat{O}_a(t)}\). \\

\noindent \textbf{Proof of unique sharpest observable}\\

\noindent Consider the descriptors of a qubit initialised such that the expectation values are:
\begin{equation} \label{initial_qubit}
   \braket{\hat{\mathbf{q}}_1(0)} = (0,0,A),
\end{equation}
for $0<A\leq 1$. In general, this qubit is in a mixed state. When $A=1$, it is pure, and $A$ is never equal to $0$, where the qubit would be in a maximally mixed state and have no unique sharpest observable. Intuitively, the sharpest observable of the qubit looks like it will be \(\hat{q}_z(0)\), since \(\hat{q}_z(0)\) has a non-zero expectation value meanwhile \(\hat{q}_y(0)\) and \(\hat{q}_x(0)\) have expectation value $0$, suggesting maximal non-sharpness of those observables. To prove this, we need to consider also whether some linear combination of \(\hat{q}_x(0)\) and \(\hat{q}_y(0)\) could be used to construct an observable with sharpness greater than or equal to that of \(\hat{q}_z(0)\).

Consider some general observable \(\hat{O}\), expressed as a linear combination of the qubit's $x$, $y$ and $z$ observables: 
\begin{align}
\hat{O} := a\hat{q}_x(0) + b \hat{q}_y(0) + c \hat{q}_z(0),
\end{align}
where \(a^2 + b^2 +c^2 =1\), ensuring \(\hat{O}^2 =1\). In general: 
\begin{align}
\text{Var}(a\hat{A}+b\hat{B}) = a^2 V(\hat{A}) + b^2 V(\hat{B}) + 2abC(\hat{A},\hat{B}),
\end{align}
where 
\begin{align}
C(\hat{A},\hat{B}) =: \frac{1}{2}\braket{\hat{A}\hat{B} + \hat{B}\hat{A}} - \braket{\hat{A}}\braket{\hat{B}}.
\end{align}
If $\{\hat{A}, \hat{B}\} = 0$, the latter term simplifies to \(C(\hat{A},\hat{B})= -\braket{\hat{A}}\braket{\hat{B}} \). Using this expression with the observable $\hat{O}$, the final term \(C(\hat{q}_i(0) , \hat{q}_j(0)) = 0 \) for all pairs of \(i,j\in \{x,y,z\}\) where \(i\neq j\). The resulting expression for the variance is: 
\begin{align}
\text{Var}(\hat{O}) &= a^2\text{Var}(\hat{q}_x(0)) + b^2 \text{Var}(\hat{q}_y(0)) + c^2 \text{Var}(\hat{q}_z(0)) = a^2 + b^2+(1-A^2)c^2.
\end{align}
Given that \(a^2 + b^2 +c^2 =1\), we can replace $c^2$ to give: 
\begin{align}
\text{Var}(\hat{O}) &=  A^2 (a^2+ b^2)+1-A^2.
\end{align}
Taking partial derivatives with respect to $a$ and $b$, we find that the minimum variance $\text{Var}(\hat{O})= 0$ indeed occurs for $a=0$ and $b = 0$, meaning that the variance is minimised when \(\hat{O}=\pm \hat{q}_z(0)\), as required. 

This proof also has the direct consequence that the variance of the sharpest observable is not affected by single-qubit gates applied to the qubit. Applying a single-qubit gate will map the descriptors at time $t=0$ to a linear combination of them at time $t=1$. Since we have shown that a linear combination of descriptors cannot give a greater variance than the sharpest observable, the variance of the evolved descriptor must have the same value as before the single-qubit gate was applied, though the gate can change the sharpest observable. 

Now we can prove that \ref{sharpest} is the right expression for the unique sharpest observable. At $t=0$, $ \hat{O}(t) = \hat{q}_z(0)$, as required by \ref{initial_qubit}. The variance of this observable, \(V(\hat{O}(t))= 1 - \braket{\hat{q}_{ax}(t)}^2 -\braket{\hat{q}_{ay}(t)}^2 - \braket{\hat{q}_{az}(t)}^2 \), also has the property of being invariant under single-qubit gates. Consider an example where an $x-$rotation gate, paramaterised by $\phi$, is applied to the qubit, evolving the descriptors to: 
\begin{equation} 
    \hat{\mathbf{q}}_a (1) = (\hat{q}_{ax}(0), \hat{q}_{ay}(0) \cos{\phi }  + \hat{q}_{az}(0) \sin{\phi} , \hat{q}_{az}(0) \cos{\phi}  -  \hat{q}_{ay}(0) \sin{\phi} ).
\end{equation}
with the resulting expectation values: 
\begin{equation} 
    \braket{\hat{\mathbf{q}} (1)} = (0, A \sin(\phi), A \cos(\phi)).
\end{equation}
The variance becomes \(\text{Var}(\hat{O}(t=1)) = 1- A^2 \sin^2{\theta} - A^2 \cos^2{\theta} = 1-A^2 \), and hence is unchanged by the gate. 

Given that (1) the sharpest observable is unique; (2) $\hat{O}(t)$ is the sharpest observable when $t=0$; and (3) $V(\hat{O}(t))$ is invariant under single-qubit gates, we conclude that $\hat{O}(t)$ is the unique expression for the sharpest observable of a qubit (up to a minus sign). 

\subsection{Connecting the sharpest observable and density matrix 
}
The reduced density matrix of a qubit captures the information that can be retrieved about the qubit from measurements, via quantum state tomography. Hence, it characterises all the locally accessible information of a qubit. The density matrix contains complete information of all the probabilities of measurement outcomes. Here we prove that the sharpest observable of a qubit is directly connected to the density matrix, by expressing the density matrix solely as a function of the sharpest observable and its expectation value. 

Consider the expectation value of an arbitrary single-qubit observable of \(\hat{A}(t)\) of \(\mathcal{Q}_a\), which has density matrix \(\hat{\rho}_a(t)\): \(\operatorname{tr}(\hat{\rho}_a(t) \hat{A}(t)) \stackrel{\text{def}}{=}  \braket{\hat{A}(t)}\). Now at a given time $t$, the density matrix can be expressed in terms of a qubit's observables as follows: 
\begin{equation}
   \hat{\rho}_a(t) = \frac{1}{\operatorname{tr}(\hat{1})} \left ( \hat{1} + \sum_{j\in \{x,y,z\}}  \braket{\hat{q}_{aj}(t)} \hat{q}_{aj}(t) \right ).
\end{equation}
The RHS contains the form of the sharpest observable, hence we can rewrite the density matrix as a function of the sharpest observable and its expectation value: 

\begin{equation}
   \hat{\rho}_a(t) = \frac{1}{2}(\hat{1} + \braket{ \hat{O}_a(t) } \hat{O}_a(t)).
\end{equation}
It is helpful to simplify the expression using the observable's eigenvalue projectors, namely: 
\begin{align}
   & \Pi_{1}(\hat{O}_a(t)) \stackrel{\text{def}}{=}  \frac{1}{2}(1+\hat{O}_a(t)), && \Pi_{-1}(\hat{O}_a(t)) \stackrel{\text{def}}{=}  \frac{1}{2}(1 - \hat{O}_a(t)),
\end{align}
where the subscripts denote the eigenvalues of the projectors. Then the density matrix becomes: \begin{equation}
   \hat{\rho}_a(t) = p_1 \Pi_1(\hat{O}_a(t)) +  p_{-1} \Pi_{-1}(\hat{O}_a(t)).
\end{equation}
The probabilities $p_1$ and $p_2$ are those of measuring the eigenvalues $+1$ and $-1$ respectively. Hence, it is clear from this expression that the density matrix is diagonal in the basis of the sharpest observable's eigenvectors. This provides an interesting connection between the density matrix and sharpest observable: the density matrix intrinsically selects the sharpest observable as unique. 

Our analysis gives a new significance to the density matrix, namely that it encodes the probabilities of measuring the eigenvalues of the sharpest observable. This result connects the density matrix formalism directly with the theory of locally inaccessible information. In section \ref{quantify_info}, we use this result to derive a measure for locally (in)accessible information of a qubit, building on the conclusion that the measure should be a function of the sharpest observable, and hence of \(\hat{\rho}_a(t)\). 

\subsection{Quantifying information} \label{quantify_info}

We denote the desired measure for locally accessible information encoded in a qubit $\mathcal{Q}_a$ as $I^{\text{(acc)}}{\mathcal{Q}_a}(t)$. There are now two constraints we require this measure to satisfy, precisely the ones conventionally imposed for quantum information to satisfy, generalising those of Shannon \cite{vedral2010}: 

\begin{enumerate}
\item Continuity: $I^{\text{(acc)}}_{\mathcal{Q}_a}(t)$ is a real-valued, continuous function of $\hat{O}_a(t)$ and the probabilities of measuring its eigenvalues are a continuous function of $\hat{\rho}_a(t)$.
\item Additivity: When $\mathcal{Q}_a$ and $\mathcal{Q}_b$ are not entangled with each other (such that they are in a separable state relative to each other), the total locally accessible information in $\mathcal{Q}_a$ and $\mathcal{Q}_b$ is the sum of the individual locally accessible information stored in each qubit.
\end{enumerate}

The unique quantity that satisfies both these conditions is the von Neumann entropy of the qubit at a time $t$, \(S_{\mathcal{Q}_a}(t):=\braket{\log_2  \hat{\rho}_a(t) }\) (setting $k_B = 1$ for simplicity), up to an affine transformation. We define the locally inaccessible information stored in an individual qubit \(\mathcal{Q}_a\) to be: 
\begin{equation}
    I^{\text{(acc)}}_{\mathcal{Q}_a}(t) \stackrel{\text{def}}{=}  1 - S_{\mathcal{Q}_a}(t),
\end{equation}
with the affine transformation chosen such that the locally accessible information correctly satisfies the boundary conditions when the qubit is in a pure state and in a maximally mixed state. When \( \mathcal{Q}_a\) is pure, \(I^{\text{(acc)}}_{\mathcal{Q}_a}(t) =1\), meaning exactly one bit of information is stored in a locally accessible form. Meanwhile when \( \mathcal{Q}_a\) is maximally mixed, \(I^{\text{(acc)}}_{\mathcal{Q}_a}(t) =0\), meaning \( \mathcal{Q}_a\) stores no locally accessible information. 

When \( \mathcal{Q}_a\) is partially mixed, \(0 < I^{\text{(acc)}}_{\mathcal{Q}_a}(t) < 1\), meaning \( \mathcal{Q}_a\) can store a fractional bit of locally accessible information. Physically, storing a fractional bit means that there is a fundamental constraint on the minimum error probability when copying that bit of information. This occurs because the qubit can store information about the value of the sharpest observable, but with an irreducible likelihood of error when being copied. The greater the likelihood of an error when copying the value of this observable, the less locally accessible information is stored. The likelihood of accurately copying the desired value of the sharpest observable becomes equal to that of copying an erroneous value, when approaching the limit of a maximally mixed state that stores no locally accessible information. \\

\noindent \textbf{The locally accessible information of a network}\\

\noindent We can generalise the total locally accessible information to a network of  \(\mathcal{N}\) qubits, denoted by \(I^{\text{(acc)}}_{\mathcal{N}}(t)\). For a network, by definition the total locally accessible information is the information that can be retrieved by measurements on the overall network, including from any combination of joint measurements, which makes it distinct from the sum of locally accessible information in the individual qubits composing the network. Simplifying to the case where the total network is in a pure state, the locally accessible information that can be stored in a network of $n$ qubits is:

\begin{equation*}
   I^{\text{(acc)}}_{\mathcal{N}}(t) = n.
\end{equation*}
We showed in section \ref{sec:acc_info} that the expression holds when all the individual qubits composing the network are pure states. To generalise this, we consider the system at time $t$, and then an arbitrary unitary gate being applied to the global system, evolving it to time $t+1$. 

At time $t$, an arbitrary observable of the network can be expressed in terms of the following basis: 
\begin{align} \label{basis}
 \Bigl \{ \Pi_{m=1}^n \hat{Q}_{m, \alpha_m }(t) : \alpha_1,...,\alpha_n \in \{x,y,z,0\} \Bigr \}.
\end{align}
This set contains $4^n$ basis vectors. We have expressed them using the observable \(\hat{Q}_{m, \alpha}\), which is defined as:  
\begin{equation}
\hat{Q}_{m,\alpha}(t) = \biggl\{  \begin{array}{ll}
\hat{q}_{m,\alpha}(t) \quad \text{if} \quad \alpha \in \{x,y,z\}, \\
    \hat{1} \quad \text{if} \quad \alpha = 0.
\end{array}
\end{equation}
The full basis set is then the set of all products of $x$, $y$, $z$ and unit descriptors for each qubit composing the network. This set forms an orthonormal basis, as the components are traceless, and the trace of their square is equal to the trace of the unit observable:
\begin{align}
& \operatorname{tr}\left(\Pi_{m=1}^n \hat{Q}_{m, \alpha_m }(t) \right) = 0 \quad \quad (\text{unless \(\alpha_m = 0\) for all \(m\)}), \\
& \operatorname{tr}\left((\Pi_{m=1}^n \hat{Q}_{m, \alpha_m }(t))^2 \right) = \operatorname{tr}(\hat{1}).
\end{align}
This basis is valid for the evolved state of the network of $n$ qubits at time $t+1$, after a global unitary evolution from $t=0$. We can therefore express \(\hat{q}_{az}(t)\) in terms of the operators $\hat{Q}_{m, i_m }(t+1)$, which describe the network at time $t=1$ in the basis \ref{basis}:
\begin{align}
 \hat{q}_{az}(t) = \sum_{i_1 \in \{x,y,z,0\}} ... \sum_{i_n \in \{x,y,z,0\}} \operatorname{tr} (\Pi_{m=1}^n \hat{Q}_{m, i_m }(t+1) \hat{q}_{az}(t)) \Pi_{m=1}^n \hat{Q}_{m, i_m }(t+1)  
\end{align}
There are $n$ sharp, commuting observables describing a network of individually sharp qubits at time $t$. Since each of these can be expressed in terms of the observables at time $t+1$, there must also be $n$ such observables describing the network at time $t+1$.  Hence, the network at time $t+1$ can store $n$ bits of locally accessible information. Now imagine that there is an additional Boolean observable at time $t+1$, that commutes with all the others, and can therefore store an additional bit of information. This observable must be possible to express in terms of the observables at time $t$, hence an additional observable at time $t$ could store an additional bit of information. This contradicts the result of \ref{sec:acc_info}, that the network at time $t$ can store maximum $n$ bits, and hence there can be no such observable at time $t$, therefore it cannot exist at time $t+1$ either. Therefore, under a globally unitary evolution whereby the qubits may become entangled, the maximum locally accessible information encoded in the network remains $n$ bits of information. \\

\noindent \textbf{Locally inaccessible information}\\

\noindent Now we have the expressions for the total locally accessible information in a network \(\mathcal{N}\), and the locally accessible information in a qubit \( \mathcal{Q}_a\), $I^{\text{(acc)}}_{\mathcal{Q}_a}(t)$, we can straightforwardly express the locally inaccessible information, \(I_{\mathcal{N}}^{\text{(inacc)}}(t)\), as the difference between the accessible information of the network and the sum of the accessible information of the $n$ individual qubits:

\begin{equation}
    I^{\text{(inacc)}}_{\mathcal{N}}(t) := I^{\text{(acc)}}_{\mathcal{N}}(t) - \sum_{a=1}^n I^{\text{(acc)}}_{\mathcal{Q}_a}(t).
\end{equation}
Here $I^{\text{(acc)}}_{\mathcal{Q}_a}(t)$ is the locally accessible information stored in the qubit \( \mathcal{Q}_a\), $i.e.$ the information that can be retrieved through local measurements on \( \mathcal{Q}_a\), in isolation from the other qubits in the network. The locally inaccessible information is the information that can be retrieved from measurements on the composite network, but cannot be retrieved by measurements on individual qubits of that network in isolation. 

This is illustrated by the example of a two-qubit maximally entangled Bell state, in \eqref{maximally mixed}. Here the locally accessible information of the individual qubits is zero: $ I^{\text{(acc)}}_{\mathcal{Q}_1}(t) = I^{\text{(acc)}}_{\mathcal{Q}_2}(t) = 0$ bits, however the locally accessible information of the joint network $I^{\text{(acc)}}_{\mathcal{N}}(t) = 2$ bits. Hence, the locally inaccessible information that can be stored in the network is $I^{\text{(inacc)}}_{\mathcal{N}}(t) = 2 - 0 = 2$ bits, as required by the results in Table \ref{tableofinformation}.

We can see directly from this formulation that since information in \(\mathcal{N}\) can only be stored in either a locally accessible or locally inaccessible form, and the total information stored in \(\mathcal{N}\) is a constant \(I^{\text{(acc)}}_{\mathcal{N}}(t) = n\) when it is in a pure state, then the sum of the locally accessible and inaccessible information is a constant:
\begin{equation}
    I^{\text{(inacc)}}_{\mathcal{N}}(t) + \sum_{a=1}^n I^{\text{(acc)}}_{\mathcal{Q}_a}(t) = n.
\end{equation}
This means that the total information that can be stored in the network is a constant, and can only be transformed between the locally accessible and locally inaccessible forms during evolution of the network. 

\section{Summary} \label{info_discussion}

We have quantified locally inaccessible and locally accessible information in a globally pure, two-qubit system, using Heisenberg-picture observables. We found the unique expression for locally (in)accessible information in terms of the sharpest observable of a system, and showed that this observable directly selects the basis in which the reduced density matrix of the system is diagonalised. This relationship connects the measure of locally (in)accessible information in terms of the sharpest observable with the conventional quantification of information stored in entanglement using von Neumann entropy. A natural development of this work is to consider the general multipartite case, and generalise the quantification of locally (in)accessible information to settings where the global state is mixed. One particularly interesting case study is to analyse the phenomenon of ``non-locality without entanglement" using the tools developed here, and also to consider thought experiments which exhibit the ``anomaly of non-locality" in different ways to Hardy's paradox \cite{bennett1999quantum, methot2006anomaly}.  % Quantifying Locally Inaccessible Information 

\chapter{Conclusions}

We have analysed macroscopic phenomena in the setting of universal quantum theory, with a focus on thermodynamics, measurement and locality. Counterfactuals play a fundamental role in our analysis, underlying the axiomatic formulation of an exact form of irreversibility and in characterising the transfer between locally accessible and inaccessible information. They also underlie the impossibility of perfectly distinguishing states, which plays a key role in ensuring macroscopic observations are fully self-consistent. Our results give new insights into the role of different quantum information resources, such as coherence, in enabling thermodynamic and more general transformations of quantum states. We also unify approaches to thermodynamic and more general transformation tasks from constructor theory and resource theories by analysing their implications for the same protocols. By expressing our results using qubit models, we create exciting opportunities to test the outcomes of experimental implementations, for instance with NMR, single-photons and cloud quantum computing platforms. 

In Part I, we focused on the field of quantum thermodynamics, and presented results regarding the limits of using quantum resources in the context of irreversibility, information erasure and coherence. We began by comparing the axiomatic, resource- and constructor-theoretic approaches, outlining their domains of applicability, the role of catalysts, and their essential ordering relations used to construct a second law of thermodynamics. Then we presented a quantum model for a constructor, which is an implicit entity required for performing the possible tasks defined by (ultimately dynamics- and scale-independent) constructor-theoretic principles. Using this model, we showed the consistency of constructor-theoretic irreversibility with fully unitary quantum theory. We conducted both a general analysis and one with an explicit qubit-based model for a constructor, the quantum homogenizer. Building on this, we showed that this model implies an additional cost to erasing quantum information, relevant in a quantum Maxwell's Demon protocol. Then we considered an alternative incoherent quantum homogenizer, based on a controlled-swap operation instead of a partial-swap, and found key similarities and differences when compared with the coherent version. We conducted an NMR experiment to implement a four-qubit quantum homogenizer and analyse the von Neumann entropy of the qubits. Finally we considered a case-study of extracting work from coherence, and unified conclusions from a resource-theoretic and constructor-theoretic analysis to show the impossibility of universal work extraction from coherence. We identified an interesting connection between this result and our quantum homogenization model for constructor-based irreversibility. 

In Part II, we considered the universality of quantum theory and approach from counterfactuals in the context of measurements. We analysed the contradictions that arise from treating quantum measurement outcomes as purely classical information. Distilling the key underlying features of counterintuitive and seemingly paradoxical thought experiments into a low-depth quantum circuit, we analysed the Penrose-triangle-like structure that emerges from treating quantum measurement outcomes as classical entities. We highlighted the role of non-commutativity in the emergence of this structure. The structure enables one to deterministically distinguish non-orthogonal states, violating a key axiomatic feature of quantum theory. Focusing on Hardy's paradox, we showed how the non-commutativity of operators used to retrieve locally inaccessible information has a quantitative relation to a measure of Bell non-locality, distinct from entanglement measures. We then used this as a case-study for quantifying locally accessible and locally inaccessible information stored in quantum systems, connecting this with entanglement measured by von Neumann entropy. 

Our work lays the foundation for many future research developments. For instance, the relative deterioration quantity used to classify the quantum homogenizer as a constructor could be generalised such that it can be used as a measure for any quantum protocol, rather than the quantum homogenizer specifically. It would be interesting to connect it to quantities and results analysed in resource theories, building on the connections identified and developed in our analysis of different approaches to showing the impossibility of a universal work extractor from coherence. Additionally the tools we developed for quantifying locally (in)accessible information could be extended to a broader range of settings that display Bell non-locality, and generalised for multi-partite and globally mixed systems, and ultimately to the dynamics-independent constructor-theoretic setting.  

Applying universal quantum theory to macroscopic phenomena has two parallel effects. The first is that it resolves apparent contradictions and explains counter-intuitive implications of quantum physics. This ensures that quantum theory is self-consistent in a range of contexts, demonstrating the universality of its domain of applicability. The second is to reveal new features of quantum theory. This includes progress towards answering technologically relevant questions regarding the usefulness of quantum resources for different tasks, and fundamental aspects of how quantum states carry and transfer information. Our results therefore both provide a satisfying account of what we observe in both quantum experiments and everyday life, and simultaneously expose the rich and subtle structures that emerge when quantum theory is taken to apply to all scales.  % Conclusion

%next line adds the Bibliography to the contents page
\addcontentsline{toc}{chapter}{Bibliography}
%uncomment next line to change bibliography name to references
%\renewcommand{\bibname}{References}
\bibliography{bibliography}        %use a bibtex bibliography file refs.bib
\bibliographystyle{unsrt}  %use the plain bibliography style

\end{document}